\documentclass{emulateapj}
\usepackage{graphicx}
\usepackage{epstopdf}
\usepackage{longtable}
\usepackage{multirow}
		
\slugcomment{Accepted for publication in ApJ: December $30^{th}$, 2010}

\shorttitle{NICMOS Morphology of Bright $24\mu m$-Sources}
\shortauthors{Zamojski et al.}

\def\deg{\ifmmode {^{\circ}}\else {$^\circ$}\fi}
\def\kms{\ifmmode {\rm\,km\,s^{-1}}\else
    ${\rm\,km\,s^{-1}}$\fi}
\def\ergcm2s{\ifmmode {\rm\,ergs\,cm^{-2}\,s^{-1}}\else
    ${\rm\,ergs\,cm^{-2}\,s^{-1}}$\fi}
\def\ergAcm2s{\ifmmode {\rm\,ergs\,cm^{-2}\,s^{-1}\,\AA^{-1}}\else
    ${\rm\,ergs\,cm^{-2}\,s^{-1}\,\AA^{-1}}$\fi}
\def\ergs{\ifmmode {\rm\,ergs\,s^{-1}}\else
    ${\rm\,ergs\,s^{-1}}$\fi}
\def\kmsMpc{\ifmmode {\rm\,km\,s^{-1}\,Mpc^{-1}}\else
    ${\rm\,km\,s^{-1}\,Mpc^{-1}}$\fi}

\def\spose#1{\hbox to 0pt{#1\hss}}
\def\simlt{\mathrel{\spose{\lower 3pt\hbox{$\mathchar"218$}}
     \raise 2.0pt\hbox{$\mathchar"13C$}}}
\def\simgt{\mathrel{\spose{\lower 3pt\hbox{$\mathchar"218$}}
     \raise 2.0pt\hbox{$\mathchar"13E$}}}

\def\gs{\mathrel{\raise0.35ex\hbox{$\scriptstyle >$}\kern-0.6em
\lower0.40ex\hbox{{$\scriptstyle \sim$}}}}
\def\ls{\mathrel{\raise0.35ex\hbox{$\scriptstyle <$}\kern-0.6em
\lower0.40ex\hbox{{$\scriptstyle \sim$}}}}

\newcommand{\um}{\,$\mu$m}

\newcommand{\iso}{{\sl ISO }}
\newcommand{\spitz}{{\sl Spitzer }}

\begin{document}

\title{HST/NICMOS Imaging of Bright High-Redshift \lowercase{$24\mu m$}-selected Galaxies:  Merging Properties}

\author{
Michel Zamojski\altaffilmark{1},
Lin Yan\altaffilmark{1},
Kalliopi Dasyra\altaffilmark{2,1},
Anna Sajina\altaffilmark{3,4},
Jason Surace\altaffilmark{1},
Tim Heckman\altaffilmark{5},
George Helou\altaffilmark{1}
}

\altaffiltext{1}{Spitzer Science Center, California Institute of Technology, MC 220-6, 1200 East California Boulevard, Pasadena, CA 91125; \email{mzamoj@caltech.edu}}

\altaffiltext{2}{Commissariat \`{a} l'\'{E}nergie Atomique Saclay, Service d'Astrophysique,
Orme des Merisiers, Bat.709, F-91191 Gif-sur-Yvette Cedex, France}
\altaffiltext{3}{Departments of Physics \& Astronomy, Haverford College, 370 Lancaster Avenue, Haverford, PA 19041}
\altaffiltext{4}{Department of Physics and Astronomy, Tufts University, 212 College Ave., Medford, MA 02155}
\altaffiltext{5}{Center for Astrophysical Sciences, The Johns Hopkins
University, 3400 N. Charles St., Baltimore, MD 21218}

\begin{abstract}
We present new results on the physical nature of infrared-luminous sources at $0.5<z<2.8$ as revealed by HST/NICMOS imaging and IRS mid-infrared spectroscopy.  Our sample consists of 134 galaxies selected at $24\mu m$ with a flux of $S(24\mu m) > 0.9$ mJy.  We find many ($\sim 60\%$) of our sources to possess an important bulge and/or central point source component, most of which reveal additional underlying structures after subtraction of a best-fit sersic (or sersic+PSF) profile.  Based on visual inspection of the NIC2 images and their residuals, we estimate that $\sim 80\%$ of all our sources are mergers.  We calculate lower and upper limits on the merger fraction to be 62\% and 91\% respectively.  At $z < 1.5$, we observe objects in early (pre-coalescence) merging stages to be mostly disk and star formation dominated, while we find mergers to be mainly bulge-dominated and AGN-starburst composites during coalescence and then AGN-dominated in late stages.  This is analogous to what is observed in local ULIRGs.  At $z \ge 1.5$, we find a dramatic rise in the number of objects in pre-coalescence phases of merging, despite an increase in the preponderance of AGN signatures in their mid-IR spectra and luminosities above $10^{12.5} L_{\odot}$.  We further find the majority of mergers at those redshifts to retain a disk-dominated profile during coalescence.   We conclude that, albeit still driven by mergers, these high-$z$ ULIRGs are substantially different in nature from their local counterparts and speculate that this is likely due to their higher gas content.  Finally, we observe obscured ($\tau_{9.7\mu m} > 3.36$) quasars to live in faint and compact hosts and show that these are likely high-redshift analogs of local dense-core mergers.  We find late-stage mergers to show predominantly unobscured AGN spectra, but do not observe other morphological classes to occupy any one specific region in the $\tau_{9.7\mu m}$ vs. PAH equivalent width (or Spoon) diagram.  This suggests a high degree of variation in the PAH emission and silicate absorption properties of these mergers, and possibly throughout the merging process itself.
\end{abstract}

\keywords{galaxies: evolution --- galaxies: high-redshift --- infrared: galaxies}

\section{INTRODUCTION}

Since their discovery in the {\em IRAS} all-sky survey more than 25 years ago, ultra-luminous infrared galaxies (ULIRGs, $L_{IR} > 10^{12} L{\odot}$) have been thought to represent a key evolutionary link between normal galaxies and quasars.  Early evidence suggested that they were young quasars fed and obscured by large amounts of gas and dust  funneled towards the center of a merger remnant of two gas-rich spirals \citep{Sanders88}.  Intense star formation rather than black hole accretion was later shown to be the primary energy source of most ULIRGs \citep{Rigopoulou96, Genzel98}, but black hole accretion has remained the dominant mechanism at higher luminosities \citep[$L_{IR} \gtrsim 10^{12.3} L_{\odot}$;][]{Lutz98,Veilleux99,Tran01,Farrah03,Veilleux09a}.  Ground-based as well as high resolution {\em HST} imaging, meanwhile, revealed that more than 95\% of ULIRGs originate in a merger, but that the ULIRG phase can also appear much before final coalescence \citep{Murphy96, Veilleux02}.

Simulations have yielded strong support to the picture of quasars originating in the merger of two gas-rich spirals after a phase of intense star formation and rapid black hole growth giving rise to the ULIRG phenomenon.  \citet{Barnes91} and \citet{Mihos94} showed that the merger of two equal-mass disk galaxies can rapidly dissipate angular momentum, causing the gas to fall to the center of the galaxy and create a starburst of ULIRG proportions.  They also demonstrated how the stellar component formed tails and streams much like the ones observed in ULIRGs.  Refinements in hydrodynamical simulations and the introduction of AGN feedback by \citet{Springel05} then showed how the AGN, once triggered, can expel remaining gas, quench star formation and become a true quasar within a remnant elliptical galaxy \citep{diMatteo05,Hopkins08}.  Finally, the addition of radiative transfer confirmed the exceptional infrared luminosity associated with the whole event, and in particular with the final coalescence \citep{Jonsson06,Li08,Younger09,Narayanan09}.

Although ULIRGs and quasars are extremely rare locally \citep{Soifer87}, observations at sub-millimeter, mid-IR, optical and X-ray wavelengths have demonstrated that their number and luminosity densities increase rapidly with redshift \citep{Chapman05, LeFloch05, Richards06, Hasinger05}.
In particular, the advent of the {\em Spitzer Space Telescope} has enabled sensitive and fast imaging at 24\um\ with MIPS, yielding the detection of a large number of infrared-luminous galaxies at $z<3$ \citep[e.g.][]{Perez-Gonzalez05}.  With such a rise in prominence, it becomes important for our understanding of galaxy/quasar evolution to ask whether, or how many of, these numerous high-$z$ ULIRGs are triggered through the same physical mechanisms as their low-redshift counterparts, and whether they also represent a transition towards quasars.  Given the higher gas fractions \citep{Noterdaeme09}, star formations rates \citep{Hopkins06} as well as specific star formation rates \citep{Zamojski07, Brinchmann00} of the overall galaxy population at these redshifts, quiescent star formation is expected to contribute increasingly more to the infrared luminosity of galaxies  \citep{Hopkins10a}.  Certainly at $z \sim 2$, some ULIRGs have been found to exhibit disk-like kinematics \citep{Forster-Schreiber09, Carilli10, Bothwell10}.  The answer as to what extent mergers are still necessary to explain the origin of the infrared-luminous population at higher redshifts is, therefore, unclear.

Meanwhile, \iso studies of local ULIRGs have demonstrated that mid-IR spectroscopy, through the resolution of PAH emission complexes and measurement of their strength relative to the underlying continuum, is the most effective single tool for identifying which of star formation or AGN activity is responsible for the observed mid-IR radiation of an object.  The launch of the {\em Spitzer Space Telescope} with its infrared spectrograph ({\em IRS}) sensitive to fluxes of $S_{24\mu m} \gtrsim 1$ mJy has, thus, brought a flurry of mid-IR spectroscopic surveys of bright, high-redshift $24\mu m$ galaxies aimed at addressing the origin of their infrared luminosity \citep{Houck05, Yan05, Weedman06, Yan07, Sajina07, Farrah08, Dasyra09, Desai09}.  The results of these efforts have demonstrated that, unlike sub-millimeter galaxies that are primarily powered by star formation \citep{Pope08,Menendez09}, $24\mu m$-selected objects appear to be more analogous to local ULIRGs in that they display both types of spectra (as well as various combinations thereof).  Also in analogy to local ULIRGs, the AGN contribution to their mid-IR flux increases with total IR-luminosity \citep{Sajina07, Dey08, Desai09}.

Our group carried out two such mid-IR spectroscopic programs of $24\mu m$-bright galaxies, the first of which yielded spectra for 52 objects at $z$\,$\simgt$\,1\,--\,3 \citep{Yan07,Sajina07,Sajina08}, while the second targeted 150 objects spanning a redshift range of $z$\,$\sim$\,0.3\,--\,2.5 and peaking at $z=1$ \citep{Dasyra09}.  Both were conducted in the \spitz Extragalactic First Look Survey (XFLS).  Analysis of this combined data has revealed the presence of an obscured AGN in $\simgt$\,75\% of our objects.  

The answer as to whether the infrared luminosity of these $24\mu m$-bright galaxies is due to mergers, however, has, thus far, not been conclusive.  Initial morphological studies have yielded mixed results \citep{Dasyra08, Bussmann09, Melbourne09}, but part of this variation could be due to small sample sizes and selection criteria.  In an effort to address this question more fully, we have obtained HST/NICMOS imaging of a sample of 135 bright high-redshift $24\mu m$-selected galaxies that combines the previously published data for 33 sources from our first program \citep{Dasyra08} to new data for 102 sources from our second program.

In this paper, we expand the analysis of \citet{Dasyra08} and perform a systematic search of merger signatures by uncovering underlying structures, resulting in the discovery that $\sim 80\%$ of our objects are fueled by either an ongoing or recent ($\lesssim 0.5$ Gyr) merger event.  We also artificially redshift local ULIRGs to quantify the detectability of merging signatures at high redshift and to create a comparison sample for our sources.  We examine similarities and differences with local ULIRGs.  We then combine our morphological results with our IRS spectra \citep{Sajina07, Dasyra09} and the SED analysis of \citet{Sajina08} and Sajina et al. (in preparation) to investigate the link between morphology/merging and, both, the relative strength of the AGN and starburst components at infrared wavelengths, as well as the degree of obscuration.  We discuss our results in the context of other LIRG and ULIRG samples at both high and low redshift.  Finally, we confront our observations with simulations, and discuss their consequences for our understanding of galaxy evolution.

The paper is organized such that we first describe our sample and analysis in \S 2, which we follow by a discussion of our simulated observations of redshifted local galaxies in \S 3.  We present our results in \S 4 and compare them to that of other ULIRG samples in \S 5.  We end section 5 by discussing implications for the theory of galaxy evolution, and conclude with a summary (section~6).  We use a $\Lambda$CDM cosmology with $H_{0} = 70 \mbox{ km s}^{-1} \mbox{ Mpc}^{-1}$, $\Omega_{m} = 0.3$ and $\Omega_{\Lambda} = 0.7$.

\section{OBSERVATIONS \& ANALYSIS \label{sec:observations}}

\subsection{Sample \label{sec:sample}}

Our full sample consists of 134 galaxies with $24\mu m$ fluxes above 0.9 mJy selected from the Extragalactic First Look Survey.  It draws from two separate programs whose IRS spectra were presented in \citet{Yan07,Sajina07} and \citet{Dasyra09}.  We have obtained HST/NIC2 F160W ($H$-band) images for 33 sources from our first program and 102 sources from our second program.  We rejected one source that we suspect of being gravitationally lensed.  All data were taken in MULTIACCUM mode, and we obtained,  for each object, four to eight dithered images with exposure times ranging from $640$ to $672$ seconds for a total exposure time of $2560$ to $5376$ seconds per object.  The data for the first 33 sources were presented in \citet{Dasyra08}.

In parallel with our NICMOS observations, we acquired WFPC2/F814W images of patches of the FLS field visible to that instrument at the time of our primary observations.  Eight of our objects fall in one or another of these pointings.  An additional twelve of our objects fall in the central $0.12 \mbox{ deg}^2$ of the FLS field that has been imaged by HST/ACS in the F814W filter \citep{Bridge07}. 

\begin{figure}[htbp]
\begin{center}
\includegraphics[width=3.5in]{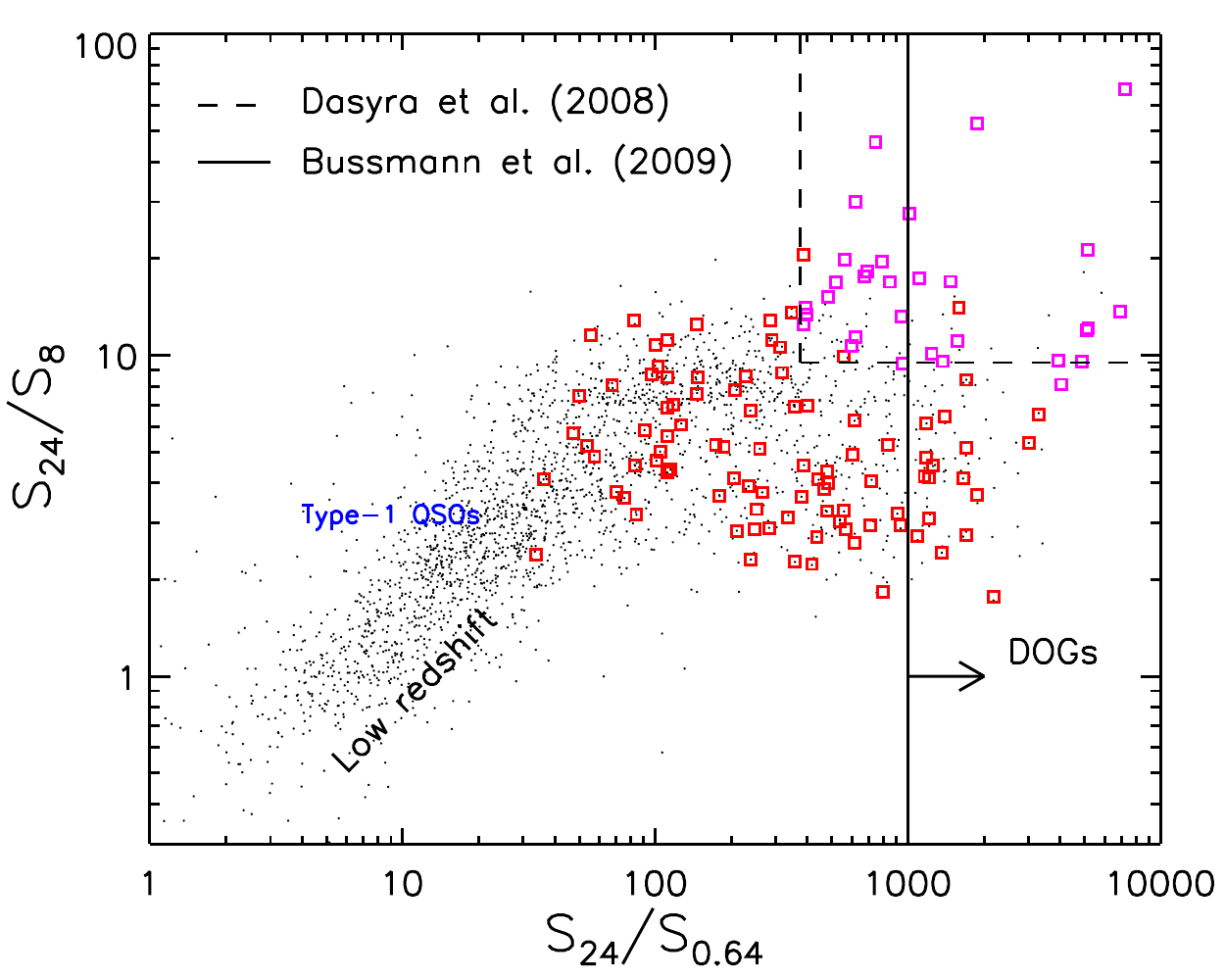}
\caption{Color-color plot comparing objects from our sample ({\em red} and {\em magenta squares}) to the general population of bright $24\mu m$-sources ({\em dots}) in the FLS field.  {\em Magenta} represents objects drawn from our first program and {\em red} from our second.  Dashed and solid lines represent selection criteria used in \citet{Dasyra08} and \citet{Bussmann09} respectively.}
\label{fig:selection}
\end{center}
\end{figure}

\begin{figure}[htbp]
\begin{center}
\includegraphics[width=3.5in]{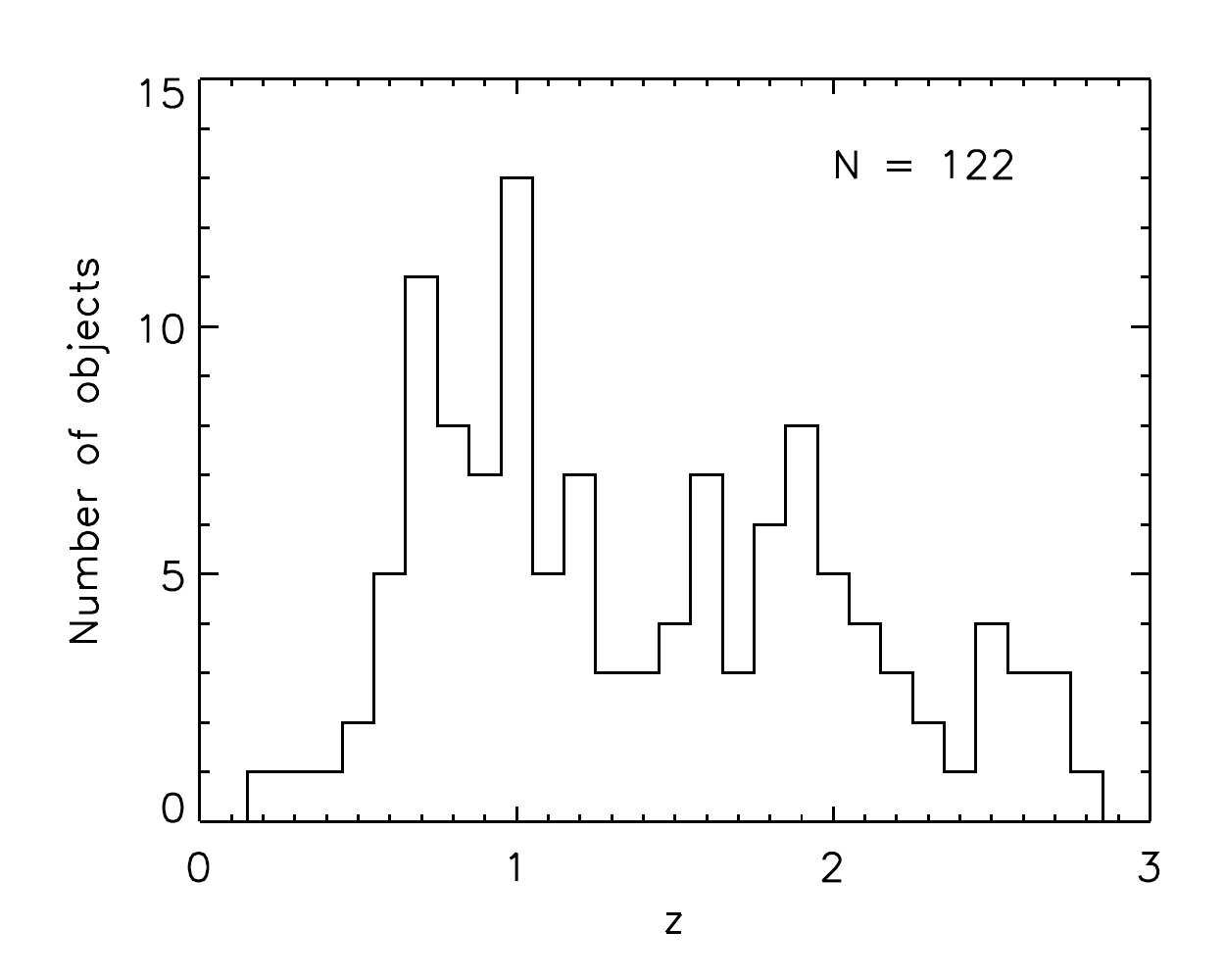}
\caption{Reshift distribution of our sample.  We miss redshift estimates for twelve of our objects.}
\label{fig:zdist}
\end{center}
\end{figure}

Galaxies in both our programs were chosen randomly from the entire set of $S(24\mu m) > 0.9$ mJy sources in the FLS main field using, in the first case, a color selection of $\nu f_{\nu} (24\mu m) / \nu f_{\nu} (8\mu m) \gtrsim 3.16$ and $\nu f_{\nu} (24\mu m) / \nu f_{\nu} (R)  \gtrsim 10$, and in the second case, a sampling rate varying according to their $R$-band magnitude, going from 1 in 10 for objects with $20 < R < 22$ to 1 in 3 for objects with $22 < R < 24.5$, while avoiding objects already targeted in our first program.  The strong weight put on $R$-band faint objects ensures that a large fraction of them lie at high redshifts ($z \sim 1 \textendash 2$), thus preventing our sample from being populated at $\sim 70\%$ by low redshift ($z < 0.3$) sources as is the case for the general $S(24\mu m) > 0.9$ mJy population.  Our initial cut on the $24\mu m / 8\mu m$ color was made to isolate objects with either strong PAH emission or a steep mid-IR continuum \citep{Yan07}, but this bias is compensated for in our combined sample as shown in Figure~\ref{fig:selection}.  Figure~\ref{fig:selection} illustrates the distribution of our sample in $S_{24\mu m} / S_{8\mu m}$ vs. $S_{24\mu m} / S_{0.64\mu m}$ color-color space in comparison to the general population of bright $24\mu m$-sources in the FLS field, and Figure~\ref{fig:zdist} shows its distribution in redshift.

In this paper, we utilize redshift measurements extracted from the IRS data using the method described in \citet{Sajina07}.  When available, we use more precise redshifts derived from optical or near-infrared spectra instead \citep[$\sim 20$ objects;][]{Sajina08}.  We miss redshift estimates for twelve of our galaxies.  These are objects that show absolutely featureless mid-IR spectra, and that have not been observed spectroscopically in the optical/NIR either.  Among the ones we do have lines and features for, our lowest redshift object is found to be at $z=0.24$ and our highest one is at $z=3.48$, but most of our sources fall in the range $0.5 \textendash 2.8$ with a main peak at $z\sim 1$, and a secondary peak at $z\sim 2$.

Using ancillary far-infrared, sub-millimeter and radio observations, \citet{Sajina08} and Sajina et al. (in preparation) were able to constrain the SED and derive full ($3\textendash 1000\mu m$) infrared luminosities for 112 objects in our sample.  Their analysis tells us that most of our sources possess luminosities ranging from $10^{11.5}$ to $10^{13} L_{\odot}$, and that, because we have a flux-limited sample, higher redshift objects also possess higher luminosities, such that most objects at $z \ge 1$ have an $L_{IR} \ge 10^{12} L_{\odot}$ and nearly all objects at $z \ge 1.5$ have luminosities above $10^{12.5} L_{\odot}$.  We discuss the relation between redshift and luminosity in more detail in section~\ref{sec:redshift_evolution}.

\subsection{Control Sample \label{sec:control}}

\begin{figure}[htbp]
\begin{center}
\includegraphics[width=3.5in]{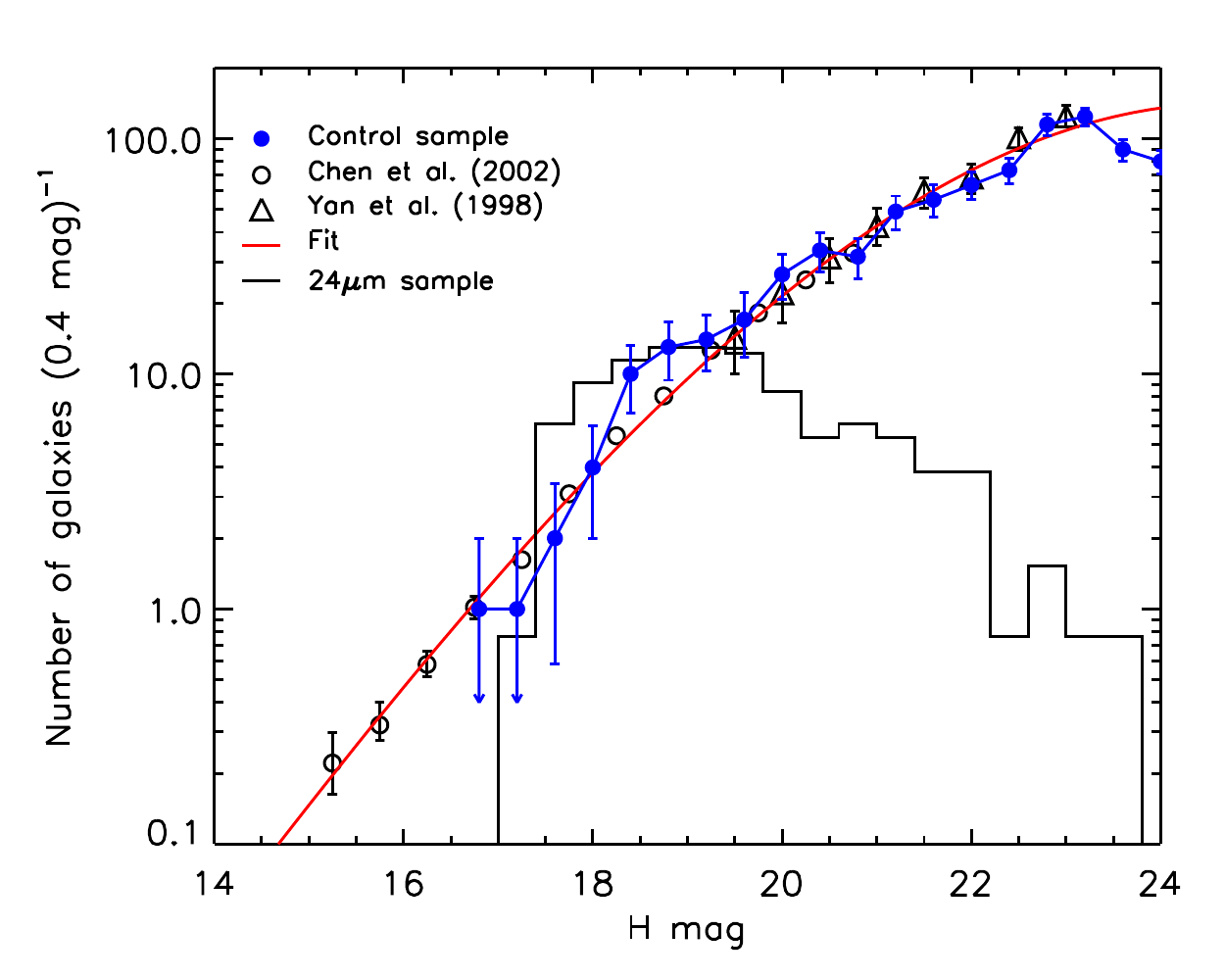}
\caption{Number of galaxies per bin of 0.4 magnitude in our control sample ({\em blue curve and points}).  For comparison, the expected number of galaxies from the number counts of \citet{Chen02} and \citet{Yan98} are shown in {\em open circles} and {\em triangles} respectively, along with a fit to those points ({\em red curve}).  Uncertainties are shown only when larger than the symbol size.  The histogram shows the magnitude distribution of our $24\mu m$-selected galaxies.  We observe an excess in our number counts at magnitudes where our $24\mu m$-selected sample peaks.}
\label{fig:ncounts}
\end{center}
\end{figure}

We use galaxies present around our $24\mu m$-selected sources in the NICMOS images as a control sample for our morphological classification.  Figure~\ref{fig:ncounts} shows the number of galaxies we find in our control sample in bins of 0.4 magnitude.  For comparison, we also plot the number of galaxies we would expect to find in our ensemble of images based solely on the number counts of \citet{Chen02} and \citet{Yan98}.  Most recent studies such as those of \citet{Metcalfe06} and \citet{Retzlaff10} arrive at nearly identical numbers.  Our control sample shows an excess at $H \sim 18 \textendash 20$, where our sample of $24\mu m$-selected galaxies peaks.  We interpret this as representative of an over-density in the environment of $24\mu m$-sources.  The remainder of our galaxy counts, on the other hand, agrees remarkably well with published values.

It is important to note that in counting galaxies, we {\em exclude} all parts associated with the $24\mu m$ source, and retain only the random component (that is the surroundings).  There is ambiguity, in a handful of cases, as to whether some distant objects are associated with their neighboring $24\mu m$-galaxy or not.  We chose to include only half of them in our number counts, but since there is no more than two such cases in any given bin (many have none), this never amounts to more than one count.  All are folded into our uncertainties though.  In our accounting, we also exclude all point sources with magnitudes brighter than $H = 22$ in order to avoid contamination by stars.  Such a cut, however, also rejects extra-galactic point sources (QSOs).  It is therefore likely that we are slightly underestimating our counts, though again, this can only be a small effect.     

\subsection{Data Reduction \label{sec:datareduction}}

We used a modified version of the standard NICMOS reduction pipeline created by V. Fadeyev, and that properly corrects for the effect of cosmic rays on neighboring pixels.  This routine is described in \citet{Fadeyev06}.  Following the standard procedures, we used the {\it calnica, nicpipe}, and {\it biaseq} routines to perform the basic reduction steps, with bias equalization, on each of our images.  We then ran {\it pedsub} to remove the \textquotedblleft pedestal effect \textquotedblright between the four quadrant of the images.  For images affected by the {\it South Atlantic Anomaly}, we then ran {\it saaclean} to remove the cosmic rays imprint that persists for a certain time after the telescope has left the SAA-region.  We then ran the {\it pedsky} routine to perform a first sky subtraction.

These first standard steps correct most instrumental signatures and bring the background to an average of zero.  However, at the end of this process remaining residual patterns are often present in many images.  Our next step was, thus, to create a \textquotedblleft supersky \textquotedblright image by median averaging images that still carried a strong residual pattern in their background, and smoothing that median image with a ring median filter with inner and outer radii of two and ten pixels respectively.  At the end, we used 25 images to construct this supersky image.  We then, for each image, found the best-fit factor by which to multiply the supersky such that, when subtracted from the image, minimizes the dispersion in that image; and proceeded to subtract the scaled supersky.  Finally, we performed a background column subtraction on images that presented vertical features usually caused by electronic ringing and streaking of bright sources.

We then combined all the dithered images of each object into a final mosaic using the {\it dither} package part of the STSDAS external package in IRAF.  We first {\it drizzled} each final NICMOS image onto separate, but aligned grids, sampled at half the size of a NIC2 pixel, using a square kernel (or \textquotedblleft drop \textquotedblright) of 0.8 times the NIC2 pixel size.  We then combined the drizzled images into one final image using a $3\sigma$ rejection around the median.

The surface brightness detection limit in our final images is $\sim 23.9 \mbox{ mag arcsec}^{-2}$, given our choice of extraction parameters, which consists of smoothing with a $3\times 3$ pixels pyramidal kernel and requiring that at least 36 pixels above $1.5 \sigma$ of the noise, whose RMS is $22.5 \mbox{ mag arcsec}^{-2}$, be connected to one another.  This detection limit corresponds well to the surface brightness we are able to pick up by eye.  Accounting for surface brightness dimming, it is equivalent to an intrinsic surface brightness of $\mu = 20.9 \mbox{ mag arcsec}^{-2}$ at redshift one and $\mu = 19.6 \mbox{ mag arcsec}^{-2}$ at $z=1.7$.

\subsection{Profile Fitting \label{sec:profile_fitting}}

We used the GALFIT package \citep{Peng02} to perform a series of fits to the profile of all of our objects.  We first fitted our objects with a one-component sersic profile with free sersic index.  For those objects with a sersic index, $n$, greater than 1 that were sufficiently extended and whose residuals (or image) suggested the presence of a disk, we repeated the fit with a two-component bulge (de Vaucouleurs) plus disk (exponential) profile.  We use this bulge-to-disk decomposition to classify our objects into disk-dominated, $B/D < 1$, and bulge-dominated, $B/D > 1$, galaxies.

Because many of our objects consist of a bulge surrounded by an envelope of tidal streams, the addition of a disk component to the fit often results in a large fraction of the flux of those objects to be assigned to that component.  This is unphysical since these tidal envelopes do not follow an exponential profile.  In those cases, we instead assume that the bulge is well represented by the sersic profile and that the disk component is negligible.  Objects that are too small to be meaningfully decomposed into bulge and disk components, on the other hand, we classified using the value of their sersic index only.  Comparison of the fitted sersic index with the bulge-to-disk ratio of the more extended objects in our sample indicates that a $B/D$ of~1 roughly corresponds to a sersic index of~2.3.  We therefore adopt $n=2.3$ as the limit between bulge and disk-dominated galaxies.  

Many ($\sim 20\%$) of our objects display a bright PSF profile at their centers, suggestive of the presence of a significant point source.  Most of them turned out very high sersic indices in their one-component fits.  We, therefore, re-fit those objects for which the one-component, sersic profile, yielded an $n>4.5$, with a two-component, PSF plus sersic, profile.  Four galaxies that looked like they might have a point source turned out indices of $n \sim 4$, only with small effective radii, and were therefore not re-fit.  Some of our largest galaxies, we were able to successfully fit with a three-component, PSF+bulge+disk, model.  We also added a PSF component to the fit of some of our galaxies when the one-component and two-component fits showed significant residuals and those residuals largely disappeared when the PSF component was added.  In all cases, we always constrained the magnitude of the PSF component (when present) using only a PSF+sersic fit, and then proceeded to fit the bulge and disk components, when possible, maintaining the PSF component fixed.  We then used the bulge-to-disk decomposition or sersic index of the host to determine, when possible, whether these objects were disk or bulge-dominated.  For a number of objects, however, the host galaxy was too faint compared to the central PSF and too small for the sersic profile to be reasonably constrained.  The profile of those objects was therefore left {\em ambiguous}.

Many of our objects have companions nearby. In such cases we always fitted the main optical counterpart to our $24\mu m$ source and its companions simultaneously.  The classification of all our objects as either bulge or disk-dominated can be found in Appendix~B, along with their magnitude, size, and morphological class.  Morphological classes are described in the following section.

\subsection{Morphological Classification \label{sec:morphclassification}}

Locally, $\sim 99\%$ of ULIRGs visually appear to be ongoing mergers or merger remnants \citep{Veilleux02}.  Analogously, a large number of our objects show signs (or possible signs) of merging.  In order to establish a solid comparison with local ULIRGs we chose to use the same classification scheme as \citet{Veilleux02}.  This scheme, originally proposed by \citet{Surace98}, classifies objects according to their stage of merging, following a merging sequence.  It is useful in that it allows to study changes in galaxy properties as a function of merging phase, and thus uncover how these properties change, statistically, throughout a merger event.  We present this scheme in section~\ref{sec:morphscheme}.  We then introduce, in section~\ref{sec:conf_classes}, confidence classes which we use to group objects according to how confident we are about their merging nature.  We show examples of objects from each of our confidence classes.  Finally, we briefly mention automated classification techniques in section~\ref{sec:automated}.

\subsubsection{Classification Scheme \label{sec:morphscheme}}

In this paper, we use the classification scheme of \citet{Surace98} and \citet{Veilleux02} which partitions objects into the following classes, following the progression of a merger:

I. {\em First Approach} or {\em Pair} \textendash~This category refers to galaxy pairs that have not passed through each other yet.  In this phase the two objects still retain their own characteristic morphology (albeit maybe with some perturbations).

II. {\em First contact} \textendash~At this stage, the two objects overlap, but tidal streams and debris have not formed yet.

III. {\em Pre-merger} \textendash~After their first encounter, the two galaxies will usually emerge again (except in very particular low-speed, co-aligned configurations where they could merge during first pass), but this time tidal tails and bridges will have formed, and their respective morphology will be highly disrupted.  Double-nucleus systems showing tails, bridges or morphologies suggestive of a recent encounter fall in this category.

IV. {\em Advanced merger} \textendash~At this stage, the two nuclei are undistinguishable meaning they have either coalesced or are on the verge of doing so.  Trails of stars that have been tidally stripped or other debris from the merging process are readily visible after, and sometimes even before, subtraction of the main, smooth component of the galaxy.  This is also a phase of rapid bulge growth.  At high redshifts, tidal features can sometimes appear as extended lopsided disks.  In those cases, we require that the main component be a bulge in order for the object to be classified as an advanced merger, so that not to be contaminated by disturbed, asymmetric or clumpy disks whose origin might not lie in a merger \citep{Forster-Schreiber10}.

V. {\em Old merger} \textendash~ In this phase, tidal tails and streams have faded out, but the galaxy still shows signs of past events in the form of residual asymmetries and/or clumps, that are detectable through subtraction of the smooth profile.  Because signatures typically associated with mergers have disappeared from the objects classified in this category, they cannot be known for sure to have experienced a recent merger.  We require them, however, to be bulge or PSF-dominated, so that their observed underlying structure, which must be important since still detected at $z \gtrsim 1$, when put together with a bulge-like galaxy, is most easily explained through the merger scenario.

\citet{Veilleux02} further split phase III into {\em close} and {\em wide} binaries.  We do not make that distinction, although we would have about equal numbers in each class.  The largest separation we observe among our {\em pre-mergers} is 35 kpc (in MIPS 298).  \citet{Veilleux02} also split phase IV into {\em diffuse} and {\em compact} mergers, but the lower physical resolution of our data does not allow us to make that distinction either.  On the other hand, we find four {\em triplets} in our sample and have a category for them.  However, because stages II, III and triplets all represent objects that are in the process of merging, but have not yet coalesced, we usually refer to them collectively as {\em early mergers}.

We note that, because the imagery of local ULIRGs by \citet{Kim02} and \citet{Veilleux02} has a much higher physical resolution and much higher surface brightness sensitivity than we have for our high-redshift sample, their phase IV covers a larger part of the merging process than our phase IV, as we lose the streams and tails that characterize that phase earlier than they do.  Furthermore, it is impossible for us to detect the residual artifacts of a merger to the level they do to define their phase V.  In our data, many of those objects would look like regular bulges (see section~3).  Our phase V objects, assuming they are truly mergers, should rather be thought of as a mix of late phase IV and early phase V objects in local terms, a kind of stage 4.5.

Although many of our objects are mergers (cf. \S~\ref{sec:mergfrac}), our sample does also contain a number of regular-looking unperturbed galaxies.  On top of the six (including triplets) merger categories described above, we therefore also have:  {\em face-on} and {\em edge-on spirals} when either spiral arms or an edge-on disk are readily distinguishable, and {\em regular bulges} when no particular features are visible and the galaxy has a bulge-to-disk ratio $B/D > 1$, or a sersic index $n > 2.3$.  We also have one object that appears as a pure point source in our data.

Lastly, we have a number of objects in our sample that we label {\em faint \& compact}, because they are best fit with a disk-ish ($n \sim 1 \textendash 2$) profile, yet are much smaller and fainter than regular spirals\footnotemark ~and show no signs of either spiral arms or of an extended disk.  Most of them show, instead, stage V type of residuals.  We show in section~3 that some local ULIRGs can indeed appear as faint \& compact at high redshift.  They are typically phase IV objects that possess diffuse tidal tails too faint to be visible at high redshifts, and dense cores often intersected by dust lanes, rather than fully formed bulges.  Because of that fact, we often group these {\em faint \& compact} objects with phase IV galaxies proper and refer to them as {\em coalescence} objects.

\footnotetext{Faint \& compact objects have typical half-light radii and apparent magnitudes of $r_{1/2} \approx 0.15\textendash 0.35"$ and $m_{H} > 19.6$, whereas spiral galaxies in our sample have values of $r_{1/2} \approx 0.4\textendash 0.6"$ and $17.7 < m_{H} < 19.0$.}

\subsubsection{Confidence classes \label{sec:conf_classes}}

Although there is inherent uncertainty associated with morphological classification, some objects can be classified with more confidence than others and, consequently, some results are more secure than others.  It is, thus, useful to distinguish these degrees of certainty.  We do so by assigning to all of our objects a confidence level with which their merging nature can be inferred.  Although these levels of confidence can themselves be somewhat subjective, we find them, from a second, independent re-classification by another co-author (L.~Yan),  to vary by no more than one confidence level in 80\% of cases, showing that they are, in practice, reasonably well-defined.  We look below at what we can learn from each of these confidence classes.

\begin{figure*}[bhtp]
\begin{center}
\includegraphics[height=1.33in]{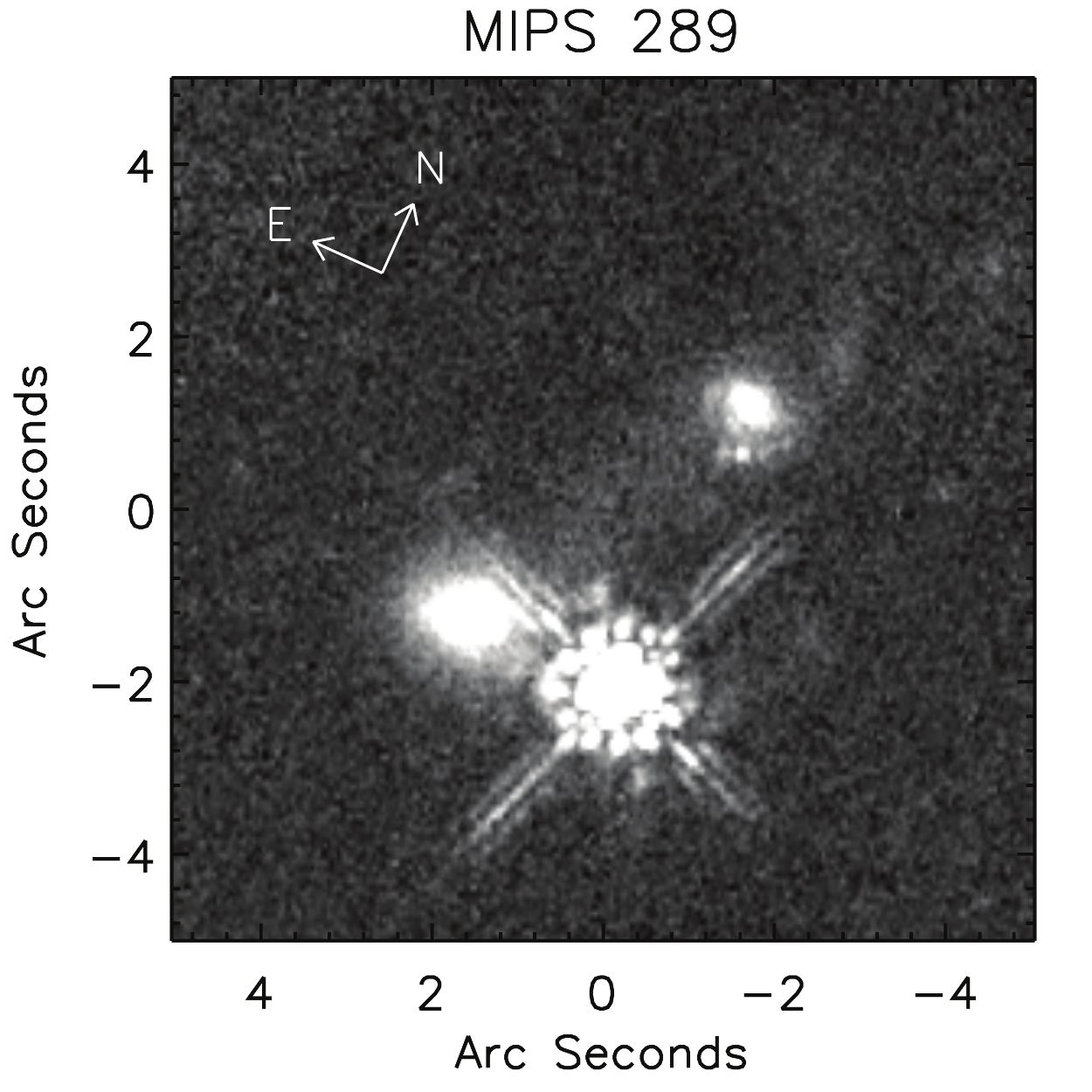}
\includegraphics[height=1.33in]{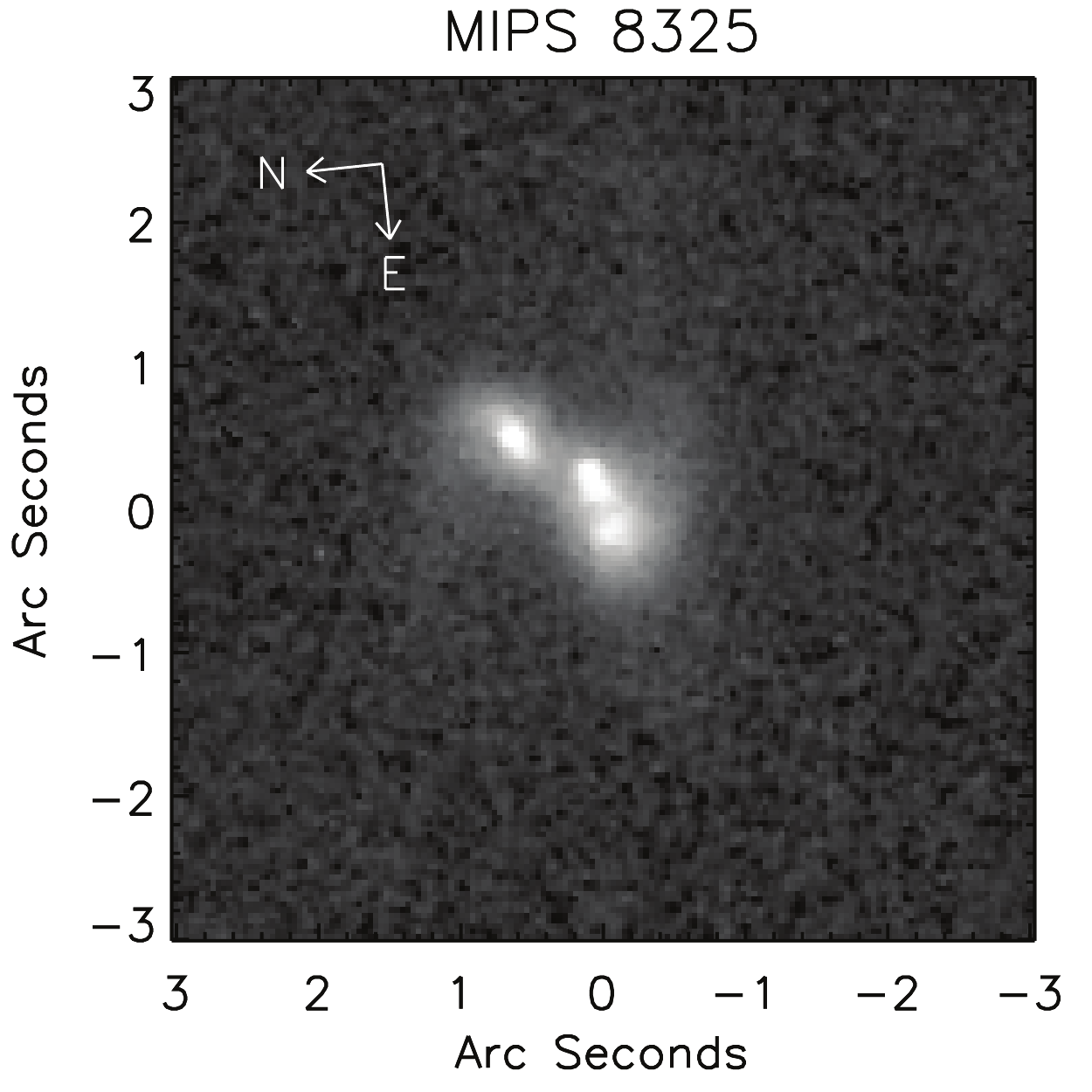}
\includegraphics[height=1.33in]{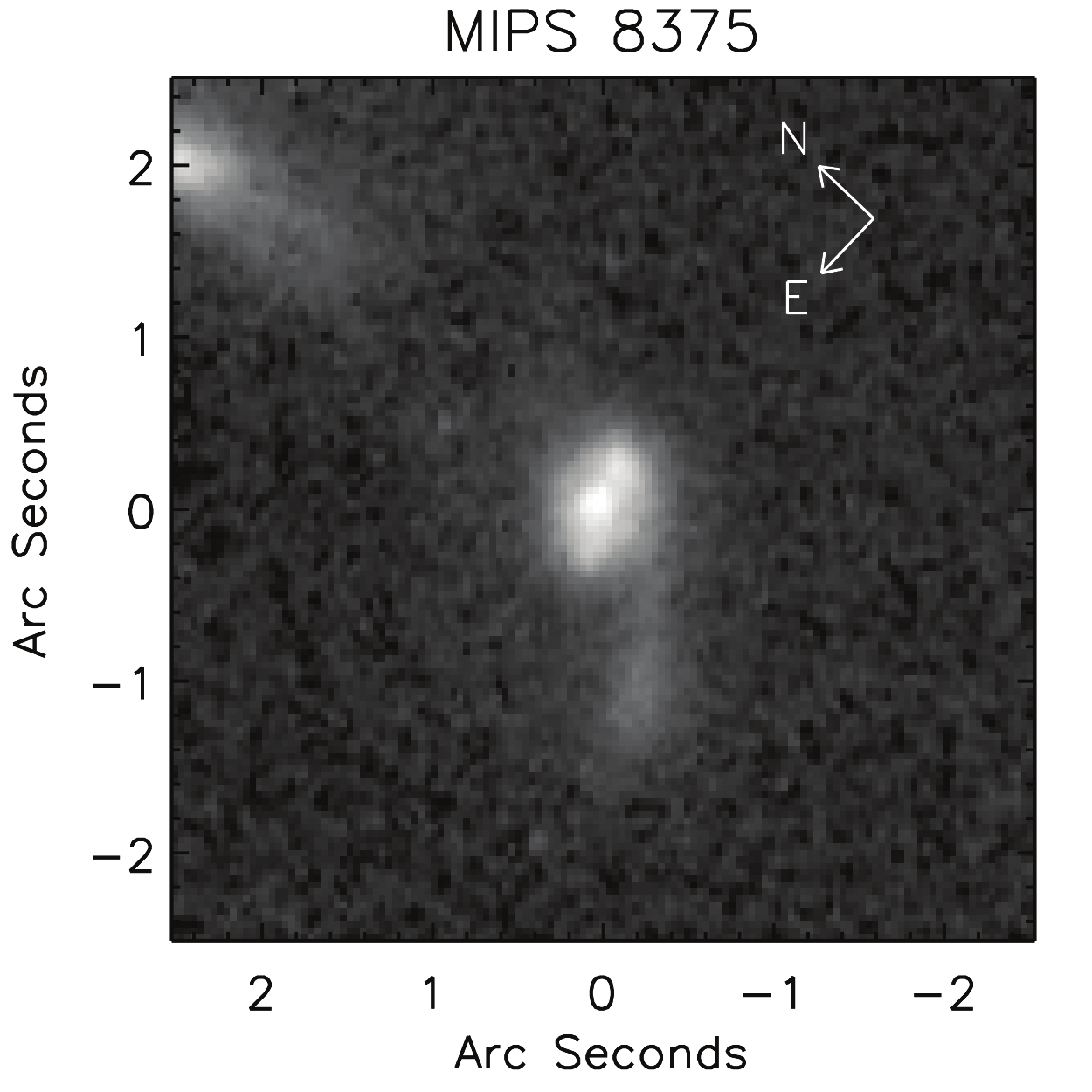}
\includegraphics[height=1.33in]{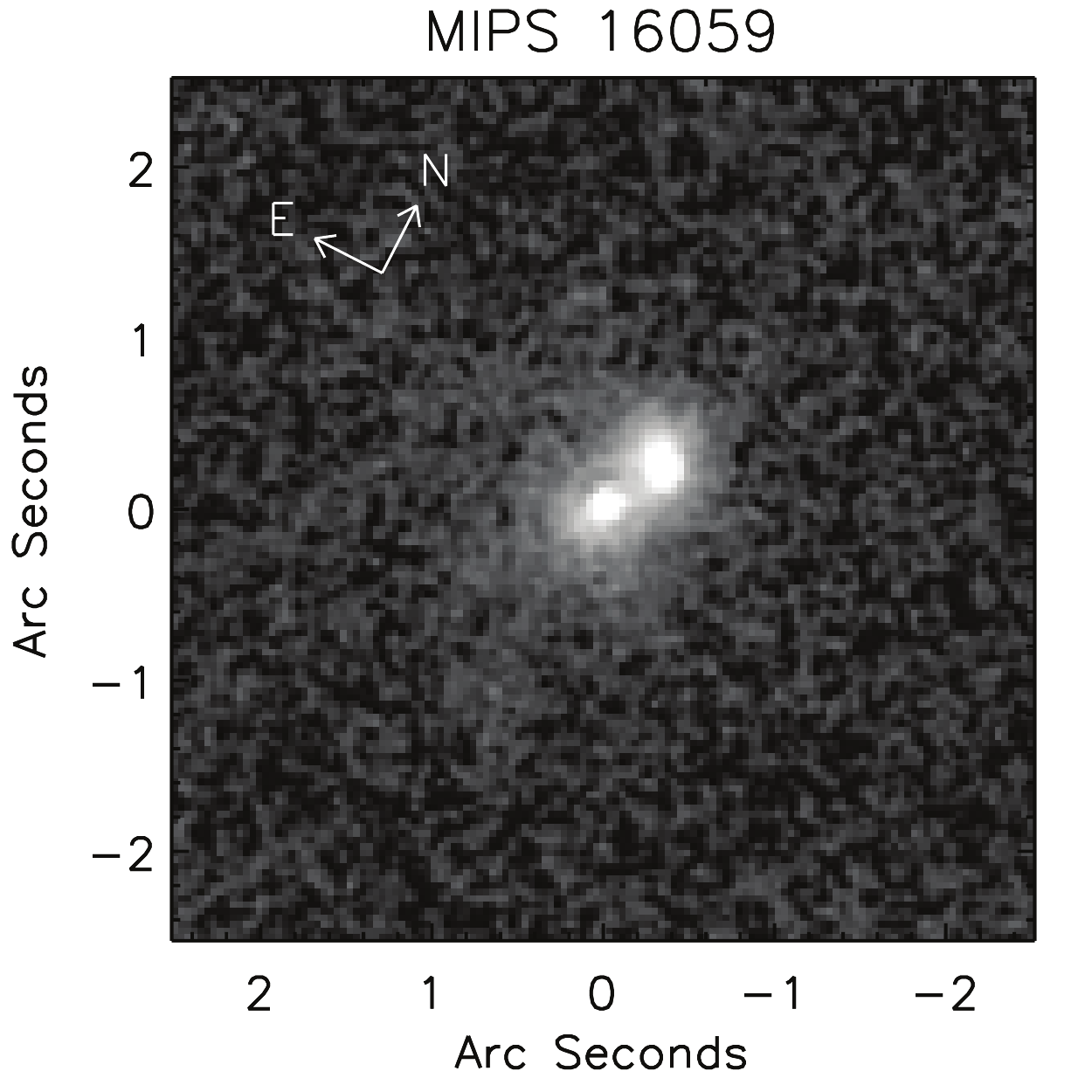}
\includegraphics[height=1.33in]{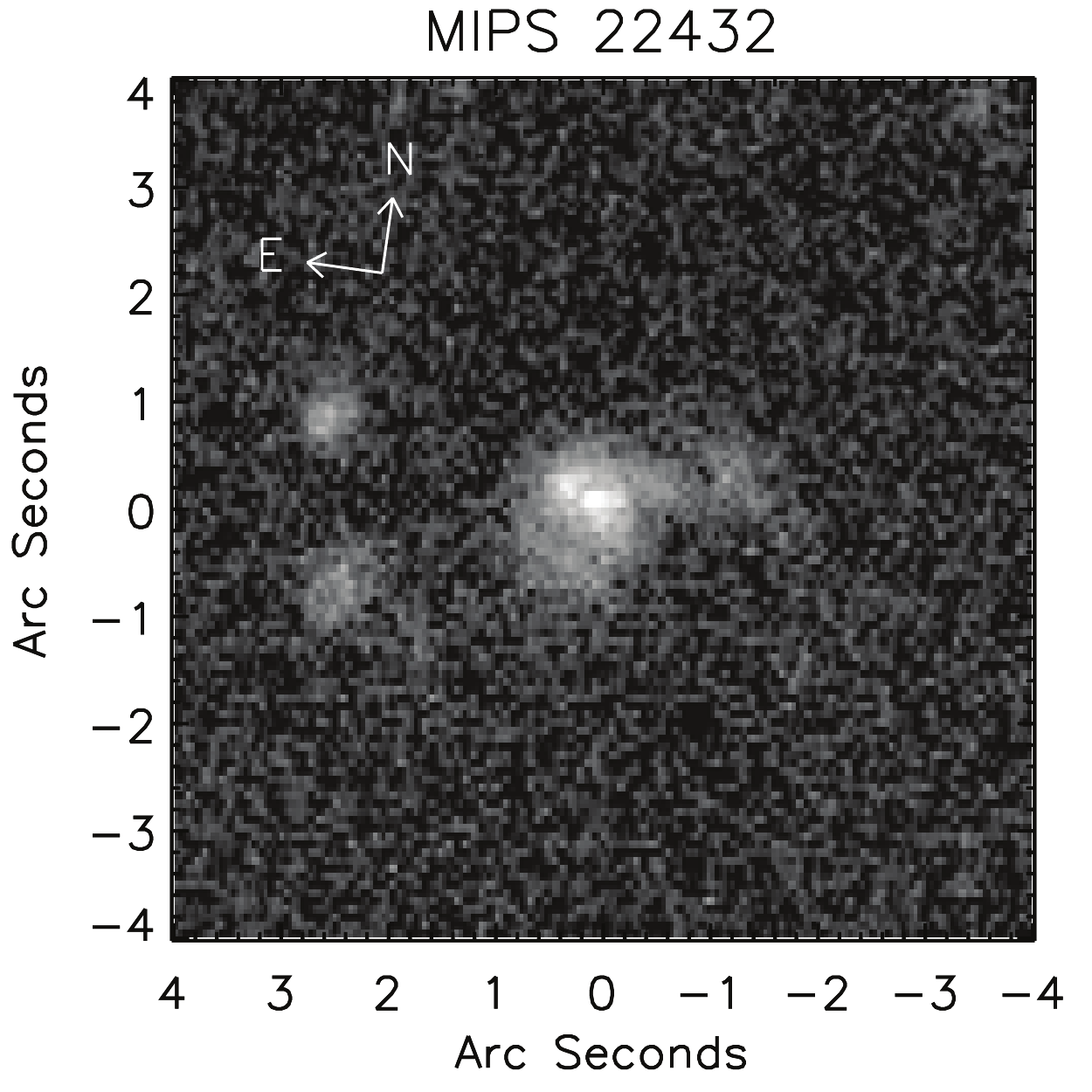}
\caption{Clear mergers (category 1). All of them are, by definition, in the early phases of merging:  four are in phase III , and one is a triplet (see text for the definition of our morphological classes).  The point source in the image of MIPS289 is a foreground star.}
\label{fig:examples1}
\end{center}
\end{figure*}

We call the most secure objects, those that can be immediately and unarguably identified as mergers at first sight, category~1.  These objects consist exclusively of two overlapping or connected galaxies that show tidal tails and streams of their interaction, as these are the only kind whose merging nature we find can be established without a doubt from visual examination alone.  A local analog would be The Antennae galaxy.  Unfortunately, we find only five objects that fit that description, in our sample.  They are all shown in Figure~\ref{fig:examples1}.  Besides containing a low number of galaxies, this first category further includes only objects in stage III of the merging process or triplets.  It thus becomes immediately clear that in order to make any progress, we need to move beyond this realm of absolute certainty.

\begin{figure*}[thbp]
\begin{center}
\makebox[3.4in]{\large ~~Close Pairs (Phase I)}
\makebox[1.7in]{\large ~~Phase II}
\makebox[1.7in]{\large ~~Triplet}\\
\includegraphics[height=1.7in]{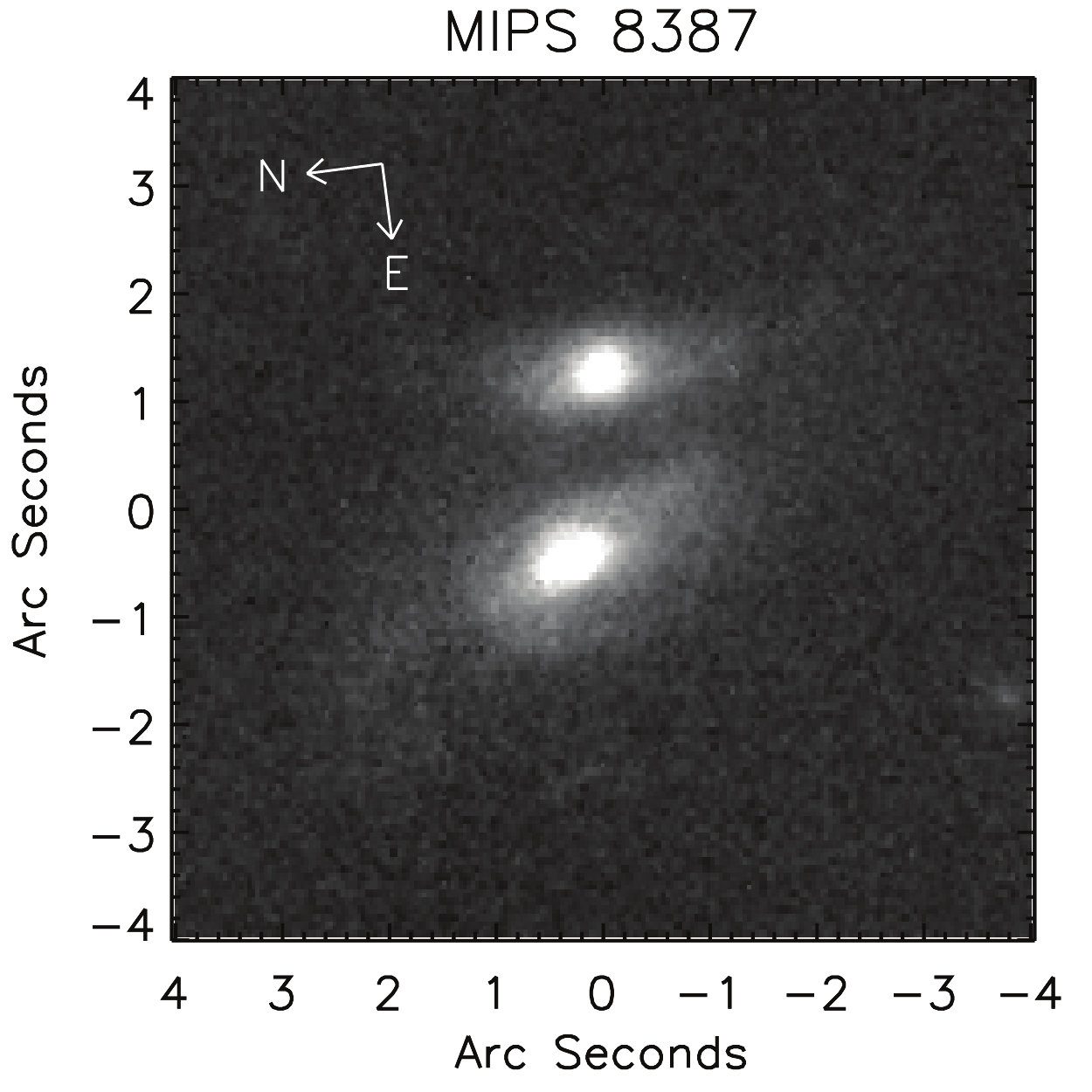}
\includegraphics[height=1.7in]{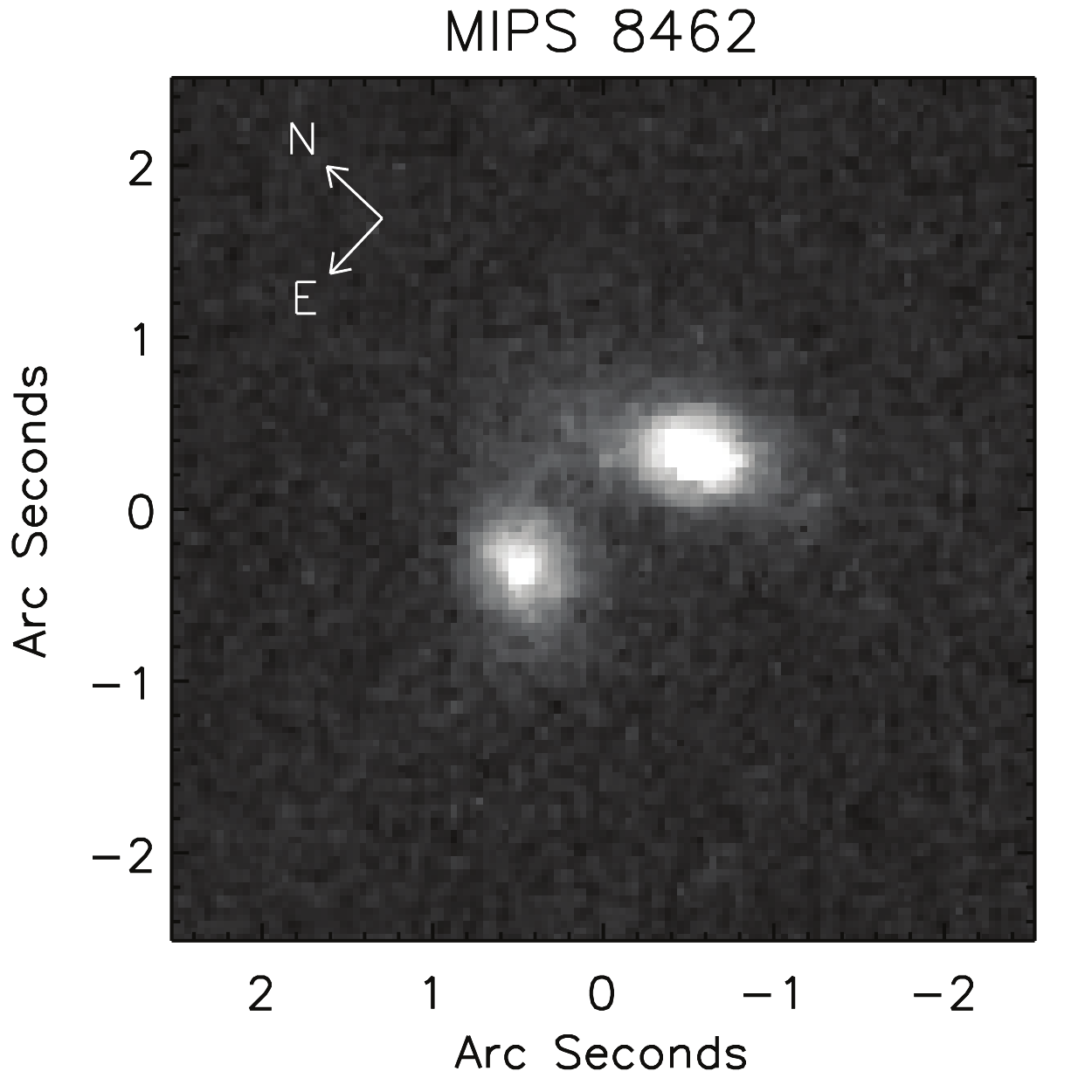}
\includegraphics[height=1.7in]{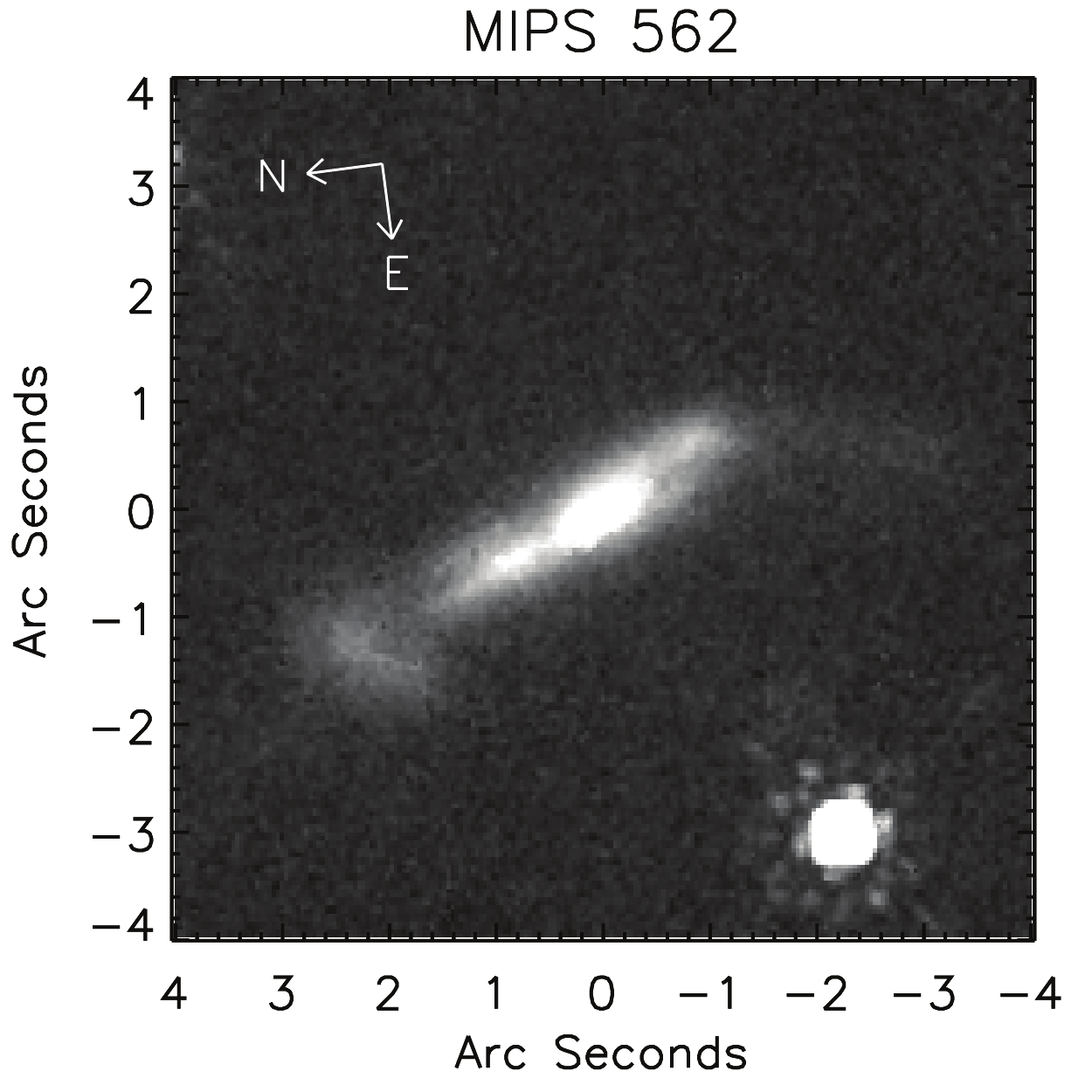}
\includegraphics[height=1.7in]{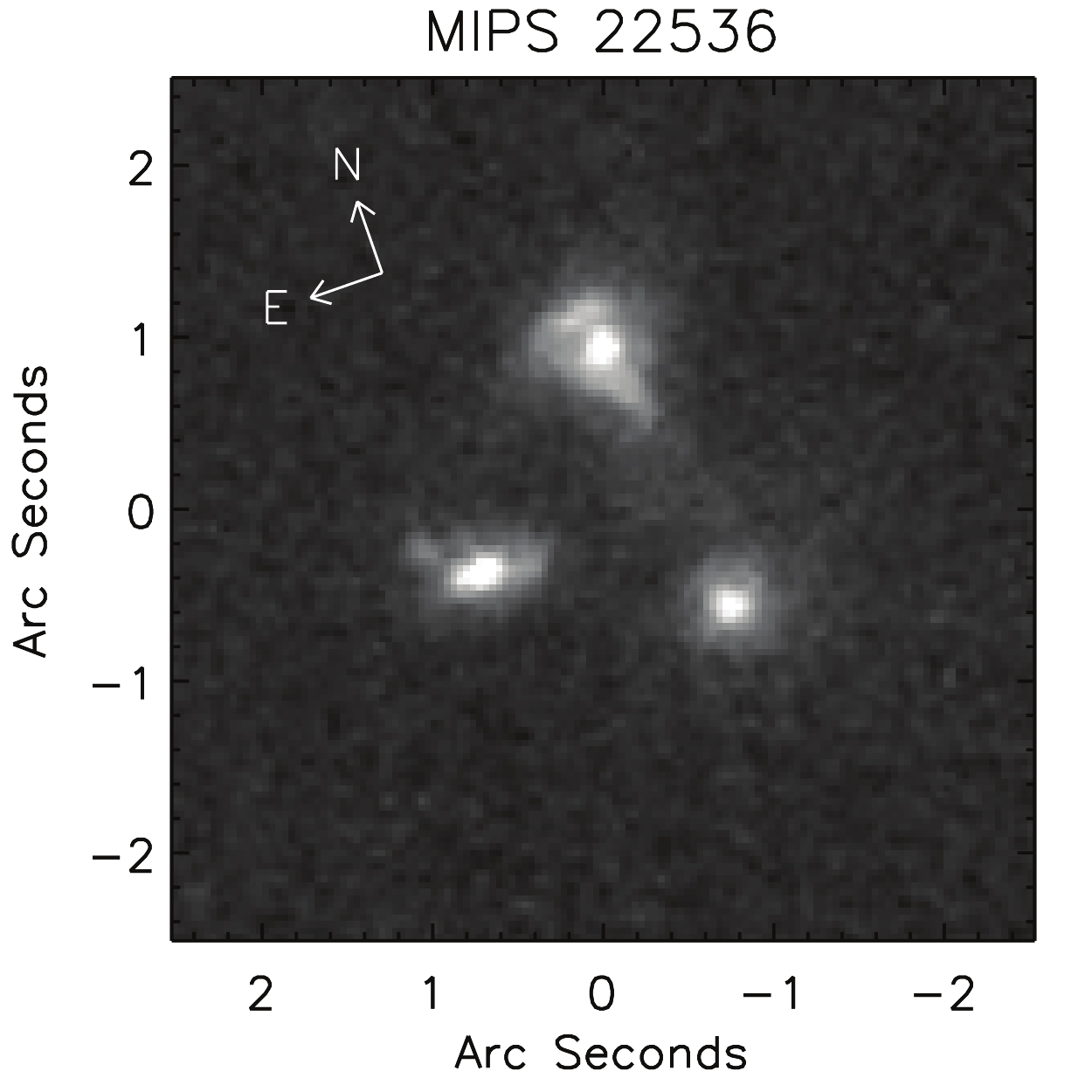}\\
\makebox[7in]{\large Pre-Mergers (Phase III)}\\
\includegraphics[height=1.7in]{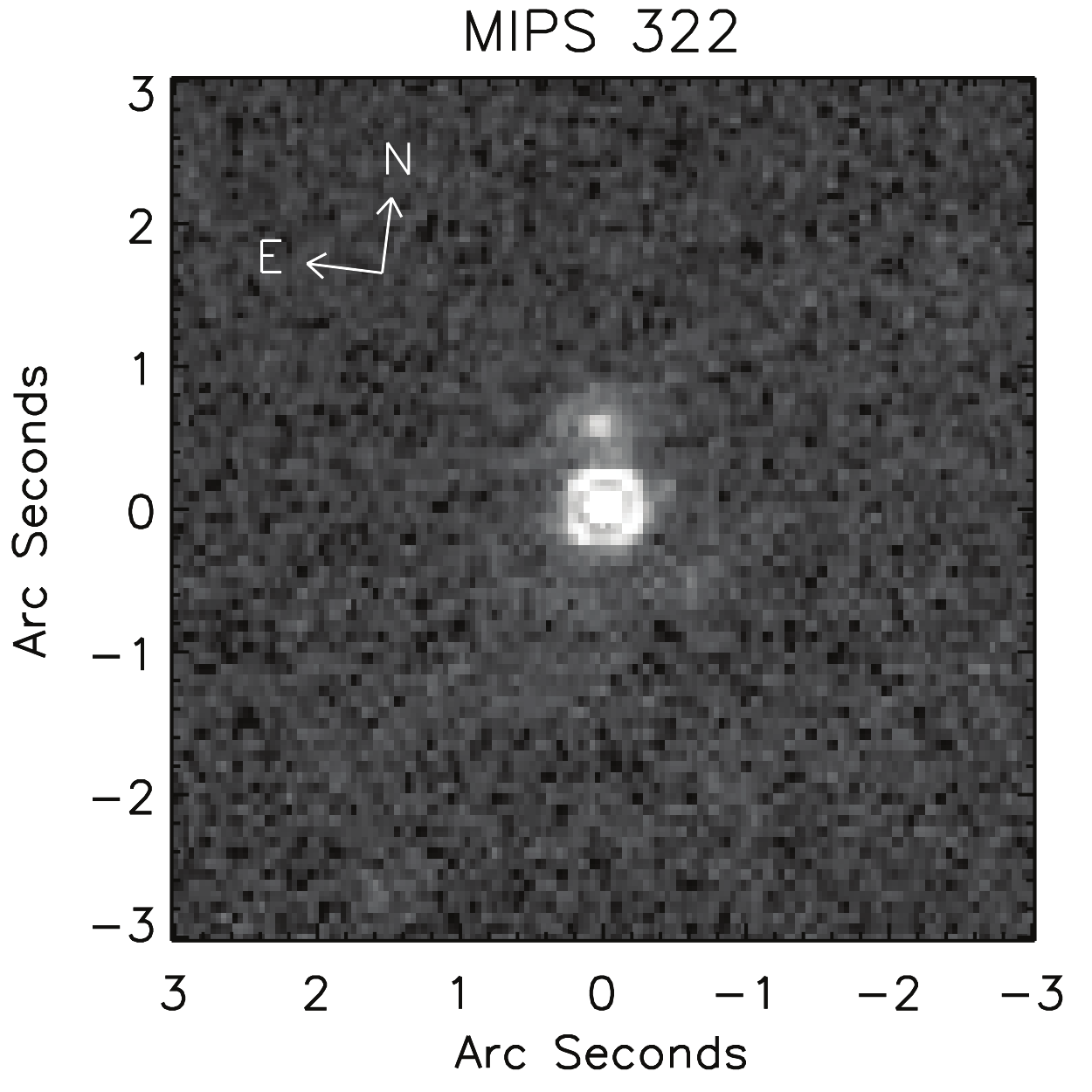}
\includegraphics[height=1.7in]{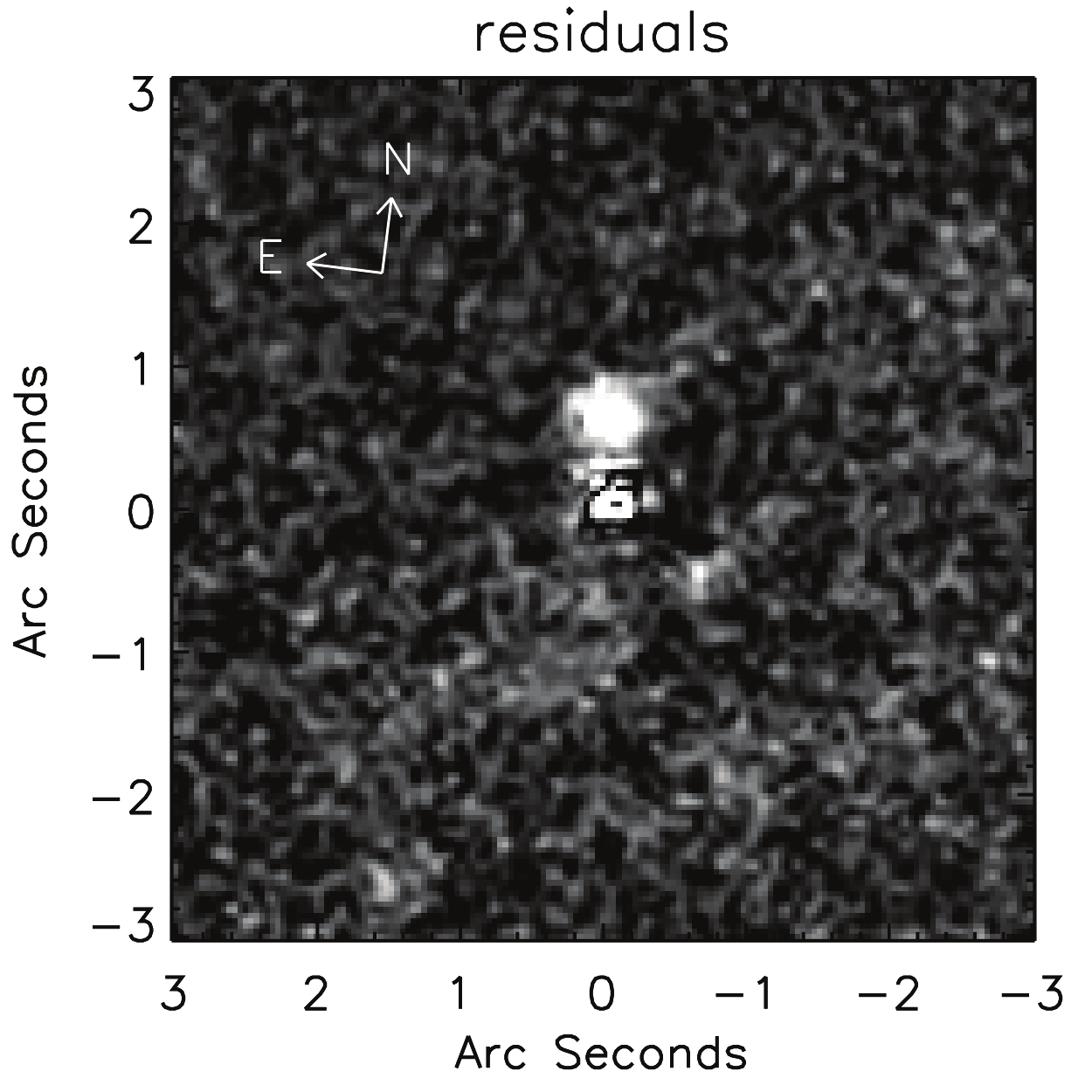}
\includegraphics[height=1.7in]{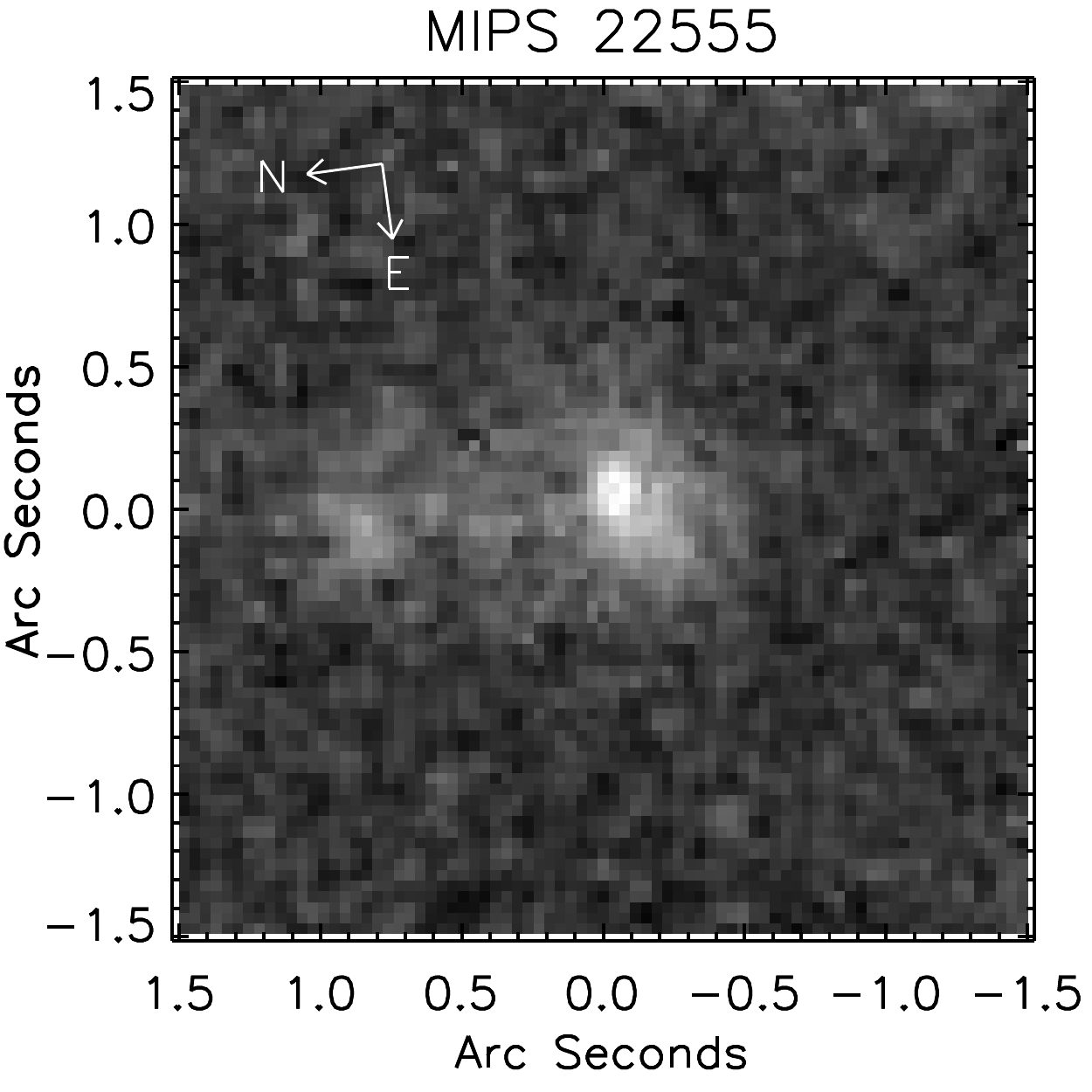}
\includegraphics[height=1.7in]{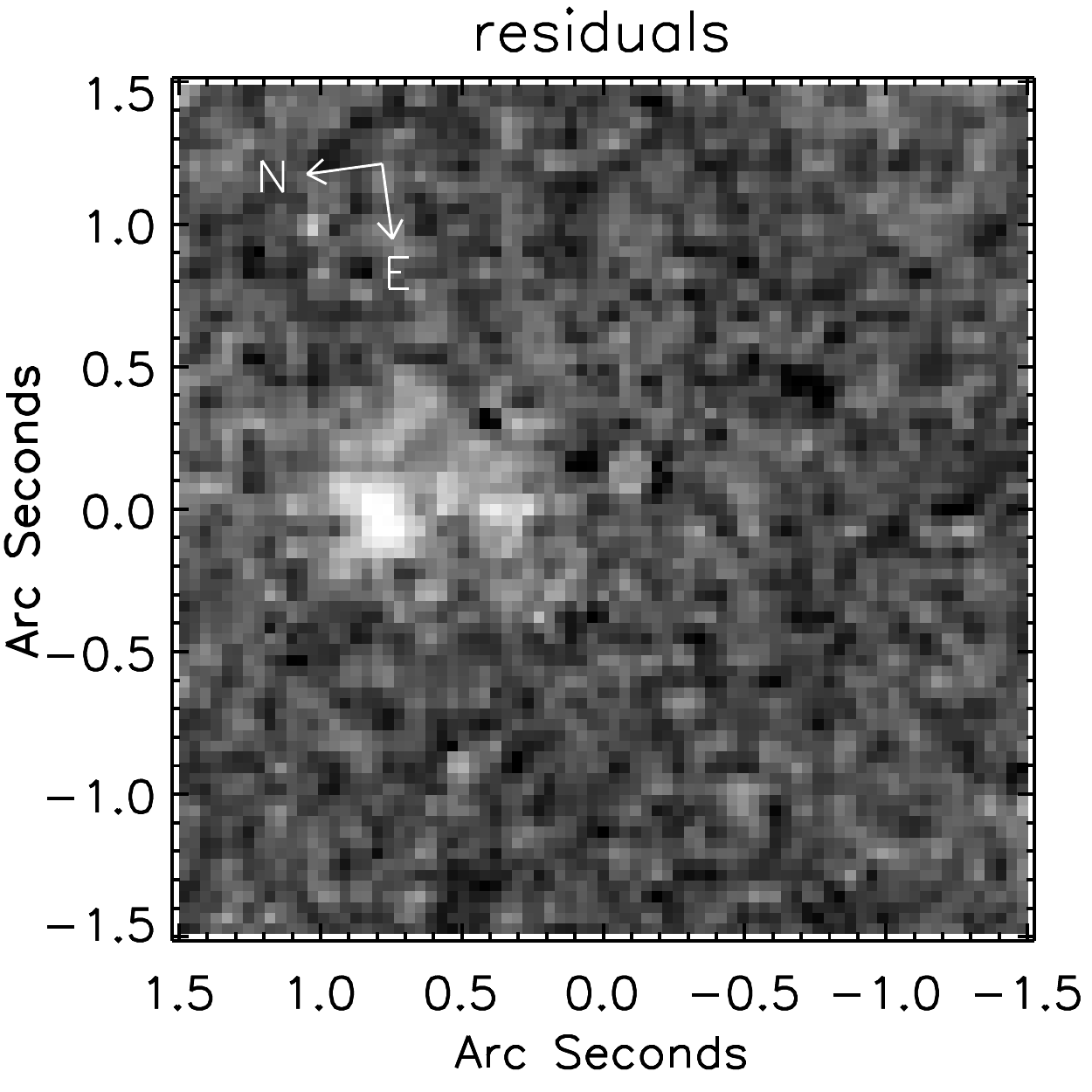}\\
\makebox[7in]{\large Advanced Mergers (Phase IV)}\\
\includegraphics[height=1.7in]{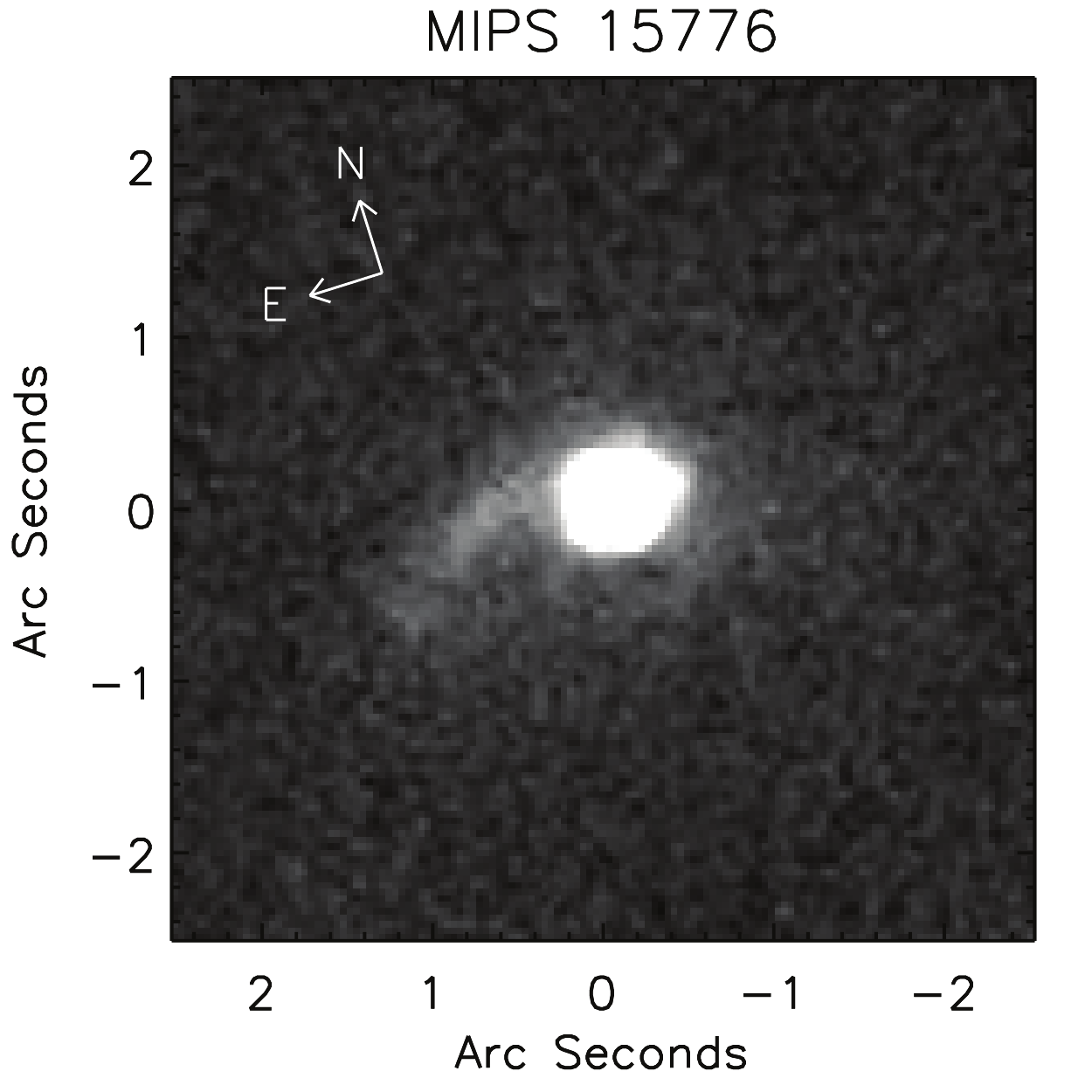}
\includegraphics[height=1.7in]{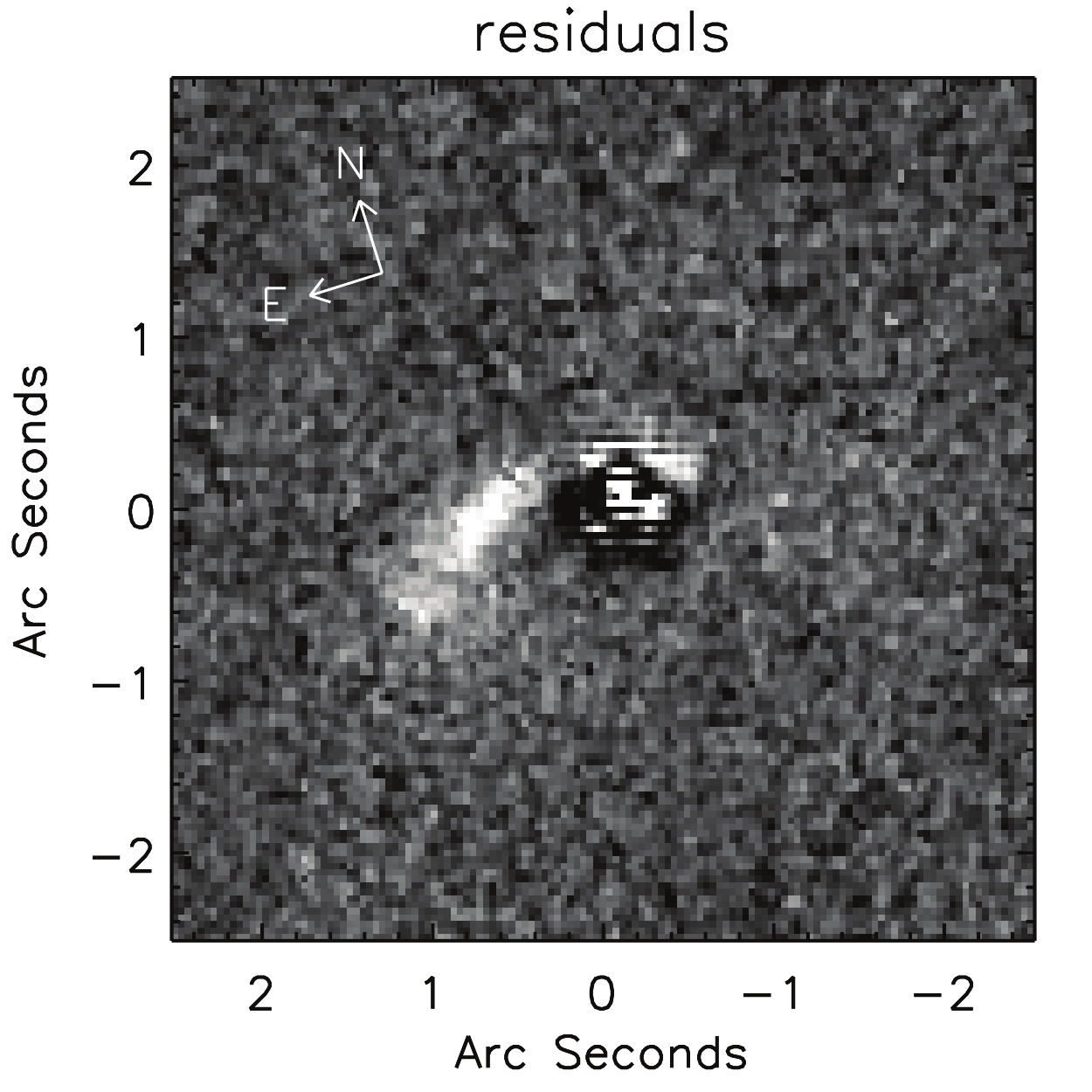}
\includegraphics[height=1.7in]{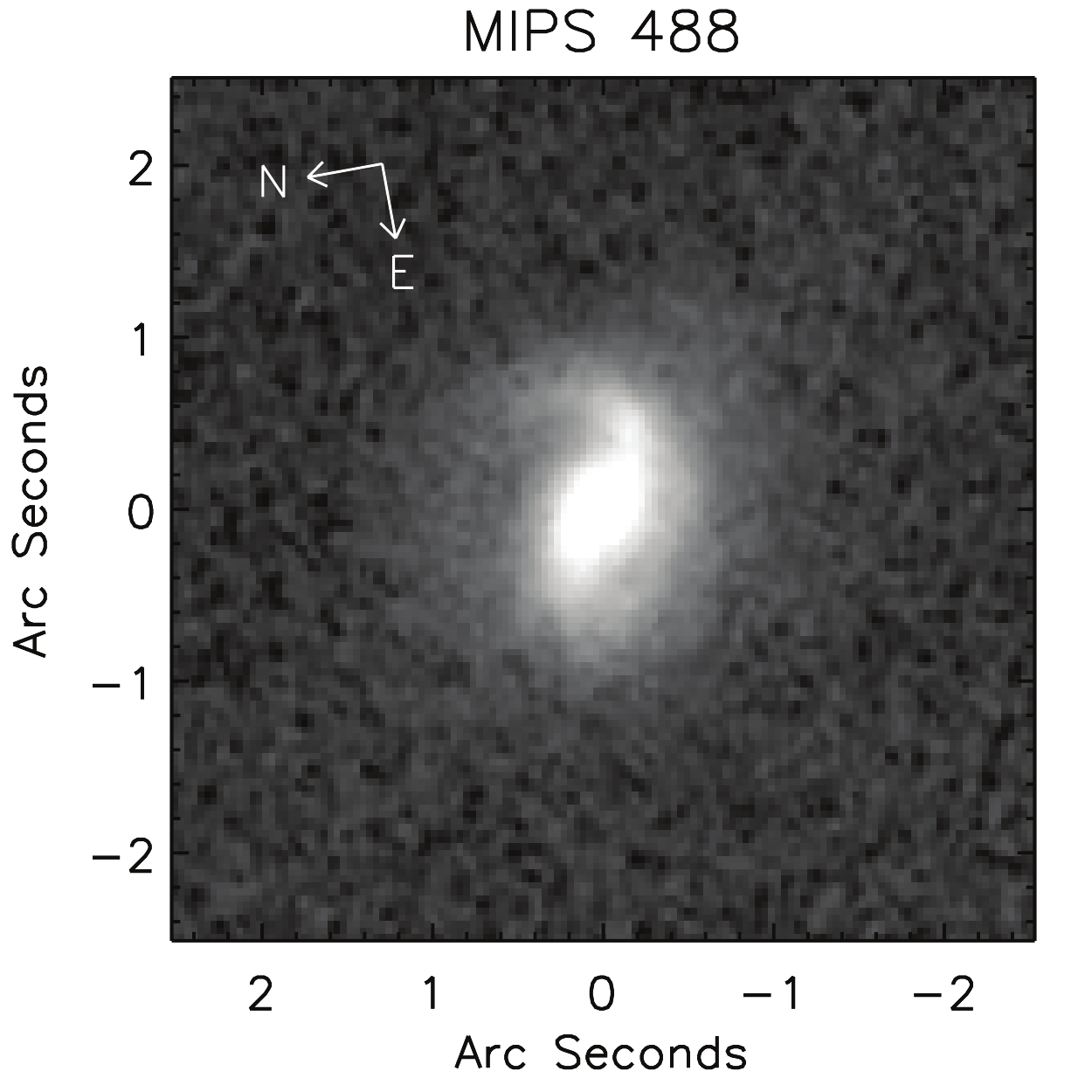}
\includegraphics[height=1.7in]{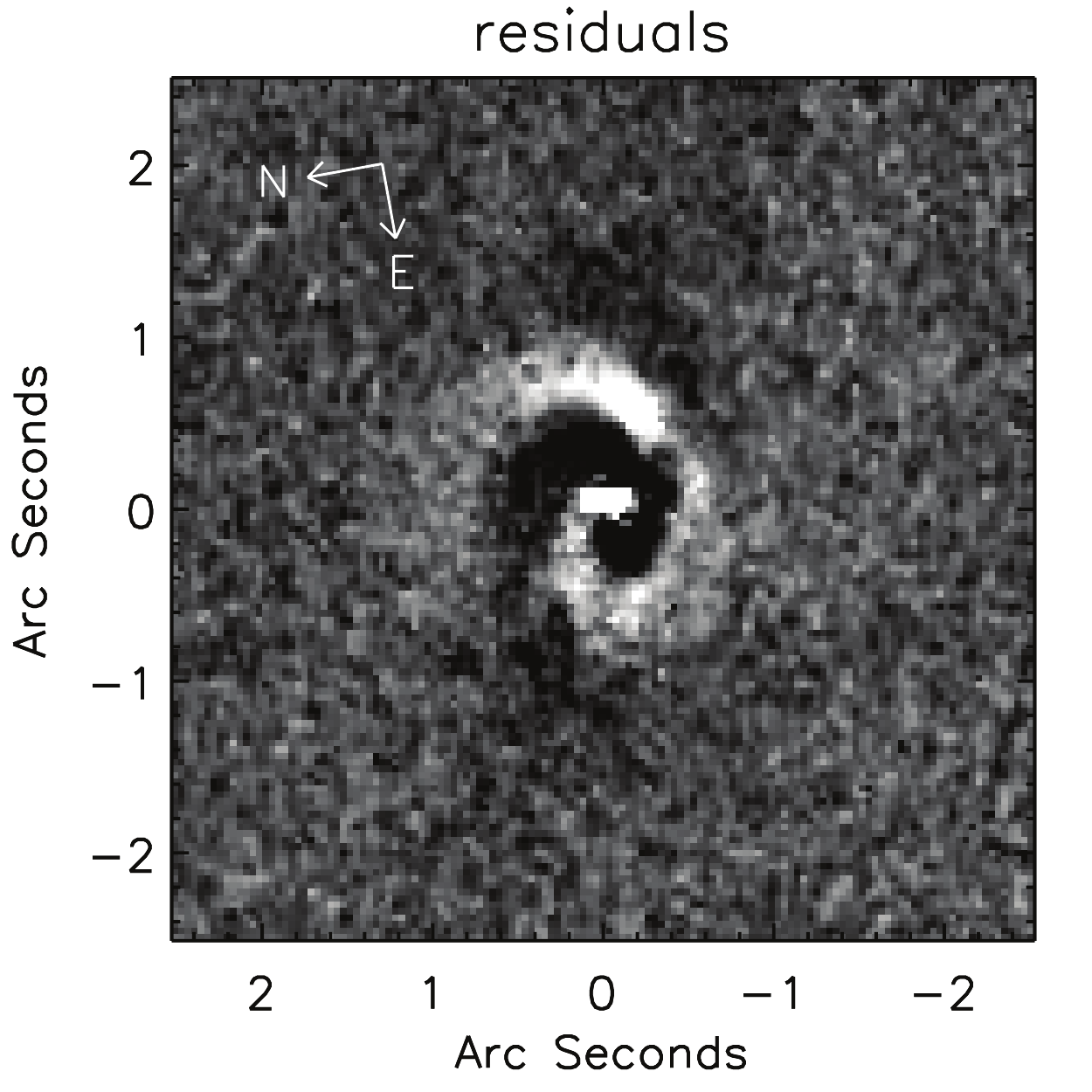}\\
\caption{Highly probable mergers (category 2) separated by morphological class.  For phases III \& IV, both the image, and the residuals after subtraction of the main component are shown.  Residuals for MIPS322 have been smoothed to better show its tidal features.  This figure also serves to illustrate the range of morphological profiles displayed by objects of those two phases.}
\label{fig:examples2}
\end{center}
\end{figure*}

We thus turn to the next best objects which we call highly probable mergers.  We include in this second category objects whose morphology is strongly suggestive of a merger event, but cannot entirely exclude other possibilities\footnotemark.  An example of which would be close pairs, since without kinematics, their merging nature cannot be asserted with $100\%$, but the proximity of the galaxies to one another (typically $< 20$ kpc in projection) and their comparable size strongly suggests that they are in the process of merging.  Another example would be advanced (singly nucleated) mergers with highly distorted morphologies and strong tidal tails, as one could imagine features like that arising in a process such as a high-speed encounter, even though they are far more typical of mergers.  Figure~\ref{fig:examples2} shows examples of different highly probable mergers.

\footnotetext{Because the exact probabilities for an object in each of our confidence classes to be a merger are unknown, we are forced to rely on approximate terminology such as \textquotedblleft highly probable\textquotedblright, \textquotedblleft strongly\textquotedblright and \textquotedblleft far more\textquotedblright.}

By including highly probable mergers, we more than quintuple our number of mergers to a total of 27, and add very little uncertainty in the process.  We also span all of the major stages of merging (four close pairs [stage I], one phase II object, seven more stage III mergers, one more triplet, and nine advanced mergers [phase IV]).  Categories~1 and~2 together, however, still represent only 20\% of our sample.

\begin{figure*}[thbp]
\begin{center}
\makebox[3.4in]{\large ~~Pre-Mergers (Phase III)}
\makebox[3.4in]{\large ~~Advanced Mergers (Phase IV)}\\
\includegraphics[height=1.7in]{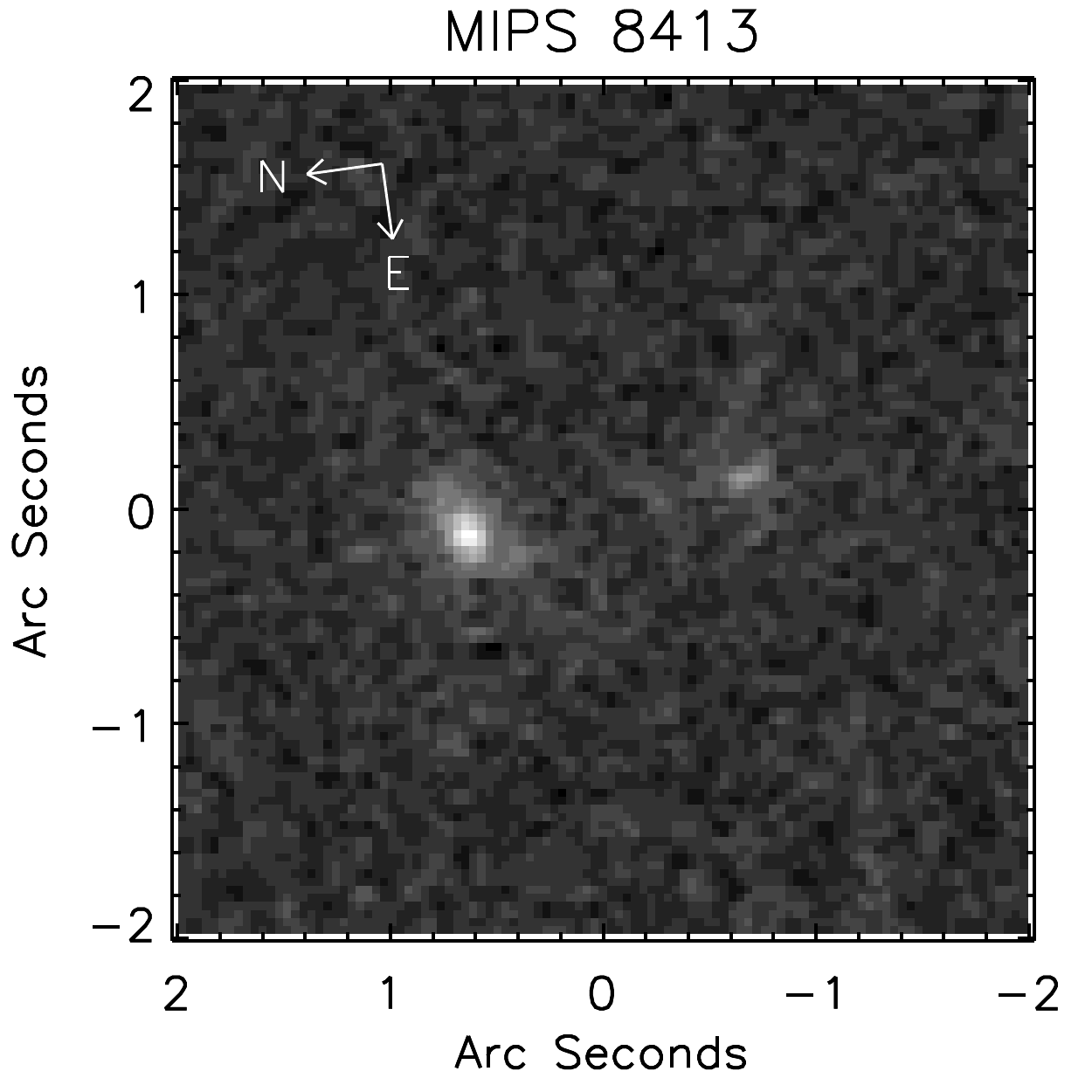}
\includegraphics[height=1.7in]{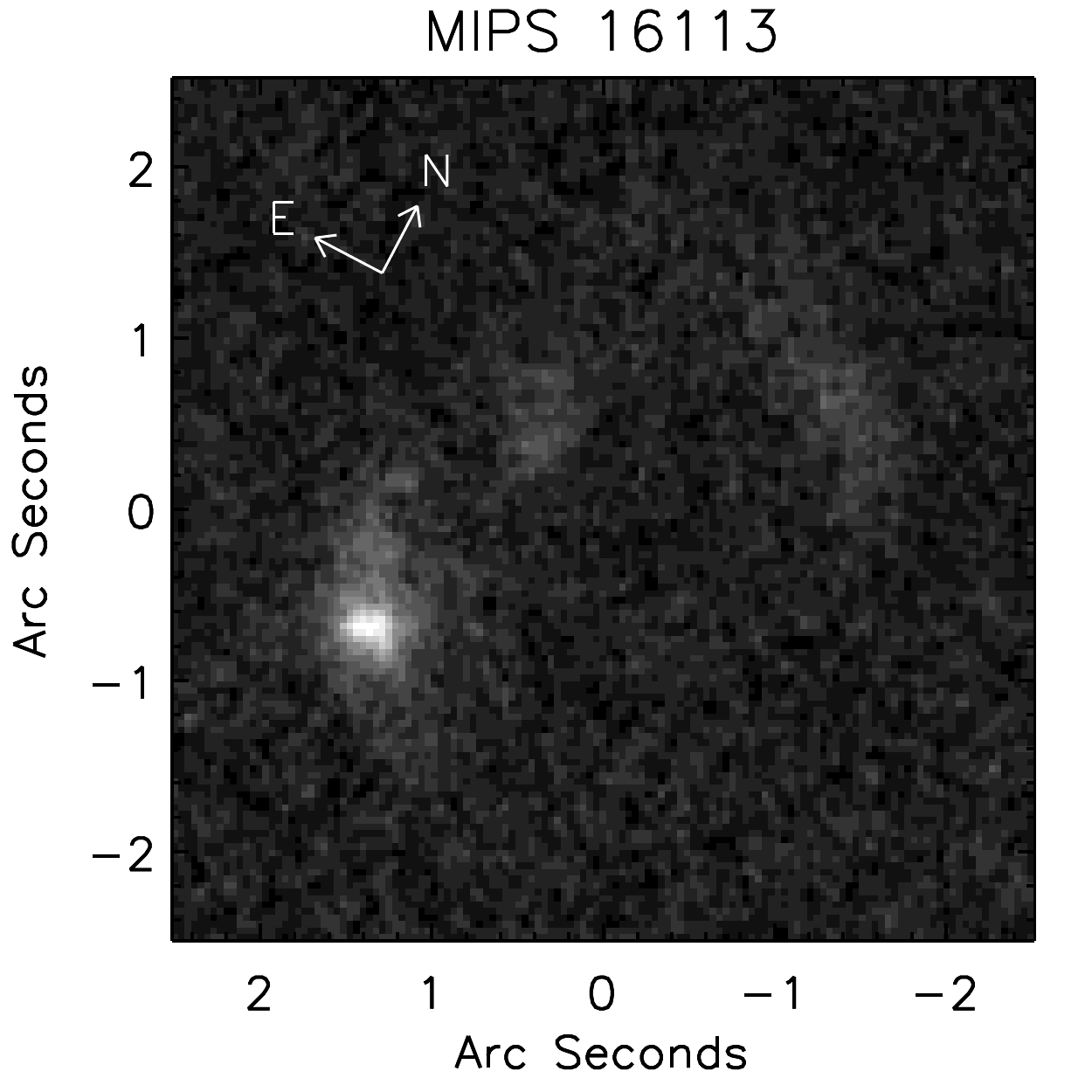}
\includegraphics[height=1.7in]{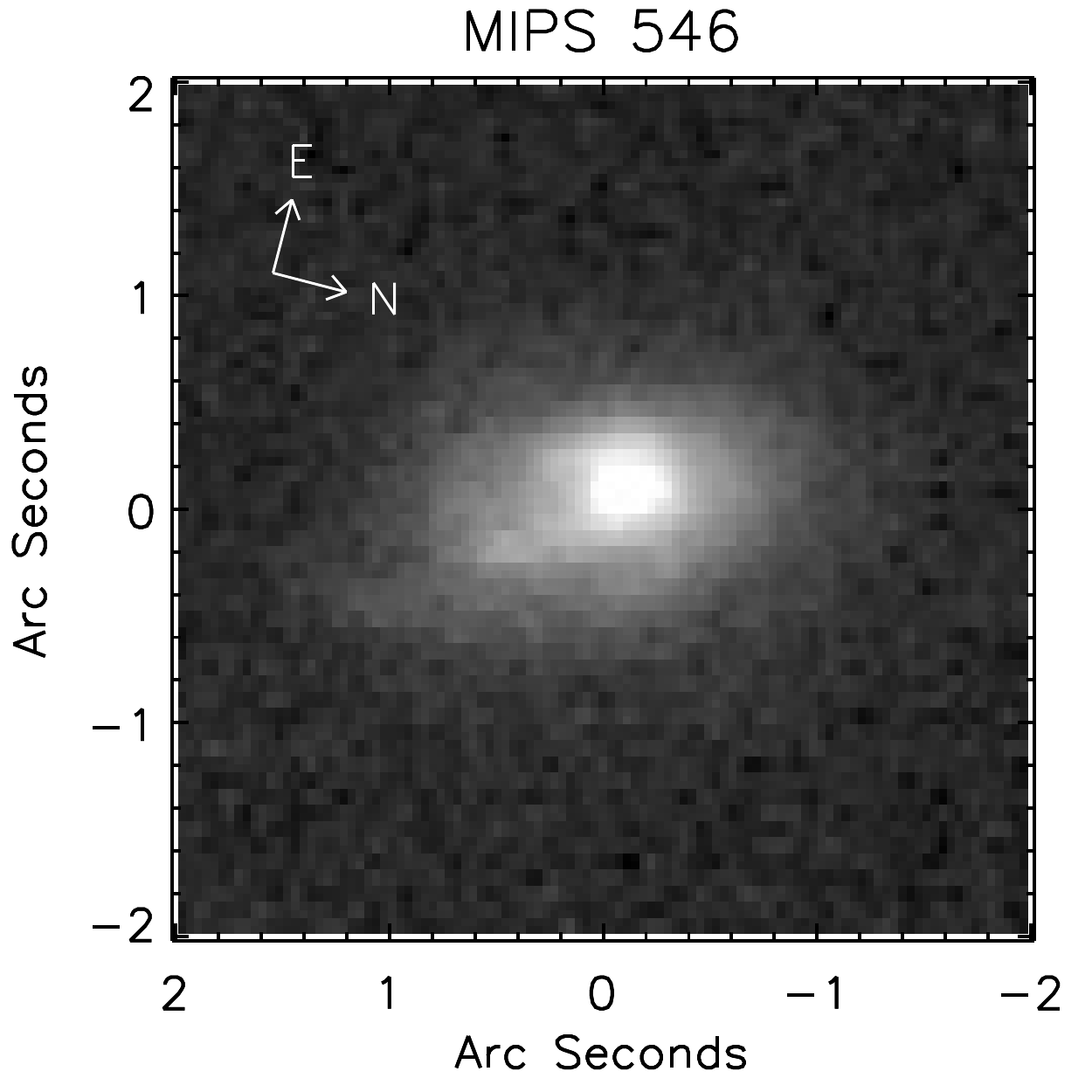}
\includegraphics[height=1.7in]{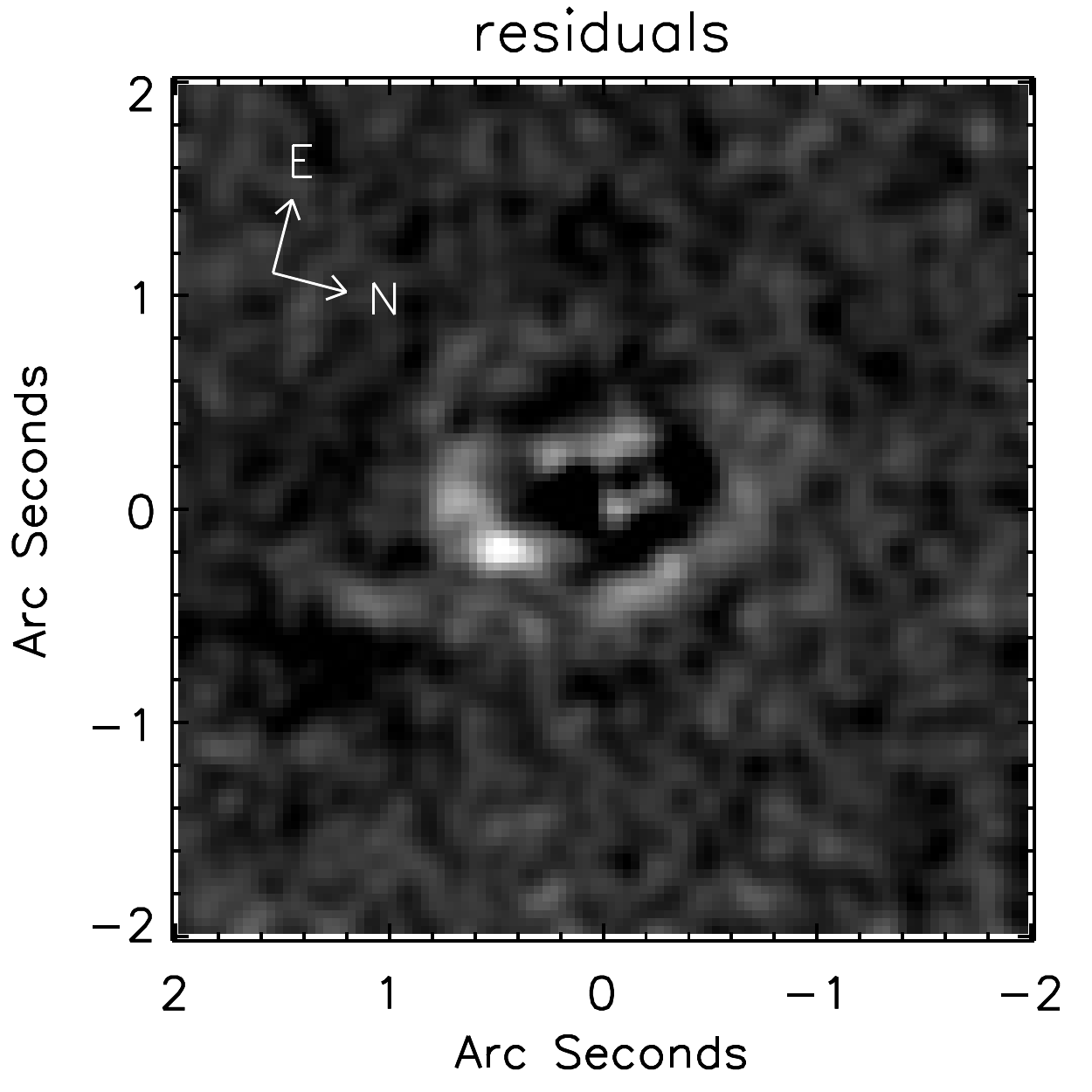}\\
\includegraphics[height=1.7in]{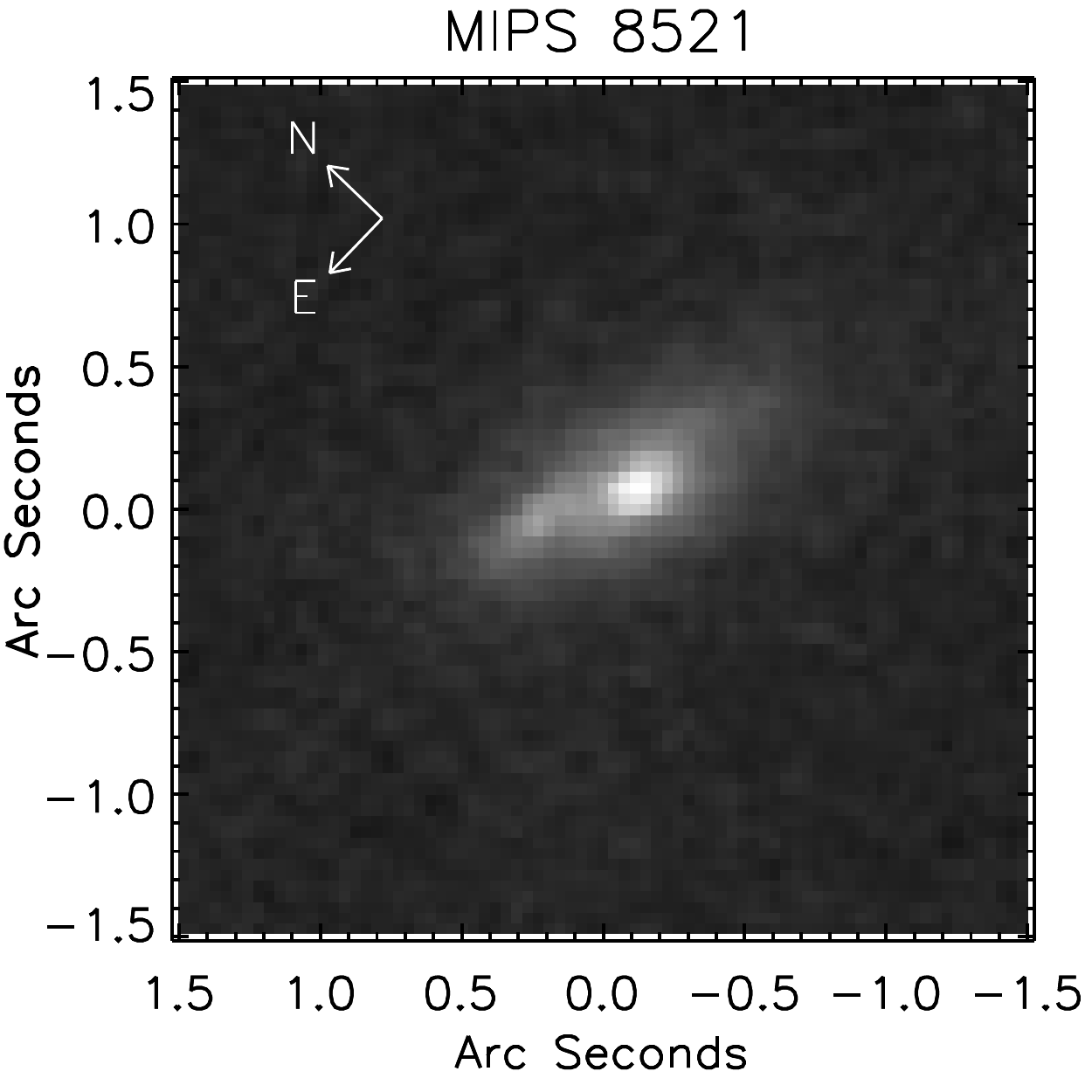}
\includegraphics[height=1.7in]{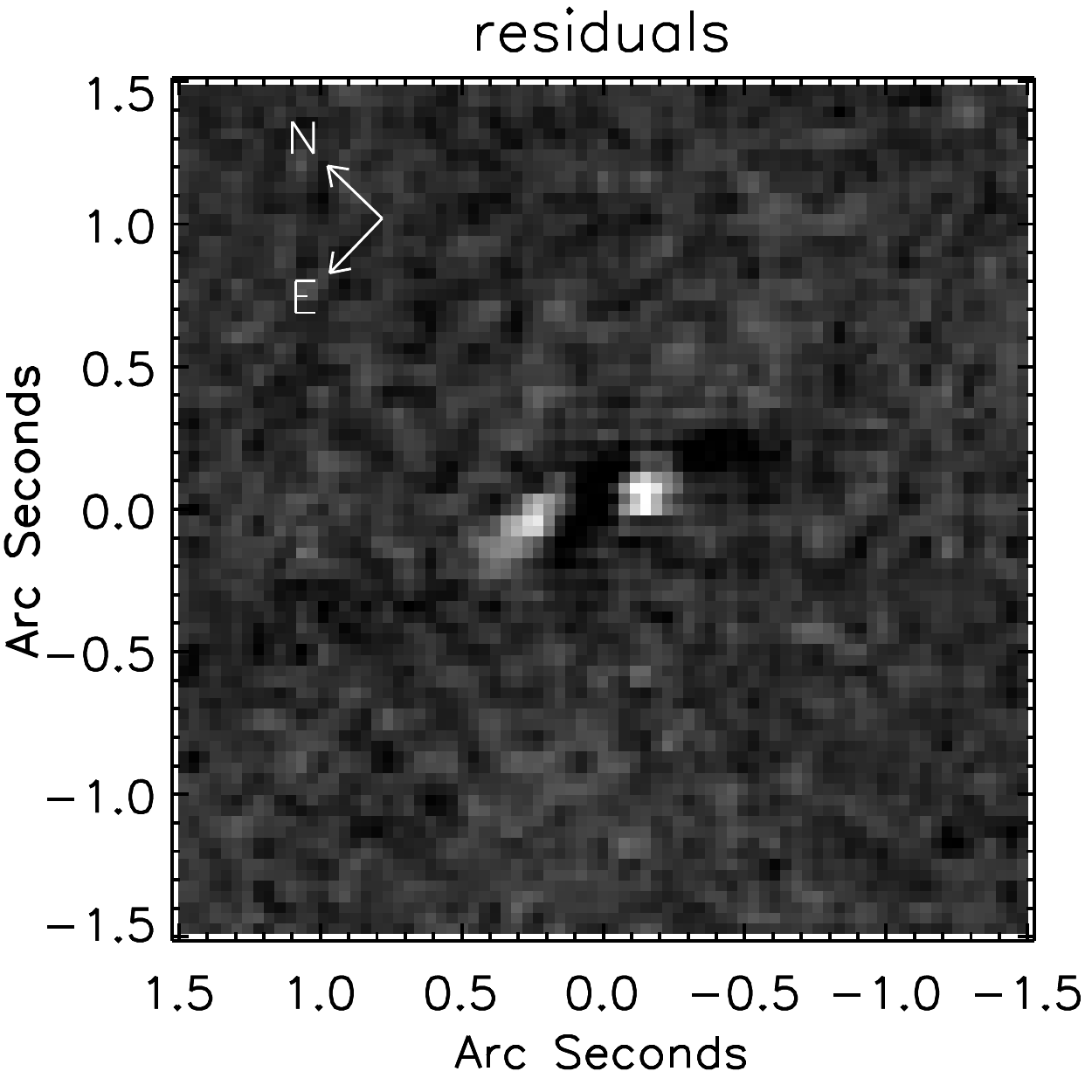}
\includegraphics[height=1.7in]{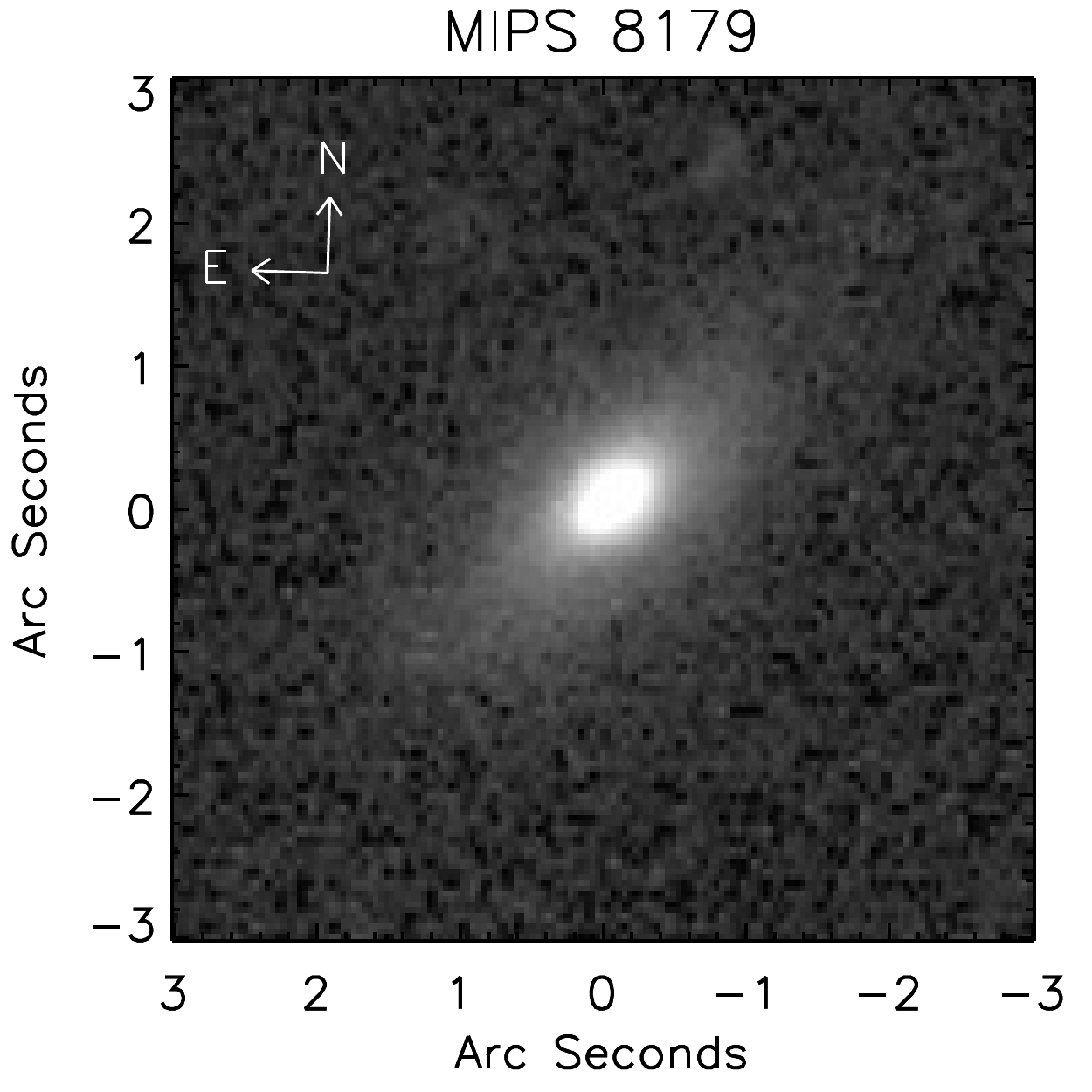}
\includegraphics[height=1.7in]{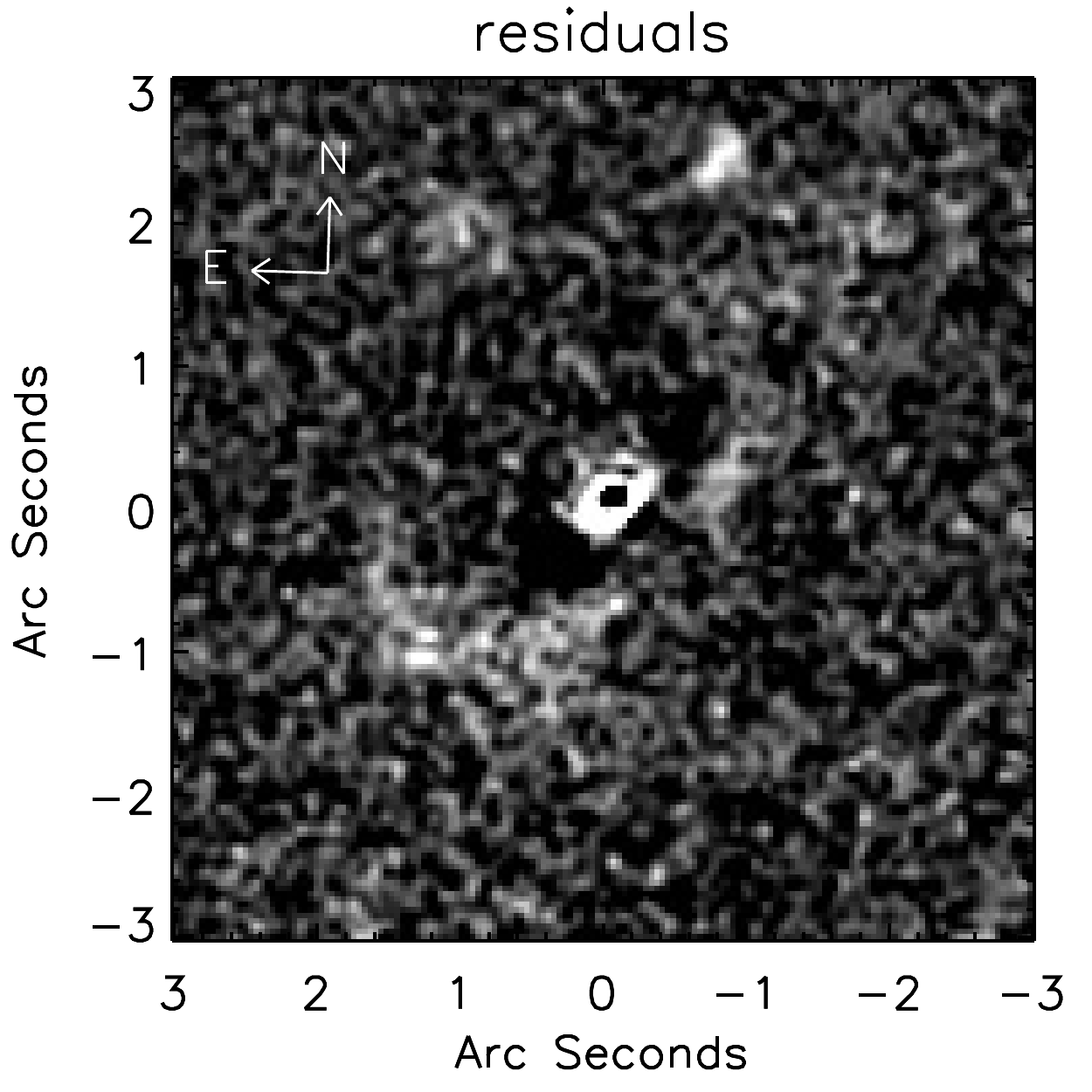}\\
\includegraphics[height=1.7in]{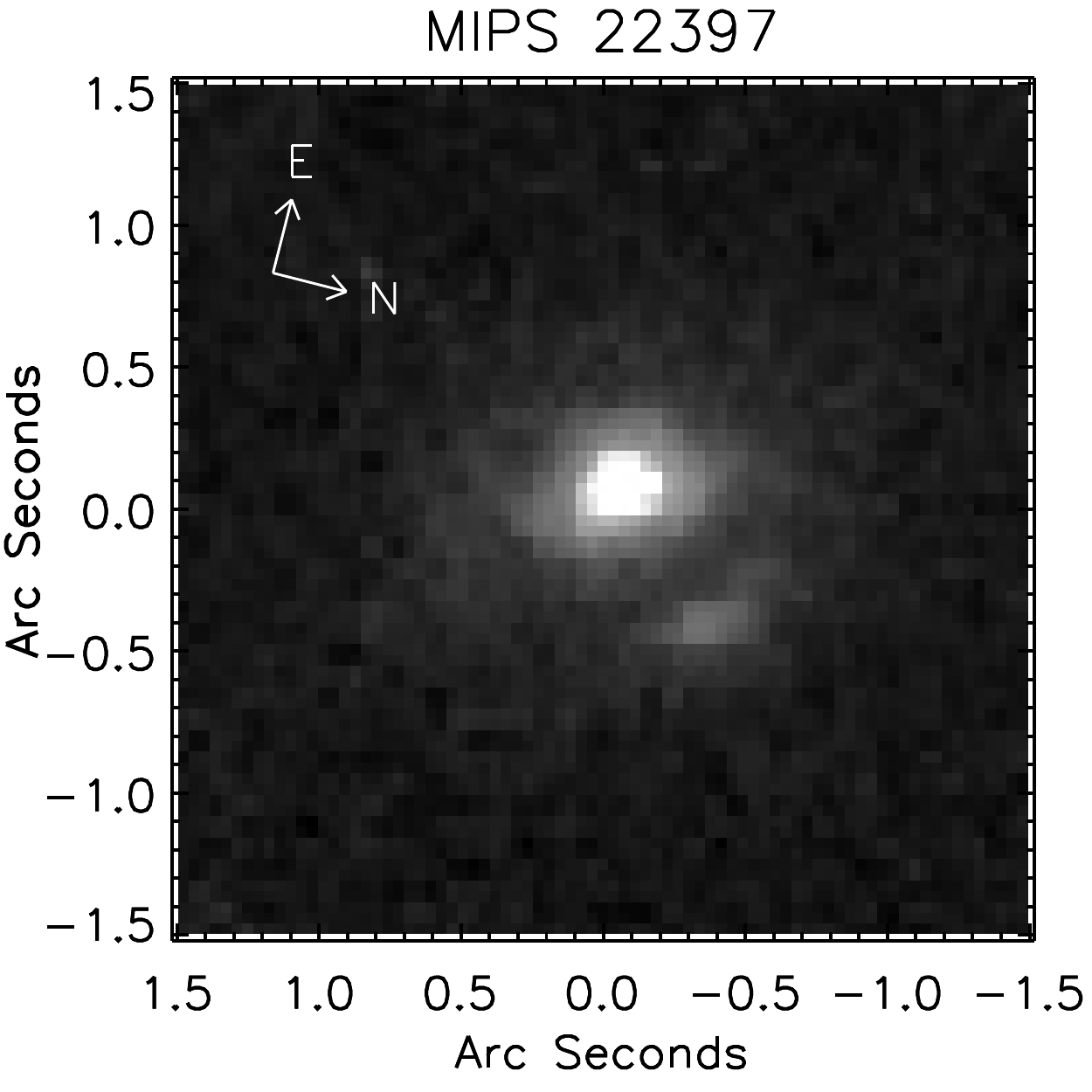}
\includegraphics[height=1.7in]{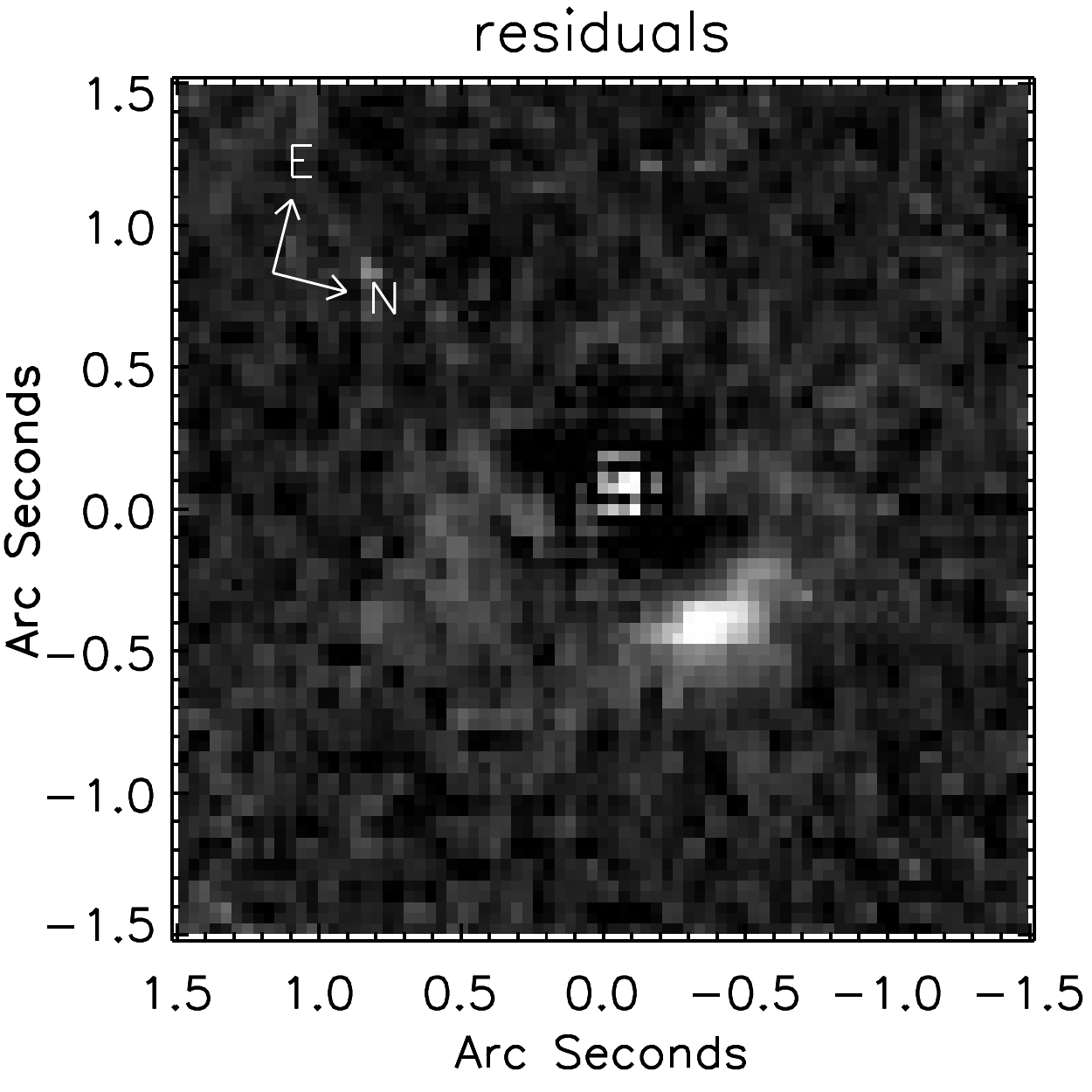}
\includegraphics[height=1.7in]{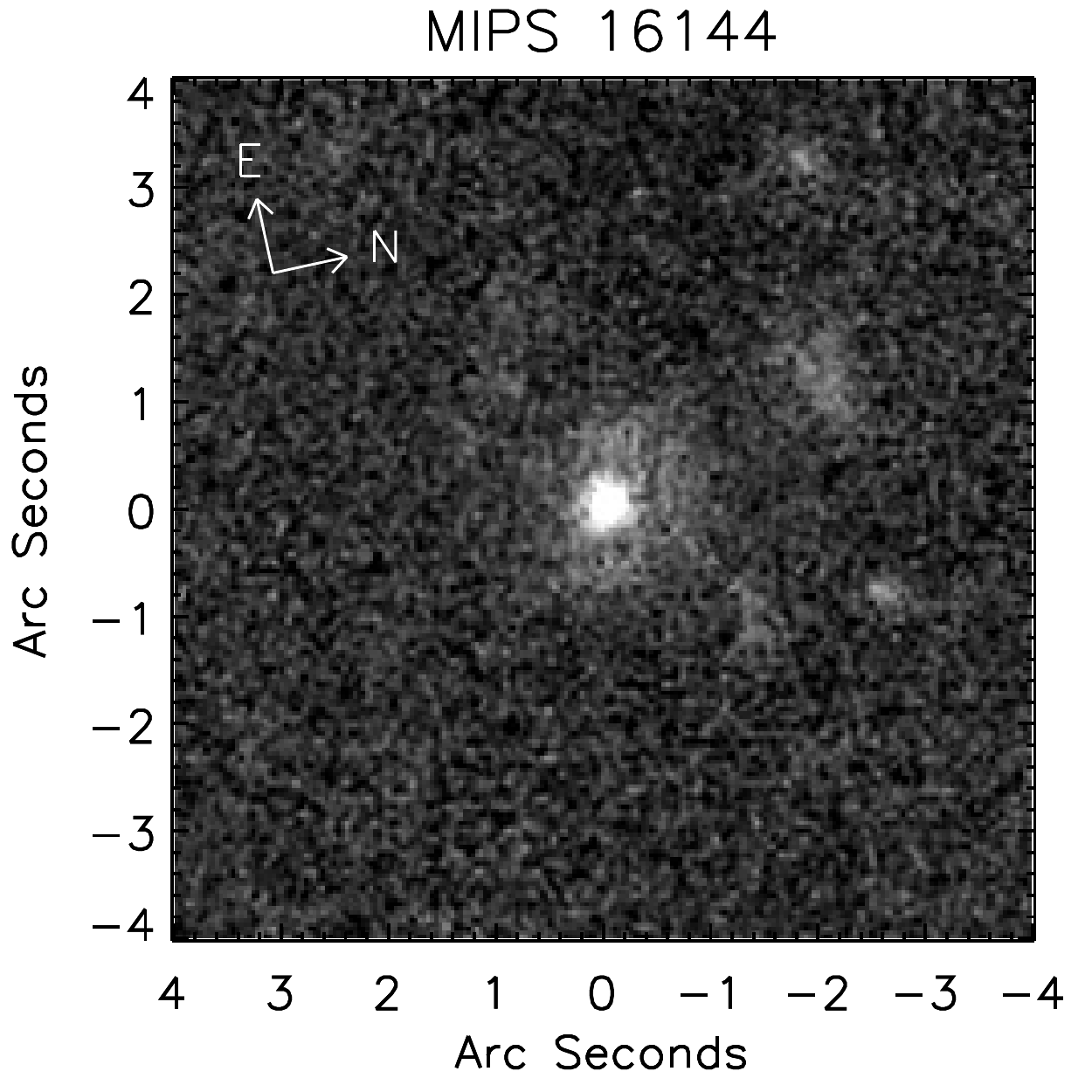}
\includegraphics[height=1.7in]{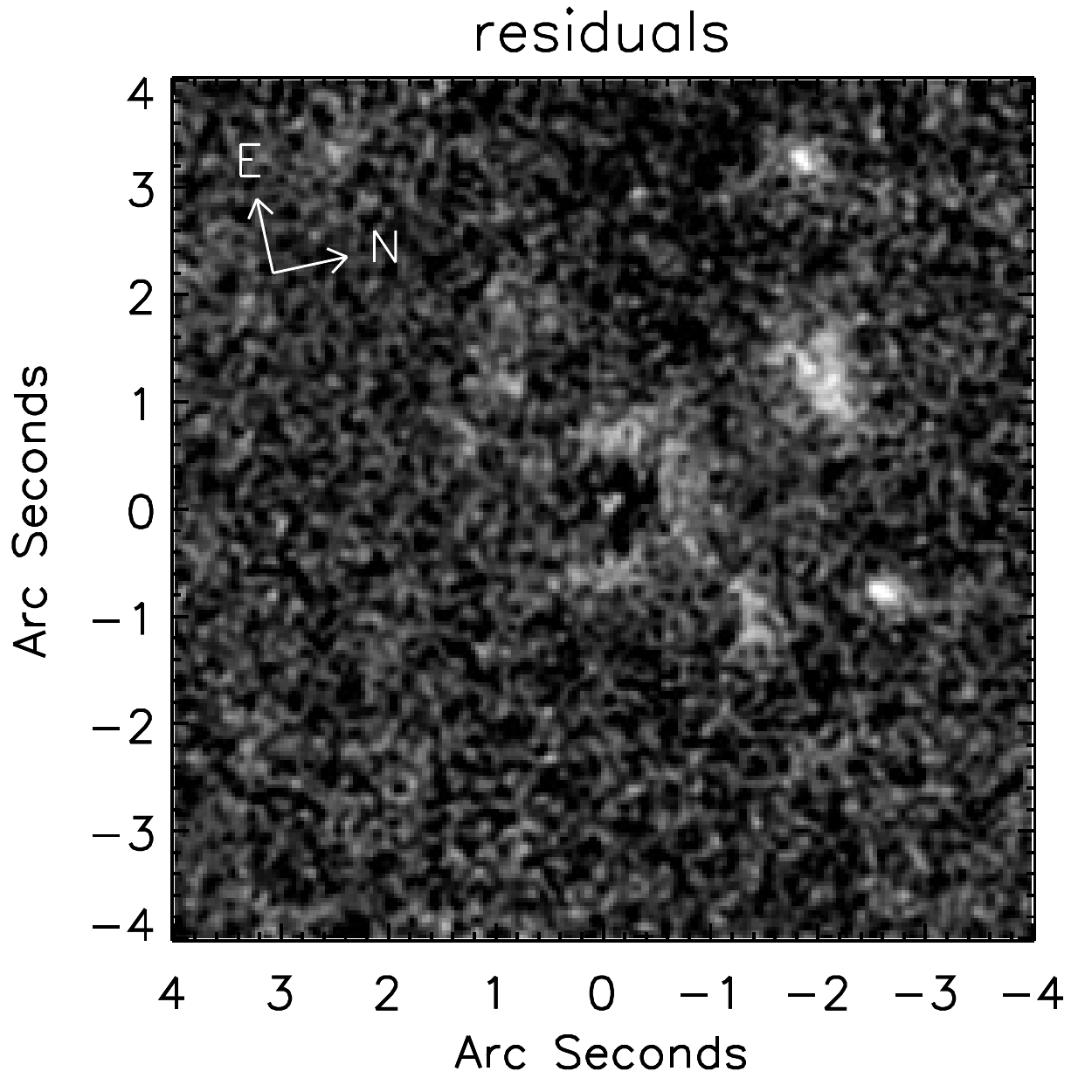}\\
\caption{Likely mergers (category 3).  Phase III objects in this category can have a higher luminosity ratio between their two components than their highly probable counterparts.  They can also be separated by a larger distance, and less obviously connected.  Phase IV objects possess tidal features characteristic of mergers, but they tend to be fainter than for those of the previous category, and can be sometimes incomplete.}
\label{fig:examples3}
\end{center}
\end{figure*}

\begin{figure*}[htbp]
\begin{center}
\makebox[3.4in]{\large ~~Early Mergers (Phase III)}
\makebox[3.4in]{\large ~~Faint \& compact}\\
\includegraphics[height=1.7in]{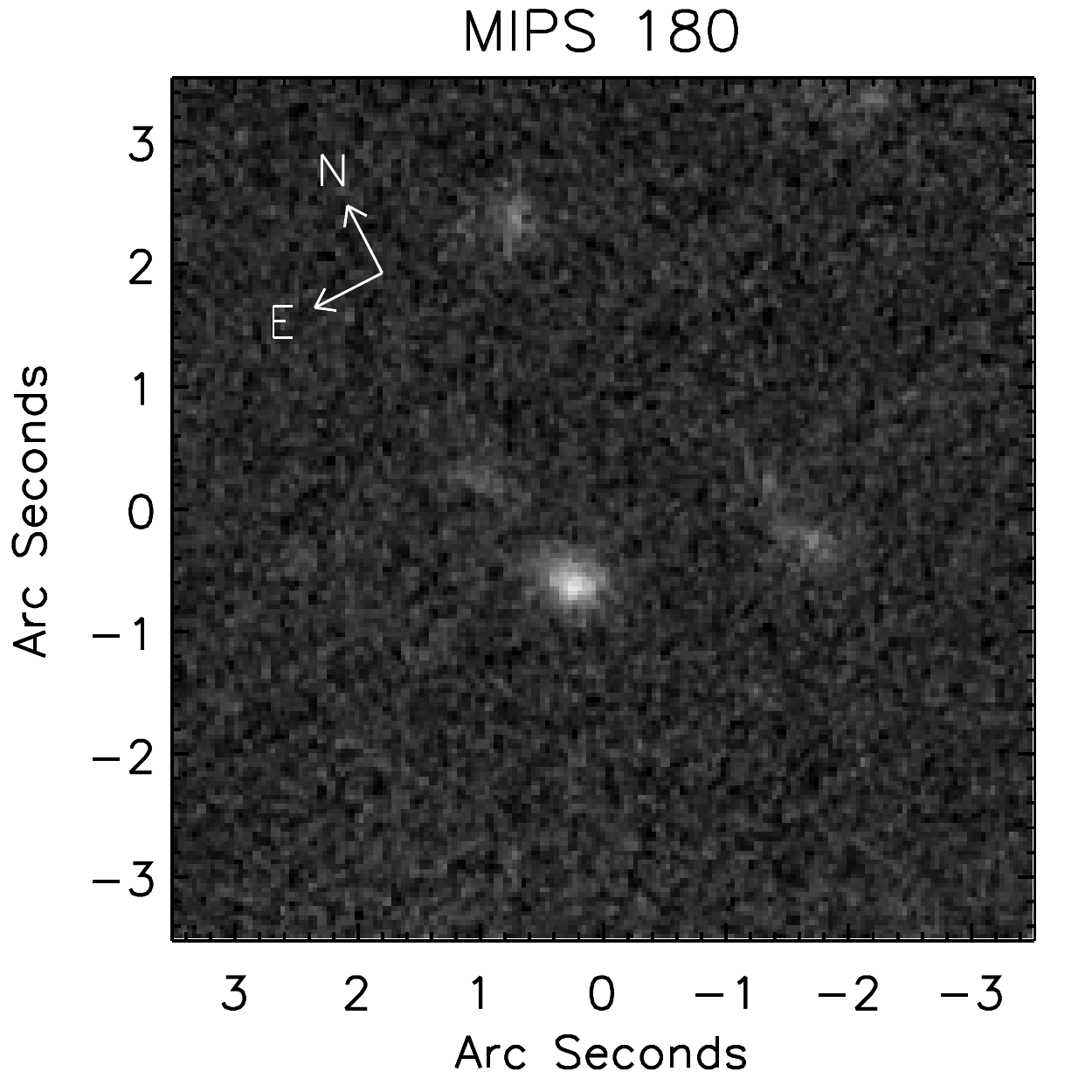}
\includegraphics[height=1.7in]{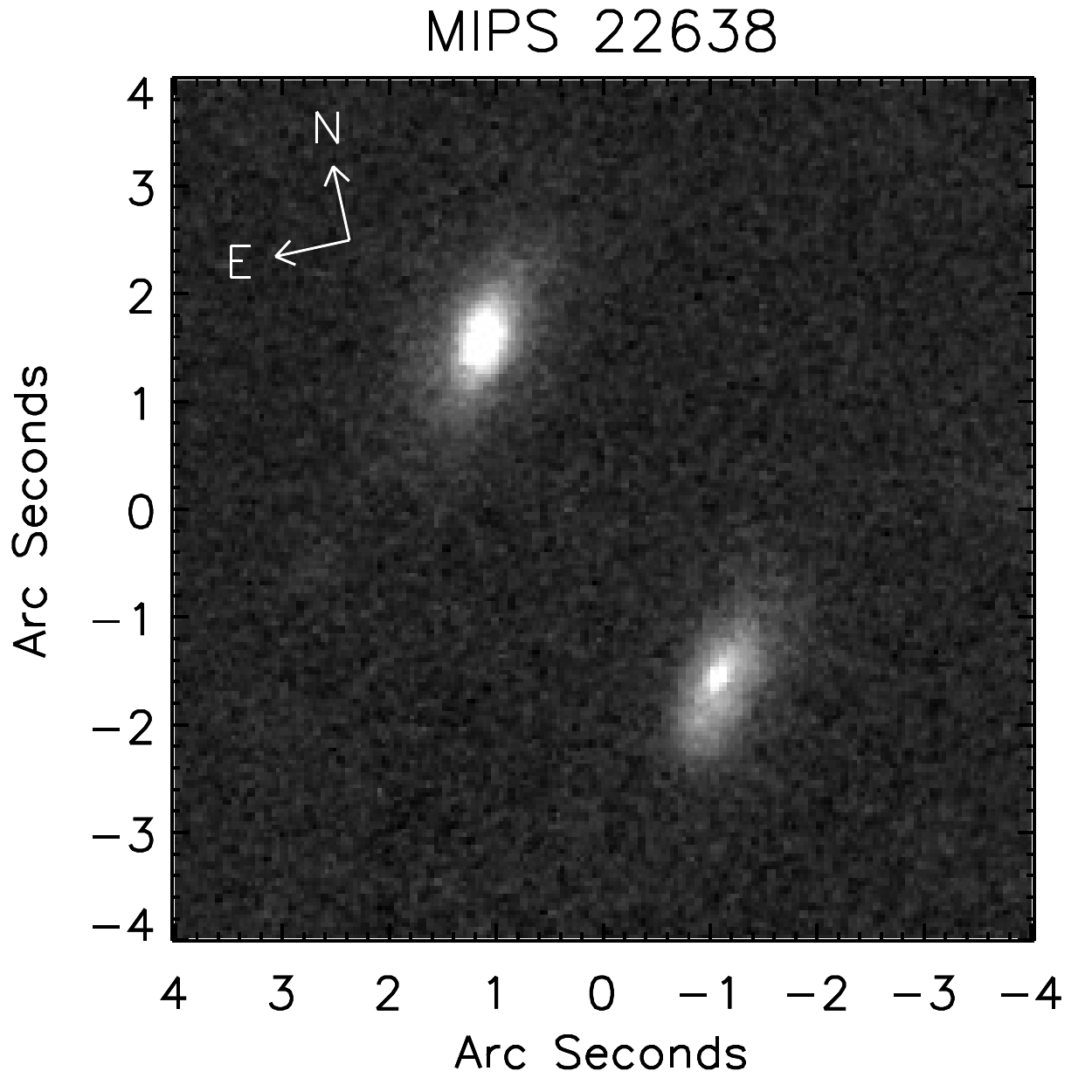}
\includegraphics[height=1.7in]{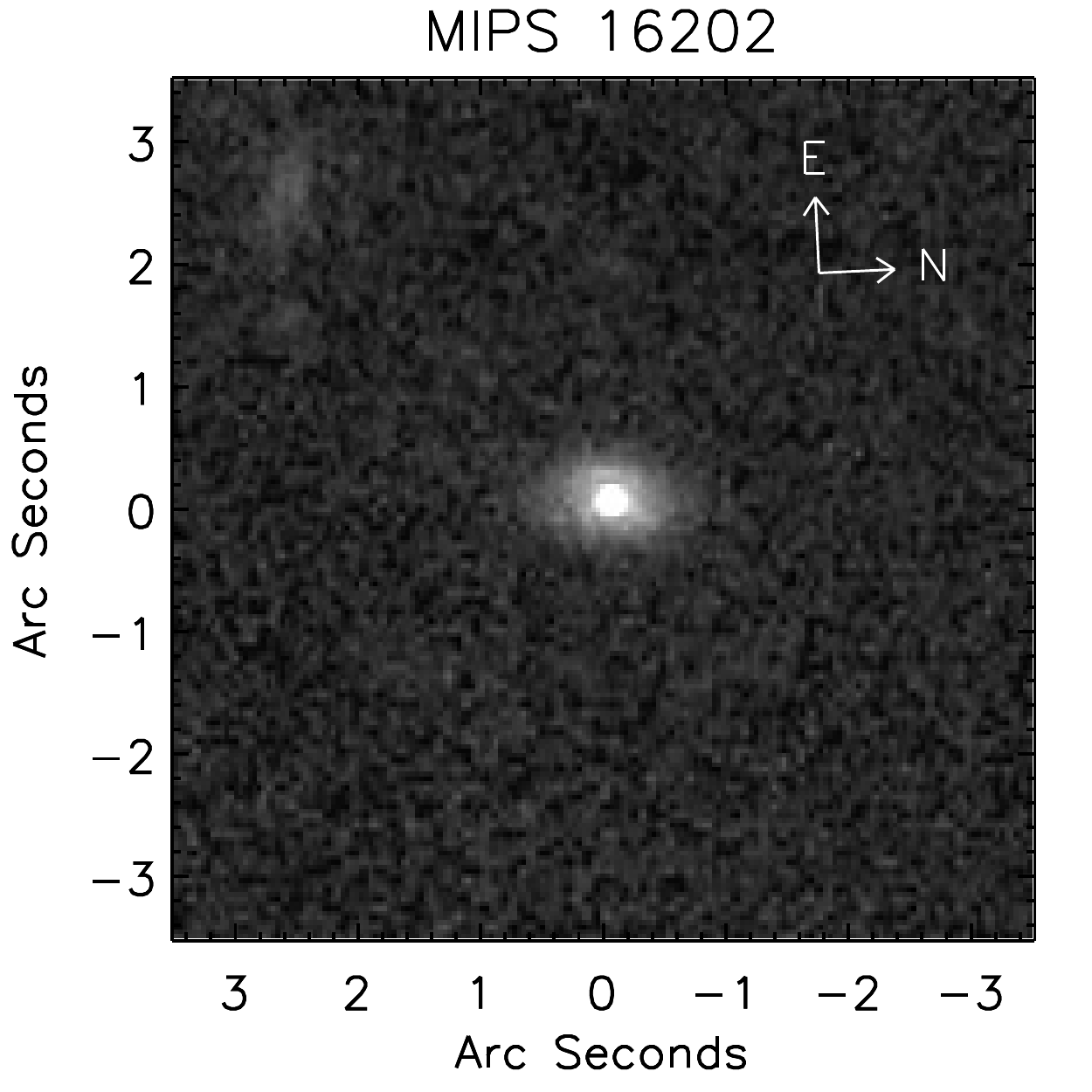}
\includegraphics[height=1.7in]{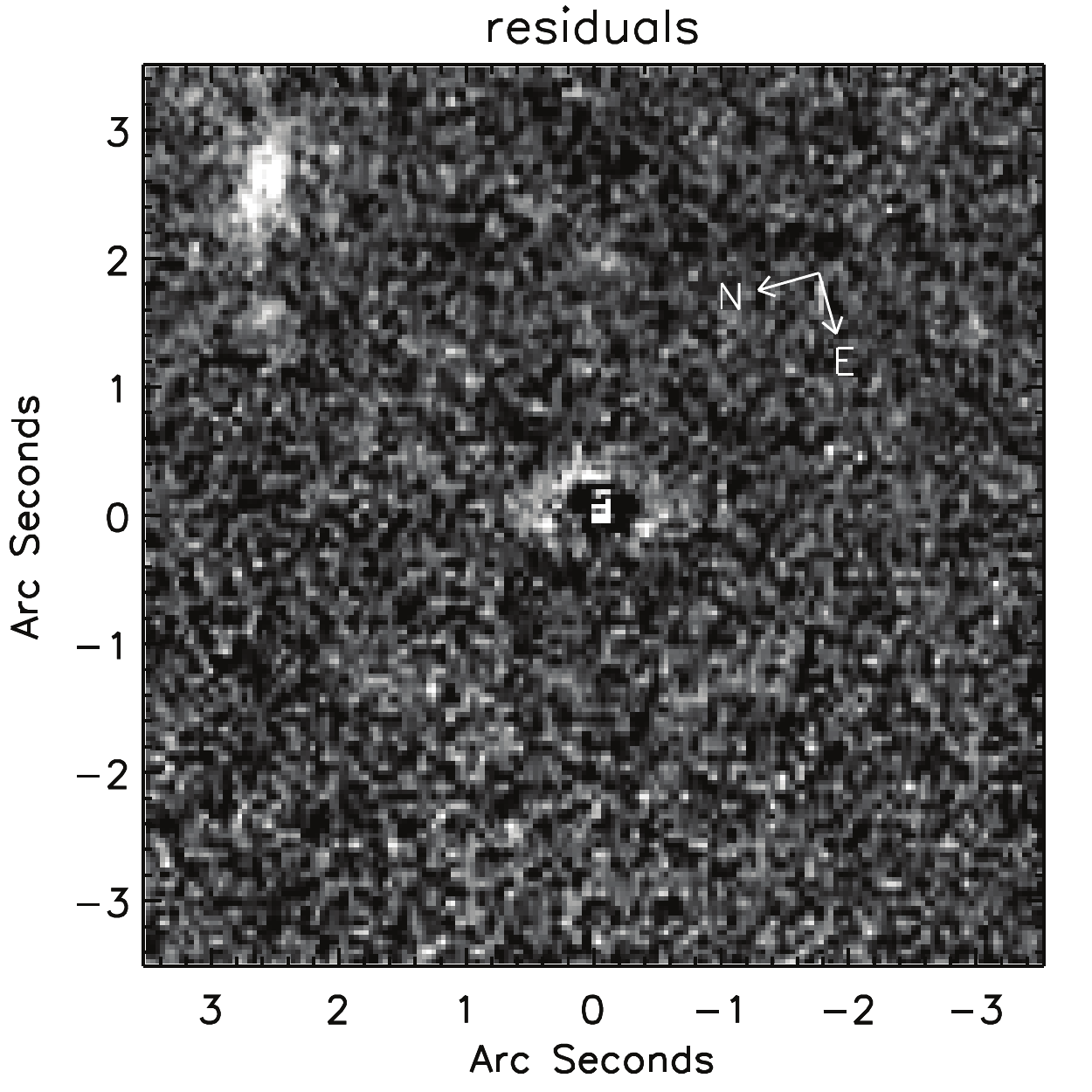}\\
\makebox[3.4in]{\large ~~Advanced Mergers (Phase IV)}
\makebox[3.4in]{\large ~~Old Mergers (Phase V)}\\
\includegraphics[height=1.7in]{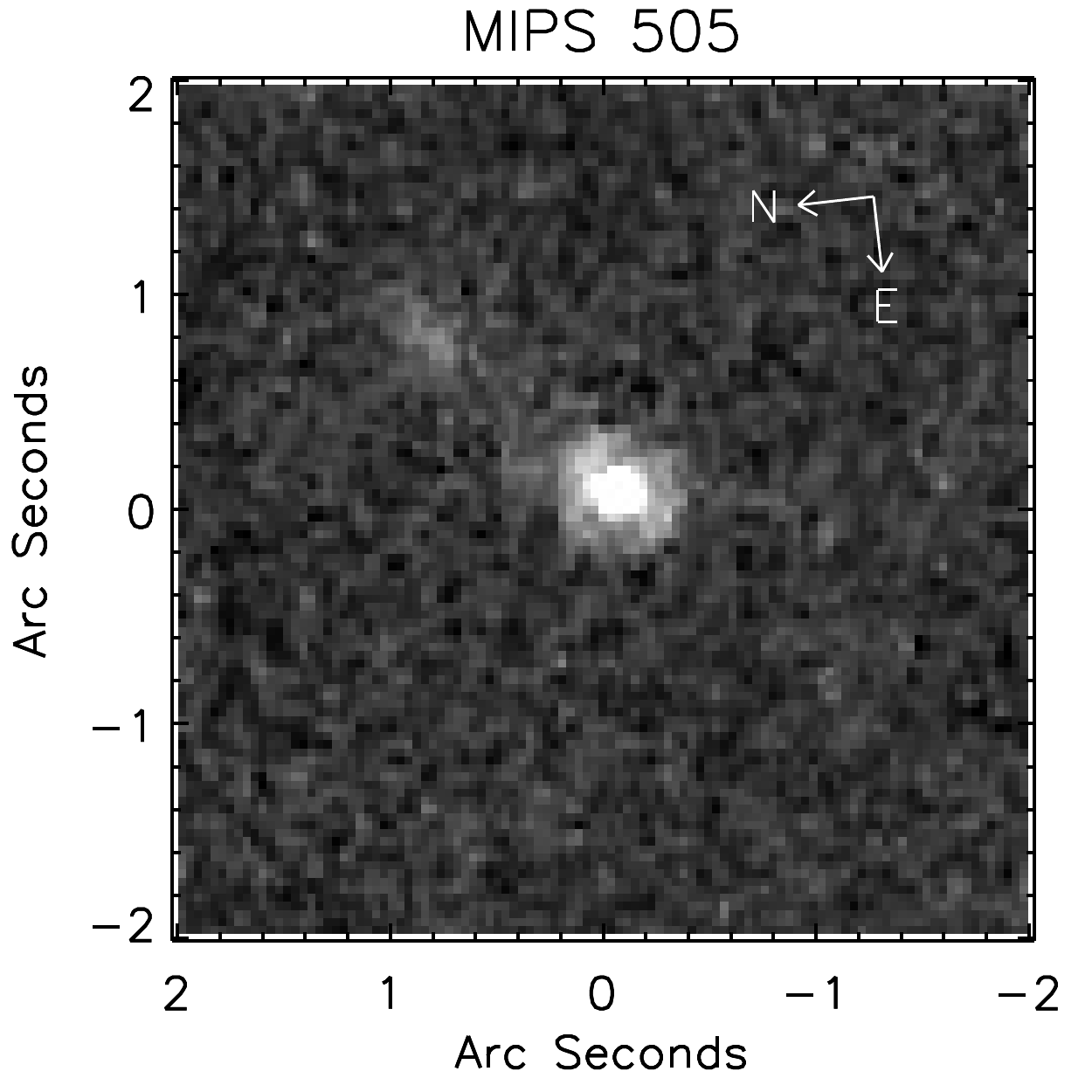}
\includegraphics[height=1.7in]{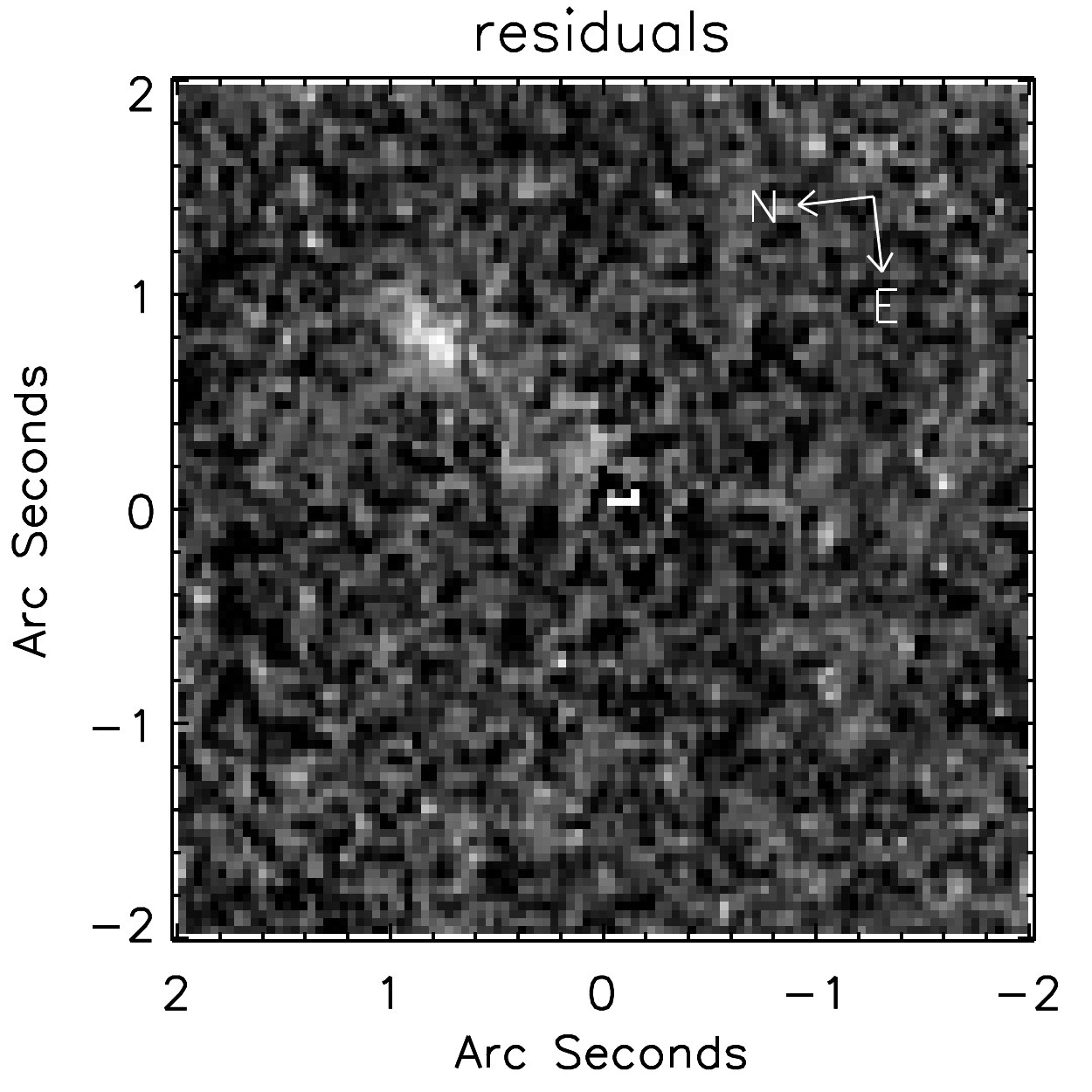}
\includegraphics[height=1.7in]{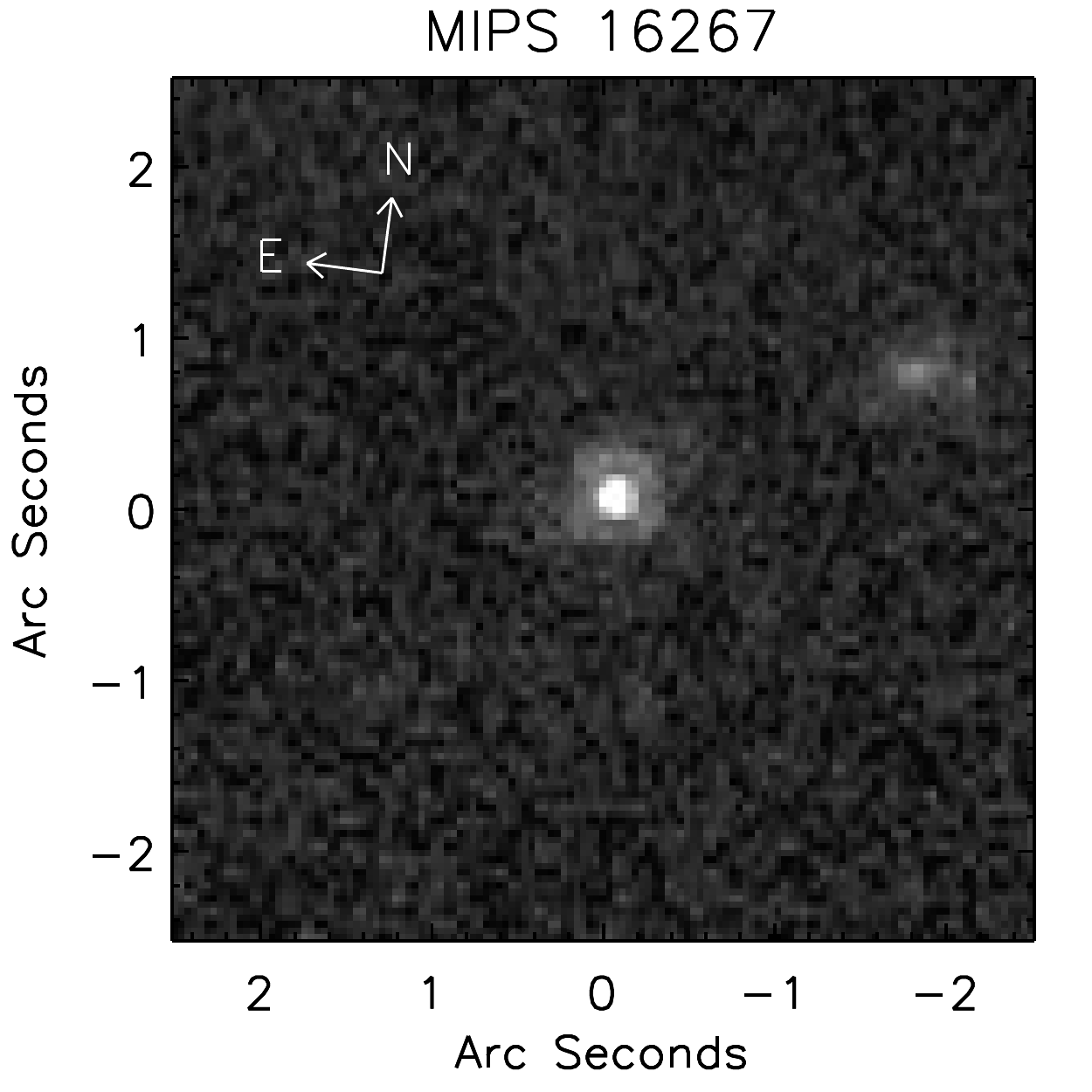}
\includegraphics[height=1.7in]{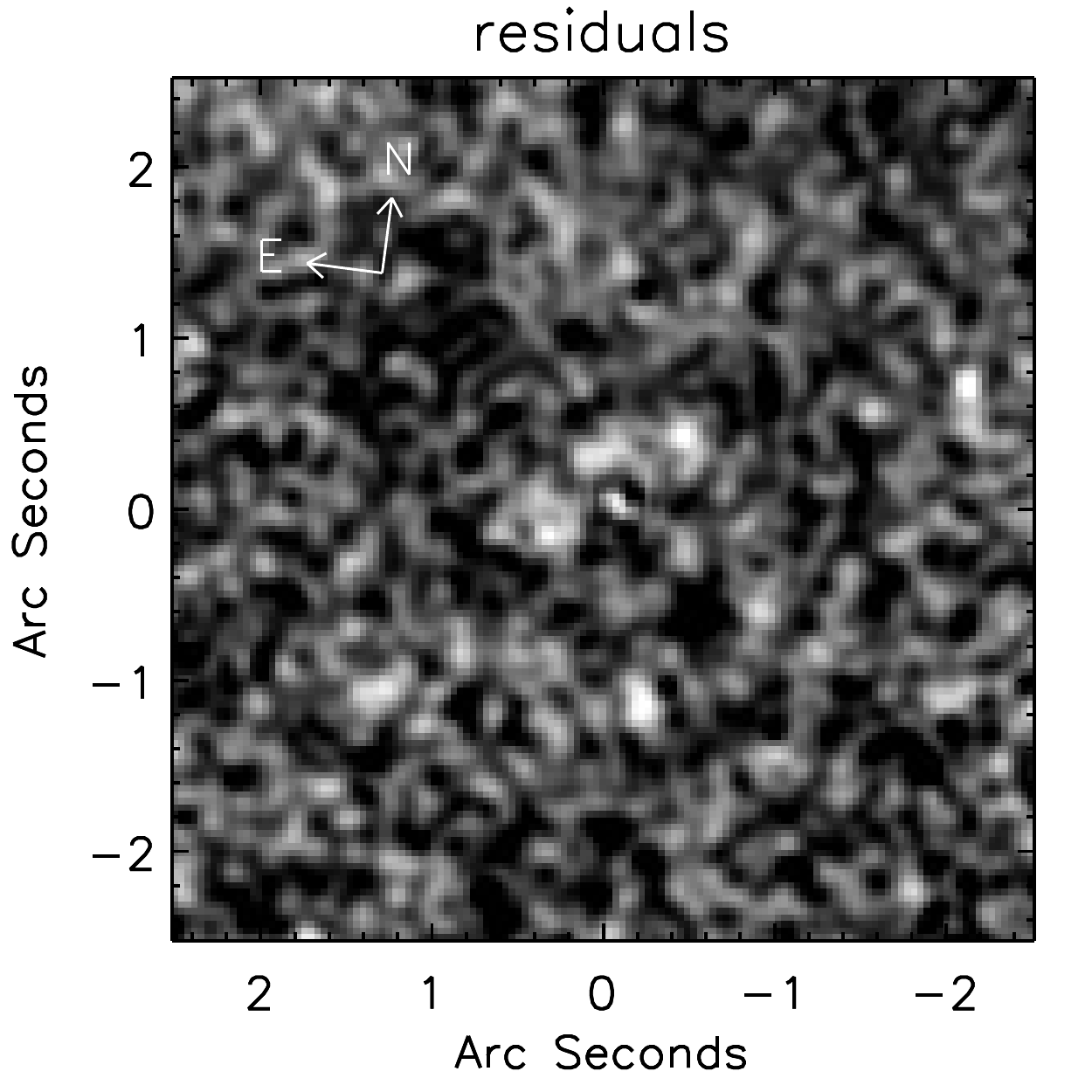}\\
\includegraphics[height=1.7in]{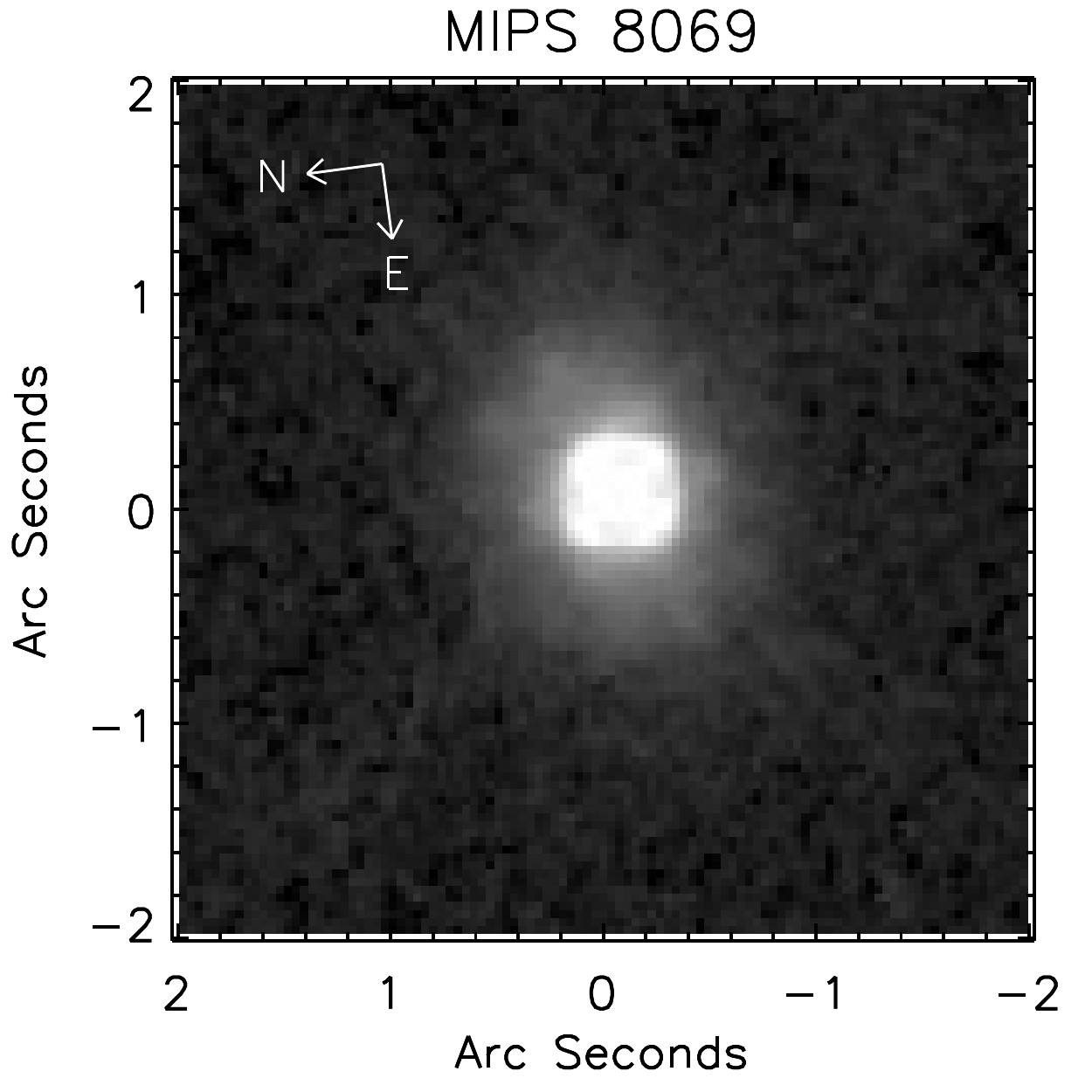}
\includegraphics[height=1.7in]{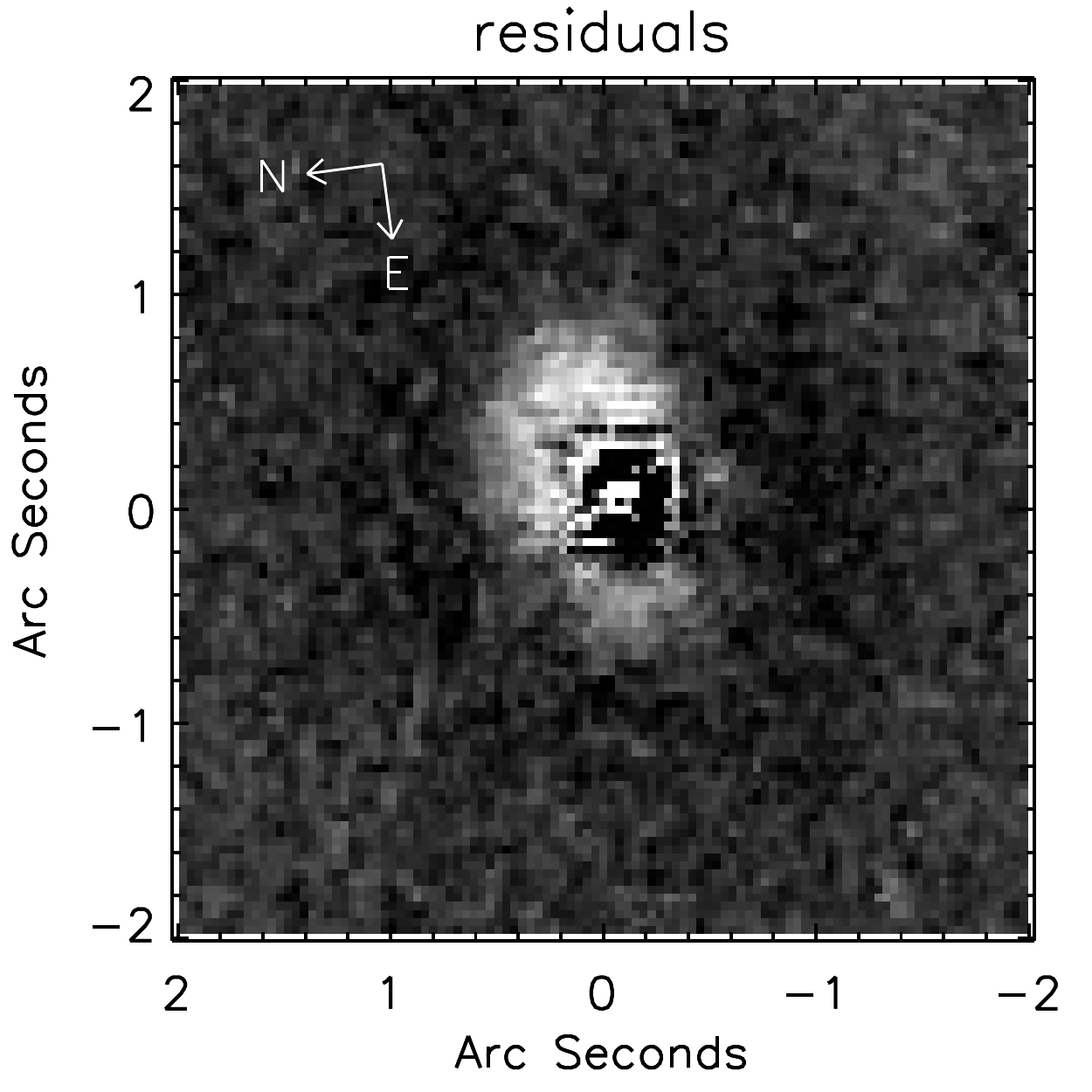}
\includegraphics[height=1.7in]{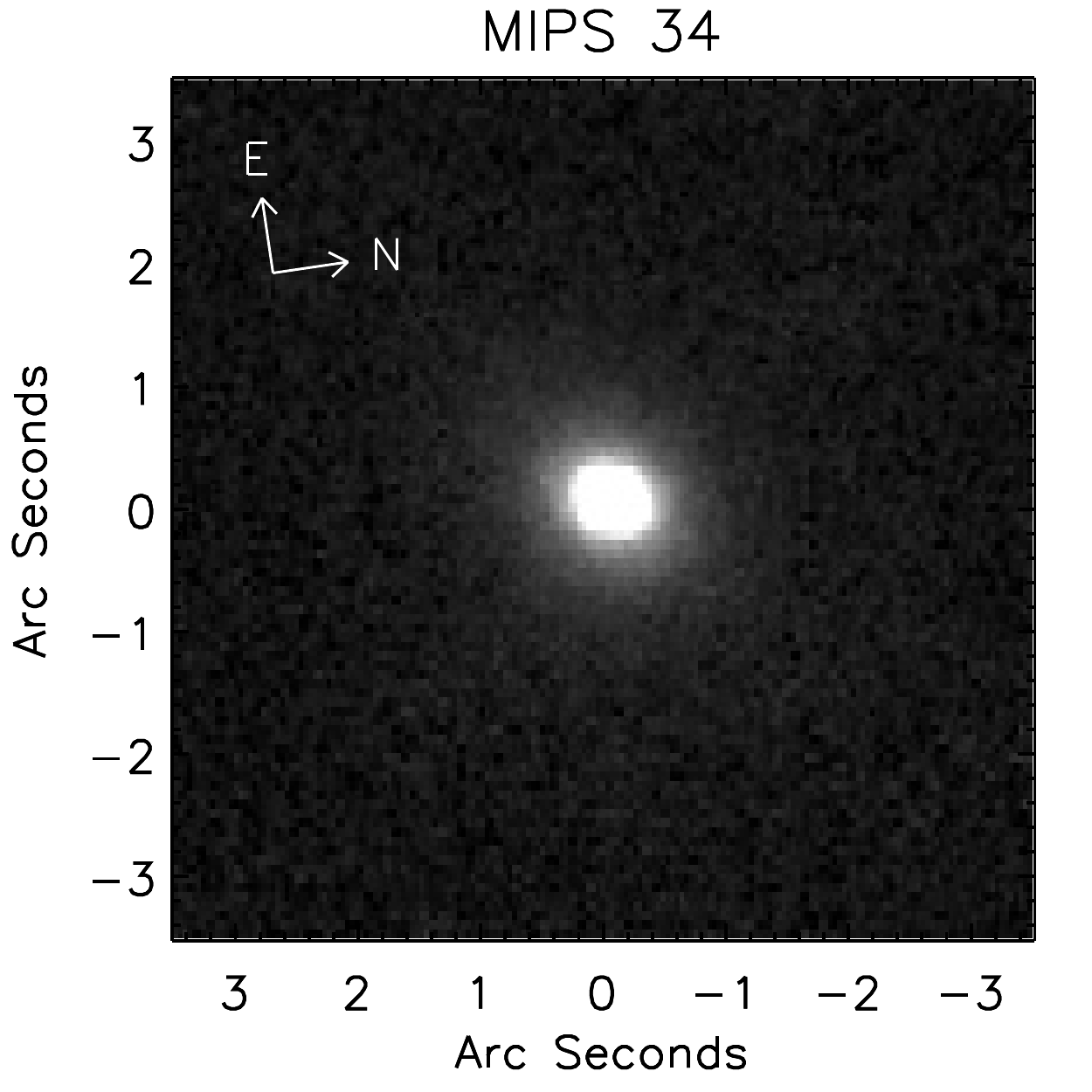}
\includegraphics[height=1.7in]{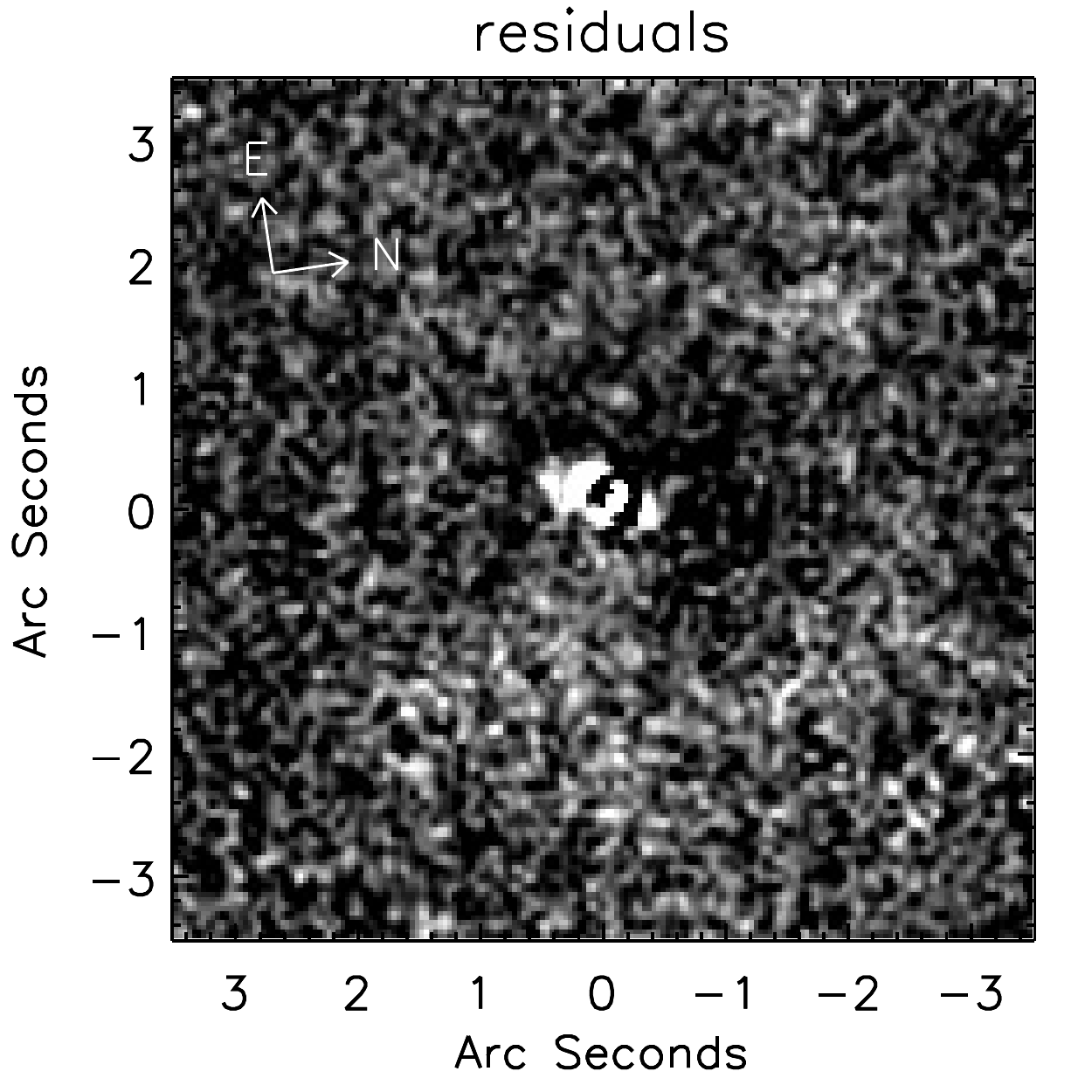}\\
\caption{Probable mergers (category 4).  Phase III objects of this category have a greater distant between their two components than likely mergers do.  The objects can also be faint, and their bridges incomplete.  Phase IV objects of this category have clear disturbances, but the shape of their features is not necessarily unique to the merging process.  Phase V objects have very weak residuals, but these appear to form tidal features typical of mergers.  Faint and compact objects have some of both.}
\label{fig:examples4}
\end{center}
\end{figure*}

Categories~3 and~4, illustrated in Figures~\ref{fig:examples3} and~\ref{fig:examples4} respectively, include, on the other hand, 35 and 33 objects each.  These objects also show merger signatures, although these progressively go down in the strength and/or shape of tidal features, the distance and/or contrast ratio between the primary object and the companion, and in the overall connectivity of the system.  The merger origin of these objects is therefore increasingly more uncertain.  It is nevertheless clear that they must contribute an important number of mergers, although we postpone more specific estimates until section~\ref{sec:mergfrac}. 

\begin{figure*}[thbp]
\begin{center}
\makebox[3.4in]{\large ~~Faint \& compact}
\makebox[3.4in]{\large ~~Phase V objects}\\
\includegraphics[height=1.7in]{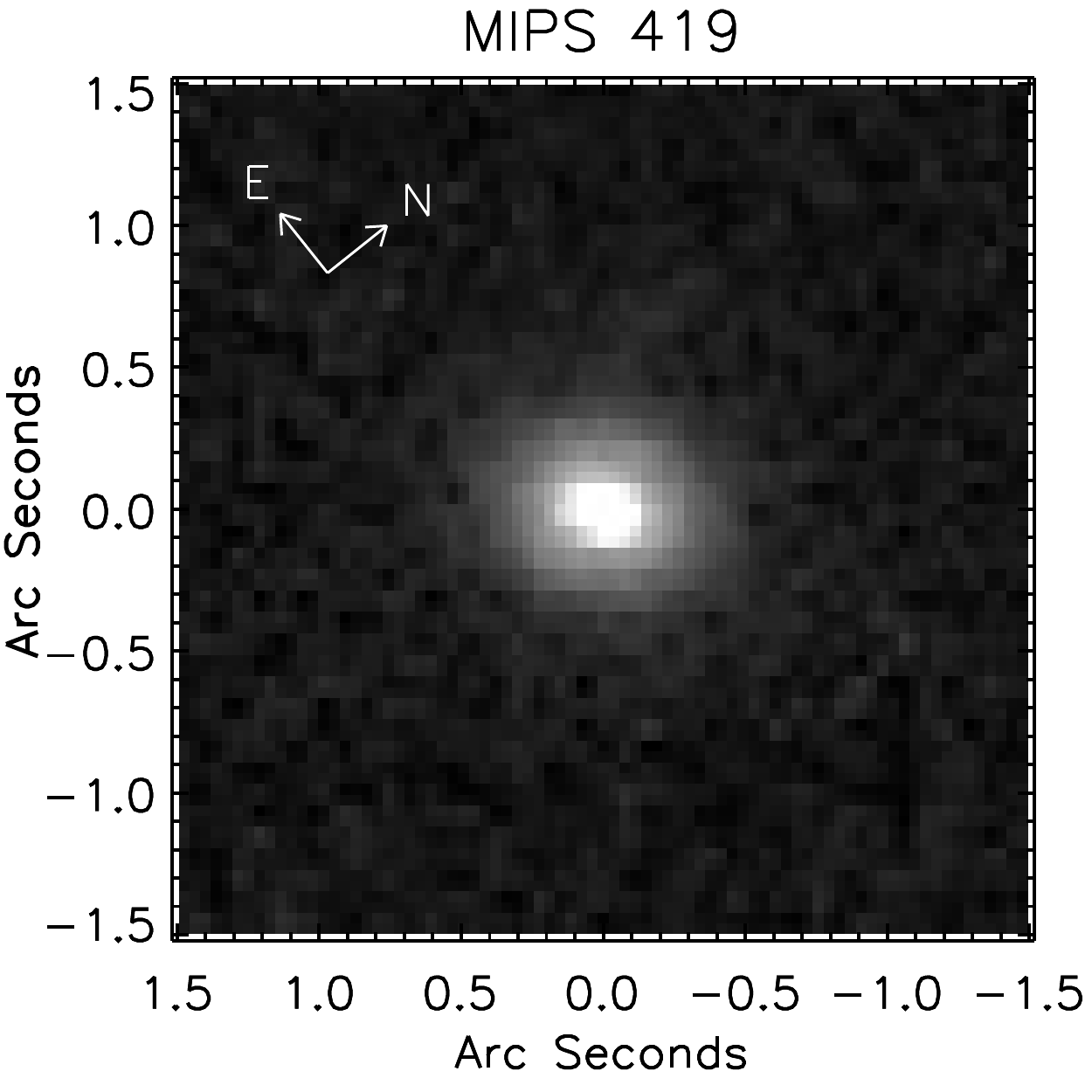}
\includegraphics[height=1.7in]{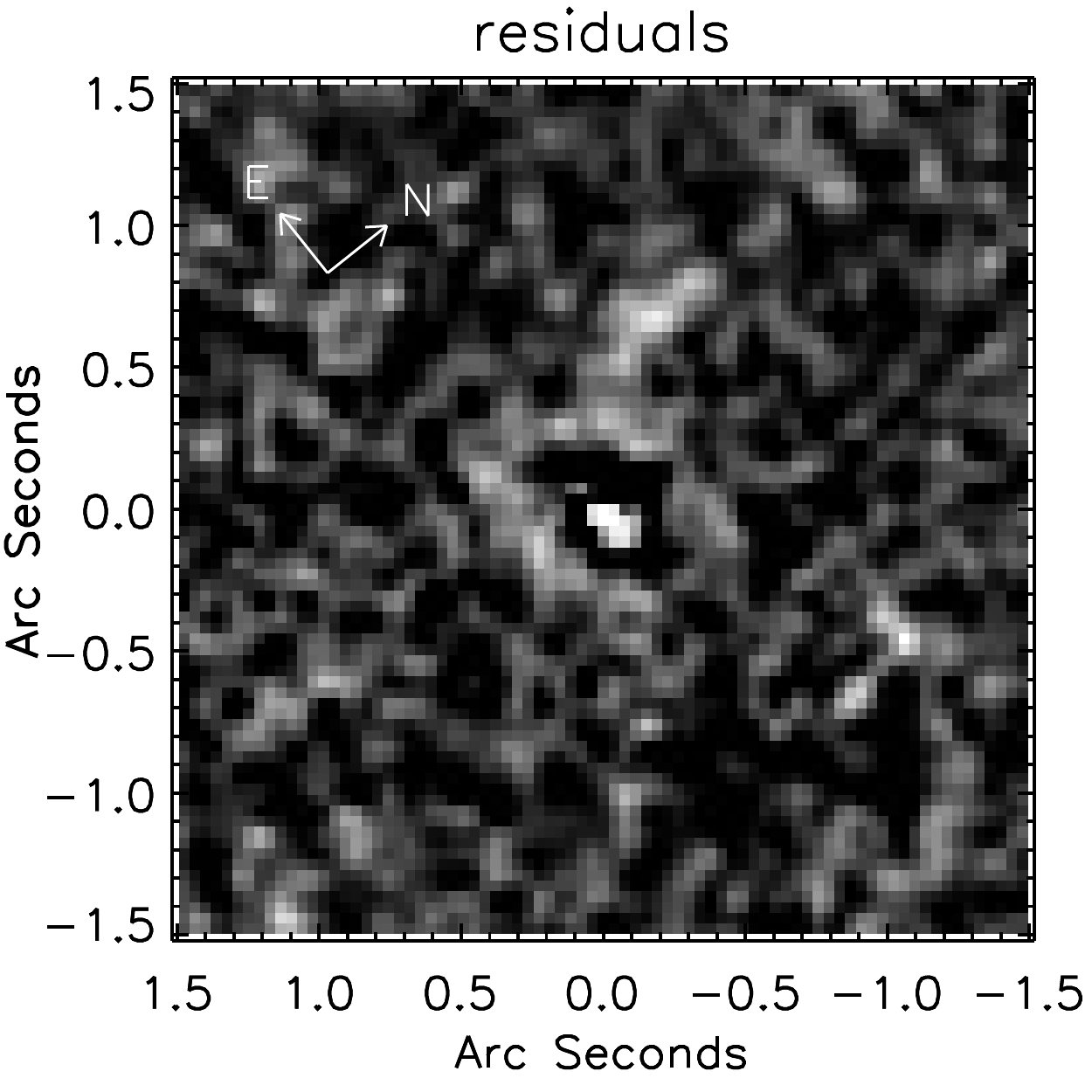}
\includegraphics[height=1.7in]{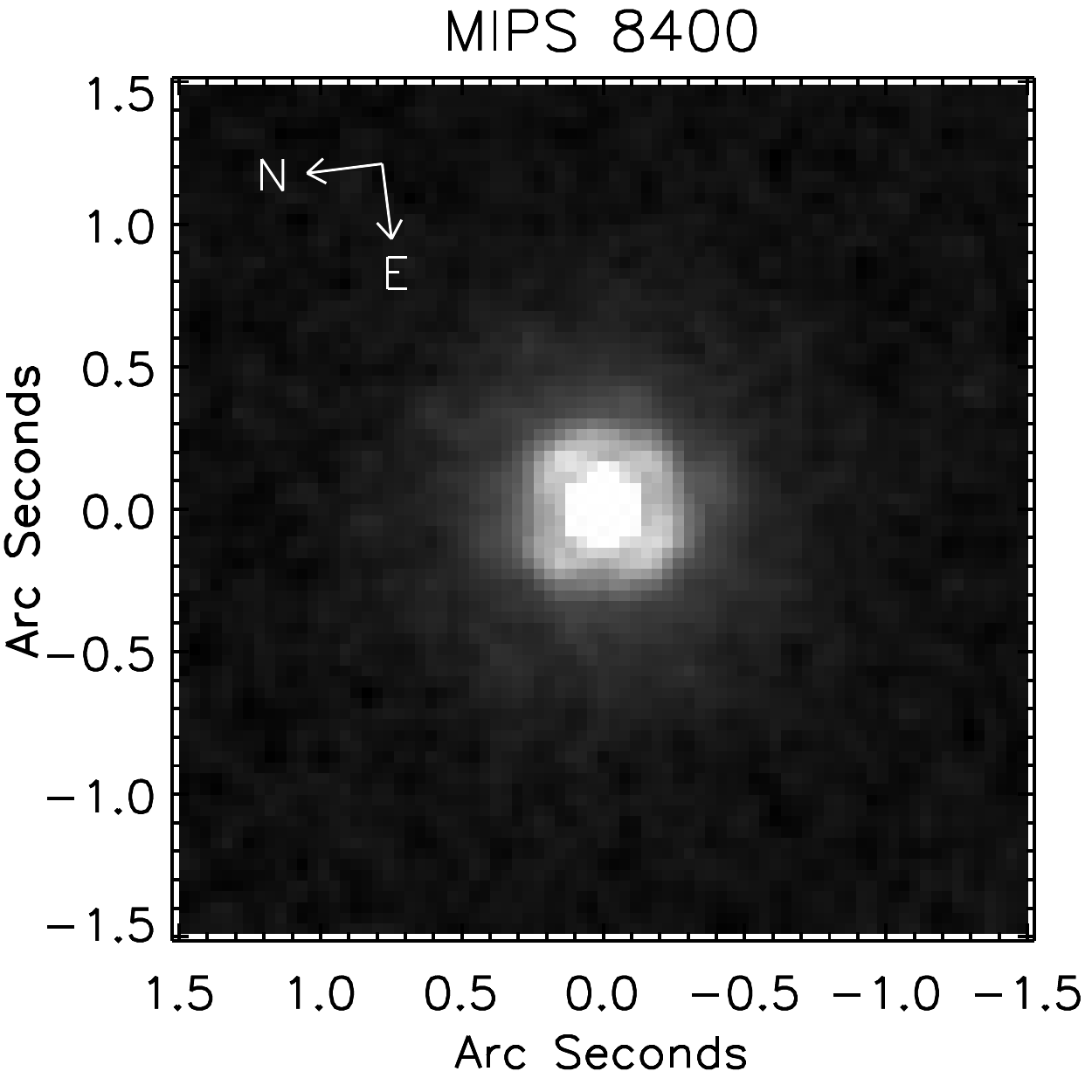}
\includegraphics[height=1.7in]{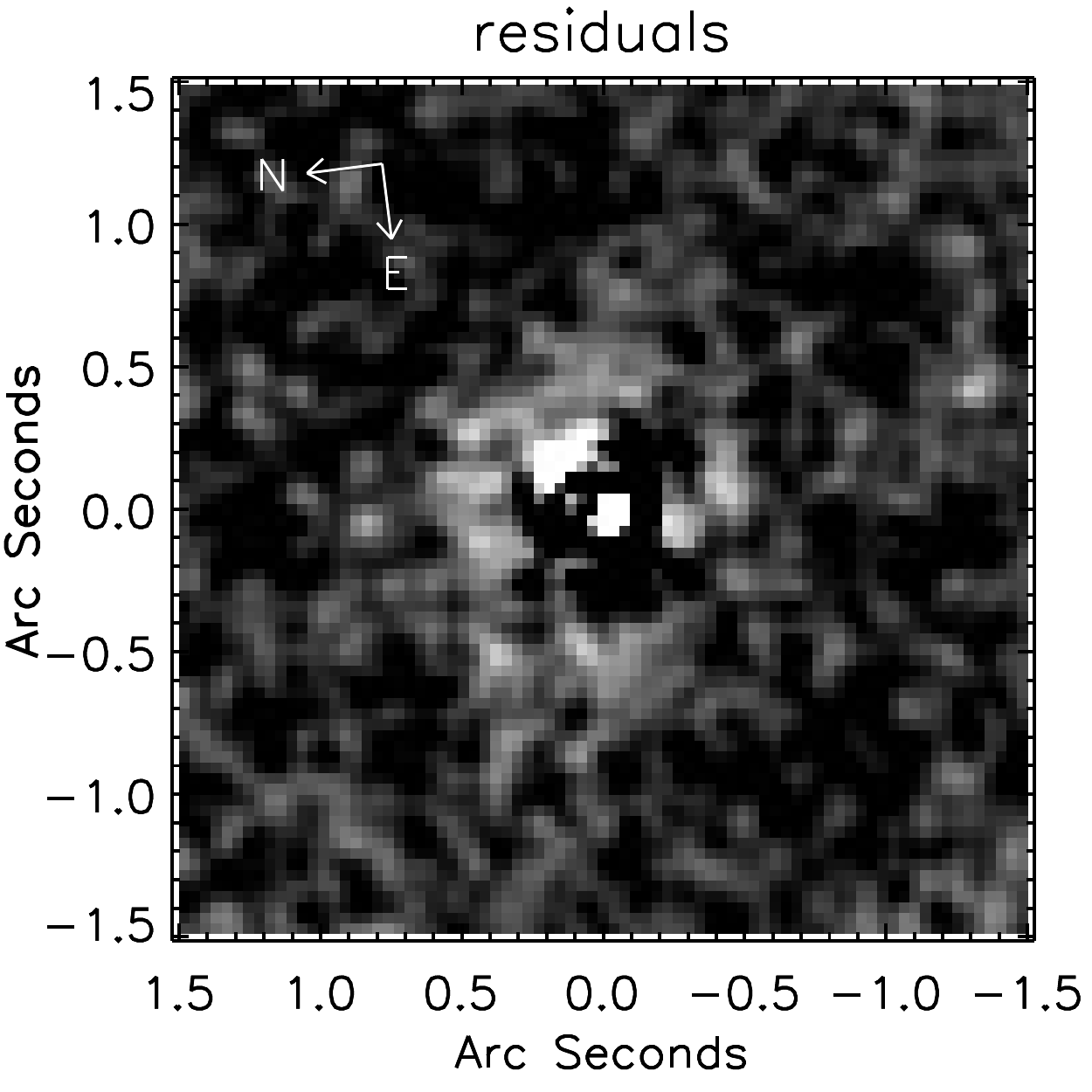}\\
\includegraphics[height=1.7in]{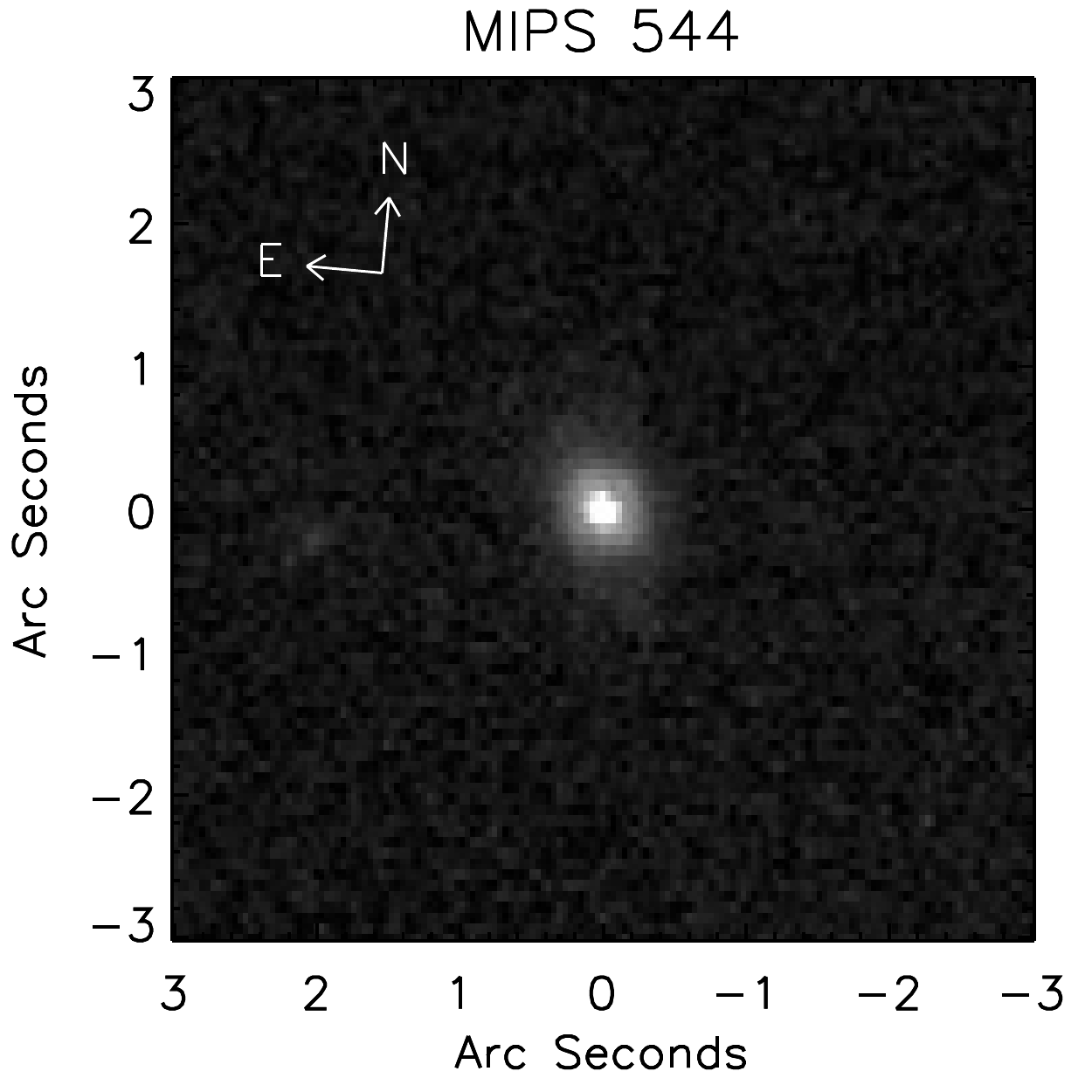}
\includegraphics[height=1.7in]{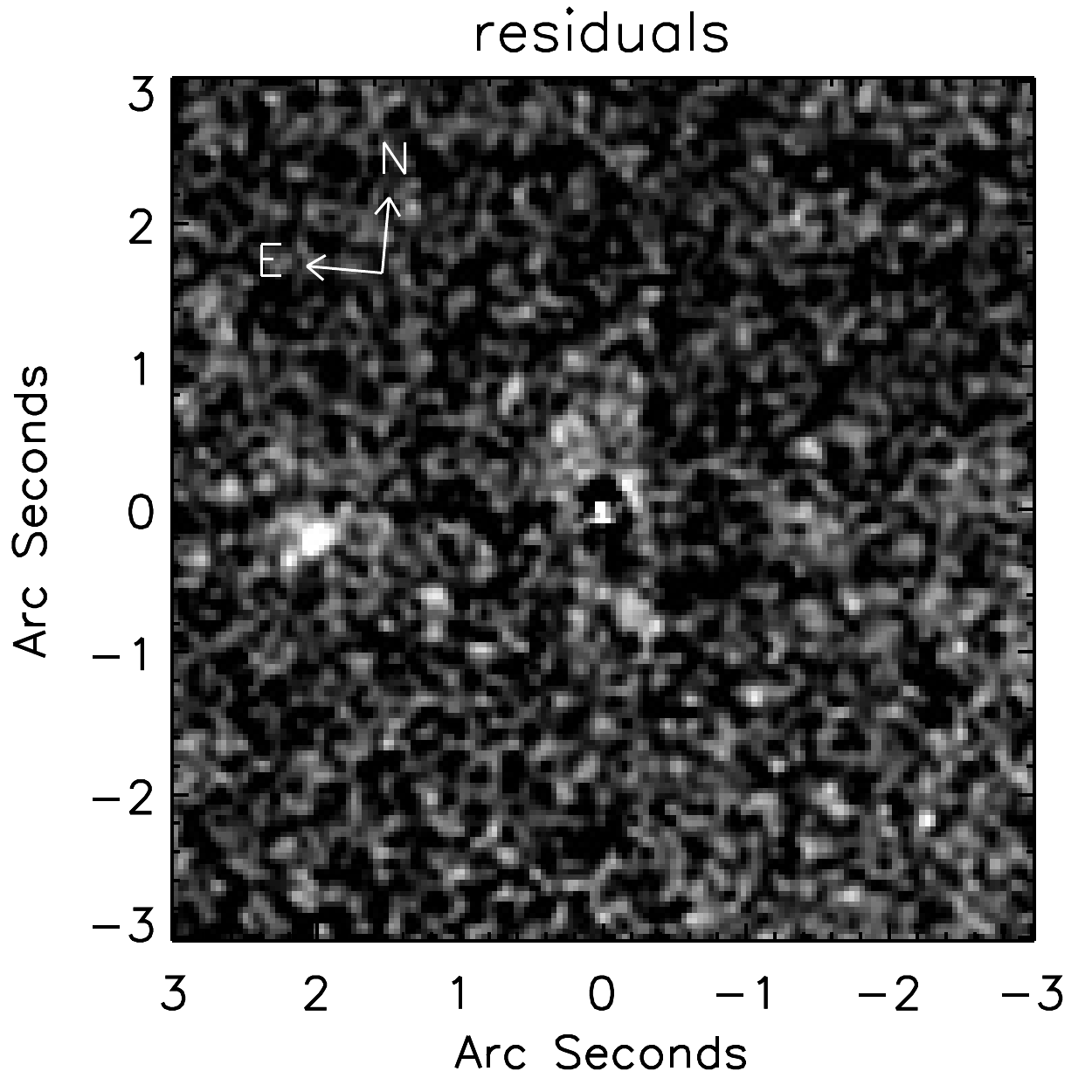}
\includegraphics[height=1.7in]{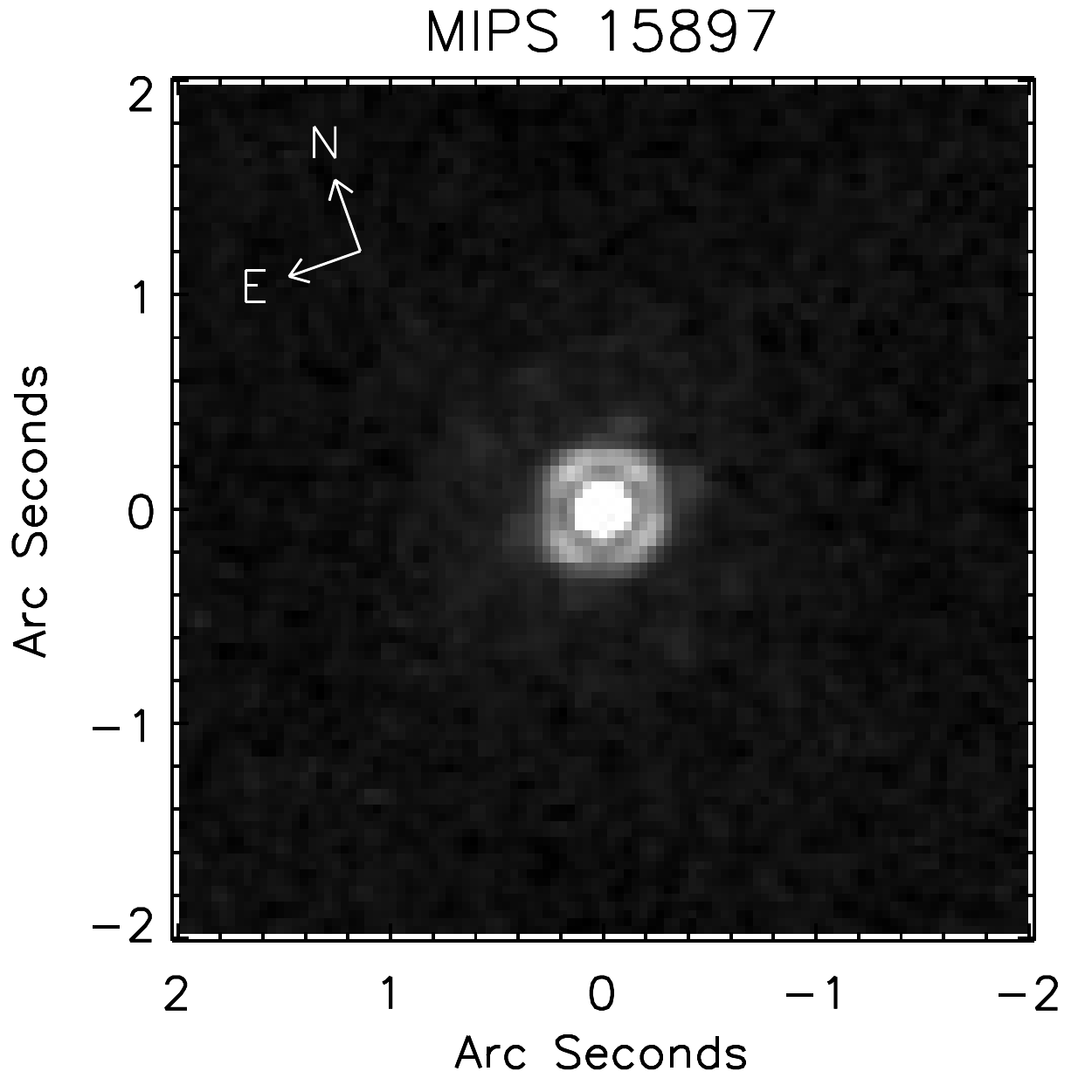}
\includegraphics[height=1.7in]{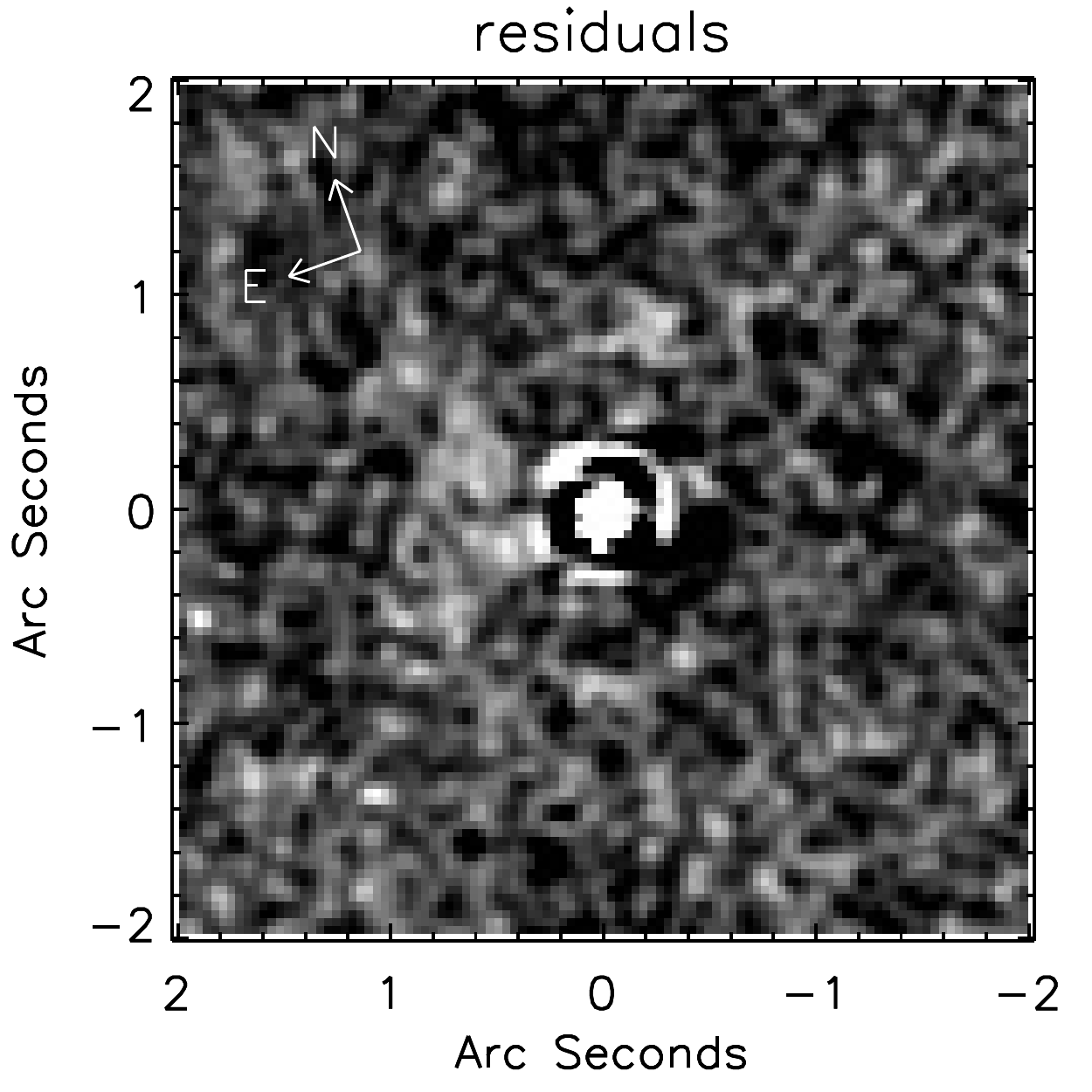}\\
\caption{Possible mergers (category 5).  They show some residuals, but one cannot tell for sure whether they originate from a merger event or not.  Residual images have been slightly smoothed to help bring out the features.}
\label{fig:examples5}
\end{center}
\end{figure*}

We then also have category~5 objects.  These are objects that no longer show companions or tidal features characteristic of mergers, but do show residuals and asymmetries after subtraction of the smooth component.  The majority of our phase V objects (the late mergers) and of our faint \& compact objects fall in this category, examples of which are shown in Figure~\ref{fig:examples5}.  Some local ULIRGs, that are clear merger remnants, also show similar features when redshifted to $z \gtrsim 1$, as demonstrated in section~3.  Therefore, although asymmetries can arise in numerous situations, we must still consider it probable that these objects have their origin in a merger event.  This is true for phase V objects as their typically high-sersic profiles combined with residuals bright enough to be detected at $z \sim 1 \textendash 2$ offers few alternatives, but also of {\em Faint \& Compact} objects as argued in \S~3.

Finally, category~6 consists of elliptical galaxies that are well-fit with a smooth sersic profile and show no residuals.  The reason why one might choose to count some of these objects towards the merger fraction originates from the fact that, as demonstrated in section~3, late mergers often lose their merging signatures in the redshifting process.  Regular ellipticals could, therefore, be hiding merging features, especially those like the ones we have in our sample that shine brightly in the mid-IR.  We thus make them into our sixth, but least confident, merger category.

The only objects that never enter our merger count are the spiral galaxies, since they are clearly isolated (that is non-merging) systems. 

We thus find that our sample spans the whole spectrum of certainty, from secure mergers to faint ellipticals without any detected features.  At the same time, we discover that most of our sample lies in the middle of that range with signatures typical of, but not always exclusively associated with, a merger event.  The most secure categories contain few of our merger candidates, and are thus highly incomplete.  The most uncertain ones, on the other hand, show very little features and are thus fairly unreliable.  In order to achieve the clearest possible picture of the importance of merging in our sample, we need to attempt to strike a balance between completeness and reliability.  We discuss where that balance might lie in section~\ref{sec:mergfrac}, together with more secure lower and upper limits.

\subsubsection{A word on automated classification techniques \label{sec:automated}}

In recent years, automated galaxy classification using measurable parameters such as concentration and asymmetry \citep{Abraham96, Conselice03b}, {\em Gini} and M20 \citep{Lotz04, Lotz08a}, or other similar combinations \citep[e.g.][]{Scarlata07, Law07, Zamojski08} have become the norm in high redshift galaxy surveys.  Using these techniques, most mergers, in optically selected samples, can be identified by their high level of asymmetry \citep{Conselice03b}.  In the case of ULIRGs, the {\em Gini}-M20 scheme provides a cleaner separation \citep{Lotz04}, although the combination of the {\em Gini} coefficient and the asymmetry parameter appears to be, overall, the most sensitive to merger signatures \citep{Lotz04, Lotz08b, Zamojski08}.

Despite being undeniably useful, especially for large samples, all of these automated classification techniques invariably miss a non-negligeable fraction of merger-induced ULIRGs, and fall short of the 99\% mark inferred visually by \citet{Veilleux02}.  In general, double or multiple nuclei systems are properly identified, but many of the singly-nucleated ULIRGs, i.e. the ones in more advanced stages of merging, as well as more distantly separated pairs, fall outside the merger cuts \citep{Lotz04, Lotz08b}.  This is due to the fact that in more advanced stages, merger signatures become increasingly faint compared to the remnant galaxy, while in distant pairs, the two objects tend to be treated separately, but not perturbed enough yet, individually, to make the cut.  In the case of advanced or late mergers, visual inspection, before and after subtraction of a smooth profile, is usually the privileged, and perhaps the only effective, approach \citep[e.g.][]{Veilleux06}.  This is analogous to quasar hosts, for which PSF-subtraction is crucial \citep[e.g.][]{Gabor09}.

In section~3, we describe simulated observations we performed on local ULIRGs redshifted to $z=1$.  Applying automated classification techniques to these redshifted ULIRGs indicates that they become even harder to separate from the rest of the galaxy population by morphological parameters, with the best criterion ({\em Gini}-asymmetry) recovering only half of our 22 redshifted local ULIRGs listed in Table~\ref{tbl:simulations}.  
In the rest-frame $I$-band, these sources have relatively smooth and broad tidal tails and bridges.  When put at high redshift, because of surface brightness dimming, these tend to get buried under the noise .  Morphological parameters then become increasingly derived from the central components which, themselves, appear smoother due to the decreased resolution.  These effects tend to draw those objects towards the same morphological space as that spanned by normal disks and bulges, therefore making mergers harder to identify.  This is what happens to three of the four sources illustrated in Figure~\ref{fig:simulations} (IRAS 22491-1808 being the only one to retain merger-type parameters at high redshift), and this despite the fact that they still show good indications of being mergers, even at $z=1$.

The same processes appear to affect our observed sample.  In addition, a large fraction of our objects possess prominent bulges whose signal overwhelms that of merging features, even in bright objects.  This is then reflected in the value of their morphological parameters, which overlap considerably with that of normal galaxies and often fall short of the merger criteria.  These objects are often at lower redshift ($z<1$) where they are seen in their rest-frame near infrared.  As a result, we find many clear mergers not making any of the morphological cuts discussed above, and, typically, only around one third of our sample would appear as mergers using these techniques.  We find that those objects that do make the cuts are really only the most obvious one:  mainly those with double or multiple nuclei.
This is extremely limitative.  We would rather include, in our study, objects in all stages of the merging process.  We therefore choose to forego, in this paper, the use of those parameters on our $24\mu m$-selected sample as we feel they do not reflect well its structural richness.  

The situation is very different for our control sample.  The latter is composed primarily of disky galaxies, and resembles other optically/NIR-selected sample for which automated techniques such the CAS system \citep{Conselice03} were designed.  It is, therefore, to no surprise that we find a much better agreement between visual and automated classifications with that sample.  A simple asymmetry cut suggests a merger fraction of 20\% in our control sample while a combination of {\em Gini} and asymmetry yields a merger fraction of 28\%.  For comparison, the merger fraction inferred visually is 25\% when counting up to confidence level~4 (section~\ref{sec:mergfrac}; Fig.~\ref{fig:mergfrac}).   Both estimates are dominated by objects with $H \approx 18\textendash 20$ as explained in section~\ref{sec:mergfrac}.  This suggests that our visual classification is indeed reliable up to confidence level~4, so that we can feel confident using it in the regime where automated techniques appear to break down, that is the one to which most of our $24\mu m$-selected sources belong.

Finally, we note that the merger fraction in our control sample is higher than most studies would suggest for a population at $z\lesssim 1$ \citep{Lotz08a, Kartaltepe07, Cassata05, Conselice03b}.  Based on the local $J$-band luminosity function \citep{Cole01}, $z = 0.15\textendash 1$ is where we expect to find most of our $H = 18\textendash 20$ sources.  We showed in section~\ref{sec:control}, however, that we do have an excess of sources in our images (that is without counting the $24\mu m$-sources) at these magnitudes compared to the average $H$-band number counts.  This excess must come from the environment these $24\mu m$-sources live in, and we therefore attribute our higher merger fraction to that over-density of galaxies around $24\mu m$ sources.

\subsection{Mid-IR Spectral Diagnostics \label{sec:mid-IR}}

The mid-infrared part of the integrated spectrum of a galaxy contains imprints of the physical processes at the origin of its infrared luminosity.  These include PAH emission complexes and silicate absorption/emission bands, as well as continuum emission from hot ($T \sim 1000 K$) dust, if present.  We discuss below how these features inform us on the nature of our sources and how we measure them.

\subsubsection{PAH Equivalent Widths}

Polycyclic aromatic hydrocarbon (PAH) molecules exist and are excited primarily in photo-dissociation regions around OB associations.  This makes them good tracers of star formation.  When excited, they reradiate their energy in a number of narrow bands at wavelengths mainly between $6.2$ and $12.7\mu m$.  These, in turn, straddle the broad peak of hot dust continuum emission that spans the entire mid-IR range.  Since, in external galaxies, the primary source of dust heating is, when present, the active galactic nucleus, the ratio of the PAH flux to that of the underlying continuum, or equivalent width, gives us a proxy for the relative importance of star formation versus AGN activity.

In this paper, we use both the equivalent width of the $7.7\mu m$ and $11.3\mu m$ PAH features as indicators of the respective contribution of the starburst and AGN components to the mid-infrared flux of our objects.  We perform our measurements using the Markov Chain Monte Carlo fitting procedure described in \citet{Sajina07} and \citet{Sajina06}.

Comparison of equivalent width measurements from our method with full SED-fitting indicates that objects with an $EW(11.3\mu m) \ge 0.8$ or an $EW(7.7\mu m) \ge 1.2$ are powered at $\gtrsim 80\%$ by star formation, whereas those with an $EW(11.3\mu m) < 0.1$ or an $EW(7.7\mu m) < 0.15$ have a $\lesssim 20\%$ contribution from star formation.  Throughout this paper, we will refer to the former category of sources as starburst-dominated, and to the latter as AGN-dominated.  Intermediate objects are given the epithet \textquotedblleft composites\textquotedblright.

Lastly, we mention that our boundary between AGN and composite systems is slightly blurred due to the fact that our upper limits of detectability of PAH features often lie around or just above the division line between these two categories.  In this paper, we choose to consider {\em all} non-detections as AGN-dominated systems.  We recognize that a few of them might actually be low-end composites, but none of our conclusions rely on knowledge of the exact ratio of the number of AGNs to low-end composites.

\subsubsection{The $9.7\mu m$ silicate feature \label{sec:tau}}

Silicate grains are ubiquitous throughout the interstellar medium.  They show a broad emission profile centered at $9.7\mu m$ when heated to temperatures of a few hundred kelvins or more, but the profile is often rather seen in absorption as colder grains screen the line-of-sight towards a hotter source.  In distant galaxies, this hot source is the galaxy's active nucleus.  In the scenario of quasar formation proposed by \citet{Sanders88}, the latter are born in the dense, highly obscured cores of coalescing galaxies.  As they light up, they start pushing away surrounding material \citep{Silk98, diMatteo05}, and  levels of obscuration are, consequently expected to drop \citep{Jonsson06}.  The presence of deep silicate absorption can therefore be indicative of this early dense phase.

In this paper, we use the optical depth of the $9.7\mu m$ silicate feature, $\tau_{9.7\mu m}$, primarily to look for these young obscured AGNs in our sample, but also as a general proxy for the level of obscuration in our objects.  It is defined as:

\begin{equation}
\tau_{9.7\mu m} = \ln \frac{f_{cont.} (9.7\mu m)}{f (9.7\mu m)}
\end{equation}

We use the method of \citet{Sajina07} to fit the optical depth of the silicate absorption feature simultaneously with the PAH and continuum fluxes used to obtain PAH equivalent widths.  This procedure assumes a screen geometry.  The optical depth at $9.7\mu m$ relates to the silicate strength often in use in the literature \citep[e.g.][]{Spoon07} through the following relation: $\tau_{9.7\mu m} = -1.4 \times S_{sil}$.

\section{Simulations \label{sec:simulations}}

A key task of any high redshift study is to establish a comparison with more nearby samples.  Because mergers and merger remnants are such defining trademarks of local ULIRGs, it is imperative, and the primary focus of this paper, to ask whether these traits are as ubiquitous among their high redshift counterparts.  Surface brightness dimming and lower resolution, however, affect our ability to detect merger signatures and identify mergers at high redshift.   We address these issues in this section using simulated HST/NICMOS observations of local ULIRGs redshifted to redshifts of one and above.

\tabletypesize{\footnotesize}

\begin{deluxetable*}{lcccccccc}
\tablecolumns{9}
\tablewidth{0pt}
\tablecaption{Redshifted objects \label{tbl:simulations}}
\tablehead{
	\colhead{Object Name\tablenotemark{a}} &
	\colhead{Instrument} &
	\colhead{Filter} &
	\colhead{Magnitude} &
	\colhead{Redshift\tablenotemark{b}} &
	\colhead{Absolute} &
	\colhead{mid-IR} &
	\colhead{Simulated} &
	\colhead{Classification after} \\
	\colhead{} &
	\colhead{} &
	\colhead{} &
	\colhead{} &
	\colhead{} &
	\colhead{Magnitude} &
	\colhead{color\tablenotemark{c}} &
	\colhead{Redshift(s)} &
	\colhead{redshifting \tablenotemark{d}}
}
\startdata
\cutinhead{\normalsize Phase III objects}
{\bf IRAS 08572+3915} & HST/WFPC2 & F814W & 15.59 & 0.058 & -21.55 & warm & 1 & III (2) \\
{\bf IRAS 14348-1447} & UH2.2m & I & 14.58 & 0.082 & -23.36 & cold & 1 & III (1) \\
{\bf IRAS 03521+0028} & HST/WFPC2 & F814W & 17.49 & 0.152 & -21.96 & cold & 1 & Irregular\tablenotemark{e}  \\
Mrk 463 & HST/WFPC2 & F814W & 13.34 & 0.0504 & -23.46 & warm & 1 & III (2)\\
{\bf IRAS 12112+0305} & UH2.2m & I & 15.05 & 0.072 & -22.59 & cold & 1 & III (1) \\
{\bf IRAS 22491-1808} & UH2.2m & I & 15.00 & 0.078 & -22.82 & cold & 1, 1.7  & III (1) \\
{\bf IRAS 23498+2423} & HST/WFPC2 & F814W & 16.81 & 0.212 & -23.50 & cold & 1 & III (3) \\
\cutinhead{\normalsize Phase IVa objects}
{\bf Arp 220} & HST/WFPC2 & F814W & ~12.25\tablenotemark{f} & 0.0185 & -22.28 & cold & 1 & F\&C\tablenotemark{g} (5) \\
{\bf IRAS 15206+3342} & HST/WFPC2 & F814W & 15.61 & 0.125 & -23.22 & warm & 1 & IV (3) \\
{\bf IRAS 02021-2104} & HST/WFPC2 & F814W & 16.29 & 0.115 & -22.48 & warm & 1 & IV (4) \\
{\bf IRAS 09039+0503} & HST/WFPC2 & F814W & 16.27 & 0.126 & -22.59 & cold & 1 & IV (3) \\
\cutinhead{\normalsize Phase IVb objects}
Mrk 273 (UGC8696) & HST/WFPC2 & F814W & ~13.16\tablenotemark{f} & 0.0382 & -22.97 & cold & 1 & IV (2) \\
IRAS 15250+3609 & UH2.2m & I & 15.19 & 0.0555 & -21.78 & cold & 1 & V (5) \\
UGC 5101 & UH2.2m & I & 14.60 & 0.0394 & -21.60 & cold & 1 & F\&C\tablenotemark{g} (6) \\
{\bf Mrk 231} & HST/WFPC2 & F814W & 12.49 & 0.0426 & -23.92 & warm & 1 & IV (4) \\
{\bf Mrk 1014} & HST/WFPC2 & F814W & 14.84 & 0.162 & -24.44 & warm & 1 & IV (3) \\
IRAS 23365+3604 & HST/WFPC2 & F814W & 14.48 & 0.0634 & -22.79 & cold & 1 & F\&C\tablenotemark{g} (5) \\
{\bf IRAS 05189-2524} & HST/WFPC2 & F814W & 13.91 & 0.0427 & -22.51 & warm & 1 & IV (4) \\
\cutinhead{\normalsize Phase V objects}
{\bf IRAS 01003-2238} & HST/WFPC2 & F814W & 17.15 & 0.117 & -21.61 & warm & 1 & Regular bulge \\
{\bf IRAS 07598+6508} & HST/WFPC2 & F702W & 14.40 & 0.149 & -24.84 & warm & 1.32 & Regular bulge \\
{\bf IRAS 13218+0552} & HST/WFPC2 & F702W & 17.24 & 0.206 & -22.55 & warm & 1.32, 1.8 & Regular bulge \\
Mrk 771 & HST/WFPC2 & F606W & 14.95 & 0.064 & -22.36 & warm & 1.67 & V (5) \\
\enddata
\tablenotetext{a}{Objects in bold are part of the IRAS 1-Jy sample}
\tablenotetext{b}{Corrected to the rest-frame of the CMB}
\tablenotetext{c}{Objects with a ratio of $f_{25\mu m}/f_{60\mu m} > 0.2$ are considered {\em warm}.  Otherwise, they are said to be {\em cold}.}
\tablenotetext{d}{Confidence levels (see \S 2.5.2) are shown in parentheses}
\tablenotetext{e}{IRAS 03521+0028, when redshifted, looks nothing like any of our objects. It rather looks like a diffuse cloud, a scaled up version of the LMC.  We therefore call it irregular, but would not have associate such morphology with a merger had we encountered it in our sample.}
\tablenotetext{f}{From \citet{Surace98} with the UH2.2m telescope.}
\tablenotetext{g}{Faint \& Compact}
\end{deluxetable*}

\begin{figure*}[htbp]
\begin{center}
\makebox[2.3in]{\Large ~~~~Original Image}
\makebox[2.3in]{\Large ~~~~Redshifted Image}
\makebox[2.3in]{\Large Residual image} \\
\vspace{0.05in}
\includegraphics[width=2.3in]{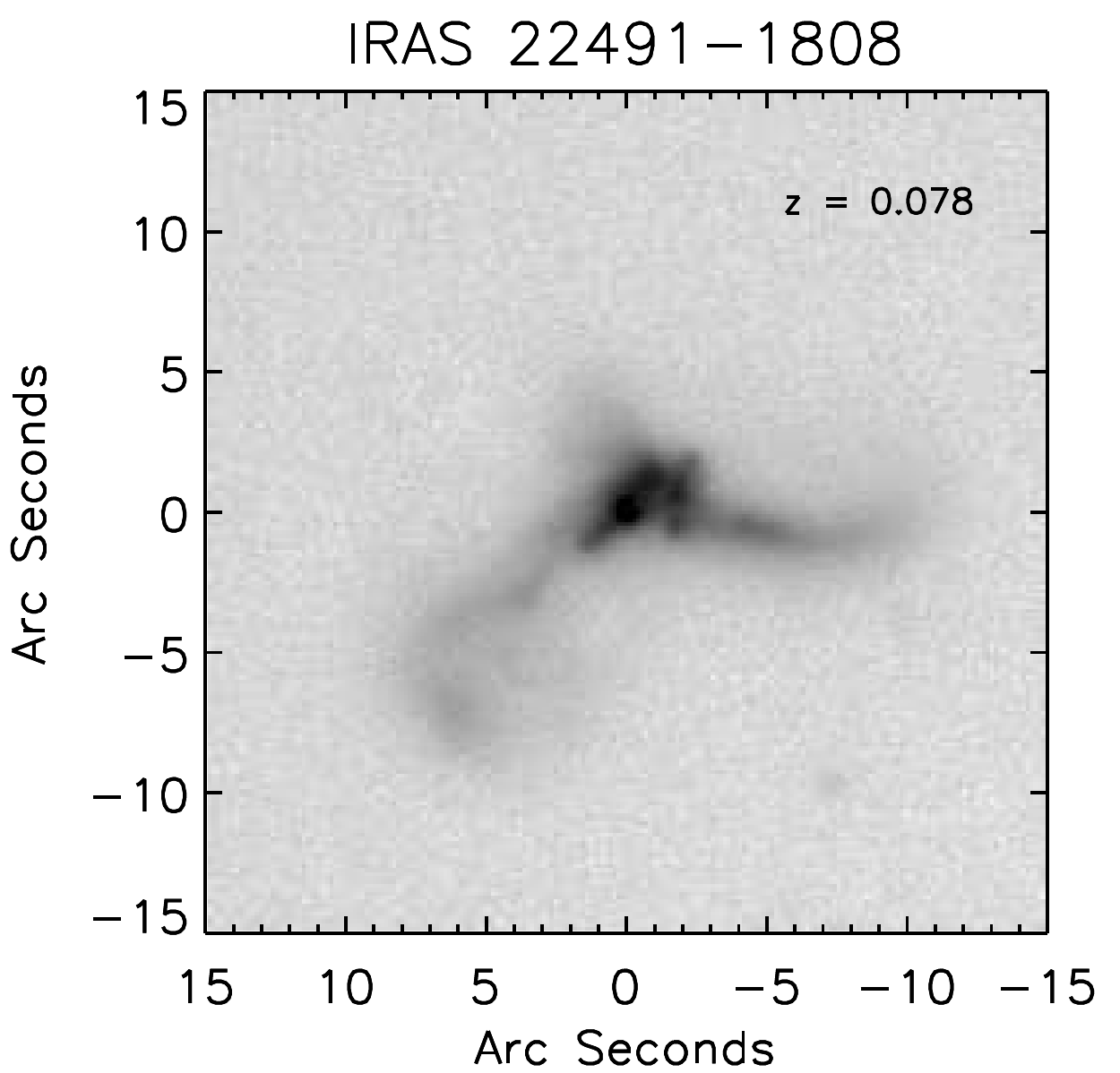}
\includegraphics[width=2.3in]{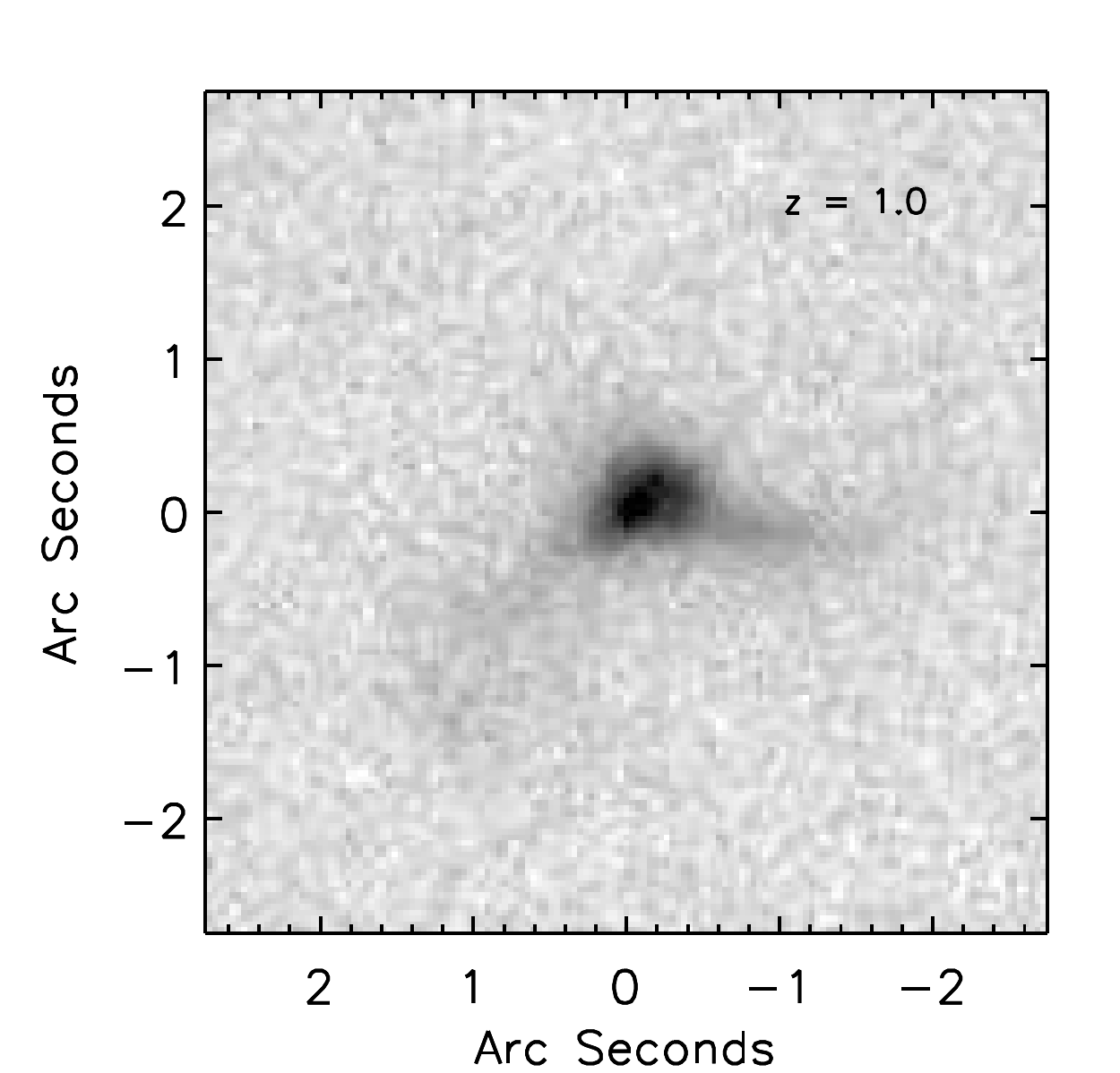}
\hspace{0.57in} \raisebox{1.2in}{\textbf{No residual image}} \hspace*{0.57in} \\
\vspace{0.05in}
\includegraphics[width=2.3in]{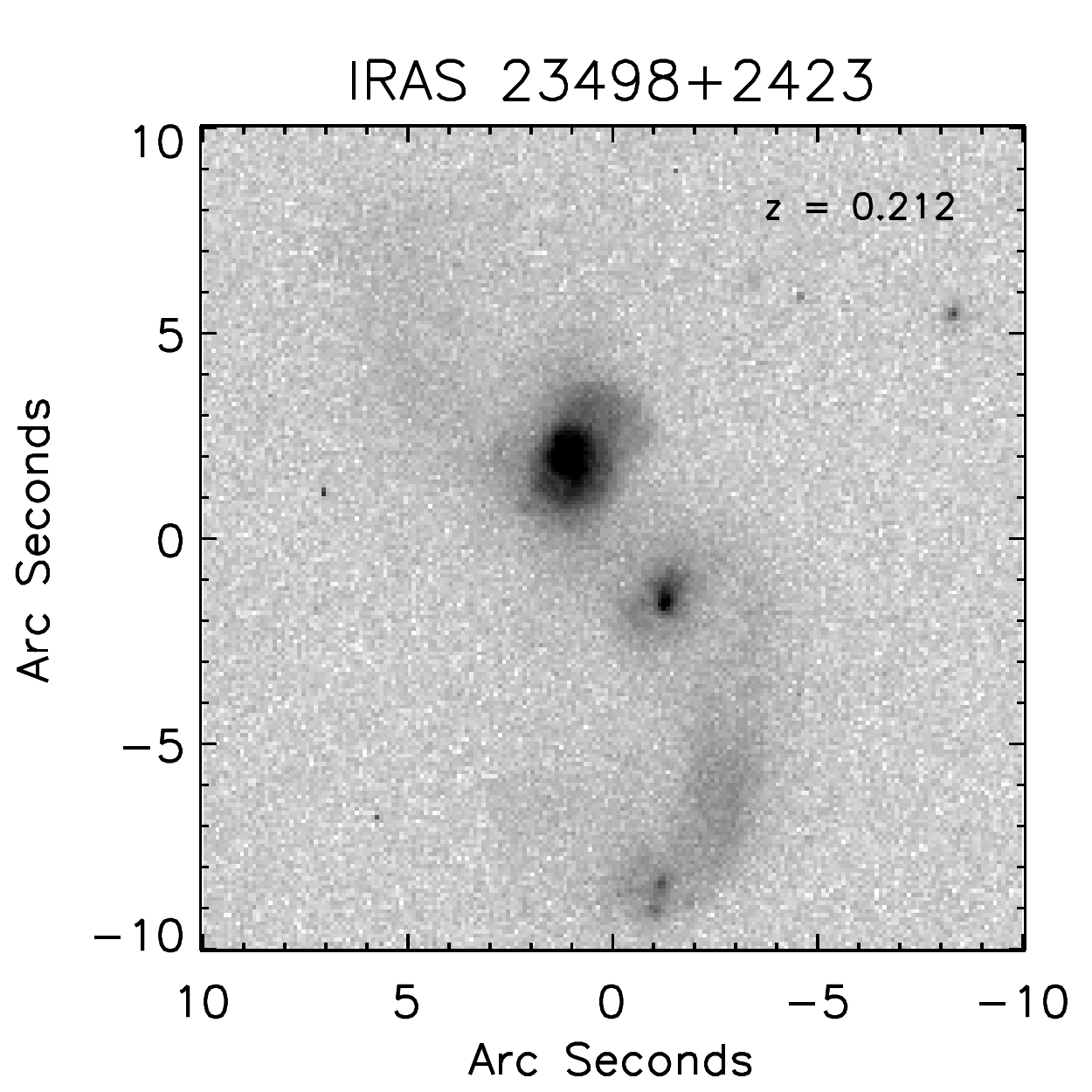}
\includegraphics[width=2.3in]{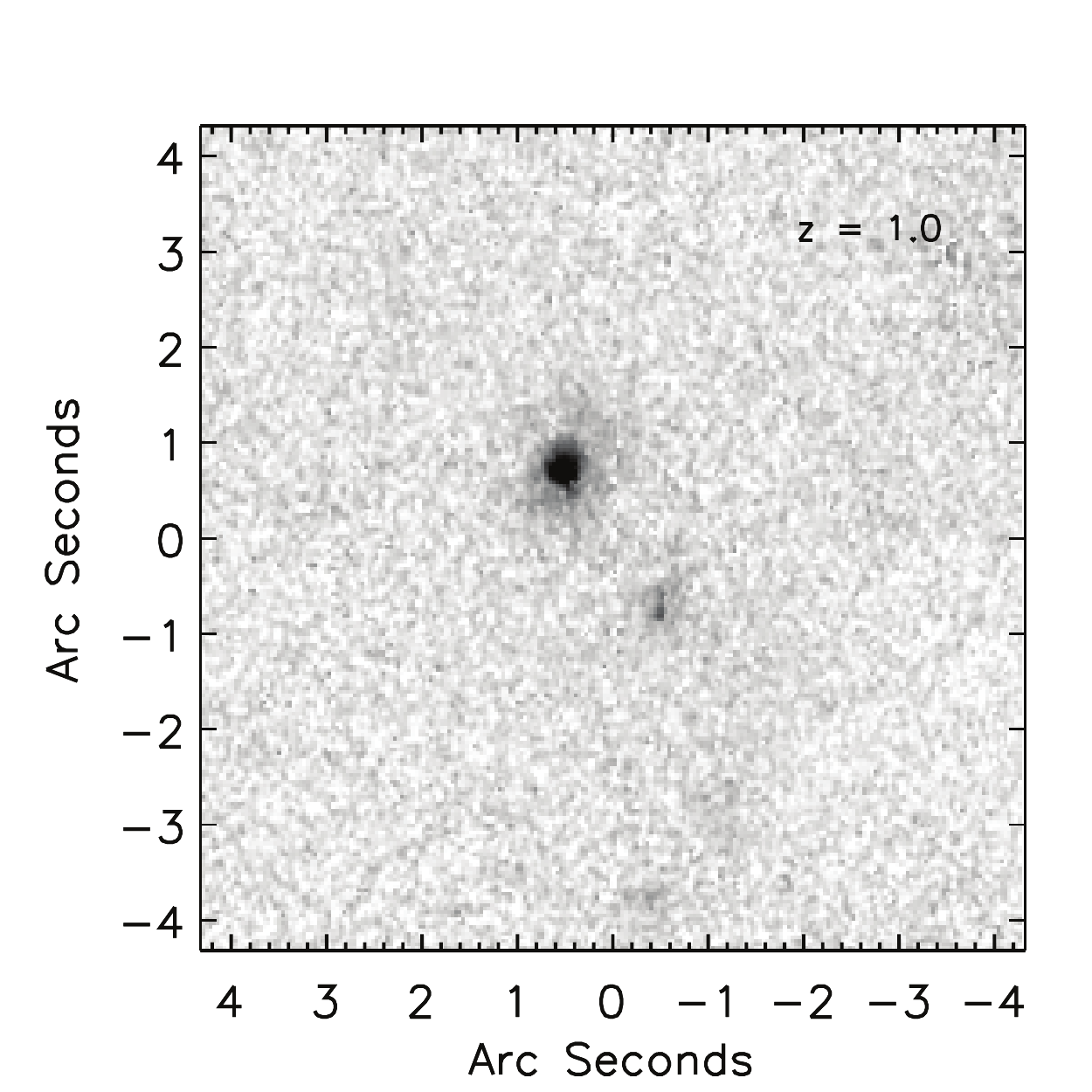}
\includegraphics[width=2.3in]{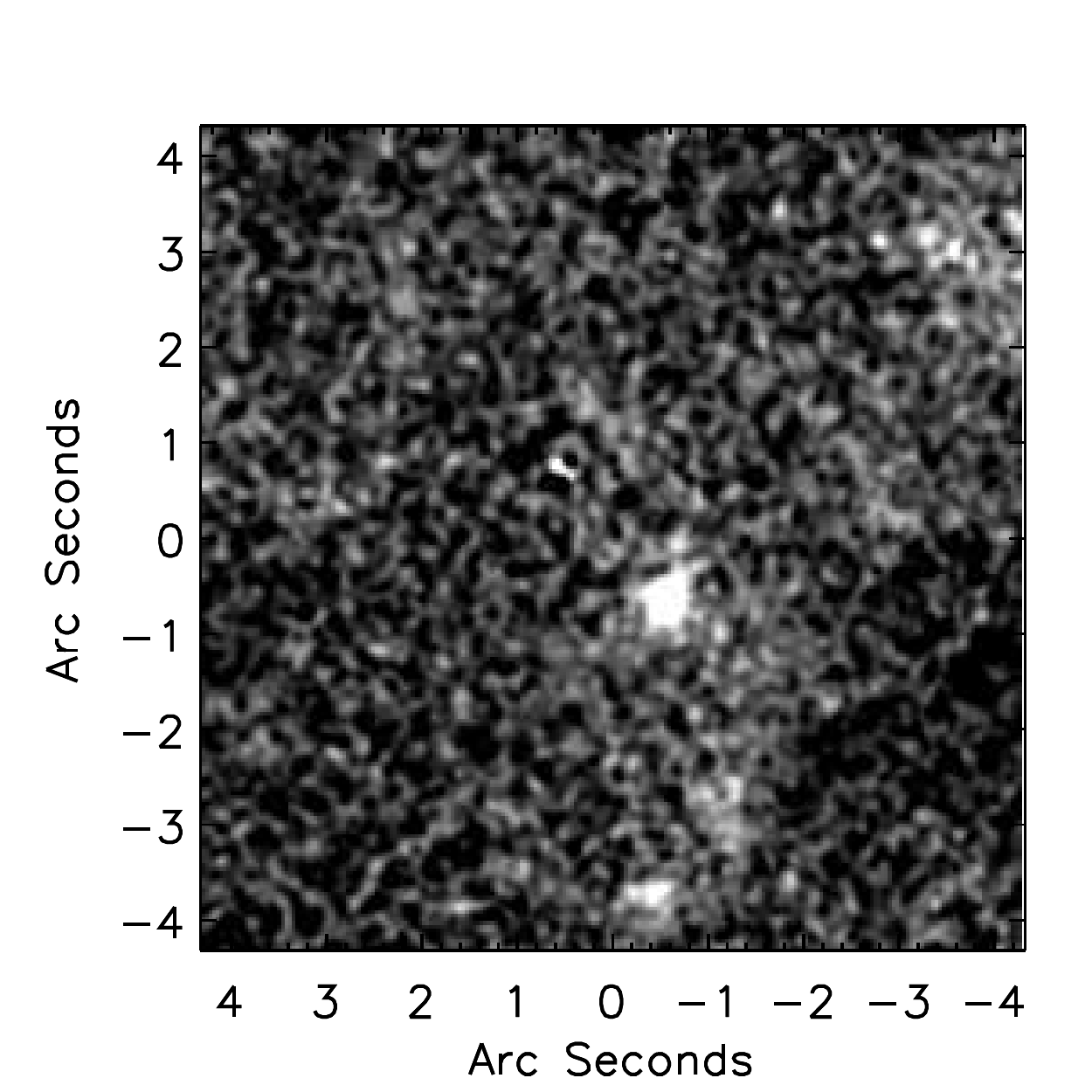} \\
\vspace{0.05in}
\includegraphics[width=2.3in]{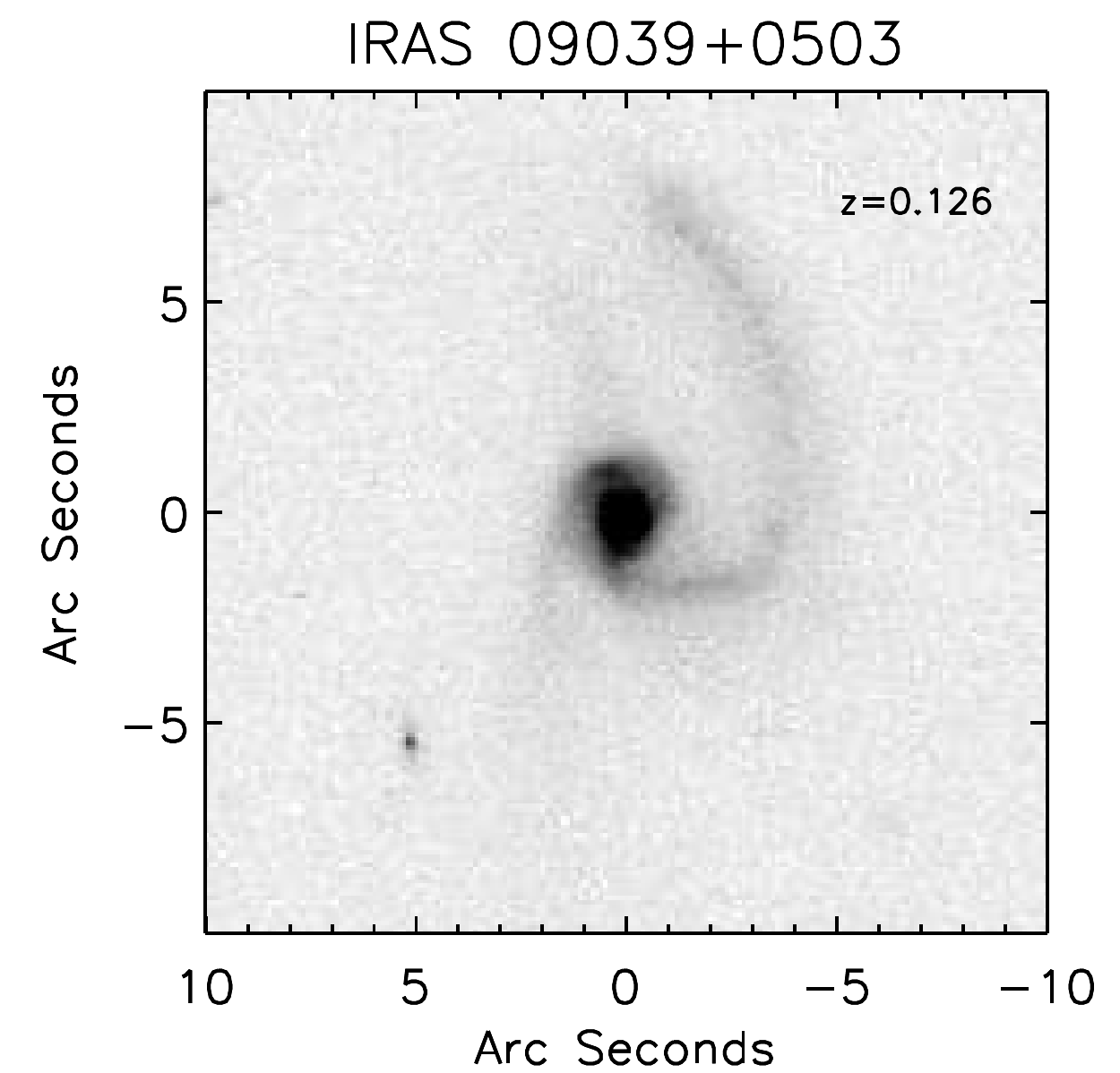}
\includegraphics[width=2.3in]{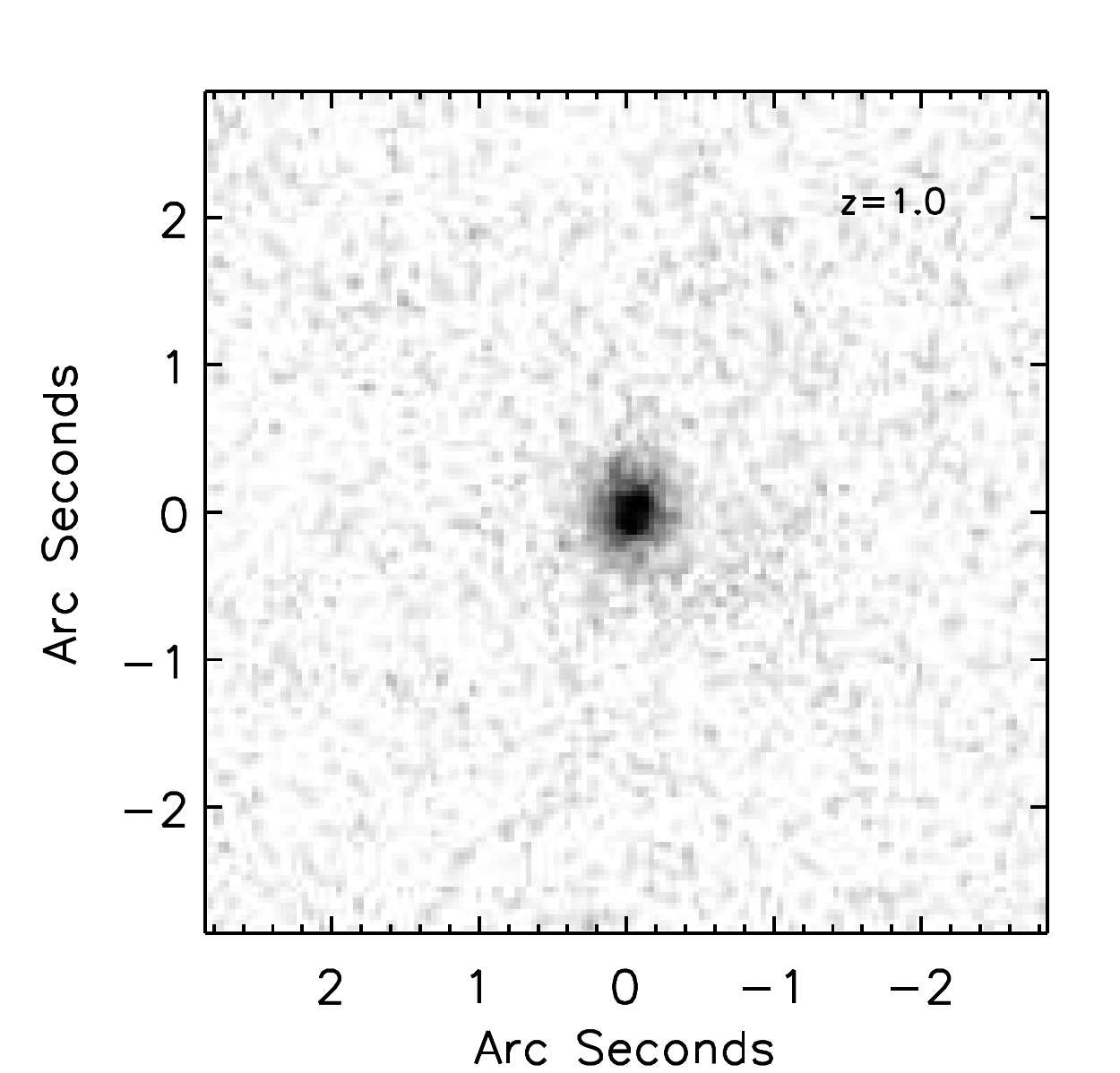}
\includegraphics[width=2.3in]{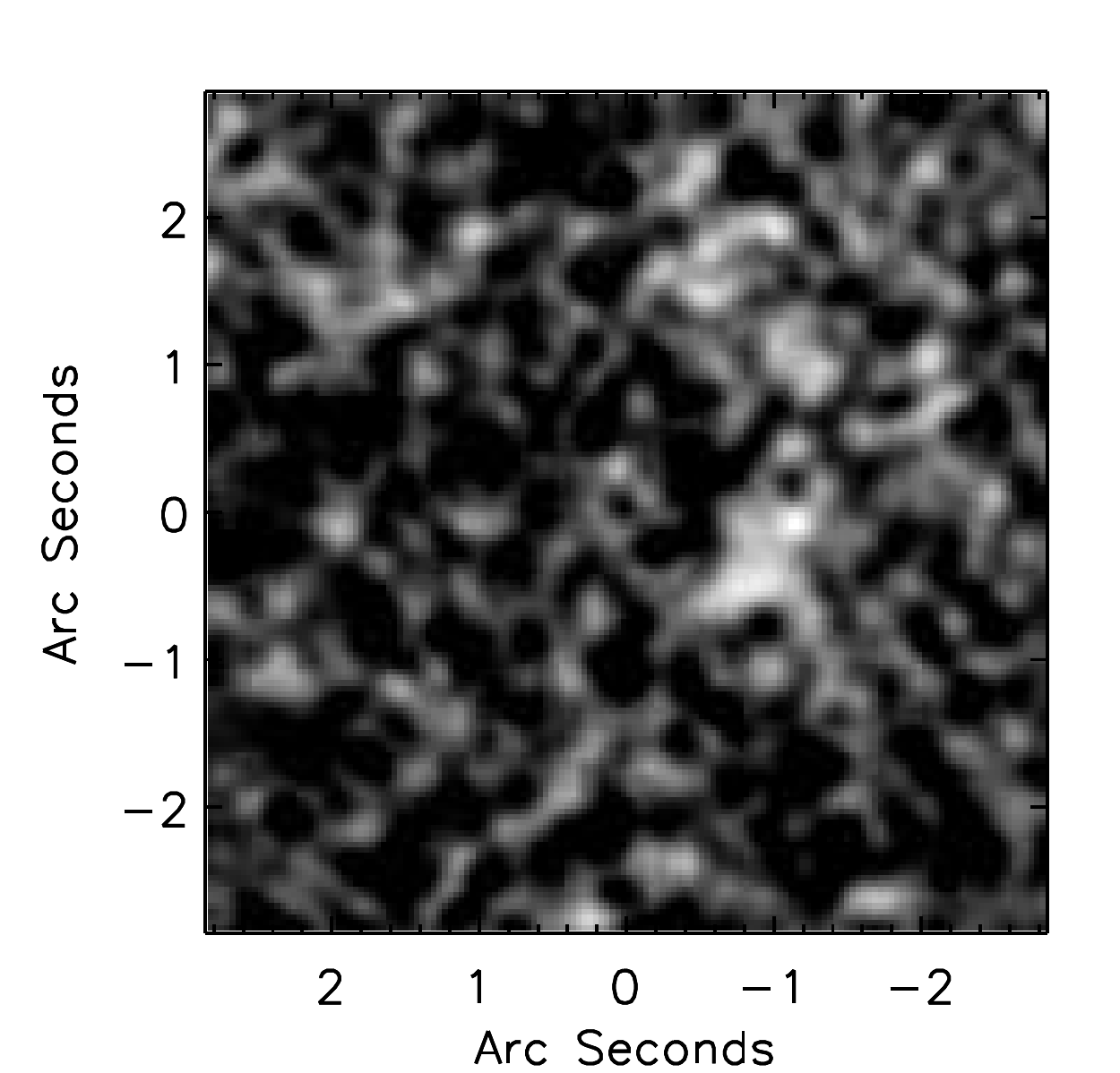} \\
\vspace{0.05in}
\includegraphics[width=2.3in]{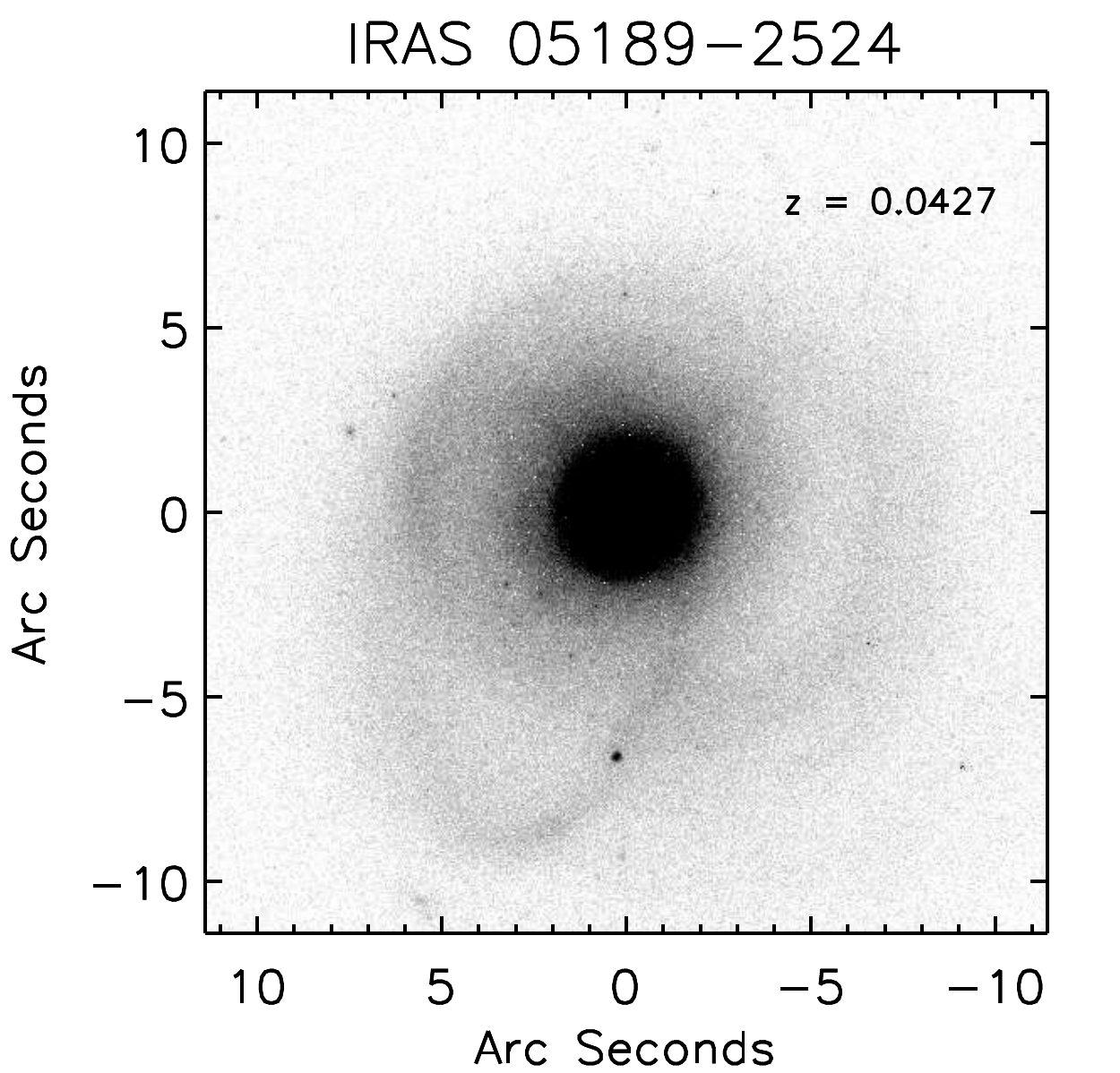}
\includegraphics[width=2.3in]{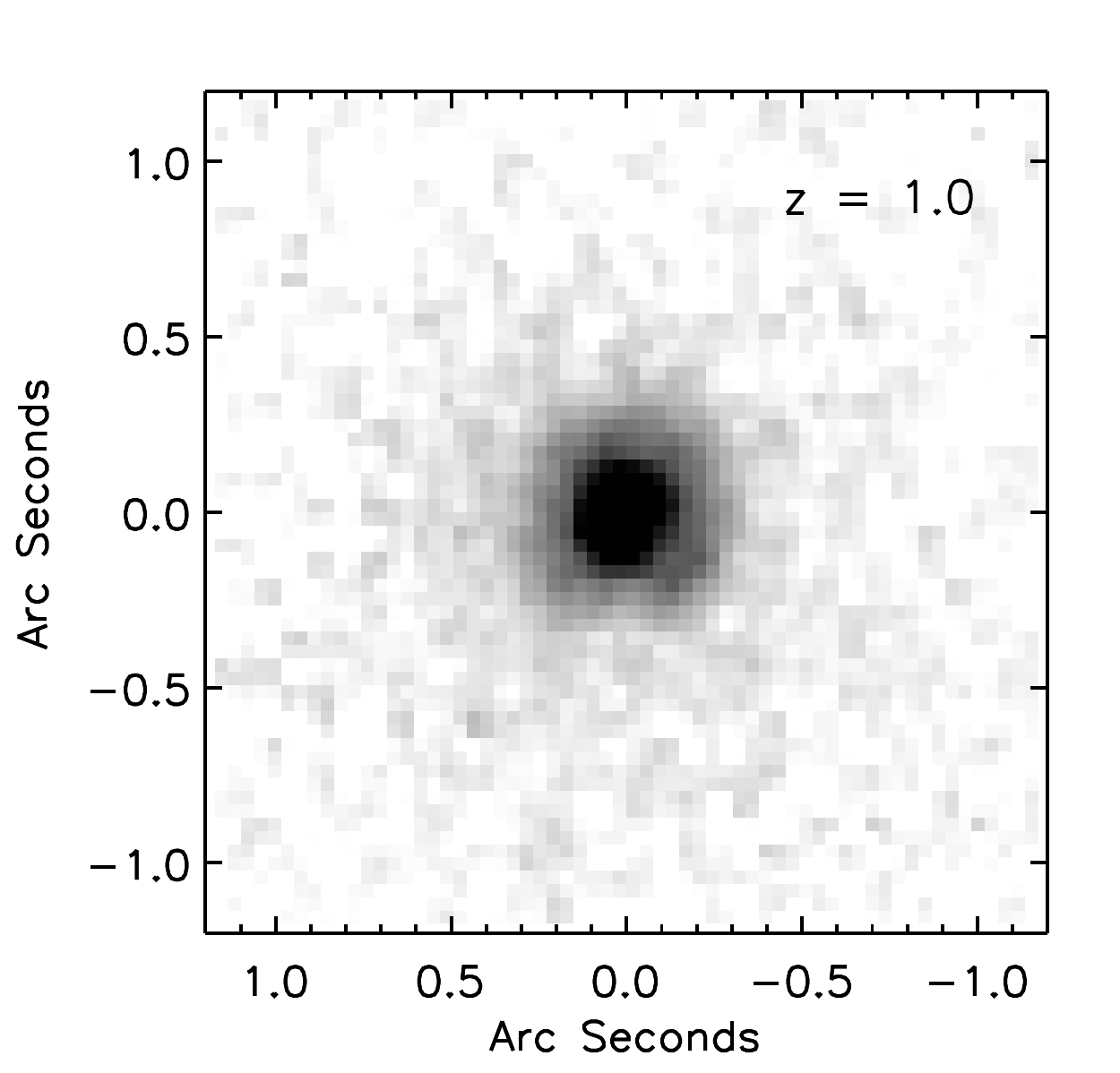}
\includegraphics[width=2.3in]{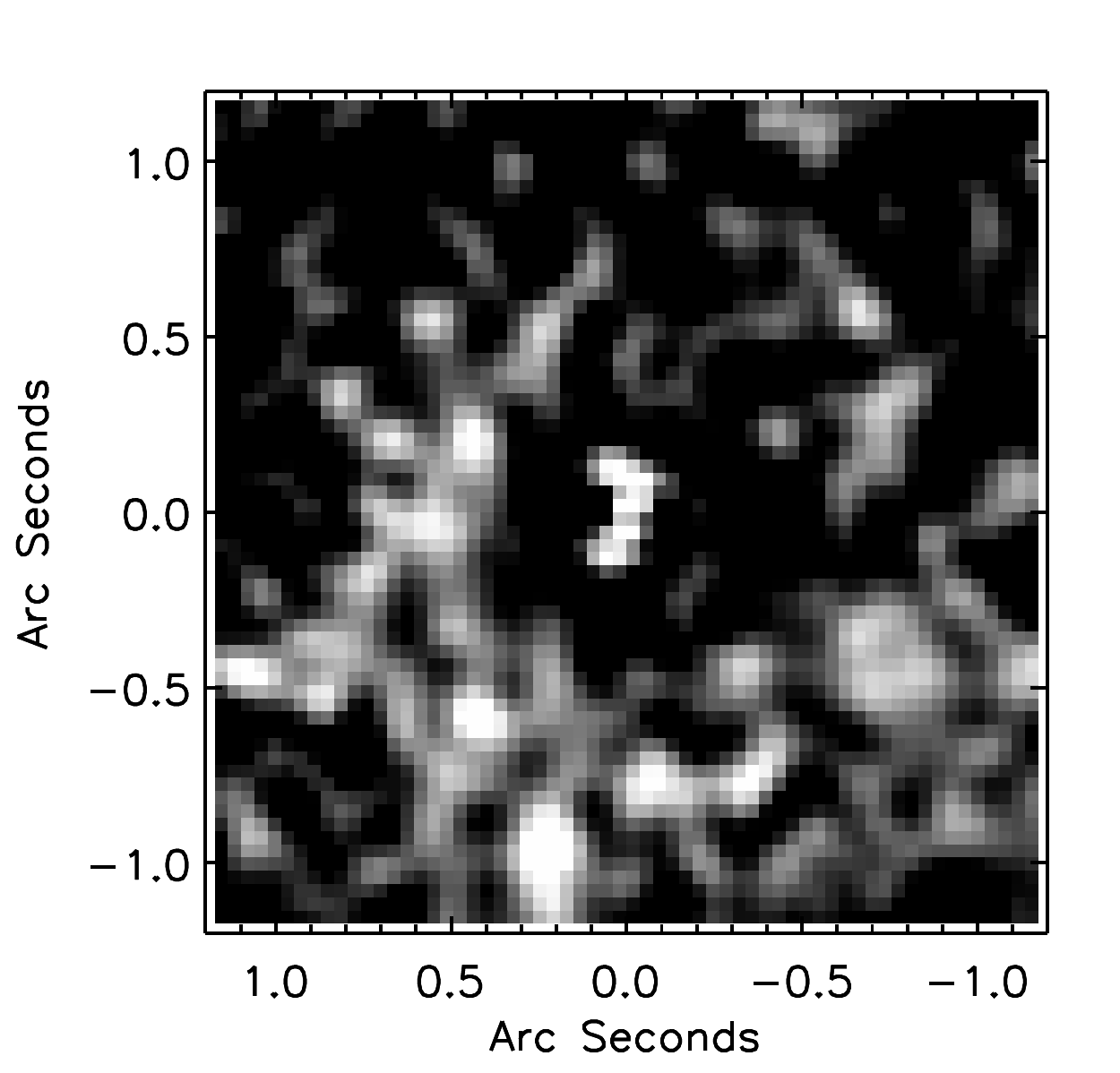} \\
\caption{Examples of simulated observations of redshifted local ULIRGs.  The color scale is inverted in residual images, and they have been smoothed to help bring out faint features.}
\label{fig:simulations}
\end{center}
\end{figure*}

We artificially redshifted a total of 22 local ULIRGs and mergers that we selected primarily from the IRAS 1-Jy sample \citep{Kim98, Veilleux02} and the sample of \citet{Surace98b, Surace00} in such a way as to cover a wide range of merger configurations and morphological types, as well as both {\em warm} ($f_{25\mu m} / f_{60\mu m} > 0.2$) and {\em cold} ($f_{25\mu m} / f_{60\mu m} < 0.2$) mid-IR colors.  We used mainly archived HST/WFPC2 F814W data when available or, otherwise, $I$-band ground-based images from \citet{Surace98} taken with the UH2.2m telescope.  Those filters were chosen to match the rest-frame band observed by the NICMOS/F160W filter at $z = 1$.  For HST images, we used the {\em TinyTim} software to model the PSF, whereas for ground-based images, we used stars directly available from the image itself.  The list of galaxies we simulated can be found in table~\ref{tbl:simulations}, along with the provenance of the data used for each object, basic information about the object such as its morphological class, redshift, magnitude and mid-IR color, and its classification after redshifting.  Figure~\ref{fig:simulations} shows examples of simulated images of local phase III and phase IV ULIRGs.  A full description of our redshifting procedure can be found in Appendix~A.

We elected to simulate most of our objects at $z=1$ for two reasons:  1) it corresponds to the peak of the redshift distribution of our sample (Figure~\ref{fig:zdist}), and 2) $I$-band images, which correspond to the rest-frame wavelengths we observe at that redshift in our data, were the ones most easily available to us.  Although we also have a significant number of objects, in our sample, at $z \sim 2$, we actually find that the merger nature of those objects is often clearer than that of the $z \sim 1$ sources.  This is most likely due to the fact that, at $z > 1.5$, objects in our sample have infrared luminosities in excess of $10^{12.5} L_{\odot}$ (see section~\ref{sec:redshift_evolution}), and are, therefore, more likely to be more major mergers.  Furthermore, we show later that, past $z = 1.5$, local ULIRGs are no longer good analogs of our bright $24\mu m$-sources, whether morphologically or in terms of their spectral properties.  It is, therefore, unclear whether local ULIRGs can be used as calibrators for our highest redshift sources even if we were to perform simulated observations of them at $z = 2$.  For these reasons, we feel that $z=1$ is the most relevant redshift to put these objects at, and that this provides us with adequate and sufficient information for our purposes.

The results of these simulations reveal that most phase III mergers are still identifiable as such at high redshifts.  They demonstrate the same to be true for two thirds of our simulated phase IV objects as well.  Table~\ref{tbl:simulations} also provides evidence, however, that our phase IV is narrower than its local homologue, and that, at least some, phase IV objects are likely to, instead, fall in the phase V category in our sample.

It also emerges from our simulations that objects we classify as {\em Faint \& Compact} at high redshift are always identified as phase IV mergers at higher resolution and higher signal-to-noise.  These objects typically possess a dense core, and are surrounded by faint and diffuse tidal structures.  At high redshifts, only the core is seen, which is what gives them their \textquotedblleft faint \& compact\textquotedblright appearance.  It is this observation that motivates our decision to associate these faint \& compact objects with the coalescence phase, together with the advanced (phase IV) mergers.  By doing so, we are further able to recover an even greater fraction of these mergers.

The larger difficulty arises with older, phase V, mergers.  We simulated high-redshift observations of four of such objects and, indeed, find that, in most cases (three out of four), the merger signature is lost in the process.  This demonstrates that our data is not very sensitive to the latest stages of the merging sequence and that late mergers could masquerade as regular elliptical galaxies.  When signatures are detected, they further consist mostly of lopsidedness or asymmetries which are difficult to link uniquely to a merger event.  We take all this into account in our estimate of the merger fraction in section~\ref{sec:mergfrac}.

Although our simulations have shown that, for the most part, local ULIRGs can still be identified as mergers at redshifts of $z \gtrsim 1$, they do not address the more difficult question as to whether, or to what extent, non-mergers could falsely pose as mergers at lower resolution and signal-to-noise.  It is, in fact, possible to imagine an alternative scenario for nearly all of our merger-identified objects, whether it is chance superposition accompanied by coincidentally misleading distortions giving the impression of an early merger, or fly-by interactions giving rise to tidal tails and posing as an advanced merger, or yet the presence of an undetected dwarf companion causing the primary object to display a lopsided profile and asymmetry characteristic of phase V objects, are all within the realm of possibility.  We argue, however, that such serendipitous configurations, at least among phases III and IV, must occur far less frequently than actual mergers, most particularly in bright $24\mu m$-selected sources, and that a full quantification of the level of contamination is, therefore, unnecessary at this stage.  Instead, we chose to assign a confidence level to each of our objects and quote our merger fraction as a function of confidence level rather than in absolute terms.  This will allow interpretation of our results to easily carry into the future as calibration of our confidence scheme, in terms of completeness and reliability, clarifies itself, and the uncertainty on the merger fraction narrows down.  We describe this approach in more details in section~\ref{sec:mergfrac}.

As a final note, we turn our attention once again to our {\em Faint \& Compact} objects, and mention that even though isolated disks that are faint and compact are known to exist at high redshifts \citep{Guzman98,Noeske06}, they tend to be very blue \citep{Phillips97}, implying that their faintness is intrinsic and not due to extreme extinction.  They would, therefore, not make it into our bright $24\mu m$ flux-limited sample.  The mid-infrared properties of our faint \& compact objects presented in section~4.5 and~4.6 would, further, be very difficult to reconcile with isolated star-forming disks.  This suggests that compact isolated disks are distinct from our {\em Faint \& Compact} sources and that contamination is not likely to be a major issue.  We nevertheless, treat these objects as mergers with a lower level of confidence because, despite the fact that all the evidence seems to paint a consistent picture, it still remains circumstantial in nature.

\section{Results \label{sec:results}}

\subsection{Merger fraction \label{sec:mergfrac}}

\begin{figure}[htbp]
\begin{center}
\includegraphics[width=3.4in]{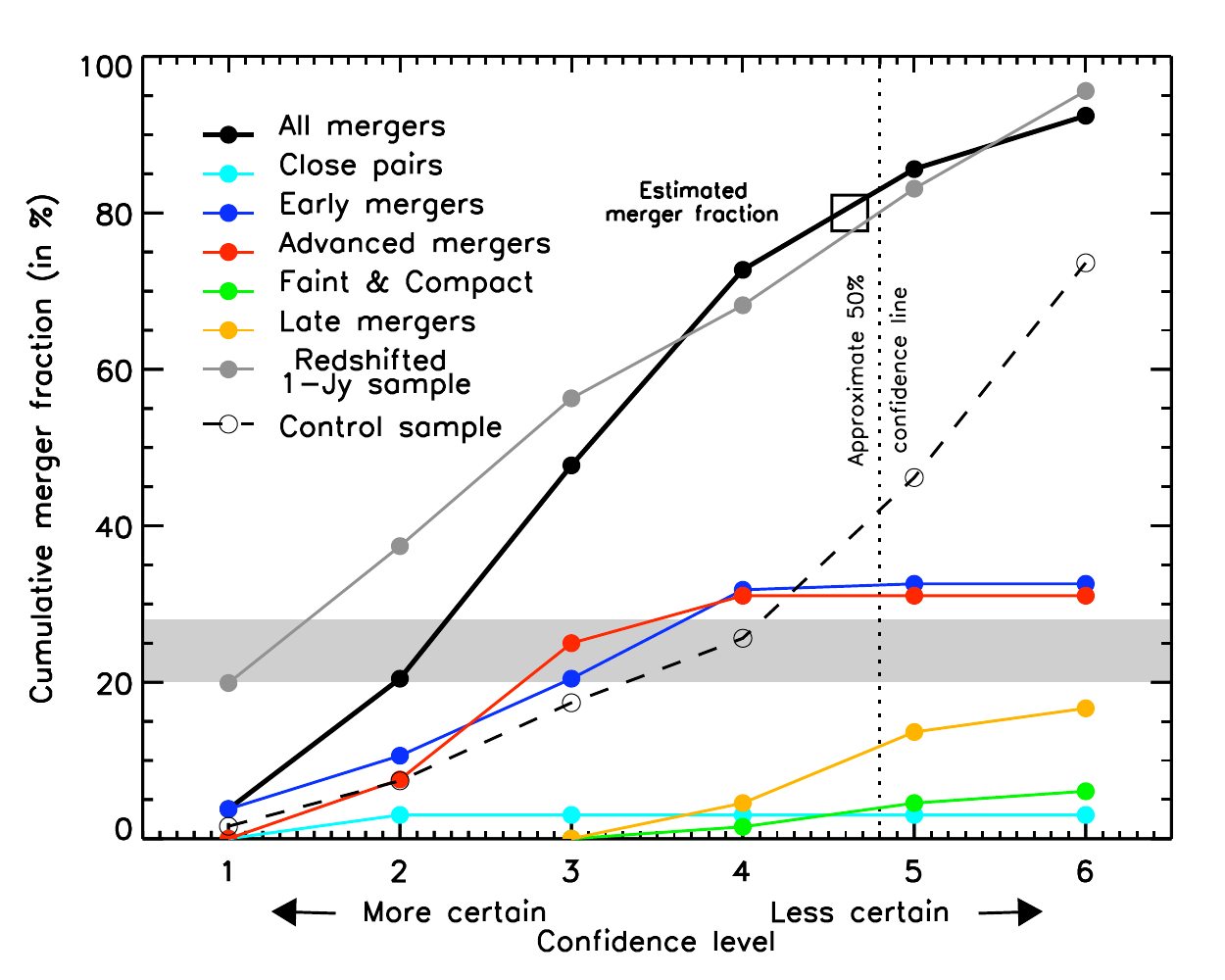}
\caption{Cumulative merger fraction from the most to the least certain confidence class.  Classes 1 through 5 are illustrated with examples in Figures~\ref{fig:examples1} to~\ref{fig:examples5}.  Class 6 are regular, but faint galaxies, and could therefore be hiding features.  Our best estimate of the merger fraction is 80\% (see text for details).  The grey area represents the range of merger fractions obtained for our {\em control} sample from automated classification techniques, illustrating that our visual classification is reliable up to confidence level~4.  Automated classification techniques fail for ULIRGs at high-$z$ as discussed in \S~\ref{sec:automated}, and hence comparisons are not shown for either ou sample or the redshifted 1-Jy sample. }
\label{fig:mergfrac}
\end{center}
\end{figure}

To describe the merger fraction of a sample in absolute terms is a perilous proposition. It is, however, the first question that arises from a data set such as ours and, therefore, needs to be addressed.  We know that, locally, interacting galaxies are readily identified visually and that high resolution observations can still reveal telltale signatures of past events even at late stages of the process \citep[e.g.][]{Surace98b, Veilleux06}.  This confidence, however, decreases exponentially with redshift as merging signatures rapidly blur and fade, and calibration of lower resolution and signal-to-noise observations against the multitude of galaxy morphologies and configurations observed in the nearby and distant Universe, despite numerous efforts \citep[e.g.][see also sections~\ref{sec:automated} and~3]{Conselice03, Lotz04, Lotz08b}, remains, to date, incomplete.  Consequently, one has to rely on visual classification where opinions might differ.  In an effort to be transparent, we decided to approach the question in steps by splitting our sample, as discussed in section~\ref{sec:conf_classes}, into confidence classes, and then accounting for the merger fraction cumulatively from the most to the least certain of them.  The results are shown in Figure~\ref{fig:mergfrac}.  We provide our own interpretation of these results below, but our method also allows them to be more easily re-interpreted in light of improved calibration constraints in the future.  

Figure~\ref{fig:mergfrac} illustrates our full accounting of the merger fraction by confidence level and morphological class.  It also shows a similar curve for the 1-Jy sample as if it were observed at high redshift with the same kind of data, as well as one for our control sample.  The curve for the redshifted 1-Jy sample was produced using the results of our simulations listed in Table~\ref{tbl:simulations} applied to the intrinsic morphological composition of the sample.  Because the magnitude distribution of objects in our control sample is wildly different from that of our $24\mu m$-sources (cf.~Figure~\ref{fig:ncounts}), a direct comparison of the merger fraction in those two samples would not be meaningful.  Instead, to derive the merger fraction shown in Figure~\ref{fig:mergfrac}, we weighted the contribution of every object in our control sample by the ratio of the number of objects within that same magnitude bin in our $24\mu m$-selected sample to its number in our control sample.  As a result, most of the weight, in the merger fraction of the random sample lies in objects with magnitudes ranging from $H \sim 18 \textendash 20$.  We use these two curves to inform our estimate of the merger fraction in our $24\mu m$-selected sample.  For example, a comparison of the cumulative merger fraction curve to that of the redshifted 1-Jy sample immediately reveals a striking similarity, from which we can already conclude that our sample is at least consistent with a high merger fraction.  More importantly, however, a high merger fraction appears to be necessary to explain the gap between our $24\mu m$-selected and control samples, the latter of which we find to be already composed of $\sim 25\%$ mergers by all measures (see \S~\ref{sec:automated}).  We argue below that, in fact, the merger fraction among our bright $24\mu m$-selected galaxies is unlikely to lie outside the $62\textendash 91\%$ range.

As a start, we can set a reasonable lower limit on our merger fraction simply by using the fact that, as demonstrated in Figure~\ref{fig:mergfrac}, our $24\mu m$-selected sample bears much closer resemblance to the redshifted 1-Jy sample than to the control sample.  We can, therefore, be confident that the merger fraction in our sample lies beyond the mid-point between those other two samples, that is above 62\%.  

In order to arrive at a more accurate estimate of the true merger fraction, however, we need to rely a little bit more on our intuition to find a balance between completeness and reliability.  This question remains, therefore, somewhat subjective and one can, naturally, form one's own opinion based on the images shown in Figures~\ref{fig:examples1} to~\ref{fig:examples5}.  However, one still needs to explain the observed morphologies of our objects, and trying to do so without invoking a merger, we more often than not found ourselves facing much more improbable scenarios.

We have two pieces of evidence that allow us to be a little more quantitative.  First, as discussed in section~\ref{sec:automated}, there is excellent agreement between our visual classification and automated classification techniques when counting up to confidence level~4, in our control sample.  (Unfortunately, because of the limitations of those techniques, this test is not applicable to our $24\mu m$-selected sample).  Then, an examination of the ACS and WFPC2 images of those $24\mu m$-selected objects that fall within the area covered by those instruments, {\em confirmed}, that is removed any doubt as to, the merging nature of many of these objects.  It even revealed the merging nature of two sources that we had thought, previously, to be disturbed disk galaxies (MIPS168 and MIPS8328).  On the other hand, it {\em never reversed} any of our NICMOS merger identifications.  This unidirectionality, along with the agreement with automated techniques in our control sample and the difficulty of explaining our observed morphologies through alternate means, all suggest that confidence levels one to four, although increasing in uncertainty, must still be primarily populated by mergers, and that contamination probably amounts to no more than a handful of objects.  We also argue, based on our redshifting analysis of the 1-Jy sample, that many, and possibly the majority, of category~5 and~6, $24\mu m$-selected objects also have their origin in an ongoing or recent merger event, and that the number of mergers in those two categories, thus, likely exceeds the number of false positives in the lower levels.  We therefore arrive at the conclusion that the correct balance between completeness and reliability should lie somewhere in the vicinity of confidence level~5, and that is where we draw the line in Figure~\ref{fig:mergfrac}\footnotemark.

\footnotetext{We note that the argument about confidence levels~5 and~6 objects does not hold for our control sample as most objects in that sample are extended disk galaxies for which we have no compelling reason to think that the presence of residuals should come from a merger more than from any normal asymmetry or structure in the disk.  The correct limit for that sample should, therefore, be drawn at level~4 and no further.  The merger fraction at that point reaches 25\%.}

Our best estimate of the merger fraction thus lies at 80\%.  We can also arrive at the same number by making the simplifying assumption that all confidence class 1 through 4 objects are mergers, while only half of the class 5 and 6 ones are.  Since all of the phase I through IV mergers belong to confidence classes 1 through 4 and almost all of the faint \& compact and late mergers belong to classes 5 and 6,  accounting for all of the former group and half of the latter also yields a similar fraction.  

In order to get a handle on the uncertainty associated with this number, it is useful to examine what our merger fraction would be under different assumptions.  For example, if we were to draw the line at category~4 and exclude all category~5 and~6 objects from the merger count, we would obtain a merger fraction of 72\%.  Counting all category~1 to~3 objects, but only half of those in categories~4 to~6, would similarly yield a 69\% fraction.  Such assumptions, although seemingly less likely, are still realistic, and cannot, therefore, be ruled out.   If we were to draw the line even further down, somewhere between categories~3 and~4, which amounts to counting all objects in confidence levels $1\textendash 3$ plus about half of those in level~4, our inferred merger fraction would be $\sim 60\%$.  60\% is also the fraction one would get by assuming a 20\% contamination in the first four categories while still rejecting all category~5 and~6 objects.  Such skepticism would imply that a large number of objects are simply masquerading as mergers, and that 40\% of our observed morphologies would have to be explained through alternate means.  This becomes very hard to justify, especially in the view of the fact that $\sim 60\%$ of our objects possess an important bulge component.  We feel that such a point of view is very much at the limit of what is reasonable to assume.   We thus find that our earlier lower limit of 62\% on the merger fraction derived from a simple comparison of our sample with the redshifted 1-Jy and control samples corresponds well to what we intuitively infer from direct inspection.  We therefore retain that number as our lower limit.

At the other end, Figure~\ref{fig:mergfrac} demonstrates that a sample of nearly 100\% merger-driven ULIRGs, namely the 1-Jy sample, can have a cumulative merger fraction curve very close to that of our sample.  It is thus not impossible that {\em all} of our merger candidates actually be mergers.  For this reason, we use the maximum value of the curve for all mergers in Figure~\ref{fig:mergfrac} as our upper limit.  It reads 91\%, and excludes only isolated spiral galaxies.  The range of reasonable values for the fraction of mergers in our sample is, thus, 62 to 91\%.

Historically, infrared-luminous galaxies have been divided between LIRGs and ULIRGs.  For comparison purposes, we thus also performed separate accountings for those two categories individually using the same dividing lines.  The numbers add up to a $57 \textendash 80\%$ merger fraction among LIRGs with a best estimate of $71\%$, and to a merger fraction between 65\% and 96\% among ULIRGs with a best estimate at 87\%.  The reason why our merger fraction is so large among LIRGs is that most of our LIRGs have luminosities just below $10^{12} L_{\odot}$, the mean luminosity being $\langle \log L_{IR}/L_{\odot}\rangle = 11.63$.  The numbers we find for ULIRGs, on the other hand, indicate that, like their low redshift counterparts, they too have their origin primarily in mergers.

\subsection{Interlude:  postulate for results to follow}

In the sections that follow, we look for correlations between various properties of our galaxies and their morphological class.  Incorporating confidence levels into all of our relations would be impractical.  We therefore choose to leave them out and include in our subsequent analysis {\em all} objects of every morphological class irrespective of their confidence level.  We argue this is justified, because the trends discussed in this paper are robust and would not be affected by the reshuffling of a few objects around morphological classes.  They focus mainly on the distinction between early and more advanced mergers which are easily distinguishable by whether the source consists of two objects/nuclei or one. Cross-contamination between these two classes of objects is therefore likely to be minimal.  Moreover, Figure~\ref{fig:mergfrac} demonstrates that early mergers and advanced mergers have a very similar curves as for their cumulative merger fraction as a function of confidence level, implying that eventual interlopers are likely to affect both categories equally (though, as argued in the last section, their number is expected to be very small).  What's more, the objects associated with the least confidence to a merger event (faint \& compact objects, and late mergers) have very uniform spectral properties, so that the conclusions remain the same for those categories irrespective of which objects one chooses to include or not.

This assumption, or approximation, we choose to pursue greatly simplifies later analysis, and allows us to retain the focus on the more physical merging sequence.

\subsection{Morphological split along the merging sequence \label{sec:morphsplit}}

Having established the overall merger fraction, we now look into the distribution of our mergers along the merging sequence (see section~\ref{sec:morphscheme} for a description of our classification).  This distribution is especially meaningful when laid in comparison with other samples.  For that purpose, we continue to use the {\em 1-Jy} morphological sample of local ULIRGs from \citet{Veilleux02}, and the results of our simulations to compare with local galaxies as if they were observed at redshift one with the same kind of data.

We show in Table~\ref{tbl:proportions} the morphological split of our sample (also represented in graphical form in Figure~\ref{fig:mergfrac}) and compare it to that of the 1-Jy sample.  We list numbers for both the intrinsic and redshifted {\em 1-Jy} sample, the percentages for the latter being obtained by applying the proportions found in table~\ref{tbl:simulations} to the sample's intrinsic morphological distribution.  Finally, we divide our sample into two redshift bins separated at $z=1.5$ and show the morphological composition of each sub-sample.  Our full sample roughly splits into $1/3$ early mergers, $1/3$ advanced mergers and $1/3$ other classes.  Among the last third, the most represented types are old mergers (14\%), isolated spirals (9\%) and faint \& compact objects (8\%).

\begin{deluxetable*} {cccccccc}  
\tablecolumns{8}
\tablewidth{0pt}
\tablecaption{Morphological content of different samples\tablenotemark{a}}
\tablehead{
	\colhead{Merging sequence} &
	\colhead{Morphological class} &
	\colhead{1-Jy sample} &
	\colhead{Redshifted} &
	\colhead{This} &
	\colhead{$z < 1.5$} &
	\colhead{$z \ge 1.5$} &
	\colhead{Unknown} \\
	\colhead{} &
	\colhead{} &
	\colhead{} &
	\colhead{1-Jy sample} & 
	\colhead{sample} & 
	\colhead{only} &
	\colhead{only} &
	\colhead{redshift}
}
\startdata
\multirow{3}{*}{Isolated objects} & Face-on spirals & 1\% (1) & 1\% & 5\% (7) & 10\% (7) & 0 & 0 \\
& Edge-on disks & 0 & 0 & 4\% (5) & 6\% (4) & 0 & 1 \\
& Irregular galaxies & \nodata & 4\% & \nodata & \nodata & \nodata & \nodata \\ \\
First approach & Close pairs (phase I) & 0 & 0 & 3\% (4) & 4\% (3) & 2\% (1) & 0 \\ \\
\multirow{3}{*}{Early mergers} & First contact (phase II) & 0 & 0 & 2\% (3) & 3\% (2) & 2\% (1) & 0 \\
& Pre-mergers (phase III) & 39\% (46) & 35\% & 27\% (36) & 12\% (8) & 46\% (25) & 3 \\
& Triplets & 4\% (5) & 4\% & 3\% (4) & 3\% (2) & 4\% (2) & 0 \\ \\
\multirow{3}{*}{Coalescence} & Advanced & & & & & & \\
& Mergers (phase IV) & 44\% (52) & 29\% & 31\% (41) & 37\% (25) & 28\% (15) & 1\\
& Faint \& Compact & \nodata & 12\% & 8\% (11) & 9\% (6) & 7\% (4) & 1 \\ \\
\multirow{3}{*}{Late mergers} & Old mergers (phase V) & 12\% (14) & 7\% & 14\% (18) & 12\% (8) & 9\% (5) & 5 \\
& Regular bulges & 0 & 9\% & 3\% (4) & 4\% (3) & 2\% (1) & 0 \\
& Pure point sources & \nodata & 0 & 1\% (1) & 0 & 0 & 1 \\
\enddata
\tablenotetext{a}{Numbers without percentages represent total number of objects.  Ellipses indicate that morphological class is not used for that sample.}
\label{tbl:proportions}
\end{deluxetable*}

Comparison between our objects and local ULIRGs reveals that not only both samples are composed, to a high fraction, of mergers, but that all phases of the merging process are represented in relatively similar proportions.  In particular, both are dominated by singly-nucleated merger remnants (coalescence and post-coalescence mergers).  We find, however, a number of subtle differences between the two samples, and address these differences below.

The main difference is in the number of isolated spirals.  Whereas all but one of the {\em 1-Jy} galaxies are mergers, we have, in our sample, 12 isolated disks (7 face-on and 5 edge-on).  We also have 11 faint \& compact objects (8 of which show residuals after subtraction of the main profile and 3 that do not).  Some of these objects could also be intrinsically isolated (i.e. not merger remnants).  This difference in the number of isolated disks is explicable by the fact that some of our lowest redshift objects are sub-ULIRG (see Figure~\ref{fig:lum_vs_z}).  

The largest discrepancy among merger classes occurs in phase V objects, for which we would have expected, from our redshifting of local ULIRGs, to find a lower fraction, and to observe, instead, many more objects with no signs at all of interaction (that is regular bulges).  However, we see quite the opposite, with phase V objects being far more common than simple bulges.  This suggests that {\em our objects carry stronger merger signatures than do local ULIRGs}.  We see that effect among advanced mergers as well when visually comparing tidal tails and streams of objects in our sample (Figures~\ref{fig:examples2} to~\ref{fig:examples4}) to those of the redshifted {\em 1-Jy} objects (Figure~\ref{fig:simulations}).  We find that those of our sample tend to be less diffuse, higher surface brightness, and readily detectable.  Among our redshifted phase IV local ULIRGs, on the other hand, only Mrk273 still shows a clear tidal tail at $z=1$.  More typical galaxies, such as those shown in Figure~\ref{fig:simulations}, display only very faint streams and tails that are hard to detect but for smoothing of the residuals.  We speculate that this effect is one more manifestation of the enhanced gas fractions and star formation rates that exist at those redshifts.

Although, overall, our sample shows a number of early mergers very close to that of the redshifted {\em 1-Jy} sample, the proportion of these objects is three times as large at high redshift than it is at $z < 1.5$.  We defer discussion of the large number of early mergers at high redshift until the next section, where we argue for a redshift evolution in the morphology of ULIRGs.  The low number of early mergers among our $z \sim 1$ objects compared to that of the 1-Jy sample, on the other hand, can probably be explained by the fact that the two samples were selected at different rest-frame wavelengths, namely at $\lambda \sim 12\mu m$ for our objects and at $60\mu m$ for the 1-Jy sample.  Many objects in the {\em 1-Jy} sample have fairly low mid-IR fluxes (those with high $f(60\mu m) / f(25\mu m)$ ratios).  Such objects are less likely to make it into our sample.  They are, however, predominantly early-stage mergers \citep{Veilleux02}.

\subsection{Redshift evolution of morphology \label{sec:redshift_evolution}}

\begin{figure}[htbp]
\begin{center}
\includegraphics[width=3.5in]{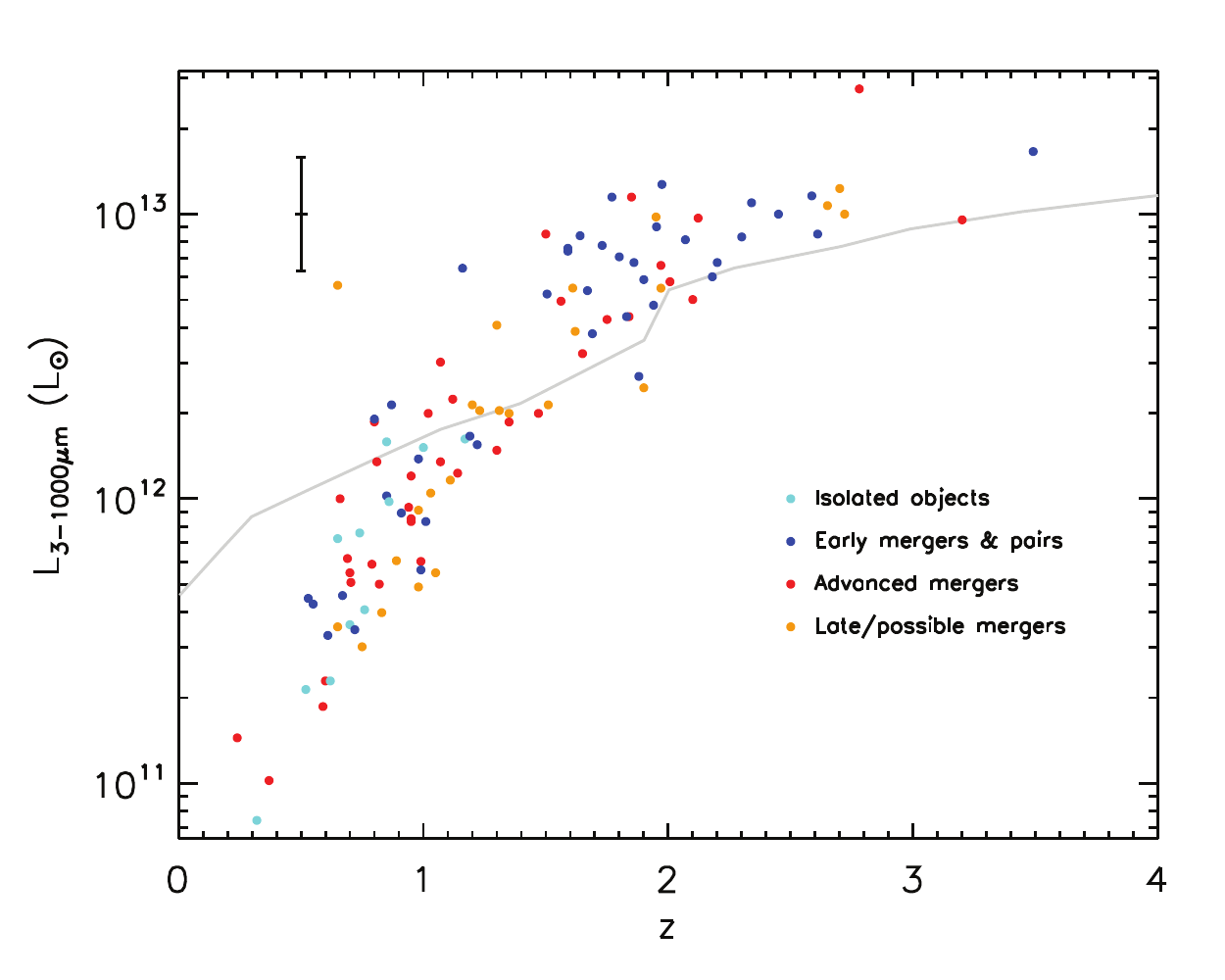}
\caption{Distribution of IR-luminosity for all objects of known redshift in our sample as a function of redshift.  The average $3\sigma$ uncertainty is shown in the upper left corner.  The light gray line represents the IR-luminosity at which star formation is predicted, from models of \citet{Hopkins10a}, to transition from quiescent to merger-driven.  Comparison between data and models is discussed in \S~\ref{sec:theory}.}
\label{fig:lum_vs_z}
\end{center}
\end{figure}

In Figure~\ref{fig:lum_vs_z} we show the total infrared ($3\textendash 1000 \mu m$) luminosity as a function of redshift for all galaxies of known redshift and luminosity in our sample.  The distribution is typical of a flux-limited sample in which the more luminous sources are at higher redshift and vice versa.  The sample seems to split naturally, though, at $z=1.5$ or $L_{IR} = 10^{12.5} L_{\odot}$, with clearly different populations on each side.  We explore the differences between those two populations, as well as with local ULIRGs, further in this section.

\begin{figure}[htbp]
\begin{center}
\includegraphics[width=3.5in]{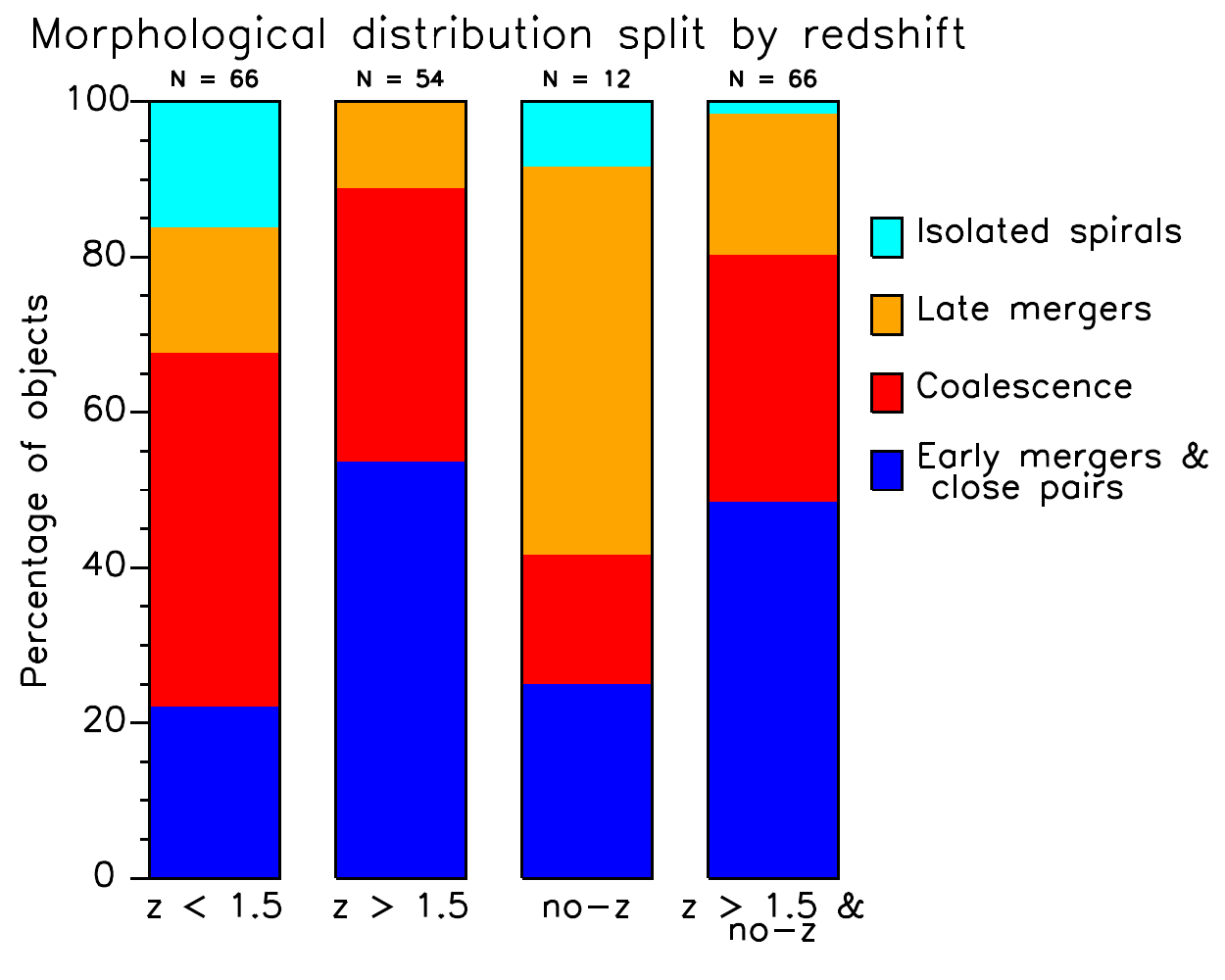}
\caption{Number of objects of each morphological class in two redshift bins.  Also distribution of morphologies for objects without known redshift.  There are more early stage mergers at high-redshift even if all objects without redshifts were put in the high-$z$ bin.}
\label{fig:morph_byz}
\end{center}
\end{figure}

Figure~\ref{fig:morph_byz} represents the morphological distribution of objects in our sample, split in high and low-redshift bins at $z = 1.5$, with the addition of a no-$z$ column for objects with unknown redshift.  It shows a clear shift from singly-nucleated mergers at low redshifts to earlier phase mergers at $z \ge 1.5$.  This is true even if all objects of unknown redshift were at high-$z$, as illustrated in the last column.  Splitting the sample in luminosity at $L_{IR} = 10^{12.5} L_{\odot}$ instead of in redshift yields identical results.

\begin{figure*}[hbtp]
\begin{center}
\includegraphics[width=3.5in]{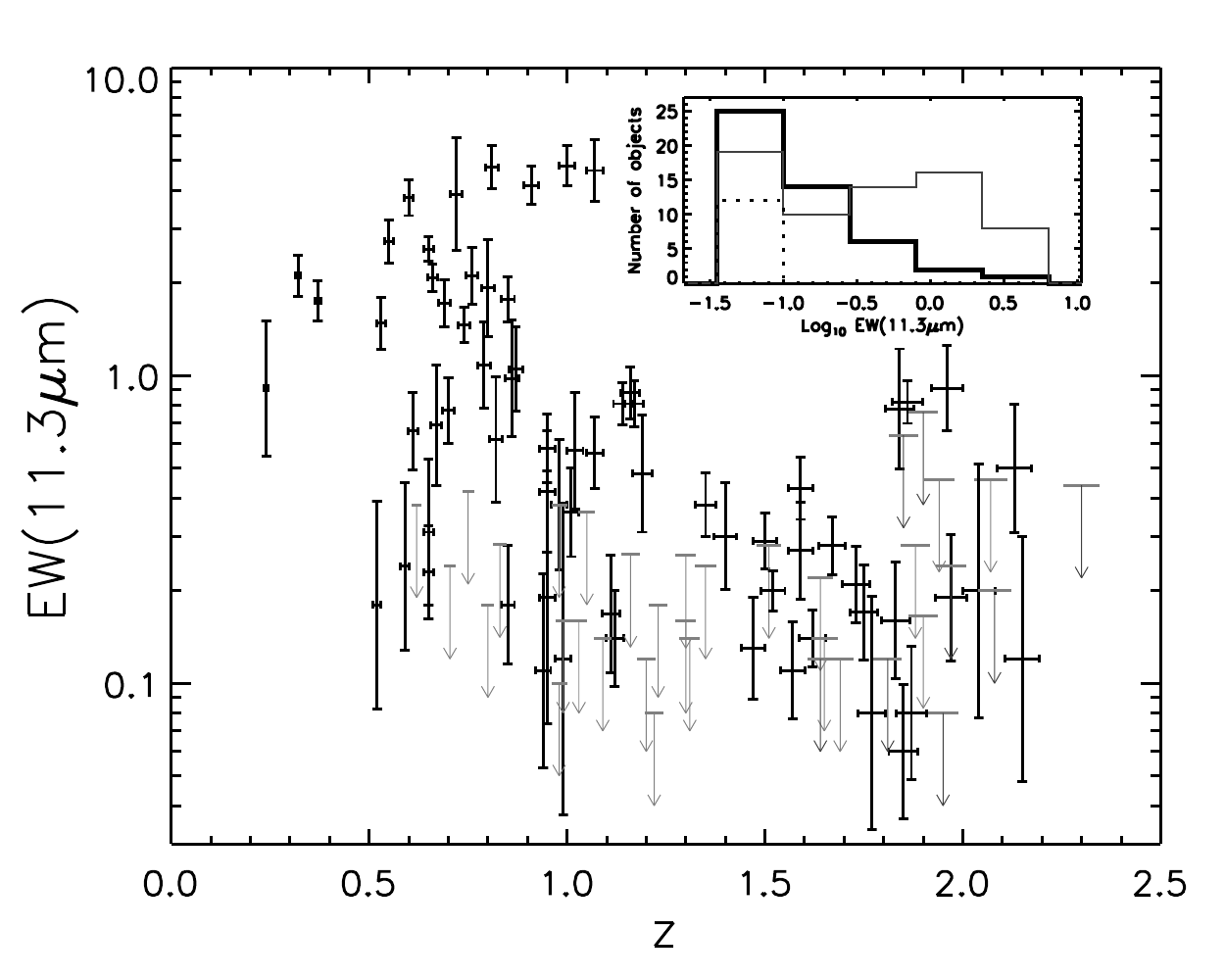}
\includegraphics[width=3.5in]{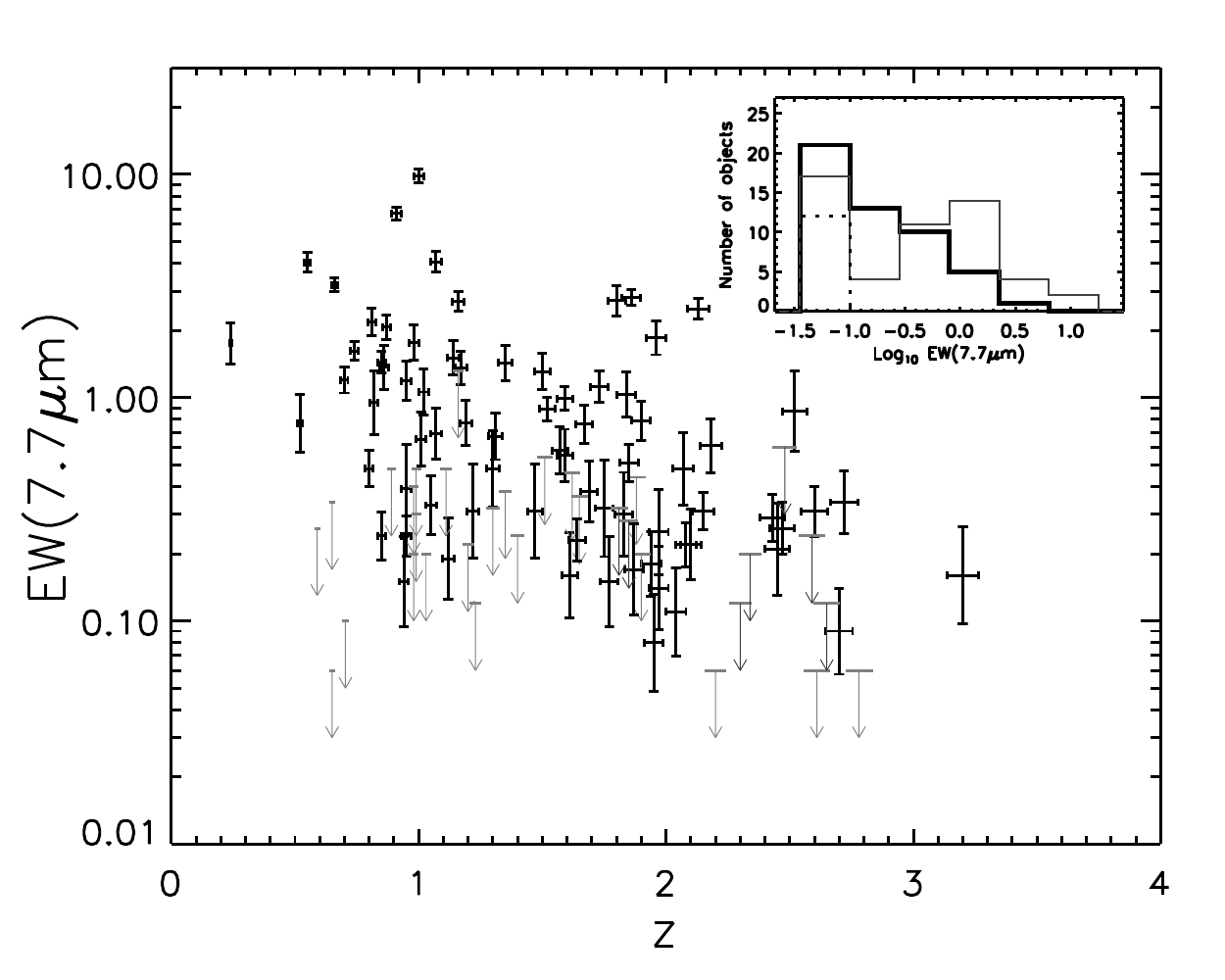}
\caption{{\em Main panels:} $11.3\mu m$ and $7.7\mu m$ equivalent widths of objects in our sample as a function of their redshift.  Objects without the proper spectral coverage are omitted from the respective plots, but every object appears at least once.  {\em Insets:}  Distribution of equivalent widths split by redshift.  The thick black histogram represents objects with $z \ge 1.5$, the thin grey one, objects with $z < 1.5$ and the dotted histogram, objects of unknown redshift.  Objects with no detected PAH emission were all added to the lowest bin.  Higher redshift objects tend to have lower PAH EWs in our sample.}
\label{fig:ew_vs_z}
\end{center}
\end{figure*}

So are we seeing redshift evolution or a luminosity selection effect?  We argue that these observations cannot be explained by luminosity, since they go opposite to the luminosity trend known locally, whereby more luminous objects tend to be in more advanced stages of merging \citep{Veilleux02,Veilleux09b}.  Neither can they be explained by an increase in the importance of star formation activity since we actually find it to decrease with redshift in our sample, as illustrated in Figure~\ref{fig:ew_vs_z} which demonstrates the smaller PAH equivalent widths of our high redshift objects compared to that of our lower redshift objects;  this, despite the fact that, at $z\sim2$, sources are selected at the rest-frame $\sim 8\mu m$ where the presence of PAH features could potentially significantly increase their observed flux\footnotemark.  From the relation between redshift and luminosity in our sample (Figure~\ref{fig:lum_vs_z}), we can, incidentally, deduce that our objects follow the same trend as found in studies of local ULIRGs where PAH EWs have been shown to decrease, on average, with luminosity \citep{Lutz98,Tran01,Veilleux09a}.  
The analogy with local ULIRGs, however, would predict that at higher luminosities and lower EWs, which, in our sample, occur at higher redshifts, we find more advanced mergers.  Since we observe exactly the opposite, we must conclude that there is intrinsic morphological evolution with redshift (which is {\em unlike} PAH EWs whose relation with redshift seems to be driven primarily by luminosity).  This morphological shift towards early mergers at $z \ge 1.5$ therefore implies that, at those redshifts, more activity, and especially more AGN activity as testified to by the low PAH EWs of those objects, occurs in early stages of merging. 
We explore possible causes for this shift in the discussion section. 

\footnotetext{The presence of $11.3\mu m$ and $12.7\mu m$ PAH features can also favor selection among lower redshift objects, but those lines typically carry much less flux than the $7.7\mu m$ complex.  In addition, the contribution of the continuum is also less important at high redshift due to bandpass shrinking.}

\subsection{Relation between mid-IR properties and morphology \label{sec:midIR_vs_morph}}

In this section we look at the relation between the physical origin of the mid-IR flux of our objects, that is whether it is from star formation or AGN activity (or a combination of both), and their observed morphology.  As mentioned in section~2, we use both the EW of the $11.3\mu m$ and $7.7\mu m$ PAH features as indicators of the respective contribution of the two phenomena.

\begin{figure*}[htbp]
\begin{center}
\includegraphics[width=1.7in]{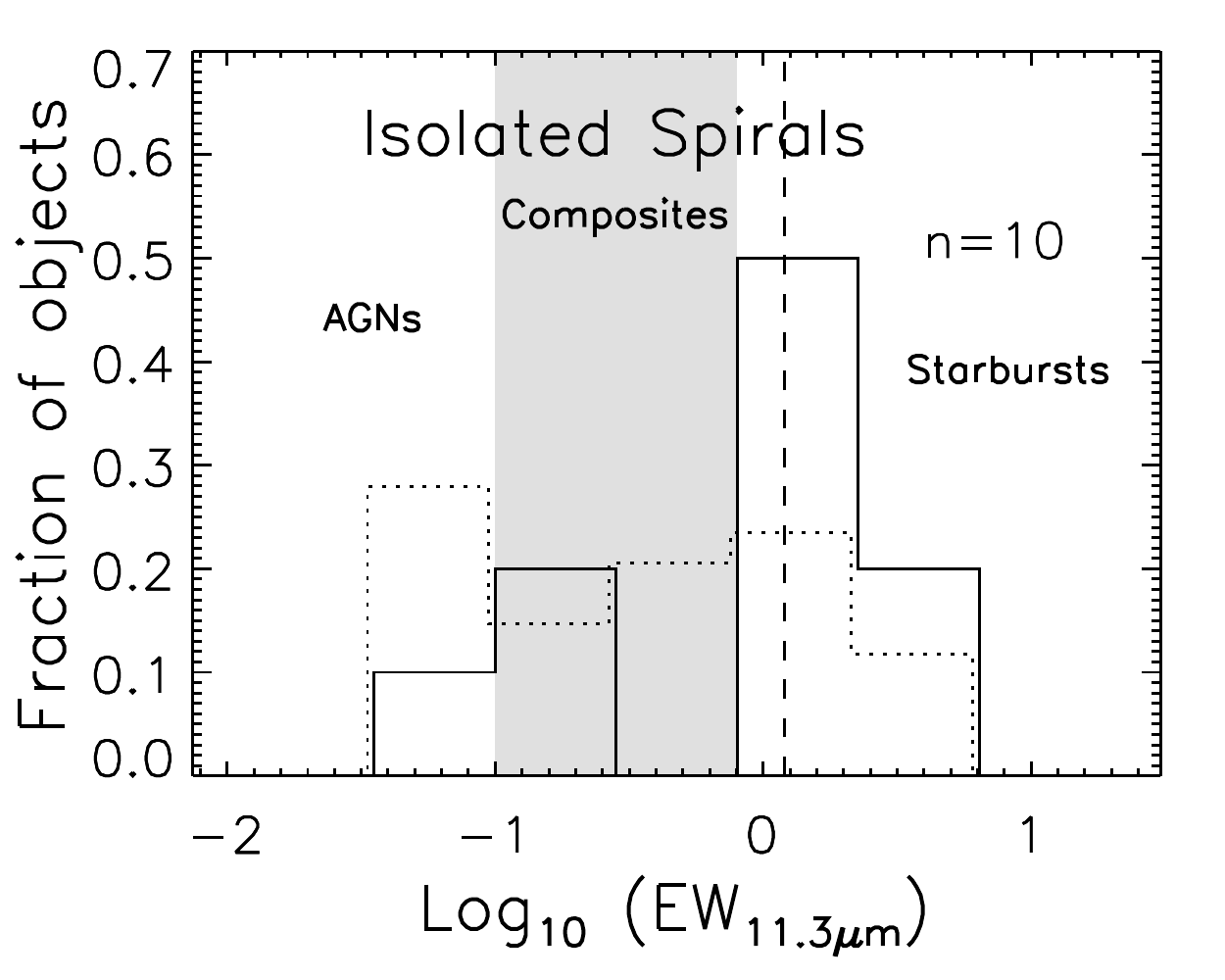}
\includegraphics[width=1.7in]{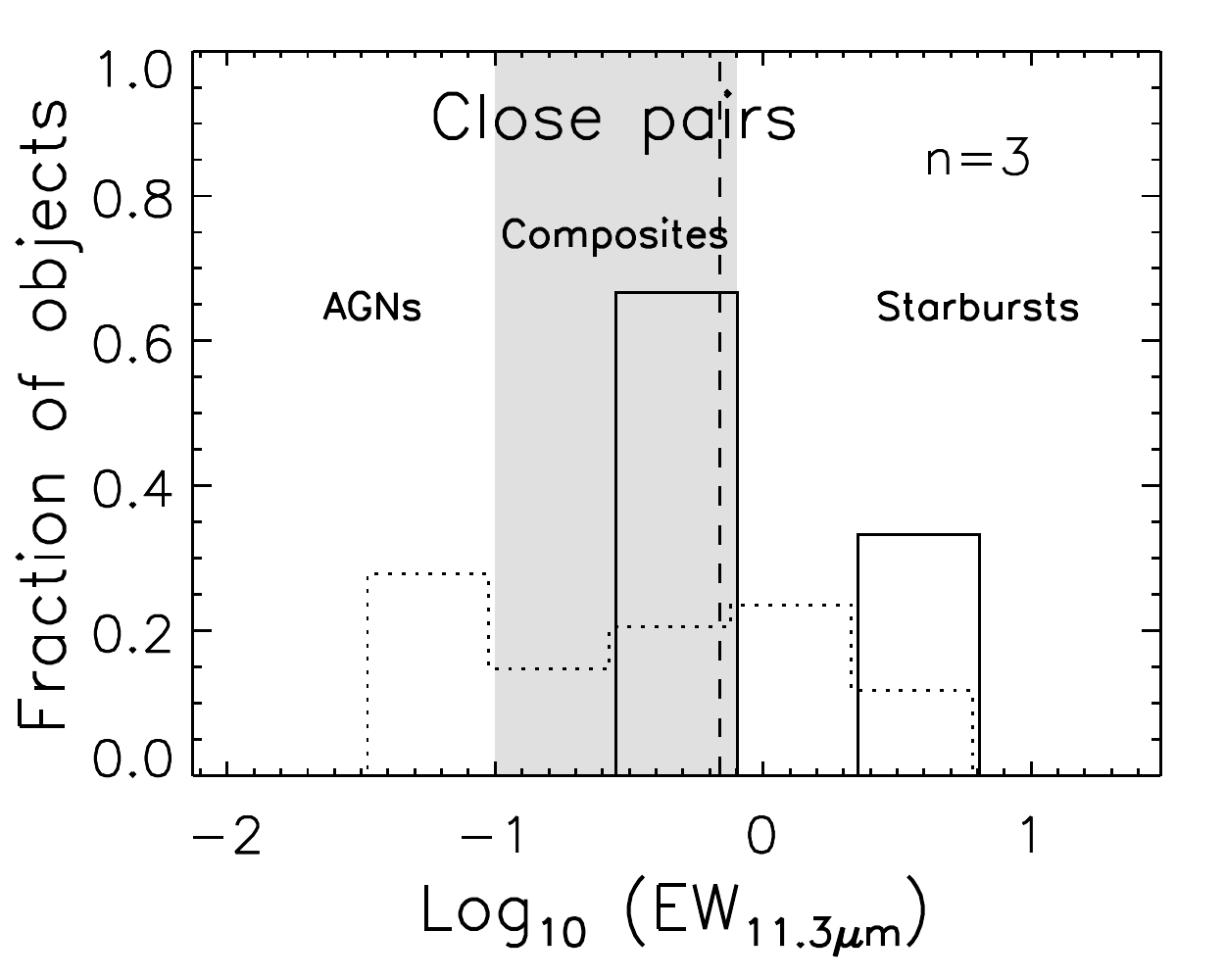}
\includegraphics[width=1.7in]{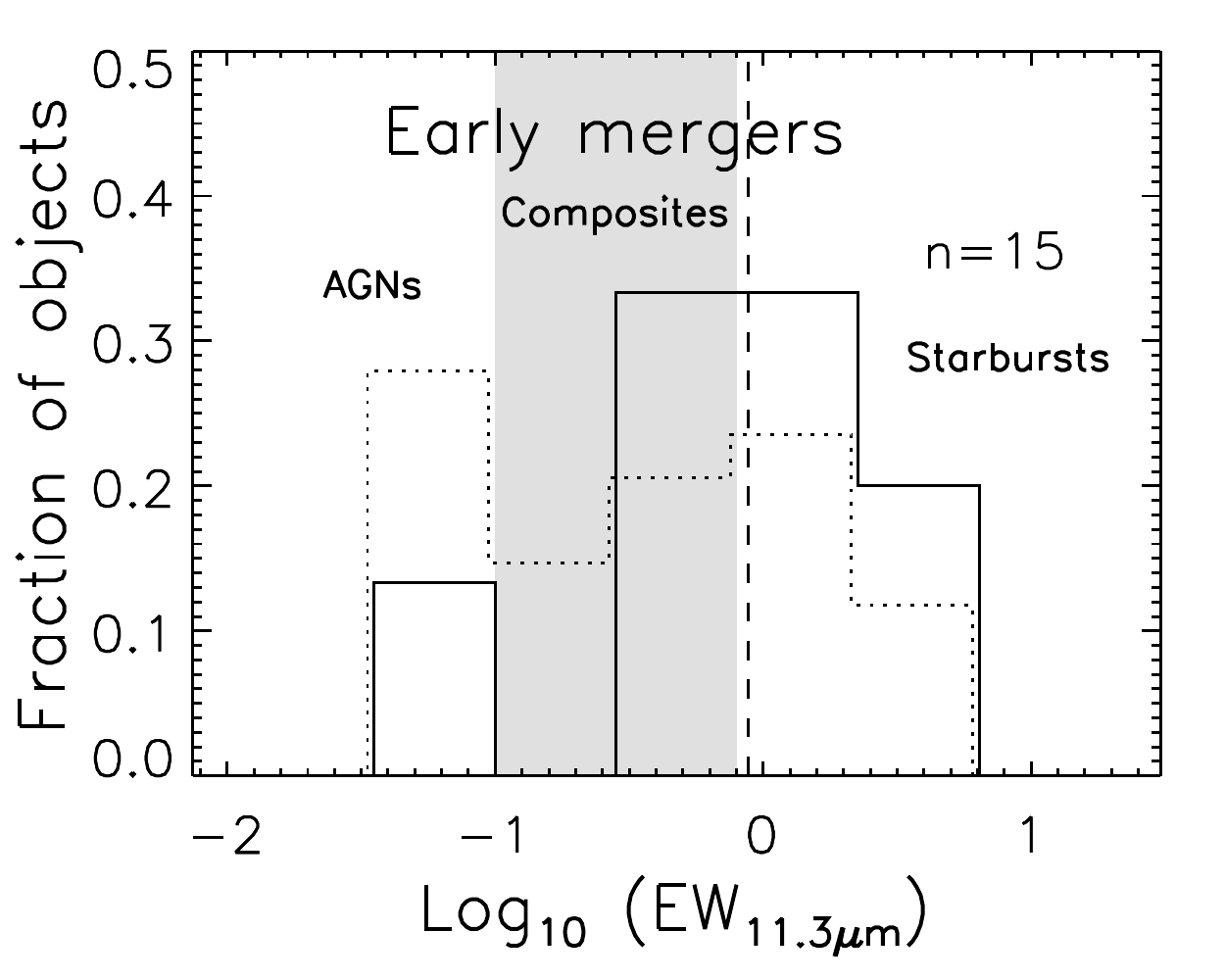}
\includegraphics[width=1.7in]{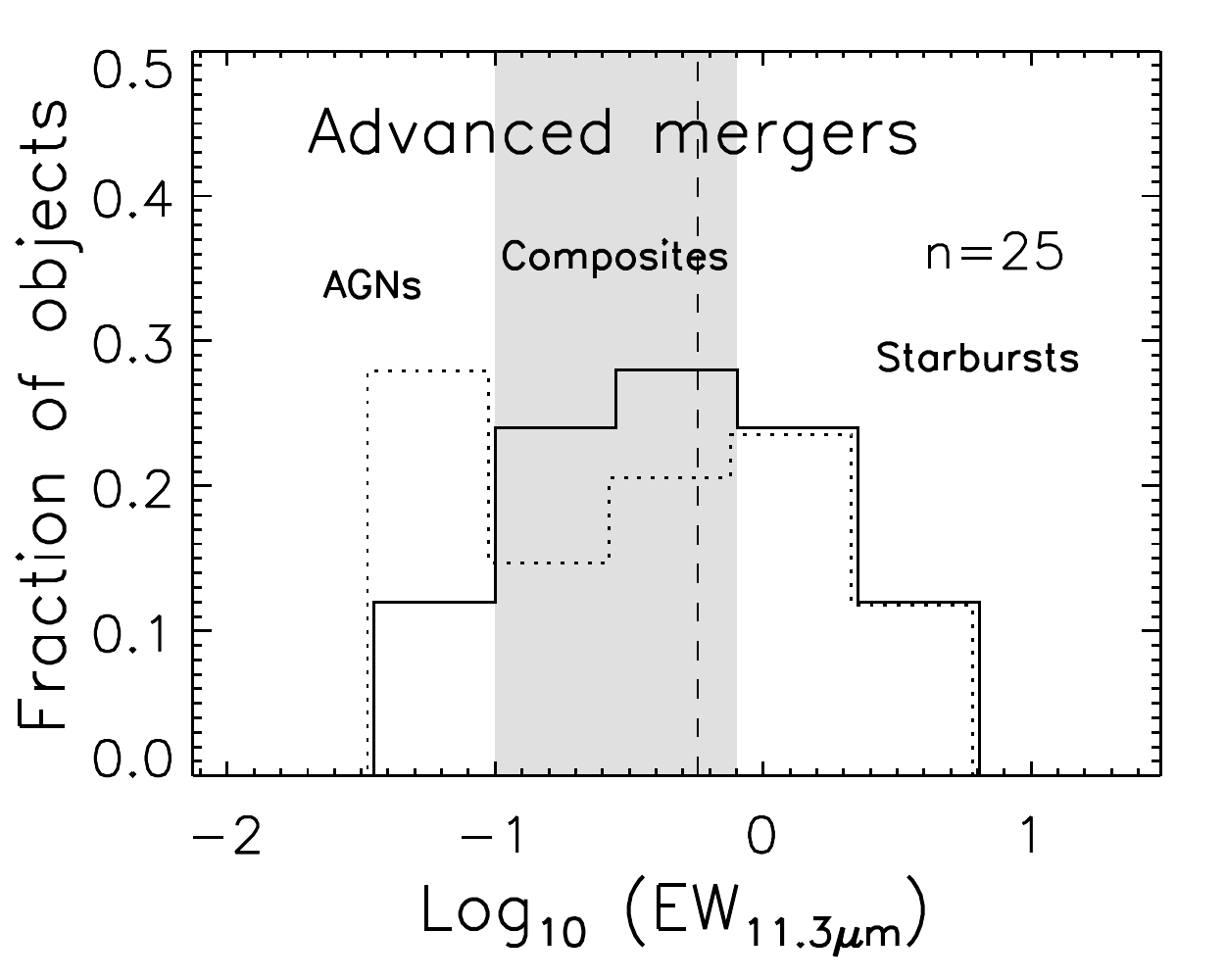}\\
\includegraphics[width=1.7in]{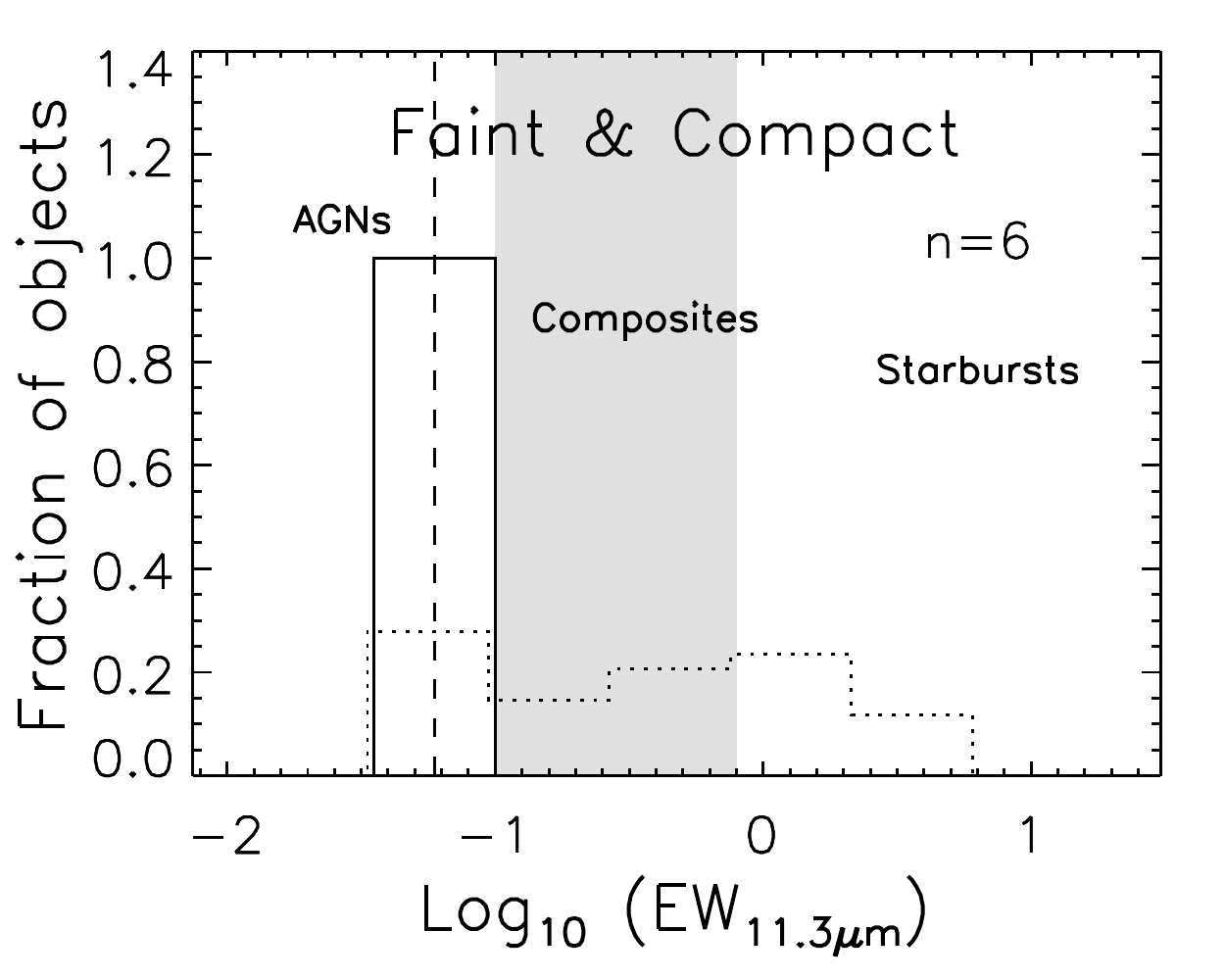}
\includegraphics[width=1.7in]{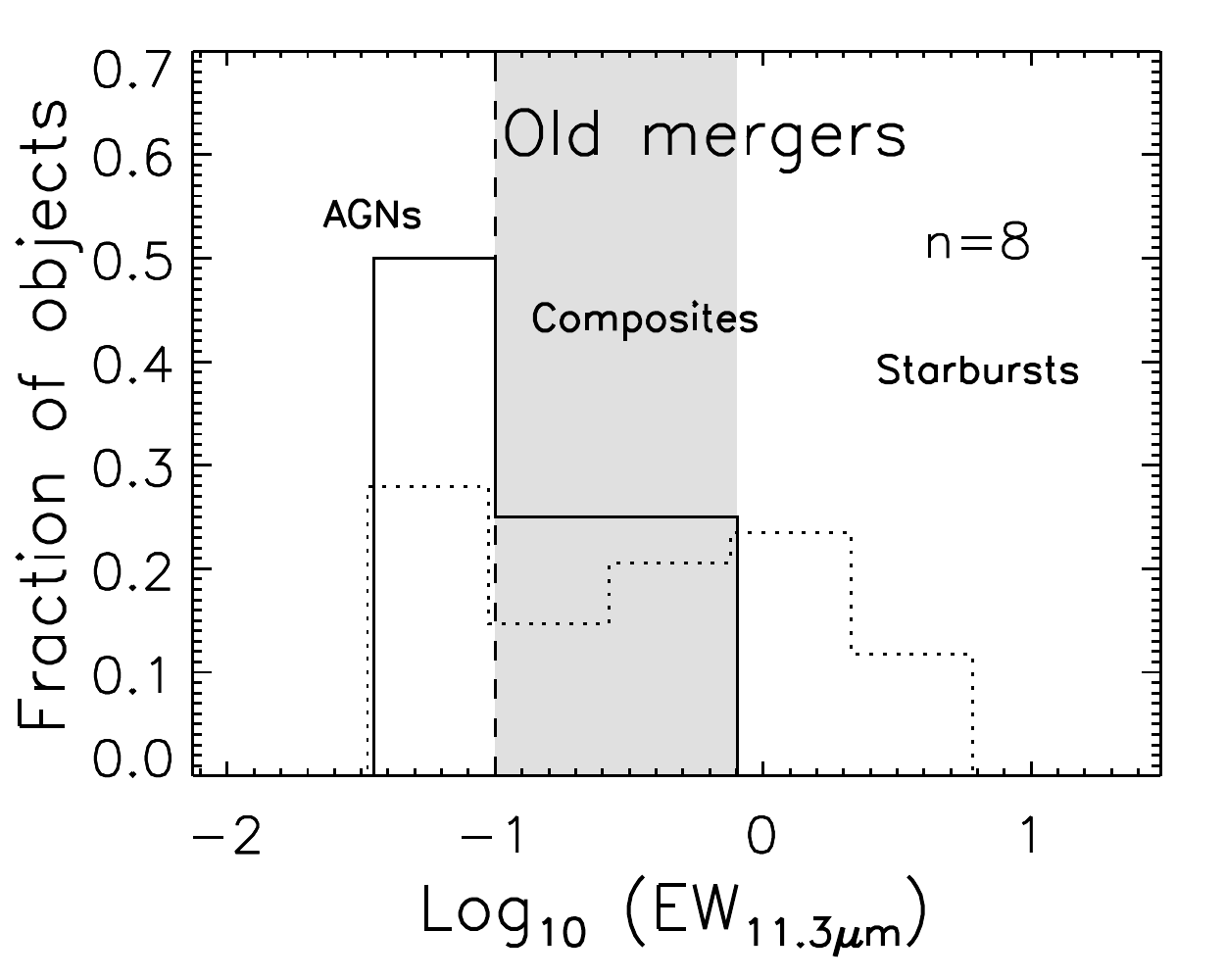}
\includegraphics[width=1.7in]{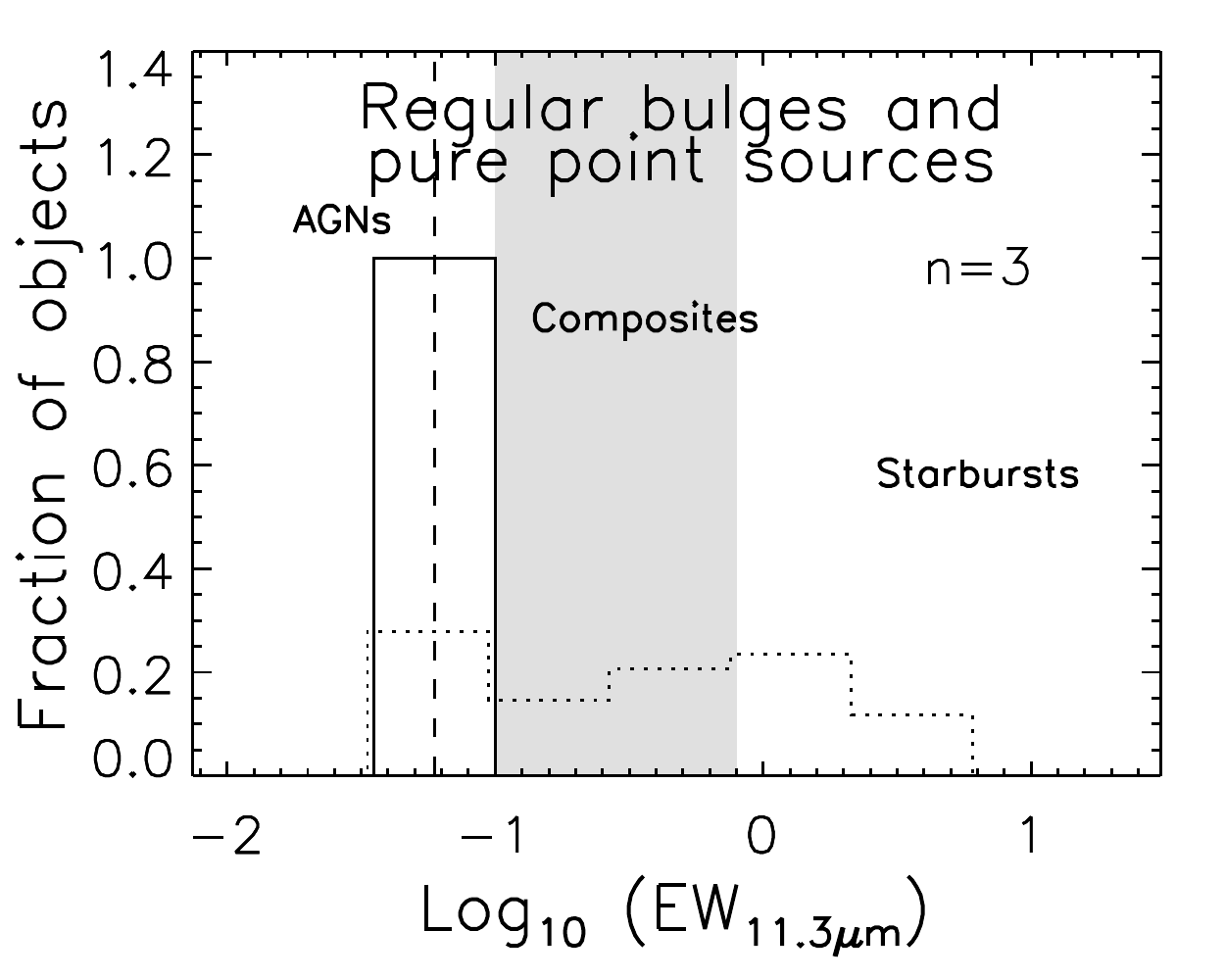}
\caption{Distribution of the $11.3\mu m$ PAH equivalent width for $z < 1.5$ objects split into morphological classes.  The dotted histogram represents the distribution of the sub-sample as a whole and has been slightly shifted for clarity.  The grey area delineates the region of AGN-starburst composite systems.  Starburst-dominated spectra fall to the right of that region whereas AGN-dominated ones populate the low-end of the distributions.  Objects without a detected feature at $11.3\mu m$ were all put into the lowest bin.  The dashed lines represent the median EW of each morphological class.  This figure demonstrates an average progression from star formation to black hole accretion along the merging sequence.}
\label{fig:ew_bymorph_lowz}
\end{center}
\end{figure*}

\begin{figure*}[htbp]
\begin{center}
\includegraphics[width=1.7in]{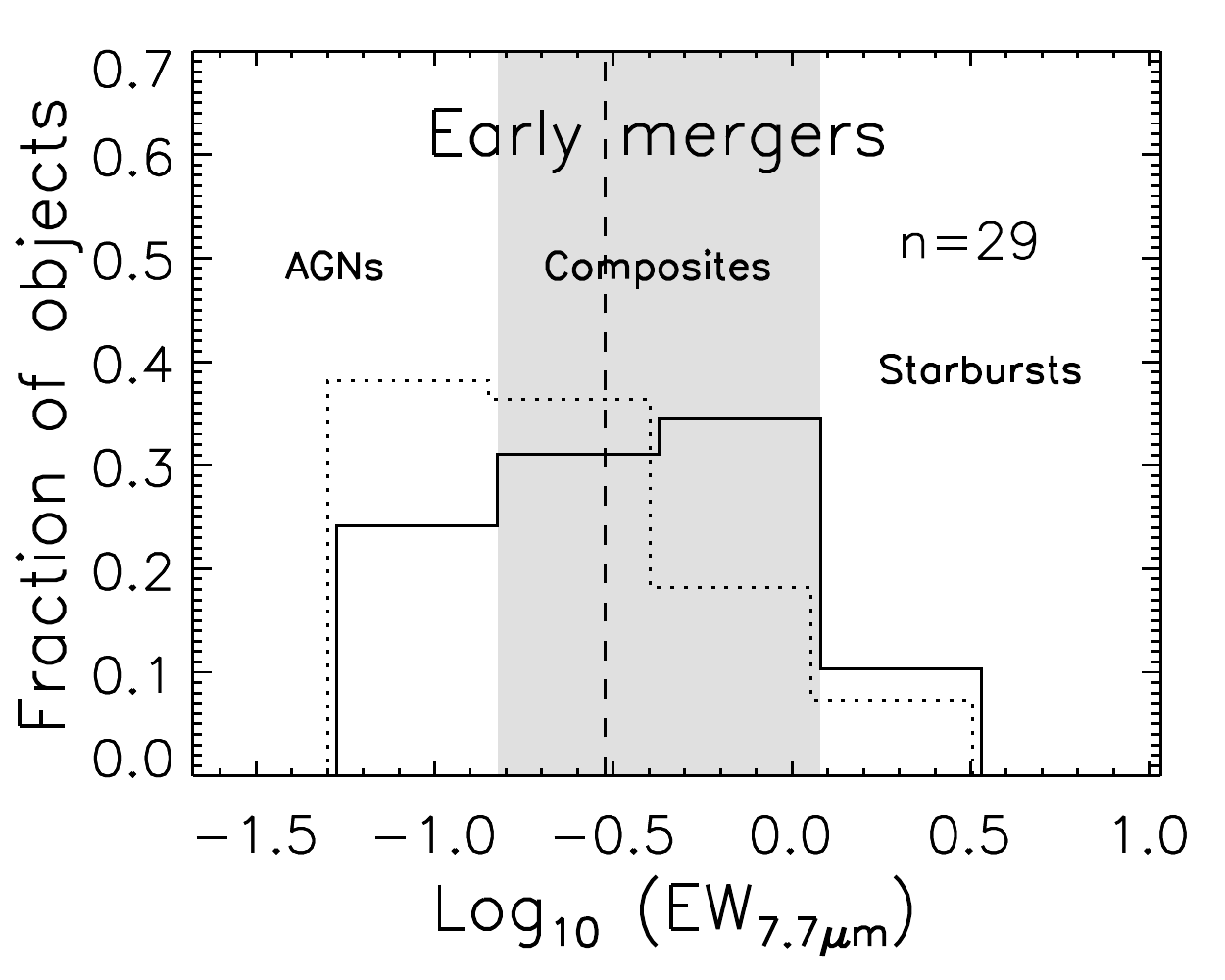}
\includegraphics[width=1.7in]{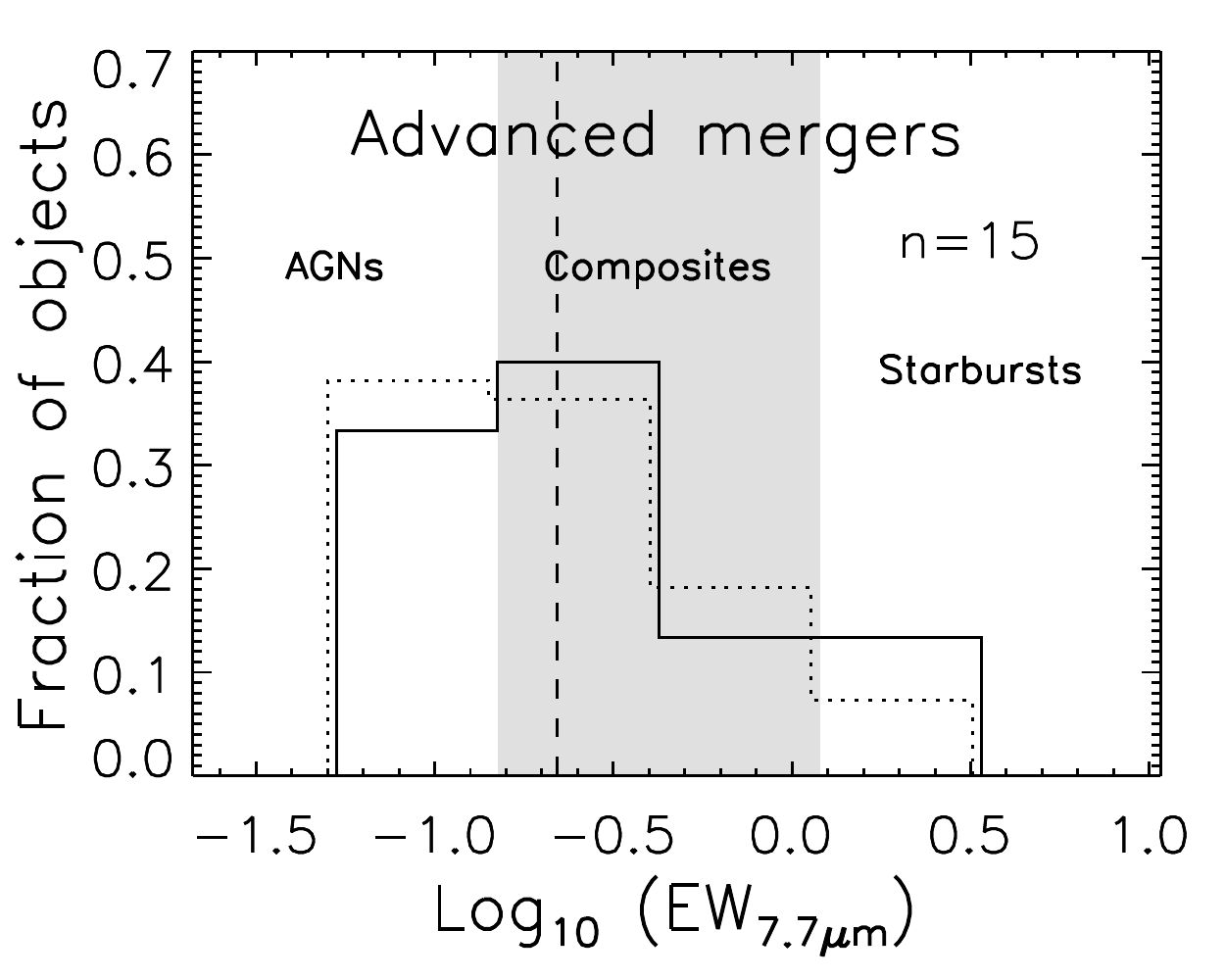}
\includegraphics[width=1.7in]{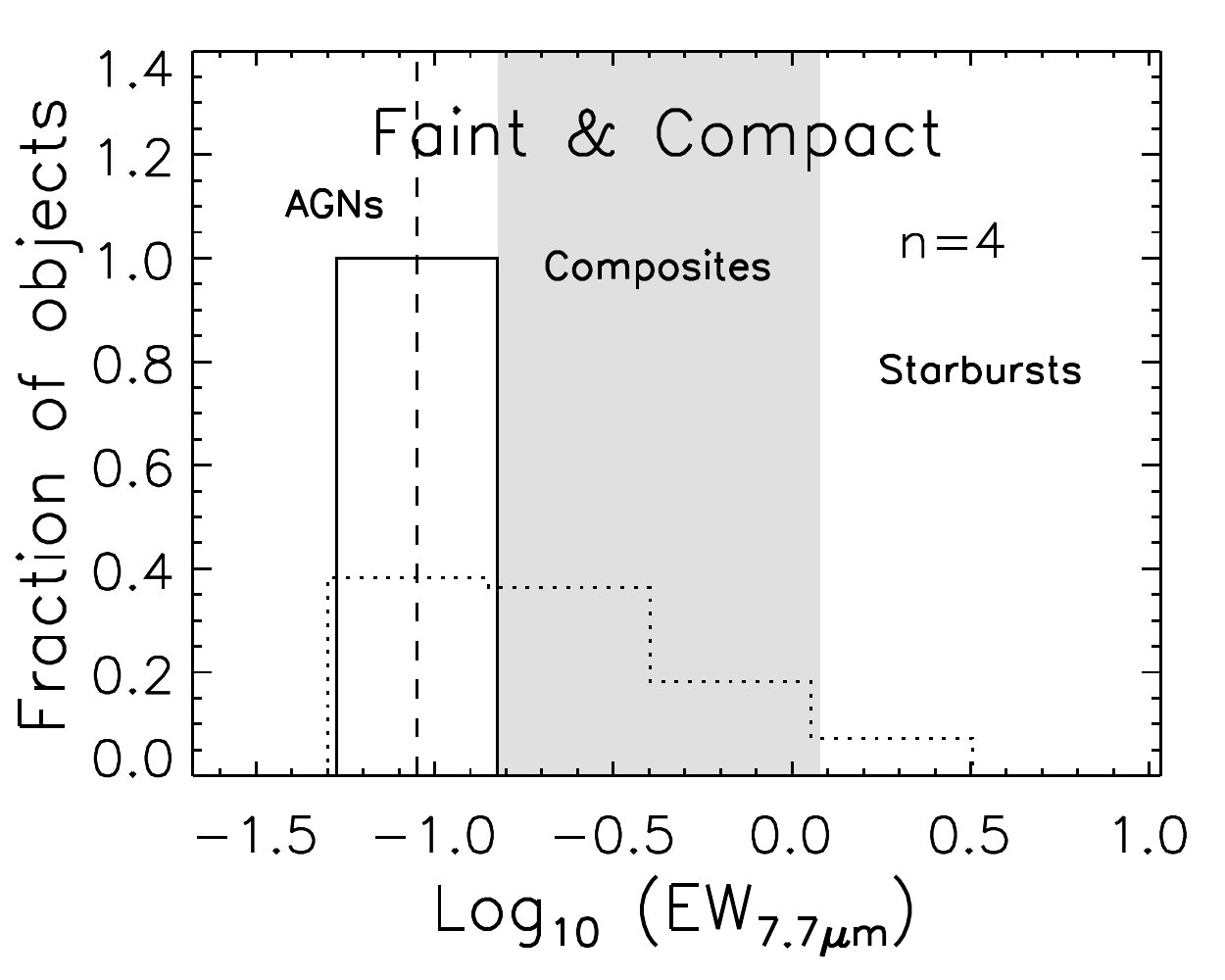}
\includegraphics[width=1.7in]{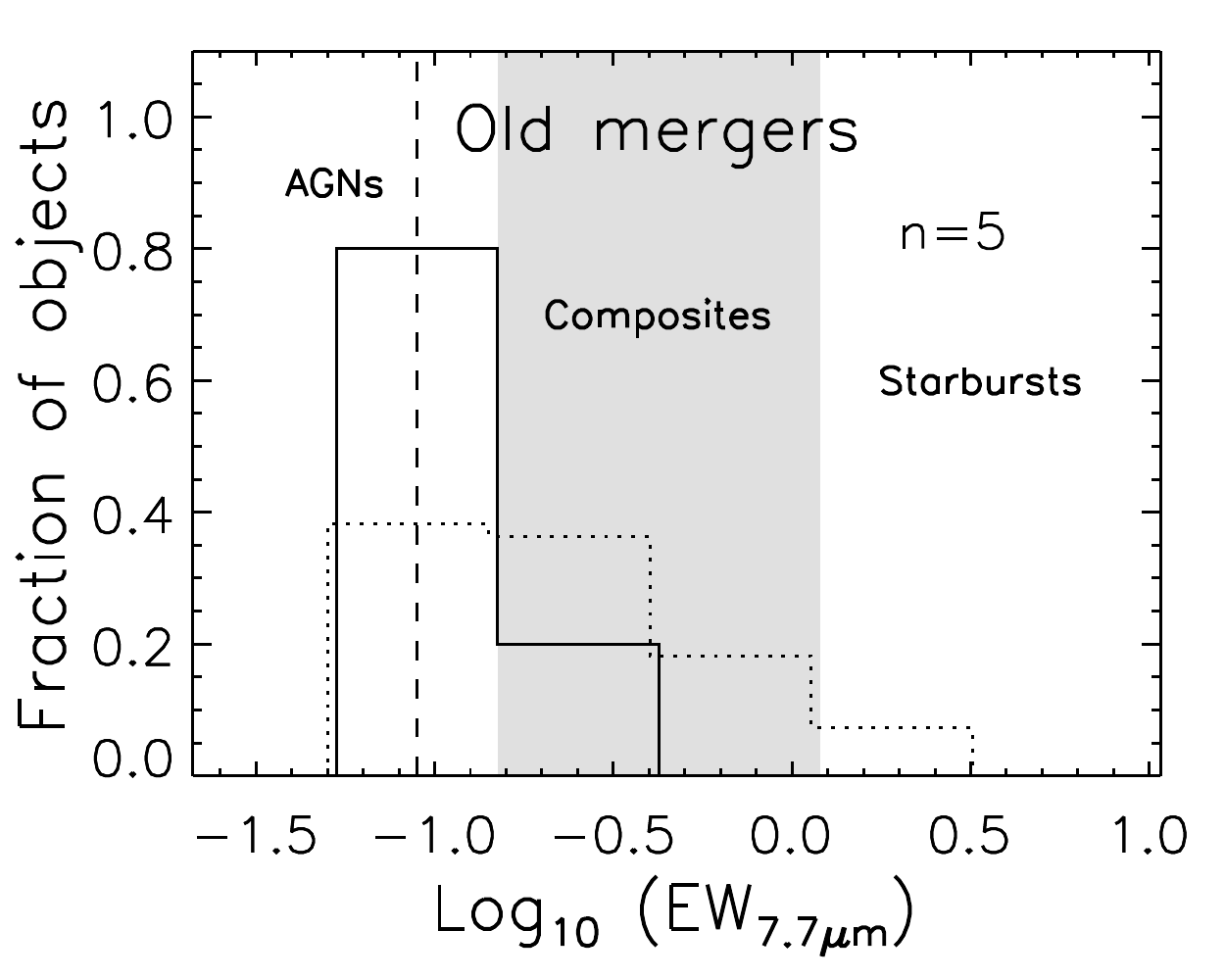}
\caption{Same as Figure~\ref{fig:ew_bymorph_lowz}, but for $z \ge 1.5$ objects using the $7.7\mu m$ PAH equivalent width.}
\label{fig:ew_bymorph_highz}
\end{center}
\end{figure*}

\begin{figure*}[htbp]
\begin{center}
\includegraphics[width=1.7in]{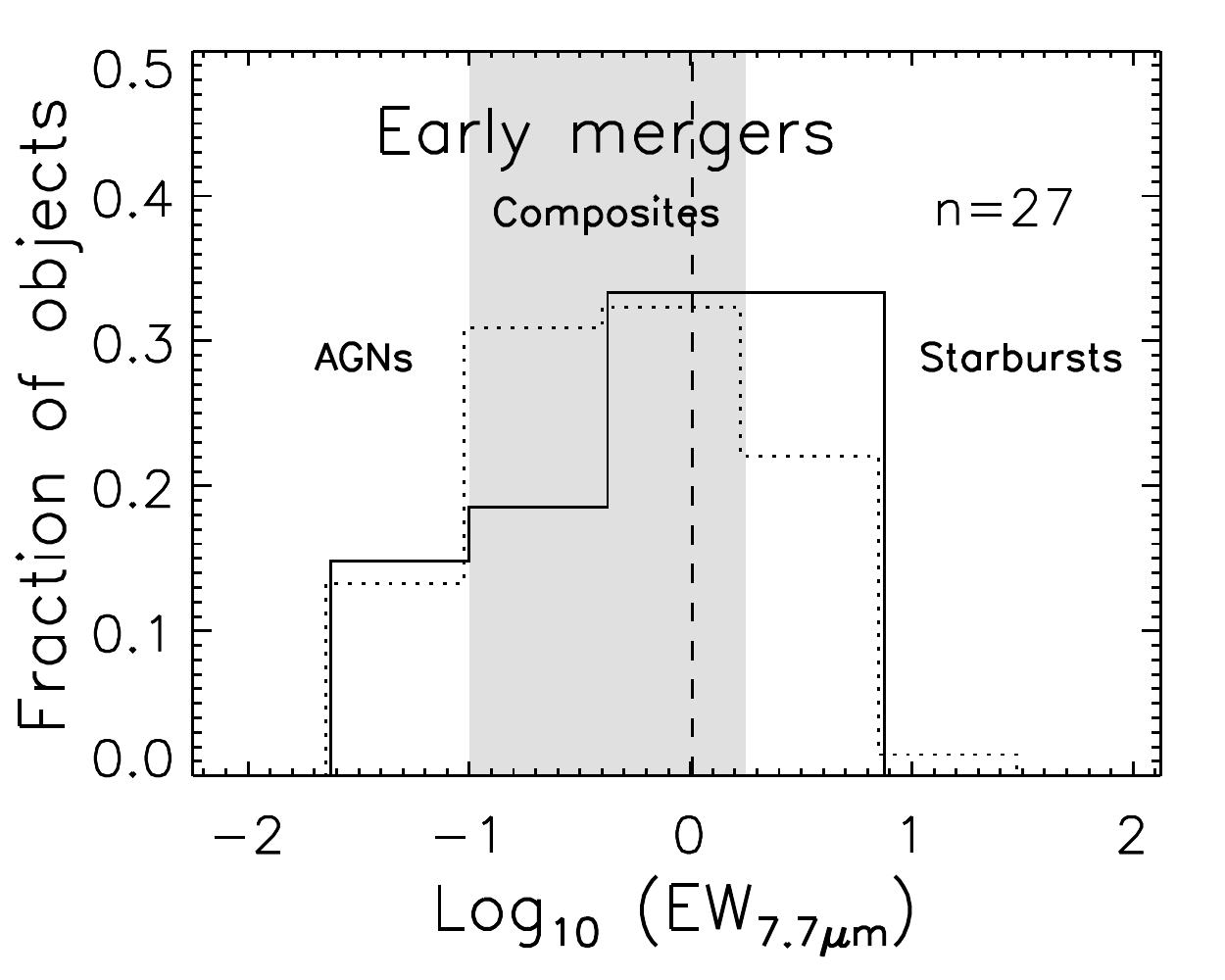}
\includegraphics[width=1.7in]{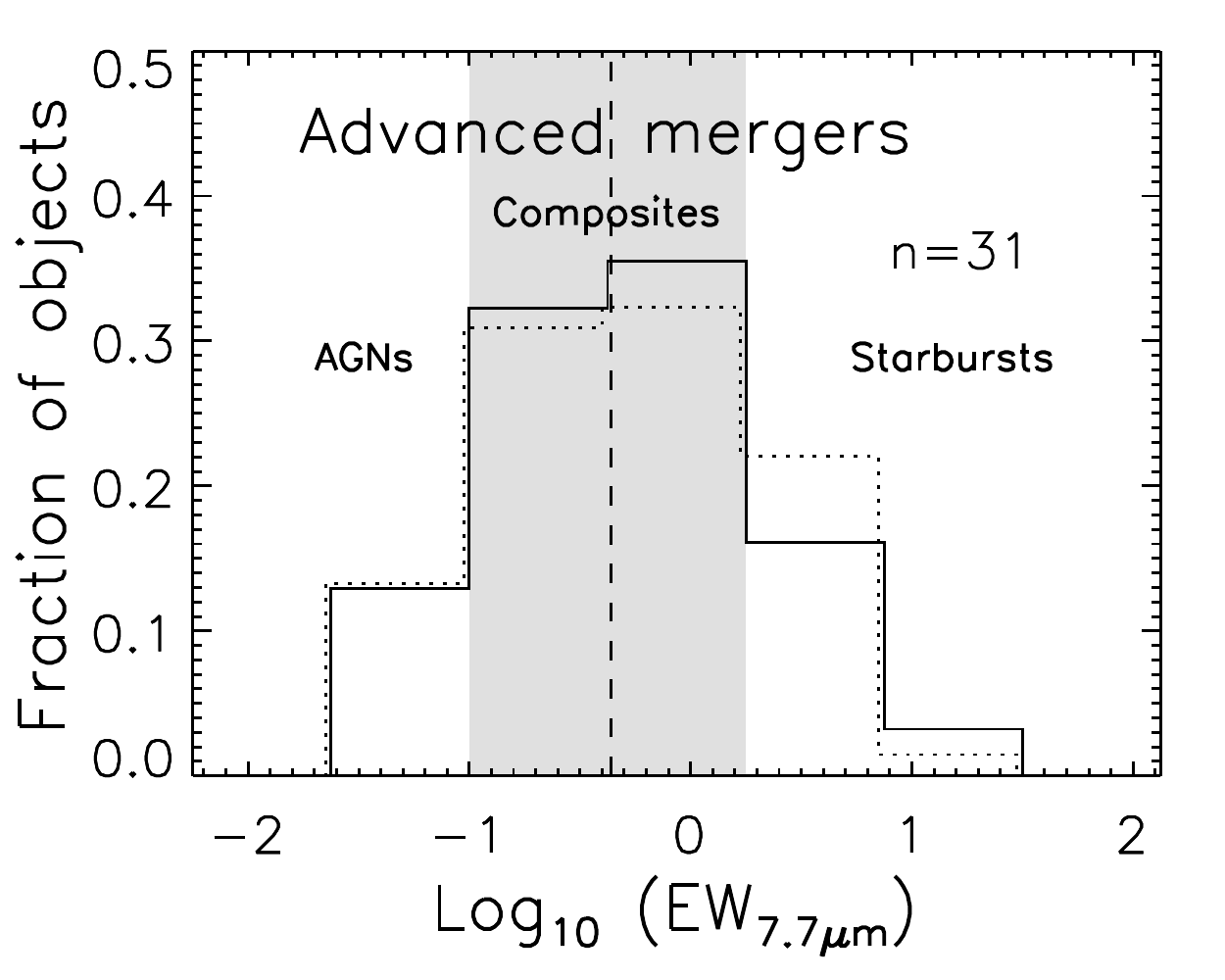}
\includegraphics[width=1.7in]{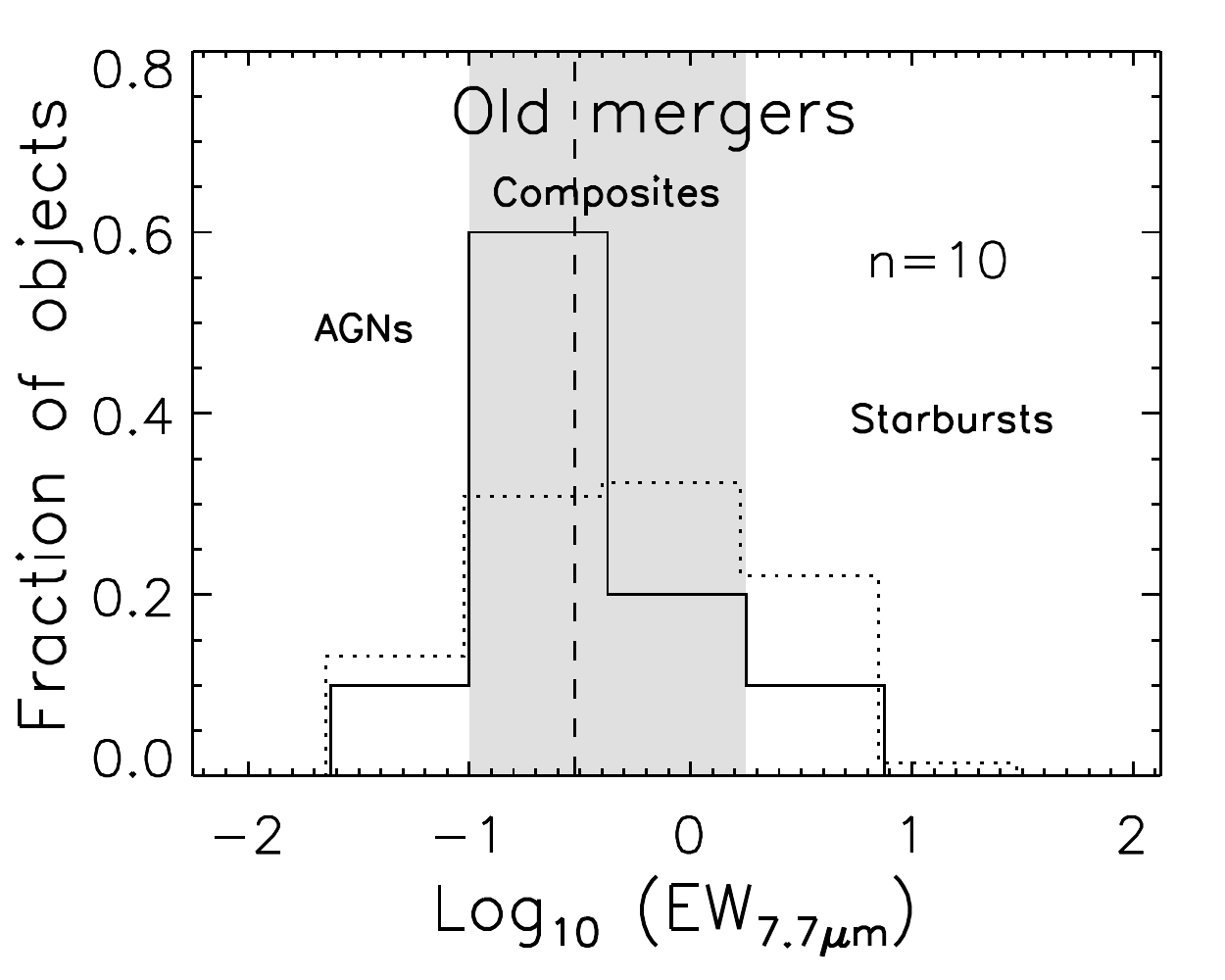}
\caption{Same as Figure~\ref{fig:ew_bymorph_highz}, but for the {\em 1-Jy} sample.  Data from \citet{Veilleux09a}.}
\label{fig:ew_bymorph_1Jy}
\end{center}
\end{figure*}

In Figures~\ref{fig:ew_bymorph_lowz} and~\ref{fig:ew_bymorph_highz} we show the distribution of equivalent widths split by morphological class for our $z < 1.5$ and $z \ge 1.5$ samples respectively.  Figure~\ref{fig:ew_bymorph_lowz} demonstrates an evolution in the median PAH EW along the merging sequence, from high PAH EWs in isolated spirals and early mergers to low EWs or no PAHs at all in old mergers and regular ellipticals.  This indicates a progression from star formation to black hole accretion activity as the merging process advances, much like the one observed in local ULIRGs \citep[Figure~\ref{fig:ew_bymorph_1Jy}; ][]{Veilleux09b}.  Just as in local ULIRGs, however, this evolution is statistical in nature:  all sources do not go through that exact sequence.  On the contrary, we observe a broad distribution of equivalent widths in most morphological classes, indicating a large variability.  We find the same trends in Figure~\ref{fig:ew_bymorph_highz}, albeit on a narrower span of the merging sequence.

Comparison of Figures~\ref{fig:ew_bymorph_lowz}, \ref{fig:ew_bymorph_highz} and~\ref{fig:ew_bymorph_1Jy} reveals that the PAH equivalent widths of the three samples are overall shifted with respect to one another.  This shift, however, is mainly a reflection of the fact that these samples cover different luminosity ranges, from $\log L_{IR}/L_{\odot} \sim 11.5 \textendash 12.5$ in our low redshift sample (highest EWs) to $\log L_{IR}/L_{\odot} \sim  12 \textendash 12.5$ among 1-Jy ULIRGs to $\log L_{IR}/L_{\odot} \gtrsim 12.5$ among our $z \ge 1.5$ objects (lowest EWs).  Our sample also differs slightly from that of \citet{Veilleux09b} in that the strongest evolution in the AGN fraction of local ULIRGs appears to be between phases III (early mergers) and IV (advanced mergers):  the progression between phases IV and V (old mergers) being weaker.  We observe, au contraire, only a small decrease in the PAH EWs of our phase IV objects compared to those of phase III, and find most of the drop to occur after (rather than during) coalescence.  We argue that the lack of a strong drop between phases IV and V in the 1-Jy sample is probably due to the selection at $60\mu m$ which highly favors composite systems over AGN-dominated SEDs, and that this effect is absent from our sample because our selection is equally sensitive to both.  Despite appearances, there is, thus, no clear indication that our objects differ much, intrinsically, from local ULIRGs in terms of their PAH properties along the merging sequence.

Focusing in more details onto specific types of objects in our sample, we can draw the following conclusions.  AGN-dominated spectra are prevalent in the later stages of the merging sequence, but can be found in almost all classes.  Starburst-dominated spectra, on the other hand, disappear after the coalescence phase, and all regular bulges are powered purely through AGN activity, consistent with the scenario in which star formation is quenched in the process of bulge formation.  Advanced mergers come in all three types of spectra, but are most highly peaked in the composite region, consistent with the idea that they are in a transition phase.

Faint \&  Compact objects are interesting in that none of them have detected PAHs, yet they all clearly show low sersic index profiles, and whether their profile is due intrinsically to a rotating disk or to dust lanes obscuring part of a nucleus, both cases require significant quantities of gas.  The reason why these objects do not show more star formation is therefore unclear.  It is possible that feedback from the AGN acts to inhibit star formation.  It is also possible, however, that star formation does occur, but that PAH features are either completely overpowered by the strong AGN continuum and/or that they are being destroyed by the radiation from the AGN.  Alternatively, the star-forming environment in those galaxies could be so hot and dense as to wipe out any PAHs, or PAH features, itself, although that would be unprecedented.  We show in the next section that these faint \& compact objects are actually highly obscured at mid-IR wavelengths, which implies a strong absorption of the radiation from a central hot source (AGN) by colder, obscuring material, or gas.  This suggests that AGN radiation might well be responsible, perhaps in combination with a compact gas distribution, for modifying the PAH properties of these sources.

\subsection{Relation between optical depth and morphology \label{sec:tau_vs_morph}}

The silicate absorption at $9.7\mu m$ is a prominent spectral feature of many local ULIRGs \citep{Hao07}.  It has been shown to arise whenever the main source of mid-infrared radiation itself is veiled by a column density of colder dust \citep{Levenson07}, and is therefore often interpreted as the signature of an obscured AGN.  The most buried of them are, in turn, seen as nascent quasars in the quasar formation scenario proposed by \citet{Sanders88}.  We compare, below, the morphology of objects with various degrees of obscuration.  In particular, we examine how morphology can help us understand the most obscured sources.  We use the $\tau_{9.7\mu m}$ silicate optical depth, as defined in section~\ref{sec:tau}, to this effect.

\begin{figure}[htbp]
\begin{center}
\includegraphics[width=3.5in]{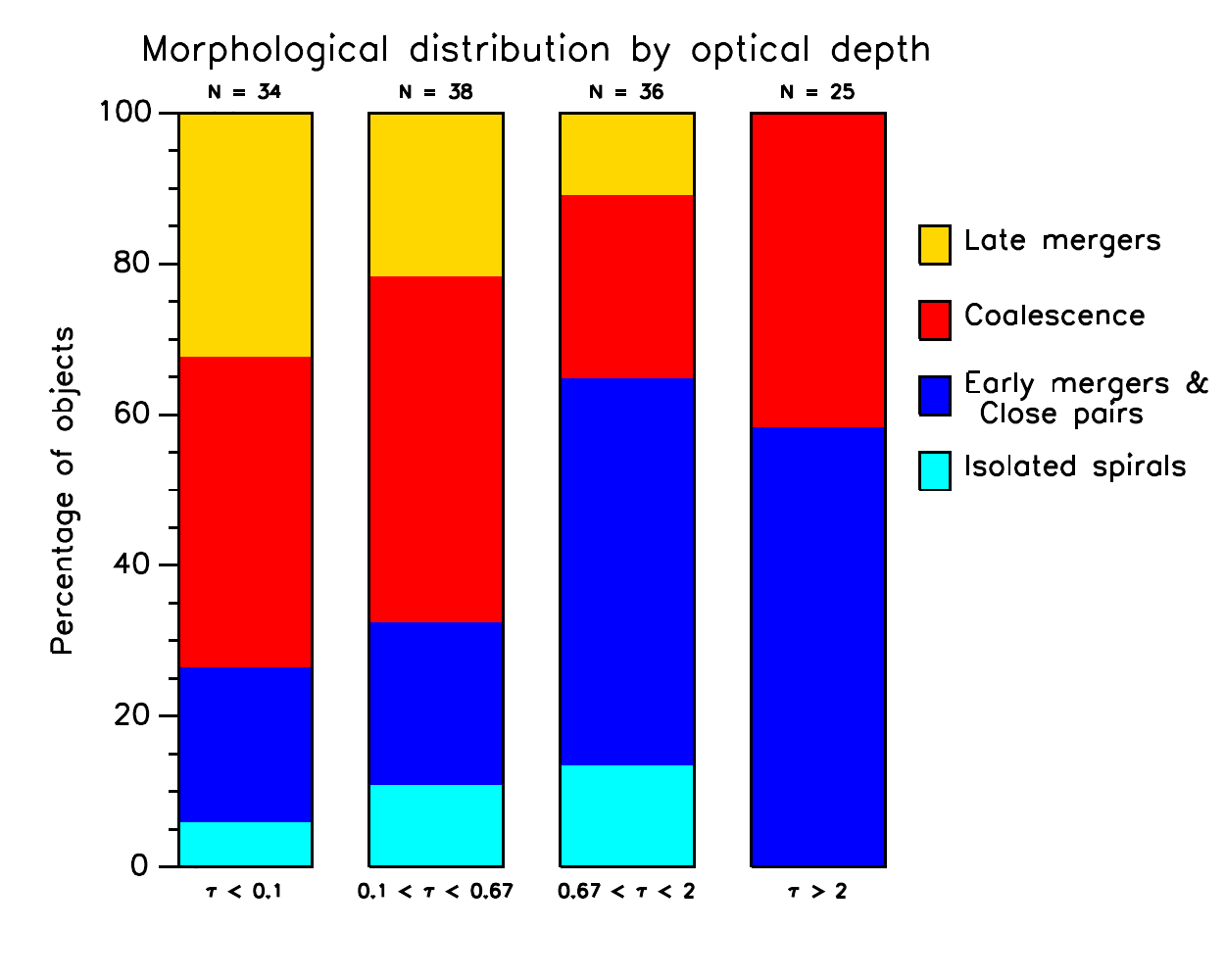}
\caption{Merger composition in four bins of $\tau_{9.7\mu m}$.  Only early and coalescing mergers populate the highest optical depths.  Late mergers have predominantly low values of $\tau_{9.7\mu m}$.}
\label{fig:morph_by_tau}
\end{center}
\end{figure}

We first examine the relation of $\tau_{9.7\mu m}$ with merging phase with the aid of Figure~\ref{fig:morph_by_tau} in which we illustrate the morphological composition of objects at different values of $\tau_{9.7\mu m}$.  Figure~\ref{fig:morph_by_tau} demonstrates that the highest optical depths are achieved in the early phases of merging and through coalescence.  Once objects reach late phases, their optical depth at $9.7\mu m$ diminishes considerably or vanishes altogether, as illustrated by the substantial increase in the fraction of late mergers at lower optical depths.  Isolated objects also do not attain the large optical depths achieved in mergers.

\subsubsection{The Spoon Diagram}

More insight can be gained by looking at correlations with more than one parameter.  One particularly useful diagram is that of $\tau_{9.7\mu m}$ versus PAH equivalent width.  \citet{Spoon07} showed that local ULIRGs lie on two distinct branches in that diagram: a horizontal branch of low optical depths spanning the whole range of EWs and a diagonal branch going from shallow silicate absorption at high EWs to high values of $\tau_{9.7\mu m}$  at low EWs.  They found the region comprised of objects with low EWs and medium optical depths, located in between the two branches, to be scarcely populated.  These observations led the authors to speculate that ULIRGs mainly evolve upward along the diagonal branch before quickly expelling their obscuring material to become quasars or Seyfert 1 galaxies on the low-EW end of the horizontal branch.  Our morphological information allows us to test these ideas.

\begin{figure*}[htbp]
\begin{center}
\includegraphics[width=3.5in]{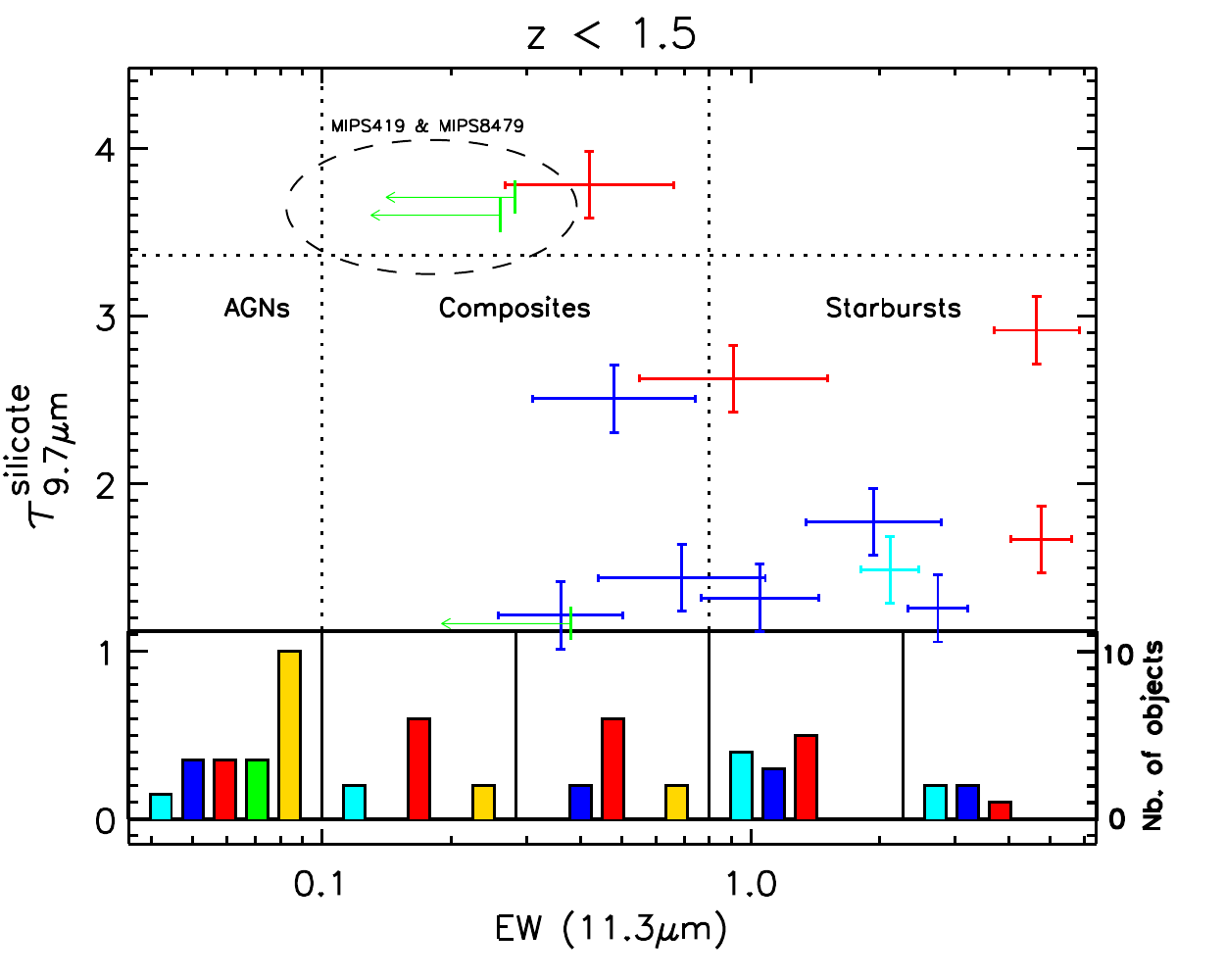}
\includegraphics[width=3.5in]{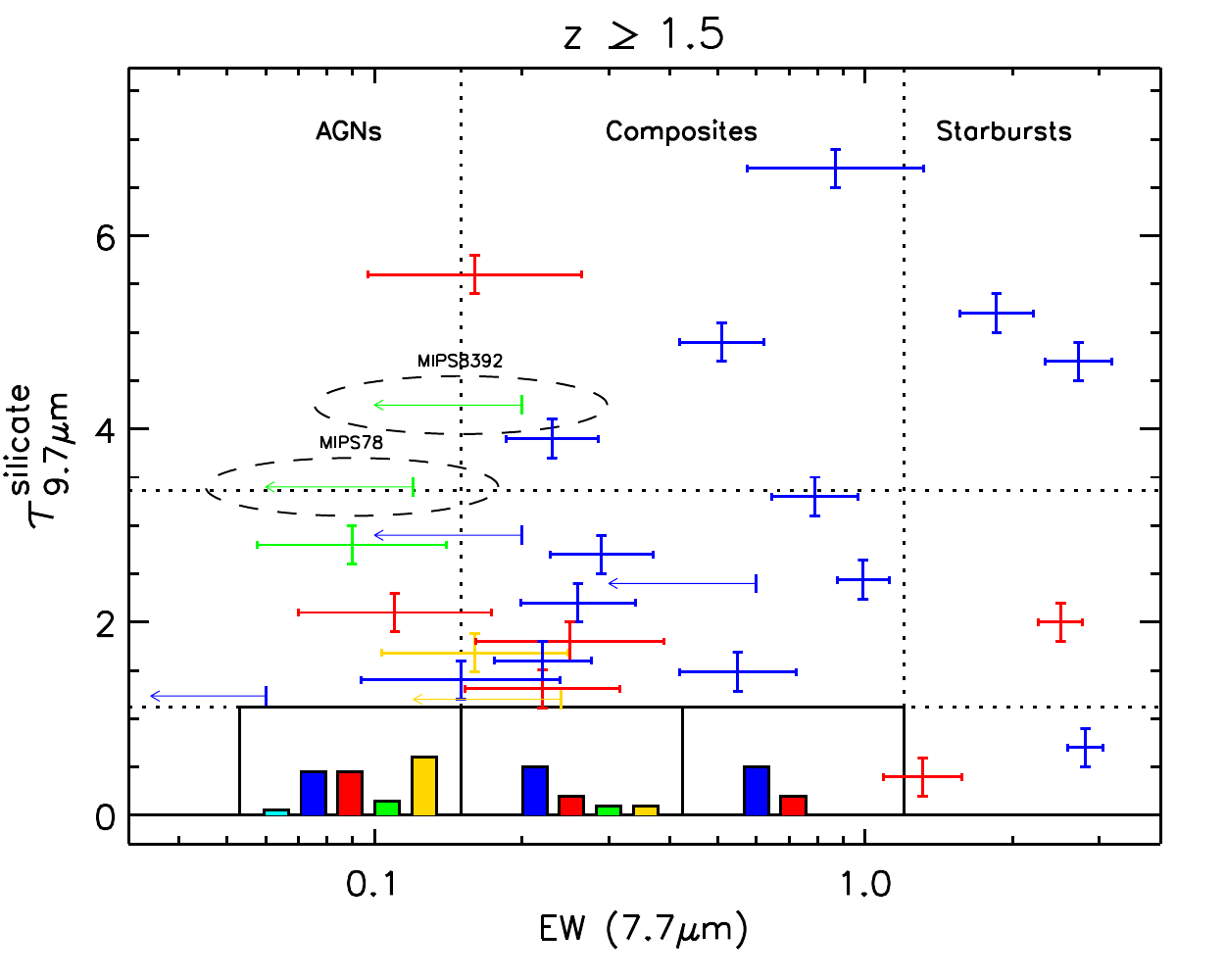}
\caption{Spoon diagram \citep{Spoon07} for our sample color-coded by morphology:  {\em cyan} for isolated spirals, {\em blue} for early mergers and close pairs, {\em red} for advanced mergers, {\em green} for faint \& compact objects, and {\em yellow} for late mergers.  Owing to the large density of points at $\tau_{9.7\mu m} < 1.12$, we show histogram distributions in bins of EW, rather than individual points.  Objects with upper limits on their EW were all placed in the AGN bin.  Objects of unknown redshift were split half-and-half between the two diagrams.  They all fall in the lower-left bin, by definition.  Bars of the histogram run from 0 to 10 objects.  Circled objects represent obscured quasar candidates.  Their morphology, IRS spectrum and SED are shown in Figure~\ref{fig:obs-quasars}.  They all possess faint \& compact morphologies.  Late mergers tend to aggregate in the lower-left part of the diagram where {\em unobscured} AGNs live.  Other morphological classes do not occupy preferentially any specific region.}
\label{fig:spoon}
\end{center}
\end{figure*}

We show the diagram of optical depth versus PAH equivalent width for low and high redshift objects in our sample separately in Figure~\ref{fig:spoon}.  Unfortunately, since our detectability is limited to PAH equivalent widths above 0.1, we do not possess the sensitivity to probe into the AGN part of the diagram where the dichotomy between the two branches occurs in local ULIRGs.  Nevertheless, we clearly detect a horizontal branch at low values of $\tau_{9.7\mu m}$, so much that we show the morphological distribution of objects along that branch in binned histograms rather than as individual points, owing to their large density.  On the other hand, the number of objects with highly obscured AGNs (top-left corner of the diagram) is rather low in our sample, with only four such candidates (circled objects in Figure~\ref{fig:spoon}), insufficient to judge of the presence of a diagonal branch.

\begin{figure*}[htbp]
\begin{center}
\makebox[6in]{\Large Obscured Quasars} \\
\includegraphics[height=1.4in]{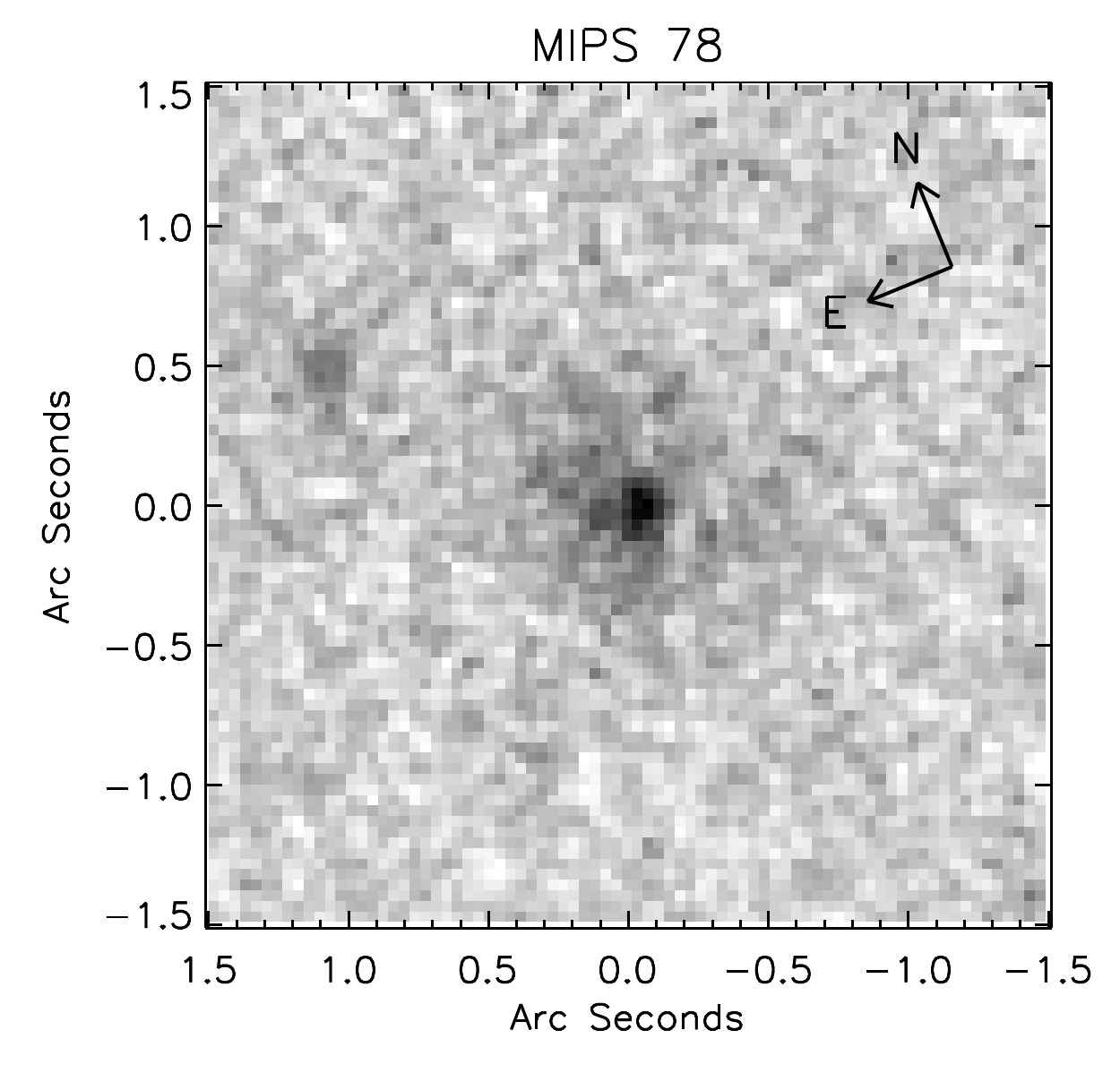}
\includegraphics[height=1.4in]{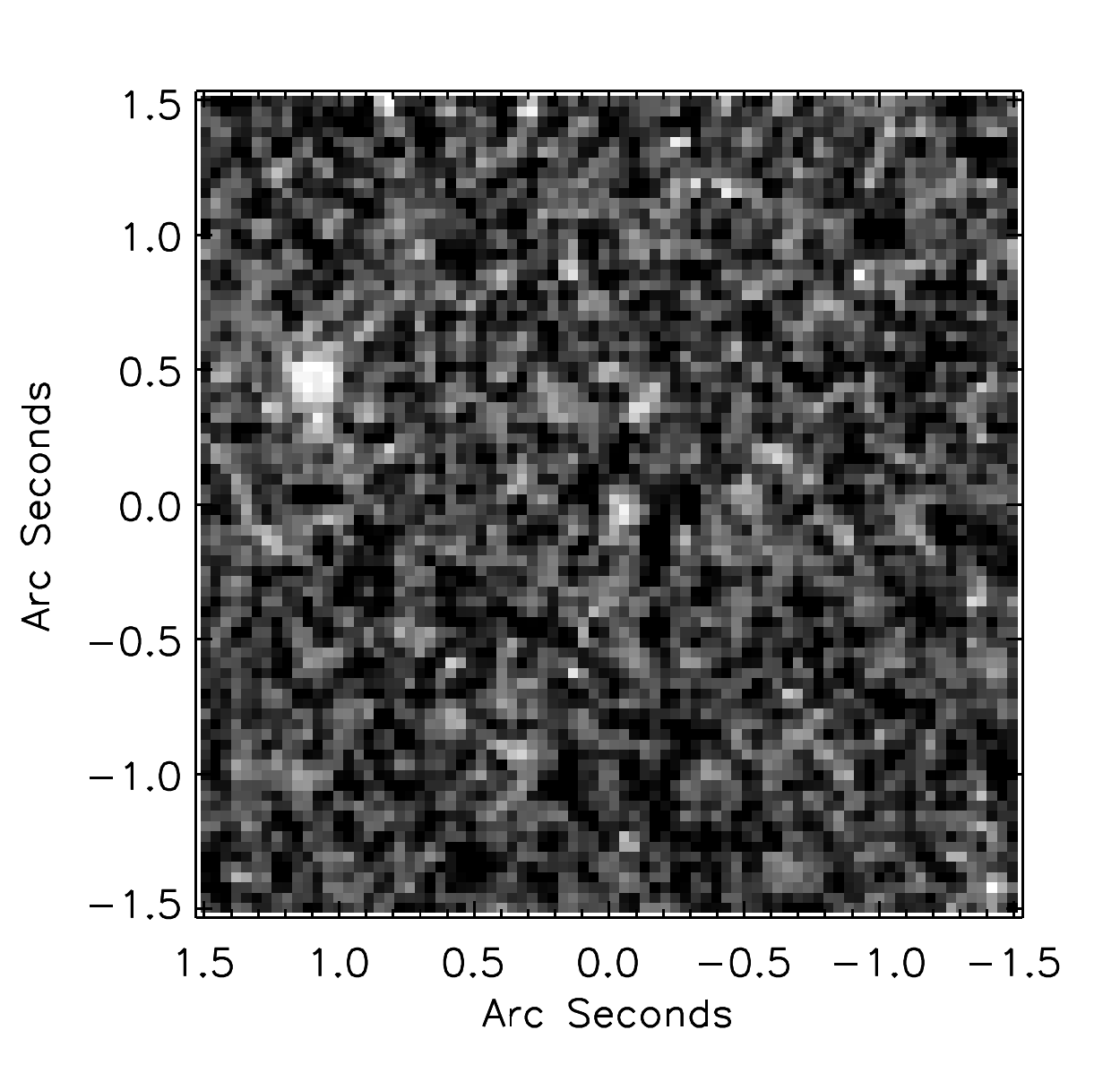}
\includegraphics[height=1.4in]{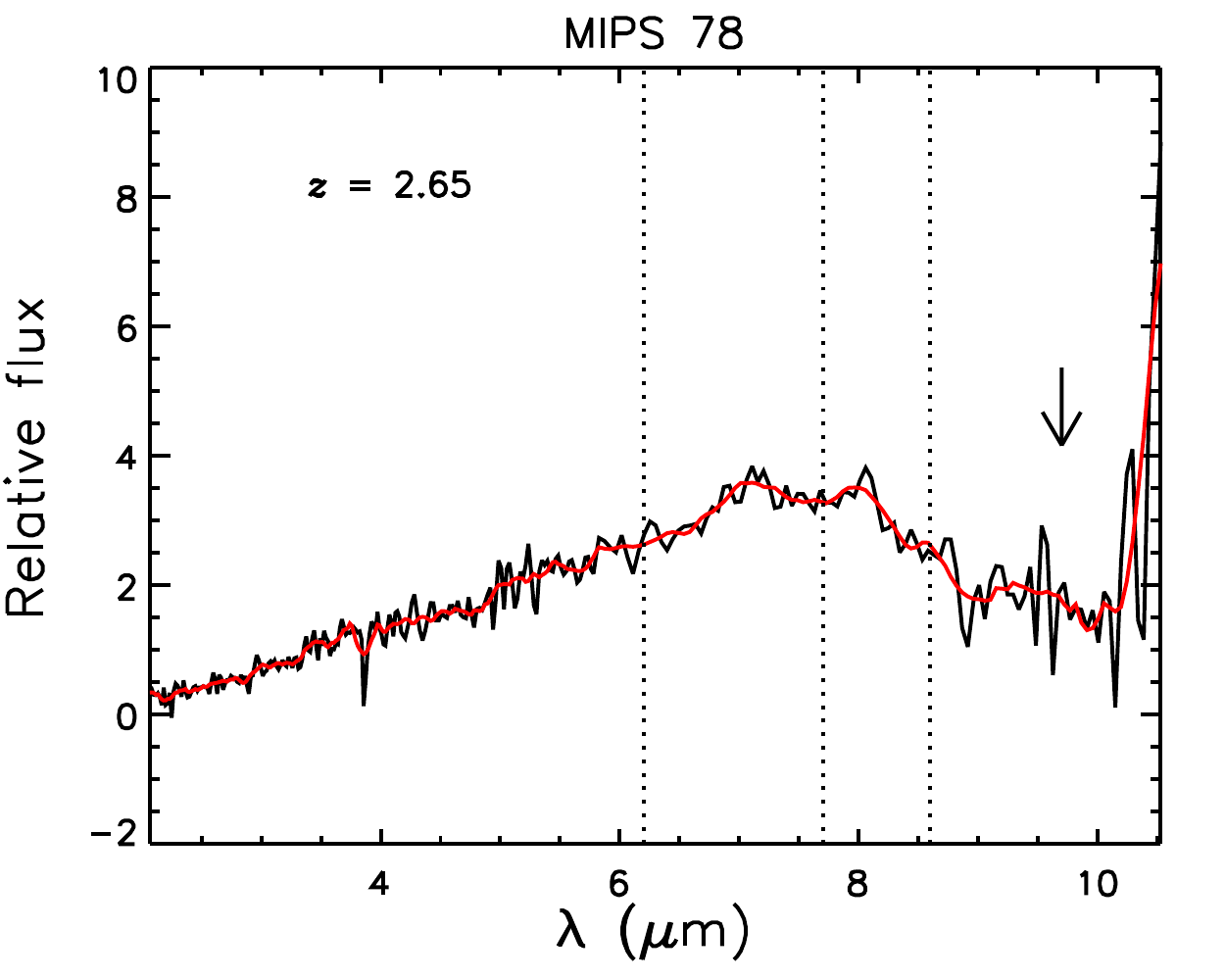}\\
\includegraphics[height=1.4in]{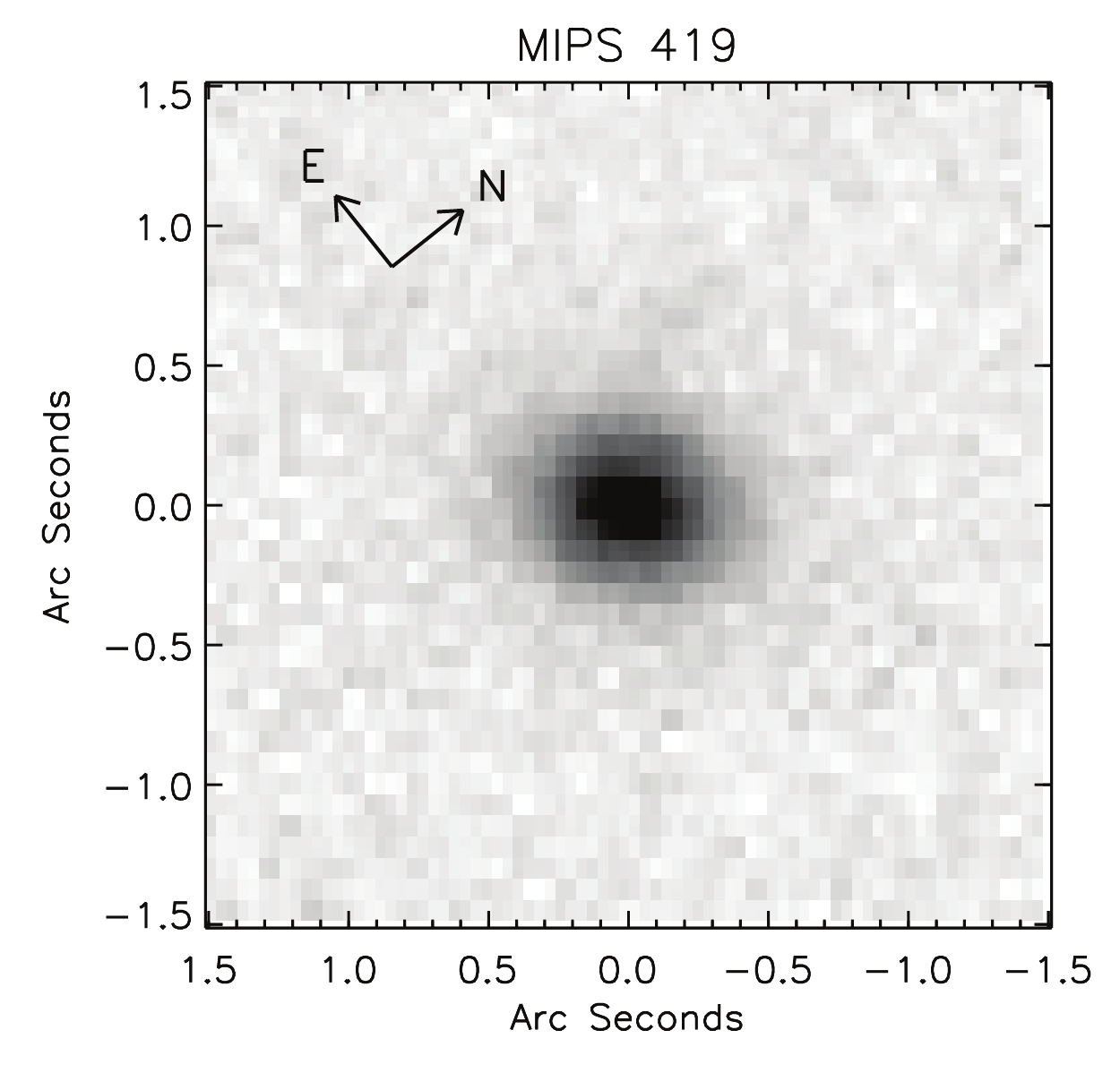}
\includegraphics[height=1.4in]{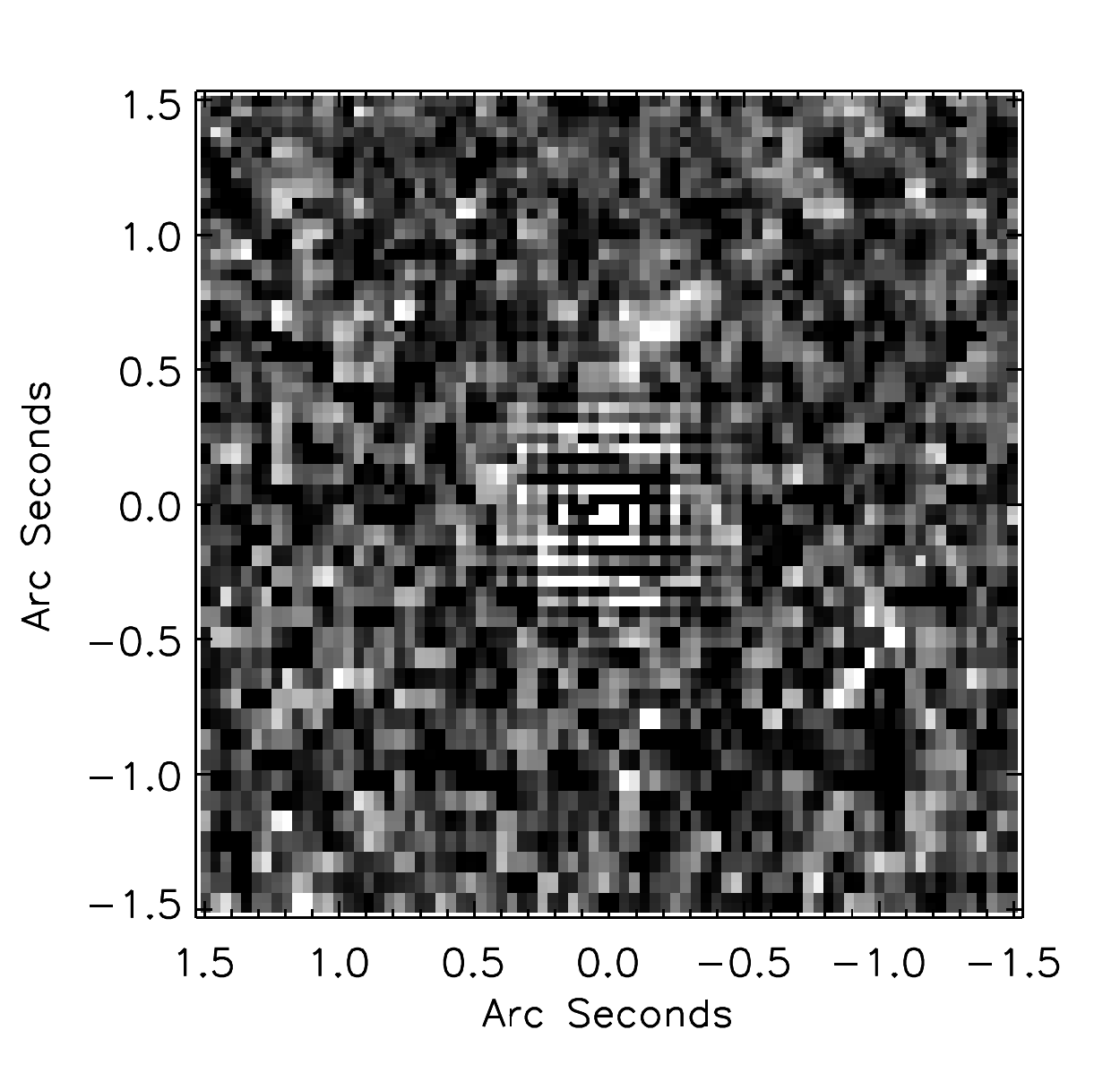}
\includegraphics[height=1.4in]{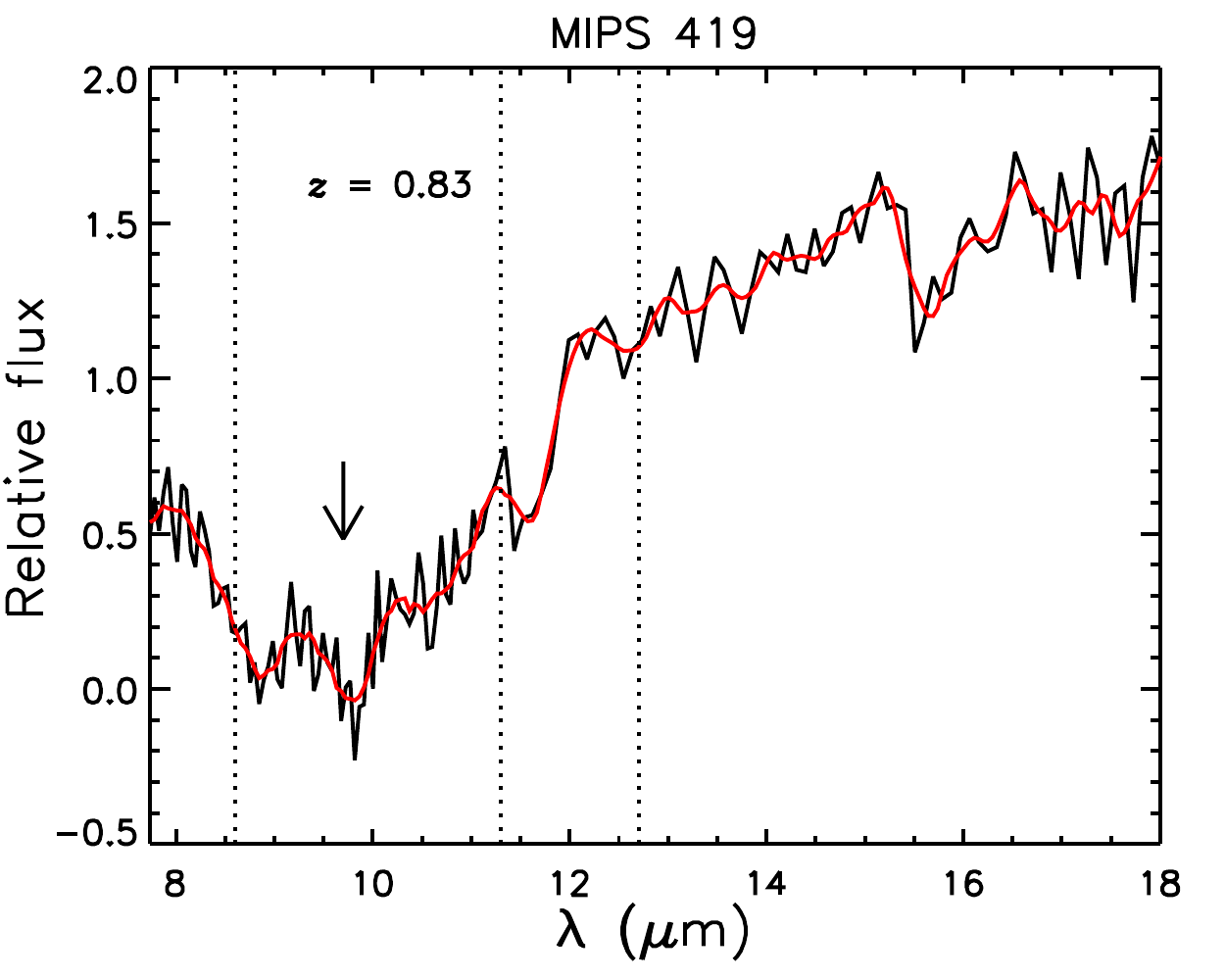}\\
\includegraphics[height=1.4in]{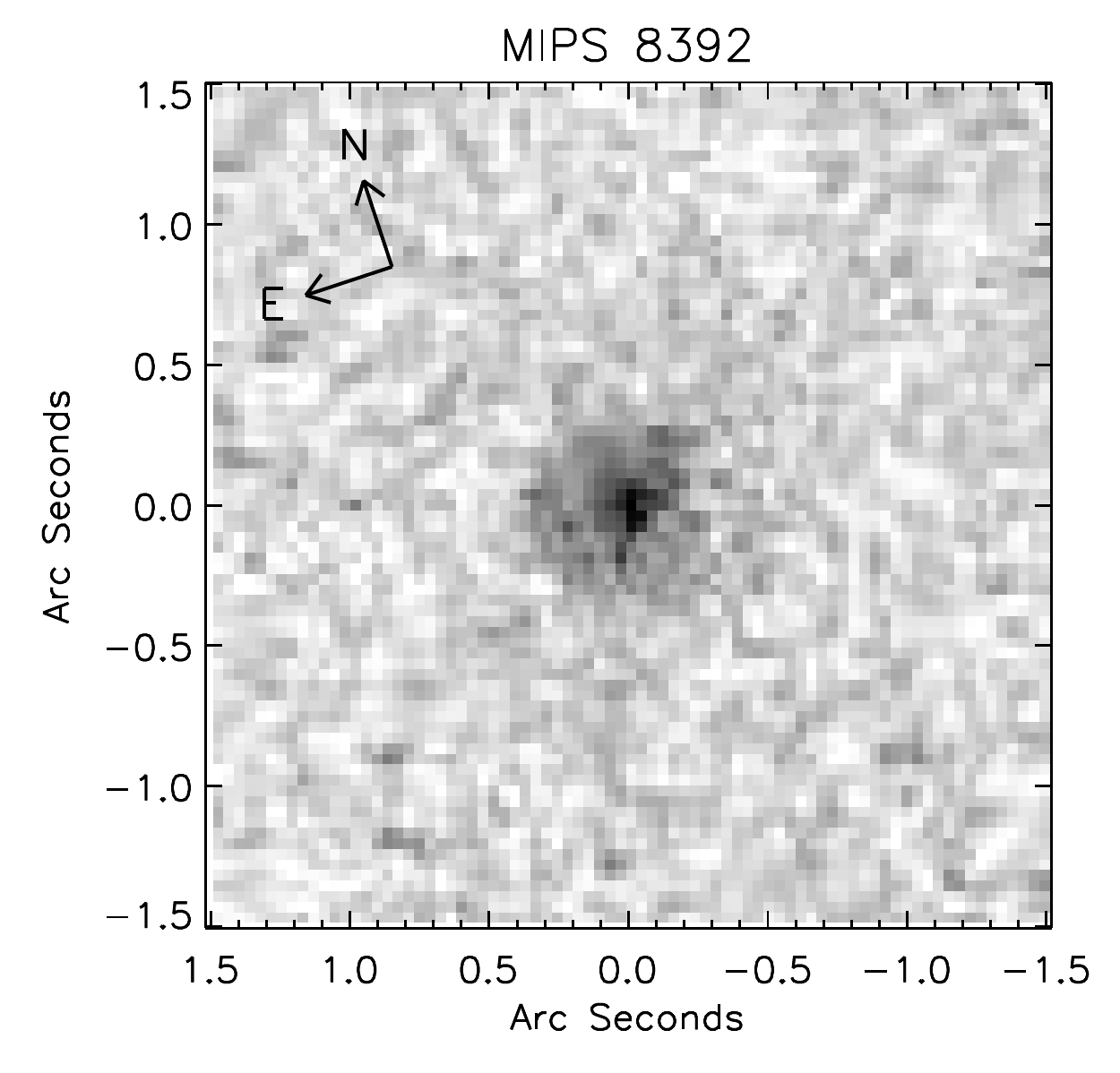}
\includegraphics[height=1.4in]{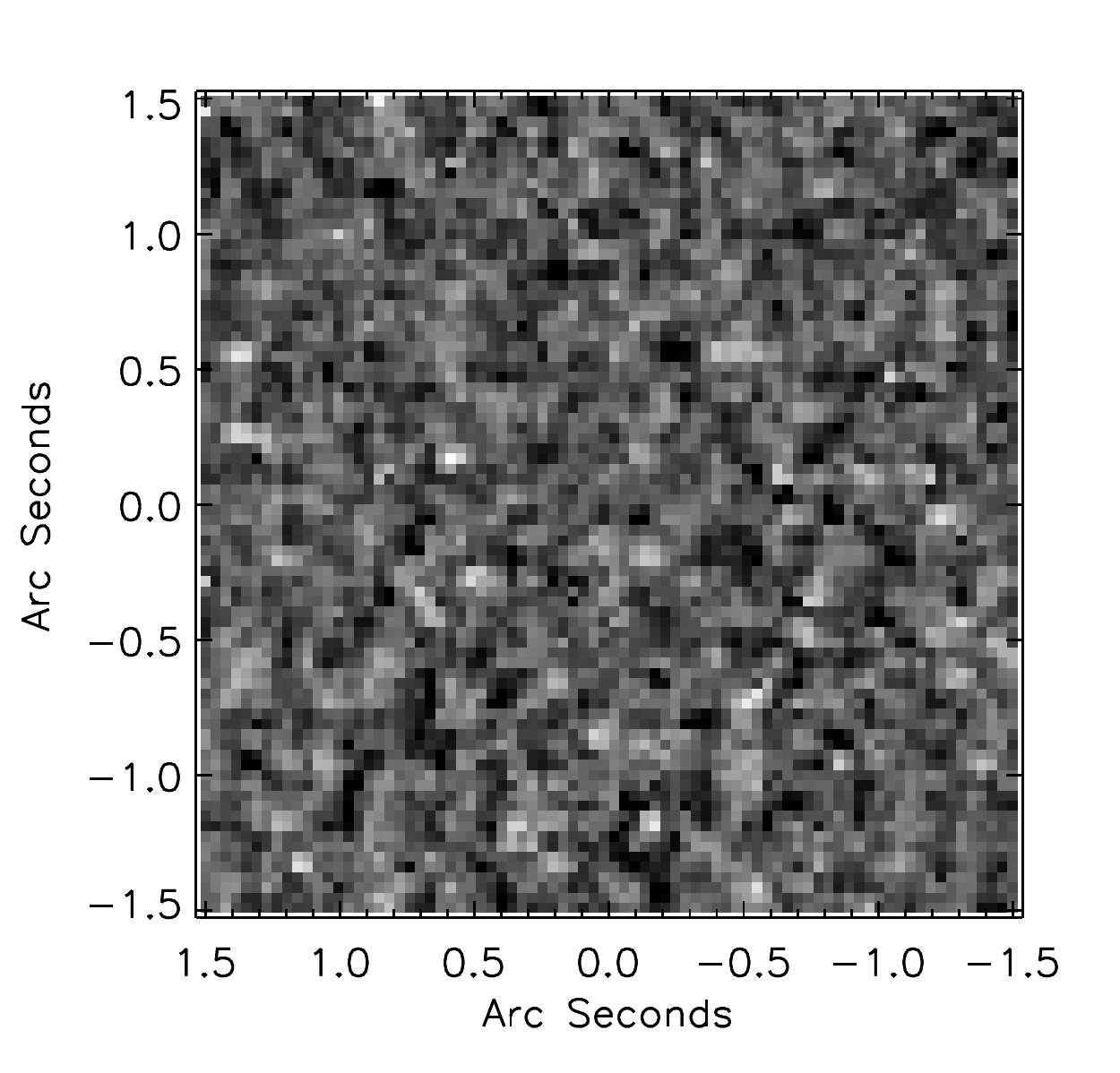}
\includegraphics[height=1.4in]{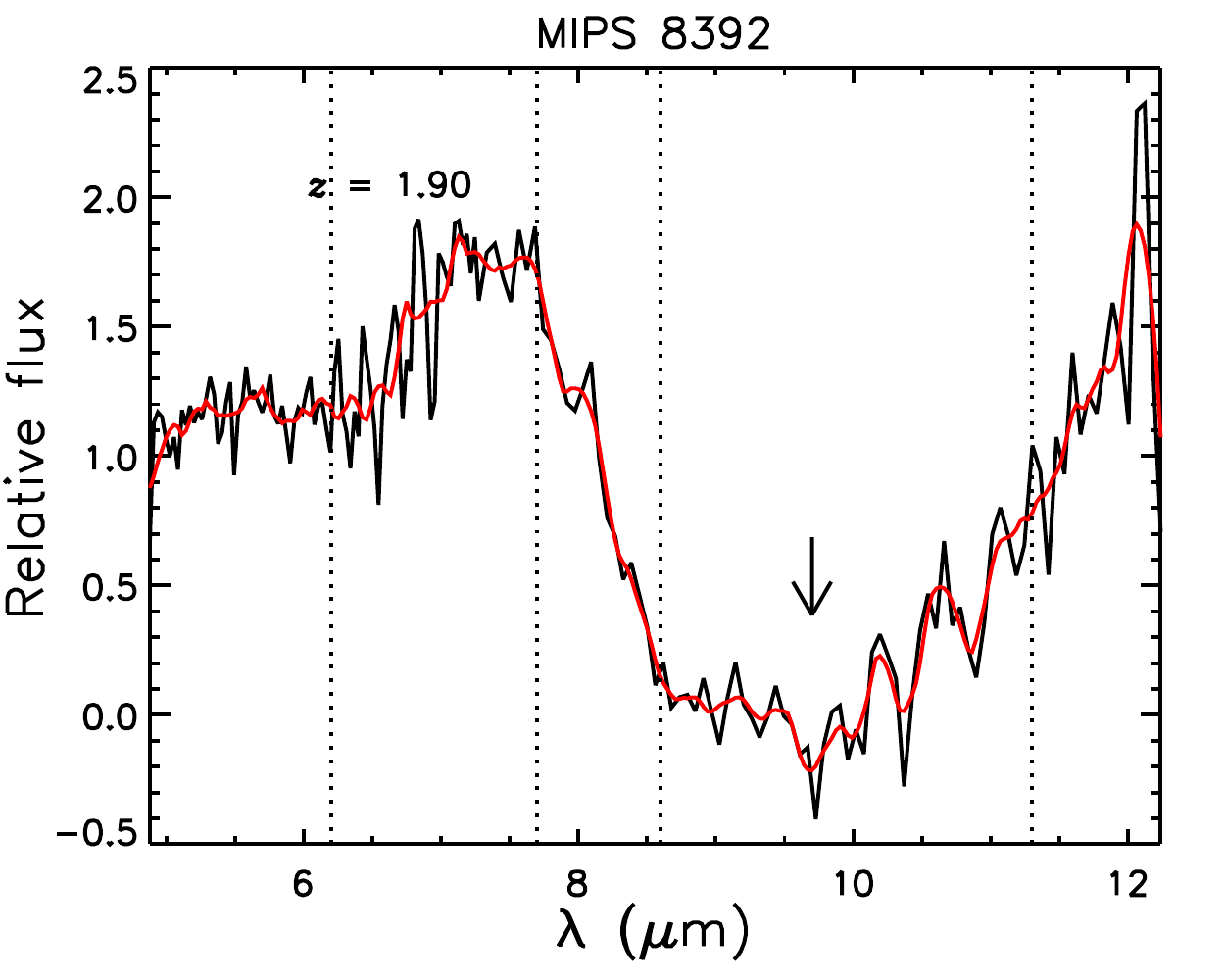}\\
\includegraphics[height=1.4in]{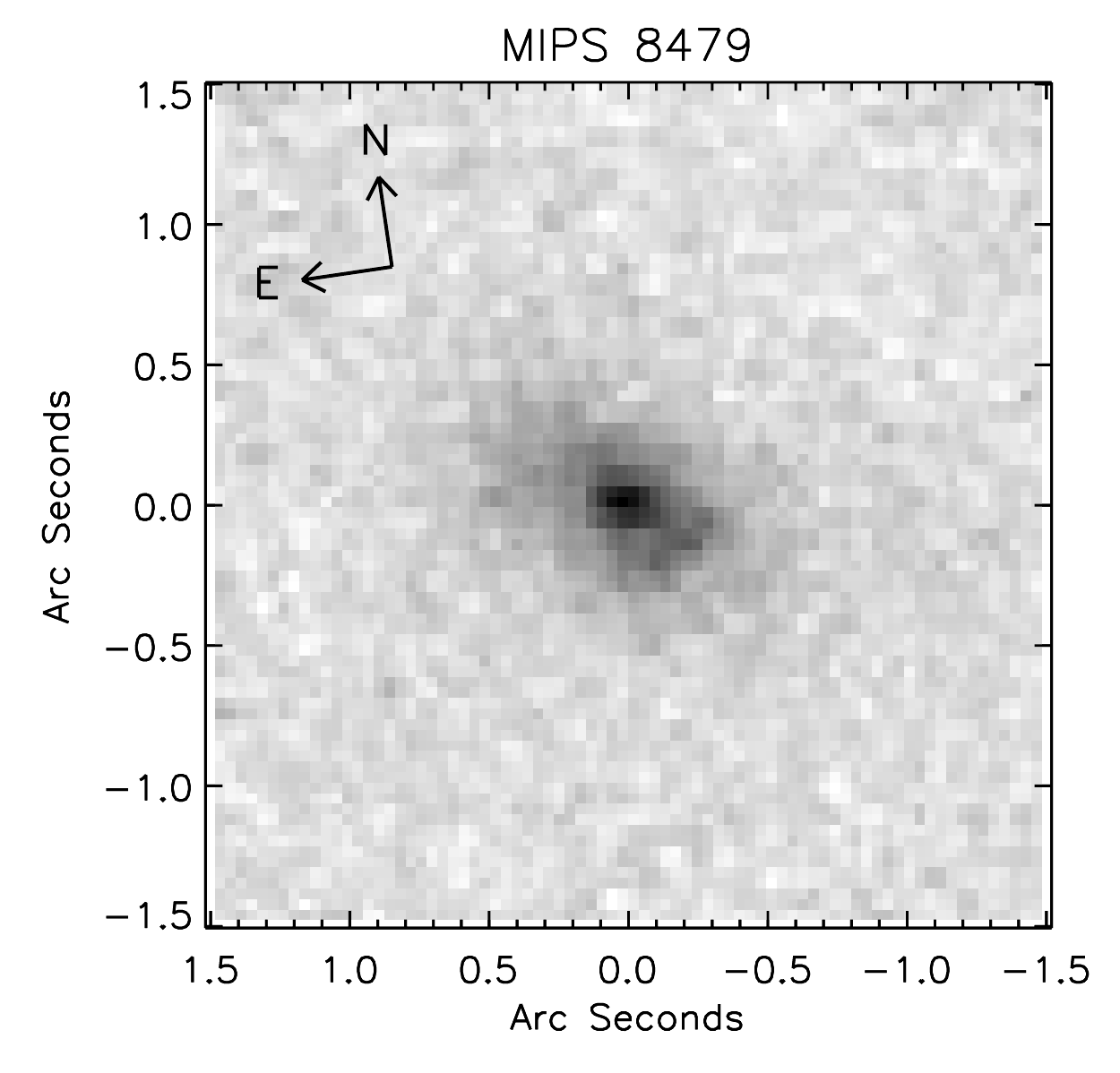}
\includegraphics[height=1.4in]{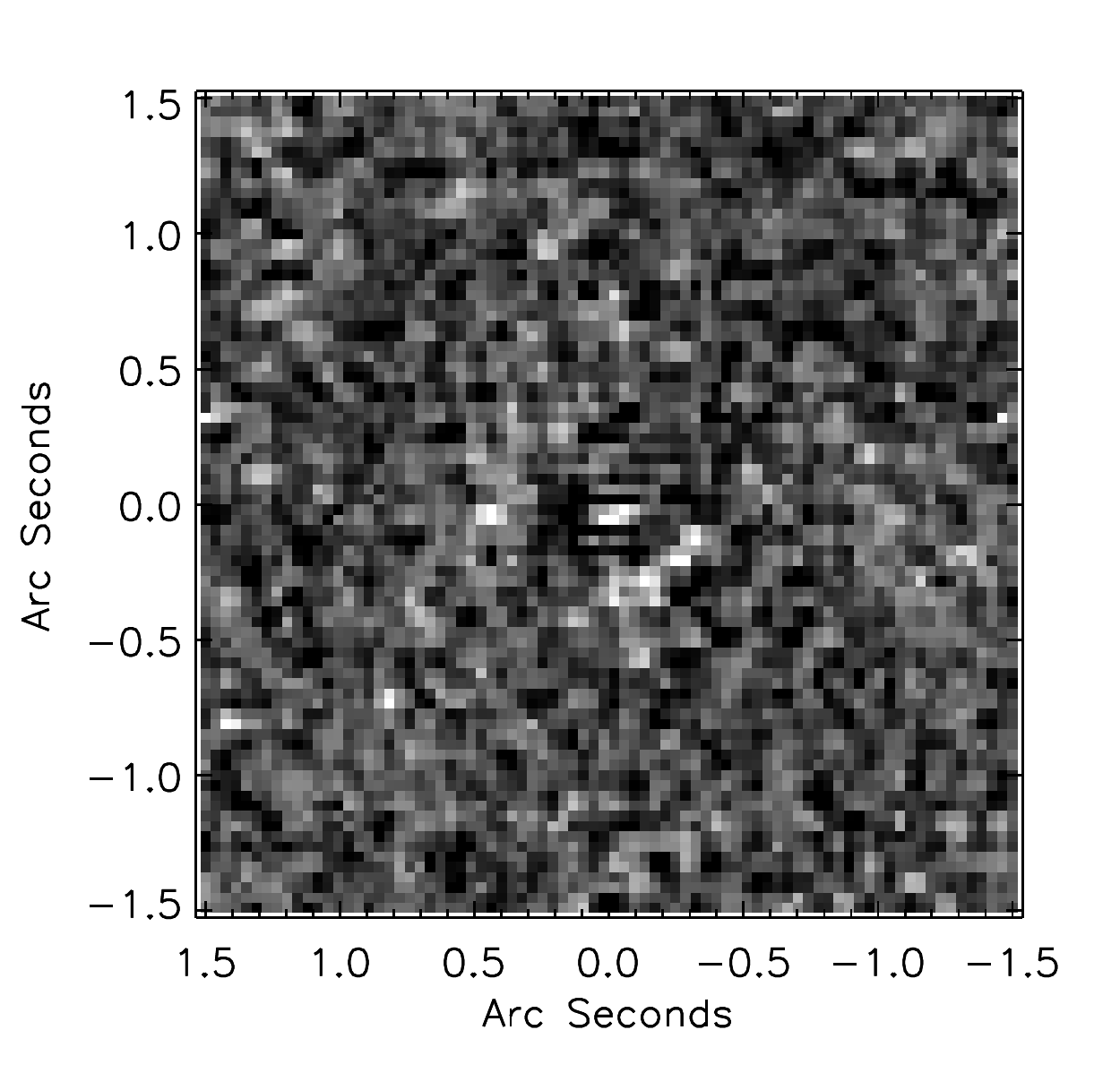}
\includegraphics[height=1.4in]{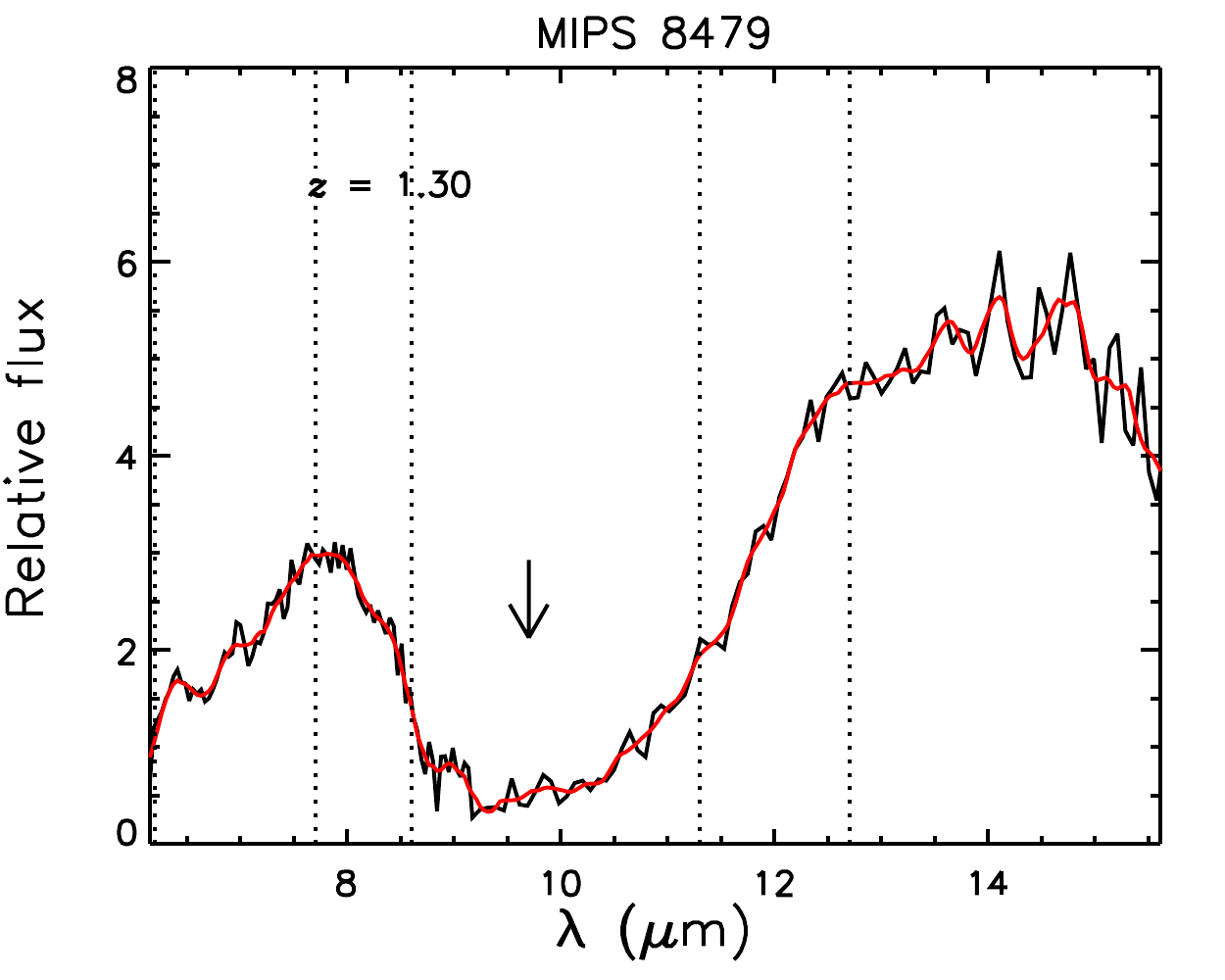}
\caption{NICMOS images, residuals after subtraction of the smooth profile, and IRS spectra (smoothed in {\em red}) of all obscured ($\tau_{9.7\mu m} > 3.36$) quasar candidates in our sample.  Images are shown in negative, whereas residuals in positive.  Dotted lines in the spectra show the location of PAH features.  The arrow marks the center of the $9.7\mu m$ silicate absorption feature.  All objects are faint \& compact and three out of four show faint residuals, the shape of which is characteristic of mergers in two of the cases.}
\label{fig:obs-quasars}
\end{center}
\end{figure*}

Despite their small numbers, our morphological analysis reveals that all of our buried AGN candidates have very specific host morphologies.  Namely, they all belong to the {\em Faint \& Compact} category.  This observation indicates that the galaxies these objects live in possess large concentrations of gas in their center but do not have fully formed bulges.  We speculate that these objects form in special merger configurations that result in the funneling of the major part of the gas and dust content into the center without much loss to stripping and tidal features.  We think that such mergers might occur in low speed encounters.

Furthermore, the highly obscured nature of these objects implies that they have not been able to clear the gas surrounding their central hot source, or active nucleus, yet and would therefore have to be very young mergers, i.e. still in the early coalescence phase.   We would expect the stellar component to begin to be randomized, and the smooth exponential profile to start to be destroyed, at that stage.  The observed sersic indices ($n \lesssim 2$) of these objects are consistent with such a process beginning to take place.  These objects thus appear to be growing their bulges and black holes simultaneously and to be perfect examples of bulge/black-hole co-evolution and of the classical quasar formation scenario of \citet{Sanders88}

Most mergers, however, do not go through a faint \& compact phase.  In fact, most objects going through coalescence show tidal tails and/or streams, as well as higher sersic indices, and thus belong to the advanced merger class.  Having already built most of their bulge, these objects will never evolve into faint \& compact objects.  Because of their tidal features, these objects are also unlikely to have evolved {\em from} a faint \& compact phase, thus making the two evolutionary paths largely independent.  Advanced mergers show, on average, shallower silicate absorption, as illustrated in Figure~\ref{fig:spoon}.  Rather than going through that obscured quasar phase, most infrared-luminous mergers thus appear to either move off and on, or about, the horizontal branch instead, with often some combination of both PAHs and silicate absorption.

Conversely, we know of at least one example of a high $\tau_{9.7\mu m}$ and weak PAH source locally that is {\em not} a faint \& compact source.  This is IRAS08572+3915, an early merger and one of the objects we simulated high redshift observations for.  Taken individually, however, both interacting galaxies in this source loose their tidal tails and extended features at high-$z$ and appear faint \& compact.  Moreover, the two bodies are only 6 kpc apart and are likely on the brink of merging and becoming a true faint \& compact source.  Nevertheless, this example illustrates that the obscured quasar phase can sometimes occur before final coalescence.

As for early and advanced mergers, Figure~\ref{fig:spoon} shows their distribution in the Spoon diagram to be fairly well mixed implying little evolution between the two phases.  The large parameter space spanned by both those categories further suggests that there are many possible paths a merger event can traverse.  It appears that the only predictable stage is the end-point of the merging sequence:  the late mergers, shown to aggregate in the lower left corner of the diagram, where both star formation and obscuration are low.

Our sample, thus, does not attest to any precise evolutionary path in $\tau_{9.7\mu m} \textendash$ EW(PAH) space, such as suggested by \citet{Spoon07}.  Rather, it seems that many avenues are possible.  Nevertheless, it appears that most objects gravitate around the horizontal branch, and only a certain kind of mergers make it to the top-left corner:  those with very dense and dusty cores that are classified as faint \& compact at high redshift.  However, all objects appear to end up in the low-$\tau_{9.7\mu m}$, low-EW corner at the end of the merging process.

\section{DISCUSSION \label{sec:discussion}}

We have shown that a high fraction, about 80\%, of bright ($S(24\mu m) > 0.9$ mJy), high-redshift $24\mu m$-sources are likely to be ongoing mergers in various stages of the process.  We further demonstrated that although our sources show a broad range of morphologies and populate the entire merging sequence, the relative number of early mergers increases substantially at $z \ge 1.5$.  We showed that star formation activity as probed by the equivalent width of PAH features, decreases, on average, along the merging sequence to the profit of black-hole accretion.  We finally demonstrated that obscured quasars, in our sample, live in \textquotedblleft faint \& compact\textquotedblright galaxies.

We mirrored our analysis on that of \citet{Veilleux02, Veilleux06, Veilleux09b} on the 1-Jy sample, and simulated high-redshift NICMOS observations of local ULIRGs in order to address the question as to whether our sources are high-redshift analogs of local ULIRGs.  To a large extent, we presented our comparison of the two samples concurrently with our results in section~4, but discuss that question in a broader context in this section.  The 1-Jy sample, however, is arguably not the best point of comparison, since the sample is selected at a substantially different rest-frame wavelength.  Furthermore, comparison with local galaxies gives us only part of the picture;  comparison with other high-redshift samples is also highly desirable.  We propose to carry such comparisons in this section.  We will discuss our results in the context of five types of infrared-luminous objects:  local ULIRGs (from the 1-Jy sample), other bright $24\mu m$-selected samples at high-redshift, sub-mm galaxies, and finally $z \sim 1$, $70\mu m$-selected objects.  Lastly, we compare our results with theoretical predictions and discuss implications for our understanding of galaxy evolution.

\subsection{Are bright high-redshift $24\mu m$ galaxies analogs of local ULIRGs? \label{sec:local_analogs}}

Our observations suggest that, up to $z=1.5$, they are analogs of, not only local ULIRGs, but also top-end LIRGs ($L_{IR} > 10^{11.5} L_{\odot}$).  This is evidenced by their large infrared luminosities ($L_{3-1000\mu m} \sim 10^{11.5}\textendash 10^{12.5} L_{\odot}$) and their high merger fraction (up to 83\% at $z < 1.5$), which are the two defining characteristics of high-end LIRGs and ULIRGs locally \citep{Murphy96,Veilleux02,Ishida04}.  Our analysis, however, has revealed that our $24\mu m$-selected sample at $z < 1.5$ tends to draw from objects in more advanced stages of merging, likely owing to the shorter wavelength selection.  When compared to redshifted galaxies from the 1-Jy sample, our objects also tend to show stronger, brighter merger signatures such as more visible tidal features and a higher incidence of detected residuals.  Spiral galaxies make for only 17\% of our objects at $z < 1.5$, thus rejecting the hypothesis that our sources are scaled up versions of lower luminosity local LIRGs \citep[which have disk morphologies in $\gtrsim 40\%$ of cases;][]{Wang06}, even though some of our spirals do show slightly elevated IR-luminosities (up to $1.6 \times 10^{12} L_{\odot}$).

At redshifts above 1.5, we observe a dramatic rise in the number of early mergers among our bright $24\mu m$ sources despite the fact that we are probing, at those redshift, only extremely high luminosities ($L_{IR} > 10^{12.5} L_{\odot}$), and despite the fact that $\gtrsim 90\%$ of these objects have PAH EWs indicative of the presence of an AGN ($\mbox{EW}_{7.7\mu m} < 1.2$).  Locally, this combination of high luminosity and low EWs can only be found among more advanced mergers.  These early mergers that represent, at $z \ge 1.5$, half of our sample, thus form a new type of sources.  Morphologically, albeit very messy, they do not appear fundamentally different from lower redshift luminous mergers at similar stages.  Their PAH EWs, however, indicate that they are primarily AGN-dominated or composite systems.  It is, therefore, an earlier triggering, and sustained fueling, of a strong obscured AGN that distinguishes them from local ULIRGs.  As we discuss in section~\ref{sec:theory}, we think this might be the consequence of the elevated gas content of these objects in comparison to that of local ULIRGs \citep{Yan10}.  The infrared-luminous aspect of galaxy evolution thus appears to proceed differently at high redshifts, in a way that is not well represented by local ULIRGs.

\subsection{Comparison with other samples of $24\mu m$-selected galaxies}

\citet{Dasyra08} presented a first analysis of a $z > 1.5$ subset of the sample used in this paper that was selected from objects in our first \spitz/IRS program.  By looking for merging pairs, they were able to place a lower limit on the number of interactions in $z\sim 2$ bright $24\mu m$ galaxies at 52\%.  We have expanded those results by performing a systematic search for merging signatures and find that the other half of our $z > 1.5$ objects all show signs of possibly being later-stage mergers (from coalescence onward).  60\% of those objects are identified with high confidence, being of confidence level ~4 or less (cf. \S~2.5.2).  

\citet{Dasyra08} further found hints that these high-redshift bright $24\mu m$ sources might be disk-dominated rather than bulge-dominated like local ULIRGs.  Our analysis reveals that this is indeed the case and that two thirds of our objects at $z>1.5$, in fact, possess disk-dominated profiles ($B/D < 0.5$ or $n<2.3$).  This can be largely attributed to the high fraction of early mergers at those redshifts, but many, more advanced, mergers also show rather low $B/D$ ratios or sersic indices.  We speculate that this is due to a combination of two effects:  high gas fractions in those objects that act to maintain a disk around the central object for longer (and perhaps indefinite) periods of time \citep[][see also \S~\ref{sec:theory}]{Robertson06}, and band-shifting that causes us to see those objects in their rest-frame $R$, $V$ or $B$ band where their disks can be readily detected.

Another type of bright $24\mu m$-selected galaxies are the so called DOGs (for dust-obscured galaxies).  These are $z \sim 2$ objects with very red observed $F(24\mu m) / F(R) > 1000$ colors.  \citet{Bussmann09} presented NICMOS (as well as ACS/WFPC2) morphologies of 31 such sources selected from the larger sample of \citet{Dey08} and chosen to have $24\mu m$ fluxes in excess of 0.8 mJy.  \citet{Melbourne09} presented NIR morphologies of 15 additional DOGs obtained through $K$-band AO observations,  eleven of which have $24\mu m$ fluxes above 0.8 mJy.  Both papers find DOGs to possess a high fraction of disk-like profiles ($\sim 90\%$ and 50\%, respectively), and argue, based on their size, sersic index, axis ratio and morphological parameters that DOGs are consistent with being late-stage mergers, transitioning from the chaotic coalescence towards becoming relaxed ellipticals.  

\begin{deluxetable*} {cccc}  
\tablecolumns{4}
\tablewidth{0pt}
\tablecaption{DOG morphologies compared to $24\mu m$-galaxies\tablenotemark{a}}
\tablehead{
	\colhead{Merging sequence} &
	\colhead{Morphological class} &
	\colhead{All $z \ge 1.5$} &
	\colhead{DOGs} \\
	\colhead{} &
	\colhead{} &
	\colhead{$24\mu m$-galaxies} &
	\colhead{}
}
\startdata
Isolated objects & All spirals & 0 & 0 \\ \\
First approach & Close pairs (phase I) & 2\%(1) & 0  \\ \\
\multirow{3}{*}{Early mergers} & First contact (phase II) & 2\%(1) & 0 \\
& Pre-mergers (phase III) & 45\% (27) & 46\% (12) \\
& Triplets & 3\% (2) & 0 \\ \\
\multirow{3}{*}{Coalescence} & Advanced & & \\
& Mergers (phase IV) & 26\% (15.4) & 19\% (5) \\
& Faint \& Compact & 7\% (4.4) & 12\% (3)  \\ \\
\multirow{3}{*}{Late mergers} & Old mergers (phase V) & 13\% (8) & 19\% (5) \\
& Regular bulges & 2\% (1) & 4\% (1) \\
& Pure point sources & 1\% (0.5) & 0 \\
\enddata
\tablenotetext{a}{After redistribution of objects with unknown redshifts in the proportions found for objects with known redshifts.}
\label{tbl:dogs}
\end{deluxetable*}

There are 35 DOGs in our sample (Figure~\ref{fig:selection}), 75\% of which are at $z \ge 1.5$.  Not counting $z < 1.5$ objects, we observe that about half of our DOGs are actually pre-mergers (phase III objects), while only half are in more advanced stages, contrary to the hypothesis that they are predominantly late-stage objects.  We agree, however, that most of them have disk-dominated profiles, and find that fraction to be 2/3.  The morphological properties of DOGs, summarized in Table~\ref{tbl:dogs}, suggest that they are not very different from the overall $24\mu m$-bright population at $z \ge 1.5$.  They only tend to draw a little bit more towards AGN-dominated morphological classes, that is faint \& compact objects and late-stage mergers.  As a consequence, we also find them to be slightly more compact, AGN-dominated and obscured than average bright $24\mu m$-selected galaxies at $z \ge 1.5$.

\subsection{Comparison with sub-mm selected galaxies}

Sub-millimeter selected galaxies (SMGs) form another type of galaxies known to lie at $z \approx 2$ \citep{Chapman05} and to be extremely luminous in the infrared \citep{Kovacs06}.  These objects are thus closely related to our bright $24\mu m$-sources.  Unlike our $24\mu m$-sources, however, SMGs possess PAH-dominated mid-IR spectra in $\gtrsim 80\%$ of cases \citep{Pope08, Menendez09}.  They would, therefore, largely populate the high-EW, high-$z$ part of Figure~\ref{fig:ew_vs_z}: the only region not occupied by our galaxies.  The two populations, thus, complement one another in covering the broader high-redshift ULIRG population.  Nevertheless, many SMGs show composite spectra, so that there is also significant overlap between the two types of sources (Sajina et al. in preparation). 

High-resolution HST observations of sub-millimeter galaxies (SMGs) at both optical \citep{Smail98, Conselice03b, Pope05, Dunlop09, Swinbank10} and near-infrared wavelengths \citep{Swinbank10} have revealed that a large fraction of them ($40\% \textendash 90\%$) show highly disturbed or multiple component morphologies indicative of merging activity.  CO kinematics have similarly confirmed the merging nature of a majority of these objects \citep{Neri03, Greve05,Tacconi06, Tacconi08, Schinnerer08,Iono09, Bothwell10}, although large star-forming disks have certainly also been found to exist among the sub-millimeter population \citep{Carilli10, Bothwell10}.  Most models also favor a merger origin for SMGs \citep[e.g.][]{Baugh05, Chakrabarti08, Narayanan09b, Narayanan10}, even though regular disk star formation fueled by cold flow accretion also appears to be a viable path \citep{Dave10}.

For our purposes, we mostly refer to the results of \citet{Swinbank10}, who published the first morphological analysis of sub-millimeter galaxies at near-infrared wavelengths.  Although they did not use the same approach towards morphological classification as we do here, we can still infer, from their $H$-band images, that 40\% of objects in their sample have two or more distinguishable components.  This makes the fraction of early mergers among SMGs very close to that of our $z \ge 1.5$ galaxies.  We can also see that their objects display many of the same kind of disturbed morphologies as ours.  We might, thus, expect the morphological distribution of SMGs not to be hugely different from that of our high-$z$ objects, especially given that CO studies, as mentioned above, have found a number of those objects to possess rather chaotic kinematics characteristic of the more advanced merging stages.  Nonetheless, we know that purely AGN-dominated objects are not found in sub-millimeter samples, and, based on the results presented in section~\ref{sec:midIR_vs_morph}, we would therefore expect faint \& compact objects, as well as very late-stage objects that appear as regular ellipticals, to be absent from SMG samples.  This does appear to be true though more detailed analysis would be needed to confirm it.  Size and structural analysis, which we postpone for subsequent publication, might also reveal more subtle differences between $24\mu m$ and sub-millimeter galaxies.

\subsection{Comparison with far-IR selected galaxies}

\citet{Kartaltepe10} have conducted the most comprehensive study of the role of mergers in far-IR selected galaxies to date using the full 2 sq. deg. of HST/ACS imaging and multi-wavelength coverage in the COSMOS field to investigate the morphological properties of a complete sample of 1500 $70\mu m$-selected galaxies.  They too adopted the morphological classification of \citet{Surace98} based on the merging sequence, enabling direct comparison with our results.  This comparison, however, is a little bit complicated by the fact that they use optical rather than NIR imaging, causing $\sim 20\%$ of their objects at $z > 1$ and $L_{IR} > 10^{12} L_{\odot}$ to be too faint to classify morphologically.  This caveat is sufficiently minor though to be circumvented, and we demonstrate below that interesting conclusions can still be drawn from a comparison of the two samples.

Visually, the $70\mu m$-galaxies of \citet{Kartaltepe10} resemble a lot more our low-redshift ($z < 1.5$) sample than either our high-redshift galaxies or the SMGs.  Since their sample includes many low-redshift objects with moderate IR-luminosities, we use, for our comparison, only those objects in their sample with a $\log (L_{IR}/L_{\odot}) > 11.5$.  We also limit our sample to $z < 1.5$ objects in order to ensure that both samples have comparable distributions in both redshift and total IR-luminosity.   Table~\ref{tbl:comparison} shows the detailed comparison by morphological class.  In their classification, \citet{Kartaltepe10} make the distinction between minor and major mergers whereas we do not.   Then, many of their sources have unknown morphologies because they are too faint to be detected in the observed optical.  For comparison purposes, we, therefore, redistributed their minor mergers, for lack of a better prior, between phases III \& IV in the same proportion as that of their major mergers, and their unknown objects equally among phase III and \textquotedblleft faint \& compact\textquotedblright objects, since these are the two classes our faint objects fall in.  We show in table~\ref{tbl:comparison} both the observed and redistributed numbers.

\begin{deluxetable*} {ccccc}  
\tablecolumns{5}
\tablewidth{0pt}
\tablecaption{Comparison of morphologies of 24 and 70$\mu m$-selected galaxies}
\tablehead{
	\colhead{Merging sequence} &
	\colhead{Morphological class} &
	\colhead{$24\mu m$-selected\tablenotemark{a}} &
	\colhead{$70\mu m$-selected\tablenotemark{b}} &
	\colhead{$70\mu m$-selected\tablenotemark{c}} \\
	\colhead{} &
	\colhead{} &
	\colhead{} &
	\colhead{(observed)} & 
	\colhead{(estimated)}
}
\startdata
Isolated objects & All spirals & 16\% & 22\% & 22\% \\ \\
First approach & Close pairs (phase I) & 4\% & 0 & 0 \\ \\
\multirow{3}{*}{Early mergers} & First contact (phase II) & 3\% & 1\% & 1\% \\
& Pre-mergers (phase III) & 15\% & 21\% & 36\% \\
& and Triplets & & & \\ \\
\multirow{3}{*}{Coalescence} & Advanced & & & \\
& Mergers (phase IV) & 34\% & 15\% & 22\% \\
& Faint \& Compact & 9\% & \nodata & 5\%  \\ \\
\multirow{3}{*}{Late mergers} & Old mergers (phase V) & 15\% & 3\% & 3\% \\
& Regular bulges & 4\% & 9\% & 9\% \\
& Pure point sources & 1\% & 2\% & 2\% \\ \\
\nodata & Minor mergers & \nodata & 16\% & \nodata \\
\nodata & Unknowns & \nodata & 11\% & \nodata \\
\enddata
\label{tbl:comparison}
\tablenotetext{a}{$z < 1.5$ objects only}
\tablenotetext{b}{Data from \citet{Kartaltepe10}, including only objects with $\log (L_{IR}/L_{\odot}) > 11.5$}
\tablenotetext{c}{After redistribution of minor mergers and unknowns.}
\end{deluxetable*}

Table~\ref{tbl:comparison} confirms the astonishing correspondence between our $z < 1.5$ sample and their bright $70\mu m$-sources.  It is remarkable that the proportions in all phases of the merging process differ by no more that $\sim 15\%$ (of the total sample), lending great support to both our results.  Nevertheless, the increase in the number of isolated spirals and early-phase mergers among $70\mu m$ sources and the corresponding decline in the number of coalescence and post-coalescence mergers when compared to $24\mu m$ sources is clear, and probably a simple consequence of the higher proportion of star-formation dominated galaxies among $70\mu m$-selected sources.

The broad correspondence between $24\mu m$ and $70\mu m$-selected galaxies suggests, as does the similarity between $z \sim 2$, $24\mu m$-selected and submillimeter-selected galaxies, that most of the observed morphological specificity in these different samples is common and intrinsic to {\em all} infrared-luminous objects, largely irrespective of whether they are powered by star formation or an AGN (although these can be responsible for small differences between the various classes).  It also suggests that this commonality springs from the origin of those objects in massive mergers.  The difference between $z \sim 1$ and $z \sim 2$ samples, on the other hand, indicates that there is redshift evolution in the morphological character of IR-luminous objects.  We argue in the next section that this difference is the result of higher gas fractions at higher redshifts.

\subsection{Comparison with theory and consequences for galaxy evolution \label{sec:theory}}

Hydrodynamical simulations of merging disk galaxies combined with radiative transfer calculations have provided us with a successful physical model of the origin of ULIRGs and QSOs \citep{diMatteo05, Narayanan09}.  These can be used, more generally, to study the outcome of any merger by varying the set of initial conditions given by parameters such as the mass ratio of the two galaxies, their gas fractions, their relative orientation and so on.  Using the statistical results of such a suite of simulations, and embedding them into a cosmological framework using an observationally motivated halo occupation distribution, \citeauthor{Hopkins08}, in a series of papers starting from (2008), have been able to calculate the role mergers play in the evolution of various global and observable properties of the galaxy population.

Most recently, \citet{Hopkins10a} calculated the IR-luminosity at which, according to their models, star formation transitions from occurring mostly in isolated galaxies to being primarily triggered by mergers.  We plotted their results on top of our Figure~\ref{fig:lum_vs_z} for comparison.  Our data agree with their models on that we find nearly all of our isolated spirals below the predicted line, indicating that their models provide an accurate upper limit as to the luminosities at which quiescent star formation usually occurs.  We also find good agreement at high redshift where nearly all of our objects fall above their line {\em and} are, indeed, found to be mergers.  We appear to disagree, on the other hand, at lower redshift where we find many merger-identified objects below their dividing line.  This, however, is not necessarily a problem, the reason being that the $24\mu m$ selection of our sample is very different from that of an $L_{IR}$ selection.  In particular, we know that our selection highly favors objects whose SED contains an obscured AGN component \citep{Sajina07,Dasyra09}.  If we adopt the results of \citet{Hopkins10b} which argue that the fueling of a strong AGN (i.e. $\dot{M}> 0.1 \mbox{ M}_{\odot} \mbox{ yr}^{-1}$ or $L_{IR} \gtrsim 10^{11} \mbox{ L}_{\odot} \mbox{ yr}^{-1}$) can only occur in mergers, then this means that we are preferentially selecting mergers, and that our sample is biased compared to the general infrared-luminous population.  Our data, thus, do not lend themselves very well to a comparison with the results of \citet{Hopkins10a} except on isolated objects, as discussed above.  They do, however, seem to support another result, namely the one argued for in \citet{Hopkins10b} which is that it is very hard to fuel a strong AGN other than through a merger event.

Another aspect of those galaxy merger simulations is that the merging of two massive gas-rich disks almost always produces an extended IR-luminous phase dominated, at first, by star formation during early stages of the merging, and then by accretion onto the central black hole after coalescence.  Our results, as well as those from observations of local ULIRGs \citep{Veilleux09a,Farrah09}, show that the above scenario is, indeed, likely to be the evolutionary path of many low to medium redshift ($z < 1.5$) ULIRGs.  However, both our observations and that of local ULIRGs also show that there is a great variety of objects at all stages of the merging process, implying many more possible evolutionary paths.  This variety has not been fairly reproduced, so far, in simulations.

At high redshifts (and high luminosities), our results indicate that mergers depart even more from that classical scenario, and that black hole accretion plays an important role, at least among our sources, already in the early stages of merging.  CO observations have demonstrated that $24\mu m$-selected galaxies at $z \sim 2$ possess larger amounts of gas than do local ULIRGs \citep{Yan10}.  We speculate that these elevated gas fractions might be at the origin of the increased AGN activity at early stages.

Another characteristic of bright $24\mu m$-galaxies at $z \sim 2$ that our observations, as well as that of others \citep[e.g.][]{Dasyra08, Bussmann09}, have revealed is their preponderantly disk-dominated profiles.  Here, we know, thanks to simulations by \citet{Robertson06} and others, that high gas fractions in major mergers can act to stabilize the disk component and that major mergers can even result in rotationally supported disk remnants provided that the progenitors are extremely gas-rich ($f_{gas} \ge 0.8$) \citep[see also][]{Springel05,Barnes02}.  The results of these simulations thus provide a natural explanation linking this large observed fraction of disk-like morphologies to the high gas content of these objects.

\citet{Robertson06} argue that the kind of galaxy assembly in which mergers result in a disk remnant is probably limited to high redshifts where gas reservoirs are much larger \citep{Noterdaeme09}.  Although it is impossible to know without detailed kinematics whether any of our $z \sim 2$ mergers exhibit such behavior, we see, based on their profile, visual appearance and merger stage, two such candidates: MIPS16144 and MIPS16122.  Both have clearly detected PAHs (EW$(11.3\mu m) = 0.5$ and 0.19, respectively).  

These two tentative disk merger remnants at $z \sim 2$, and the lack thereof at lower redshifts, combined with the overall large fraction of objects with disk profiles at those redshifts, suggest that this epoch might correspond to the end of the era of gas-rich mergers \`{a} la \citet{Robertson06}.   Interestingly, that period in cosmos history also concurs with the onset of the rapid rise in mass density of elliptical and red sequence galaxies \citep{Arnouts07,Ilbert09}.  These two observations can naturally explain each other, and point to a transition, at $z \sim 2$, from an epoch of disk-dominated galaxy evolution to an epoch of bulge formation.

\section{Summary}

Our new analysis of 134 HST/NICMOS images of bright high-redshift $24\mu m$-selected galaxies reveals a higher incidence of mergers than previously reported.  Our study is based on a flux-limited sample of objects selected at $S(24\mu m) > 0.9$ mJy, with a sampling of the $[24]-R$ color space that favors high redshift objects.  Our sample shows two broad peaks in its redshift distribution, one at $z \sim 1$ and one at $z \sim 2$.  Full SED analysis presented in a companion paper (Sajina et al., in preparation) demonstrates that most of our objects above $z=1$ have total IR-lumonisities above $10^{12} L_{\odot}$, whereas those at lower redshifts mostly have luminosities ranging from $10^{11.5}$ to $10^{12} L_{\odot}$.  Above $z=1.5$, nearly all objects in our sample have infrared luminosities in excess of $10^{12.5} L_{\odot}$.  The merging properties of our sample are summarized below:

\begin{itemize}

\item We find a high overall merger fraction, which we estimate to lie at $\sim 80\%$.  Possible values, however, range from 62 to 91\%.

\item Mergers represent 65 to 96\% of ULIRGs in our sample, but our best estimate puts that fraction at 87\%, suggesting little evolution in the dynamical origin of that category of objects.

\item Isolated spirals form 9\% of our sample.  They all lie at $z < 1.2$ and have luminosities of $L_{3\textendash 1000\mu m} \le 1.6 \times 10^{12} L_{\odot}$.

\item At $z < 1.5$, the distribution of our objects draws towards more advanced stages of the merging process when compared to local ULIRGs.  Both samples, however, are dominated by singly-nucleated (coalesced/coalescing) mergers with bulge-like profiles.

\item Comparison with simulated observations of redshifted local ULIRGs demonstrate that our objects also tend to show stronger merger signatures.  This manifests itself in brighter tidal tails and a higher detection of residuals after subtraction of the smooth component.

\item At $z \ge 1.5$, the morphological composition changes significantly with a dramatic rise in the number of early mergers:  systems in which the two merging objects are still distinguishable.

\item Still at $z \ge 1.5$, 60\% of our objects possess disk-dominated profiles which we argue is partially a consequence of their higher gas content that acts to retain a larger disk for longer periods of time during a merger.

\end{itemize}

Mid-IR IRS spectra for all objects in our sample were presented in \citet{Sajina07} and \citet{Dasyra09}.  These data give us an indication of the power source at the origin of the bright mid-IR flux of our objects in the form of the equivalent width of the PAH features they probe.  They also provide a measure of AGN obscuration through the depth of the $9.7\mu m$ silicate feature.  Combining this information with our morphological data gives us new insights on the merging process and the formation of quasars at high redshift.  It allows us to deduce the following conclusions: 

\begin{itemize}

\item Statistically, the importance of AGN activity relative to star formation increases as objects progress along the merging sequence.  At $z < 1.5$, most objects are dominated by star formation before their first pass, whereas past coalescence, most become AGN-dominated.  This trend is similar to that observed in local ULIRGs \citep{Veilleux09b}.

\item At $z > 1.5$, however, even early mergers possess, in $\gtrsim 90\%$ of cases,  a significant power-law continuum indicative of the presence of an AGN.  Combined with the fact that these have luminosities in excess of $10^{12.5} L_{\odot}$, and the rarity (or inexistence) of such objects locally, this signifies that black-hole accretion plays an important role earlier on in the merging process at higher redshifts.  We speculate that this, again, might be a consequence of the elevated gas fractions of objects at those redshifts.

\item We find that obscured quasars live in hosts with faint \& compact morphologies.

\item Our simulated observations indicate that local dense-core mergers that have diffuse tails and not yet fully formed bulges are the best candidate analogs of faint \& compact objects at high redshift.  These kind of mergers probably occur only in certain types of initial configurations.

\item We find, however, that only few objects, four in total in our sample, possess spectra characteristic of obscured quasars.  We conclude that either mergers rarely go through that phase or else it's very short-lived.

\item We find late-stage mergers to predominantly show spectra characteristic of {\em unobscured} AGNs.

\item We do not find other morphological classes to occupy any definite region in the Spoon diagram, implying that infrared-luminous mergers do not follow any characteristic trajectory, but can evolve through many possible paths.

\end{itemize}

Comparison with other galaxy samples indicates that up to $z \approx 1.5$ the properties of bright $24\mu m$-selected galaxies are very similar to that of local ULIRGs and top-end LIRGs ($L_{IR} > 10^{11.5} L_{\odot}$).  They also match well those of $70\mu m$-selected galaxies of similar luminosity and redshift.  At $z > 1.5$, we begin to see a new type of galaxy appearing in large numbers: early mergers with high infrared luminosities ($L_{IR} > 10^{12.5} L_{\odot}$) and low PAH equivalent widths (EW$_{7.7\mu m} \sim 0.3$).  These have no known local equivalent.  We also begin to see a preponderance of disk-dominated profiles, which contrasts that of local ULIRGs \citep{Veilleux06}.  On the other hand, these high-$z$ bright $24\mu m$-galaxies appear not to differ as much from other types of $z \sim 2$ infrared-luminous objects such as DOGs \citep{Dey08} or SMGs \citep{Blain02}.  We therefore conclude that, although still driven by mergers, high redshift ultra-luminous infrared galaxies are substantially different in nature from local ULIRGs and that this difference is likely rooted in their higher gas masses \citep{Yan10}.

\acknowledgements
We are grateful to M.~Elitzer and L.~Armus for insightful discussion about various aspects of mid-IR spectroscopy, as well as to J.~Lotz on aspects of automated classification.  We are very much obliged towards J.~Kartaltepe for kindly sharing her results ahead of publication and for providing us with helpful information.  We would also like to thank P.~Hopkins for providing us data from his simulations.  We are grateful to V.~Fadeyev and C.~Peng for their help with their improved NICMOS pipeline and GALFIT softwares, respectively.  We would like to extend our thanks to the referee for his/her comments that helped improve this paper.  M.~Z. acknowledges support from the HST GO grant 11142.

{\it Facilities:} \facility{HST(NICMOS)}, \facility{Spitzer(IRS, MIPS)}

\bibliographystyle{apj}
\bibliography{apj-jour,preprint_arxiv}

\appendix

\section{Redshifting Procedure \label{app:simulations}}

We describe, below, the procedure we followed to create simulated observations of local ULIRGs and mergers at high redshift.  The simulation procedure involves two main steps.  The first one consists of bringing the image to the NICMOS resolution at the desired redshift.  To achieve this, we effectively resample the WFPC2 PSF on a grid whose pixel size equals $\frac{D_{A,z} \times s_{\mbox{\tiny NIC2}}}{D_{A,0} \times s_{\mbox{\tiny WFPC2}}}$ times the original pixel size of the PSF ($s_{\mbox{\tiny NIC2}}$ corresponds to the pixel scale of the NIC2 camera in arc seconds, or $0.0756"$, and $s_{\mbox{\tiny WFPC2}}$ to that of the appropriate WFPC2 camera in arc seconds, or $0.1"$ for the WF cameras, and $0.0455"$ for the PC camera).  This gives us the intrinsic resolution of our object at that redshift.  We then deconvolve the NICMOS PSF with that resampled PSF by simply dividing in Fourier-space and applying a tapering at the edges, if necessary.  This step is the one that involves the most user interaction in order to make sure that the division in Fourier-space does not create too much ringing, or does not overly amplify the noise.

Once we have our convolution kernel in hand, we proceed to resample our image on a grid whose pixel size is multiplied by the same factor as above, and then convolve it with the WFPC2 $\rightarrow$ NIC2 kernel.  We then dim our image by the square of the ratio of the appropriate luminosity distances, plus a factor of $(1+z)$ to account for the fact that we are working with fluxes {\em per unit wavelength}, rather than bolometric quantities.  We then further multiply our image by the ratio of the sensitivities of the two cameras, expressed in erg cm$^{-2}$ $\AA^{-1}$ / DN, and the ratio of exposure times, in order to get an image in units of NICMOS counts per second.  We finally put that image onto an actual, but blank, NICMOS image obtained during the course of our observations.

We repeated that redshifting procedure four times, shifting the object around using the same dither pattern we used for our observations.  We also added a random jitter component to our shifts to reproduce observations most faithfully (though it turned out not to make much difference).  Each image was laid onto a different blank background image.  We then drizzle the four simulated images back onto one grid using the same procedure as for our actual data to produce our final simulated images.

\section{Morphological Data}
\LongTables
\tabletypesize{\scriptsize}

\begin{deluxetable}{lcccccccccc}
\tablecolumns{7}
\tablewidth{0pt}
\tablecaption{Source morphology \label{tbl:morphclass}}
\tablehead{
	\colhead{MIPS ID} &
	\colhead{Redshift} &
	\colhead{$m_{H}$} &
	\colhead{$r_{1/2}$ (kpc)} &
	\colhead{Dominant} &
	\colhead{Morphological} &
	\colhead{Confidence} \\
	\colhead{} &
	\colhead{} &
	\colhead{} &
	\colhead{} &
	\colhead{component} &
	\colhead{class} &
	\colhead{level}	
}
\startdata
MIPS34			& 0.65    & 17.84 &  2.55 & bulge & Old merger (V) & 4 \\
MIPS42\tablenotemark{a} & 1.95    &       &       & disk & Faint \& Compact & 6 \\
MIPS78    		& 2.65    & 21.96 &  2.84 & disk & Faint \& Compact & 4 \\
MIPS159   		& \nodata & 18.43 &       & ambiguous & Point source & 6 \\
MIPS168   		& 0.24    & 17.61 &  1.20 & bulge & Advanced merger (IV) & 2 \\
MIPS180   		& 2.47    & 22.08 &  2.08 & disk & Pre-merger (III) & 4 \\
MIPS184   		& \nodata & 18.12 & $\sim  5.81$\tablenotemark{b}~ & disk & Edge-on spiral & \nodata \\
MIPS213   		& 1.22    & 20.61 &  1.37 & disk & Pre-merger (III) & 4 \\
MIPS224   		& 1.47    & 18.55 &  2.83 & bulge & Advanced merger (IV) & 2 \\
MIPS227   		& 1.63    & 18.65 &  2.64 & bulge & Advanced merger (IV) & 3 \\
MIPS268   		& 1.69    & 19.90 &  1.95 & disk & Pre-merger (III) & 4 \\
MIPS289   		& 1.86    & 18.27 &  3.54 & disk & Pre-merger (III) & 1 \\
MIPS298   		& 3.49    & 22.08 &       & disk & Pre-merger (III) & 5 \\
MIPS322   		& \nodata & 19.38 & $\sim  0.96$\tablenotemark{b}~ & ambiguous & Pre-merger (III) & 2 \\
MIPS324   		& 0.95    & 18.40 &  3.36 & bulge & Advanced merger (IV) & 3 \\
MIPS331   		& 1.03    & 18.66 &  0.90 & ambiguous & Old merger (V) & 5 \\
MIPS351   		& 1.16    & 19.44 &  1.41 & bulge & Old merger (V) & 5 \\
MIPS358   		& 0.81    & 18.75 &  3.09 & disk & Advanced merger (IV) & 2 \\
MIPS369   		& 0.70    & 19.05 &  3.28 & disk & Face-on spiral & \nodata \\
MIPS397   		& 1.35    & 19.32 &  3.47 & bulge & Regular bulge & 6 \\
MIPS419   		& 0.83    & 19.61 &  1.51 & disk & Faint \& Compact & 5 \\
MIPS446   		& 0.82    & 18.37 &  3.84 & ambiguous & Advanced merger (IV) & 3 \\
MIPS464   		& 1.85    & 20.86 &  1.92 & disk & Close pair (I) & 4 \\
MIPS488   		& 0.69    & 18.13 &  3.70 & disk & Advanced merger (IV) & 2 \\
MIPS495   		& 0.75    & 19.73 &  3.00 & ambiguous & Old merger (V) & 5 \\
MIPS505   		& 1.65    & 21.12 &  0.85 & bulge & Advanced merger (IV) & 4 \\
MIPS506   		& 2.47    & 20.78 &  1.71 & disk & Pre-merger (III) & 4 \\
MIPS512   		& 0.99    & 19.33 &  1.98 & bulge & Advanced merger (IV) & 4 \\
MIPS544   		& 0.98    & 19.68 &  1.76 & disk & Faint \& Compact & 5 \\
MIPS546  	 	& 1.07    & 18.55 &  4.01 & disk & Advanced merger (IV) & 3 \\
MIPS562   		& 0.55    & 17.85 &  6.88 & disk & First contact (II) & 2 \\
MIPS7985  		& 2.78    & 18.42 &  1.26 & disk & Advanced merger (IV) & 3 \\
MIPS8069  		& 0.70    & 17.50 &  1.68 & bulge & Advanced merger (IV) & 4 \\
MIPS8071  		& 0.98    & 18.42 &  0.53 & bulge & Regular bulge & 6 \\
MIPS8098  		& 1.07    & 18.46 &  3.19 & bulge & Advanced merger (IV) & 4 \\
MIPS8107  		& 0.94    & 19.02 &  5.64 & disk & Advanced merger (IV) & 3 \\
MIPS8172  		& 1.11    & 19.10 &  0.41 & ambiguous & Old merger (V) & 5 \\
MIPS8179  		& 0.59    & 18.04 &  3.17 & bulge & Advanced merger (IV) & 3 \\
MIPS8185  		& \nodata & 19.29 & $\sim  0.58$\tablenotemark{b}~ & bulge & Old merger (V) & 5 \\
MIPS8204  		& 0.85    & 18.83 &  1.75 & disk & Pre-merger (III) & 4 \\
MIPS8224  		& \nodata & 19.89 & $\sim  1.31$\tablenotemark{b}~ & bulge & Old merger (V) & 4 \\
MIPS8226  		& 2.10    & 19.86 &  0.17 & bulge & Advanced merger (IV) & 4 \\
MIPS8233  		& 0.99    & 19.09 &  1.92 & disk & Pre-merger (III) & 4 \\
MIPS8242  		& 2.45    & 19.72 &  4.92 & disk & Pre-merger (III) & 3 \\
MIPS8245  		& 2.70    & 22.37 &  1.63 & disk & Faint \& Compact & 6 \\
MIPS8251  		& 1.94    & 20.33 &  1.34 & bulge & Pre-merger (III) & 2 \\
MIPS8253  		& 0.95    & 18.64 &  3.41 & bulge & Advanced merger (IV) & 3 \\
MIPS8308  		& 0.37    & 17.58 &  2.80 & disk & Advanced merger (IV) & 3 \\
MIPS8311  		& 1.17    & 18.93 &  2.99 & disk & Face-on spiral & \nodata \\
MIPS8315  		& \nodata & 19.86 & $\sim  1.52$\tablenotemark{b}~ & ambiguous & Old merger (V) & 5 \\
MIPS8325  		& 0.61    & 18.53 &  2.27 & bulge & Triplet & 1 \\
MIPS8327  		& 2.44    & 20.95 &  1.11 & ambiguous & First contact (II) & 4 \\
MIPS8328  		& 1.02    & 18.93 &  3.10 & disk & Advanced merger (IV) & 3 \\
MIPS8342  		& 1.56    & 20.26 &  1.21 & bulge & Advanced merger (IV) & 4 \\
MIPS8360  		& 1.50    & 18.91 &  3.33 & disk & Advanced merger (IV) & 3 \\
MIPS8375  		& 0.87    & 18.89 &  2.12 & disk & Pre-merger (III) & 1 \\
MIPS8387  		& 0.91    & 17.57 &  9.11 & disk & Close pair (I) & 2 \\
MIPS8388  		& 1.14    & 18.95 &  3.38 & bulge & Advanced merger (IV) & 3 \\
MIPS8392  		& 1.90    & 21.79 &  2.40 & disk & Faint \& Compact & 6 \\
MIPS8400  		& 1.51    & 18.44 &  1.46 & bulge & Old merger (V) & 5 \\
MIPS8405  		& 1.16    & 18.66 &  1.24 & bulge & Triplet & 3 \\
MIPS8407  		& 0.89    & 20.93 &  1.32 & ambiguous & Regular bulge & 6 \\
MIPS8413  		& 2.18    & 21.57 &  1.75 & disk & Pre-merger (III) & 3 \\
MIPS8430  		& 0.67    & 18.93 &  2.15 & disk & Close pair (I) & 2 \\
MIPS8450  		& 1.00    & 19.53 &  6.16 & disk & Face-on spiral & \nodata \\
MIPS8462  		& 1.01    & 18.82 &  3.16 & disk & Close pair (I) & 2 \\
MIPS8465  		& \nodata & 19.33 & $\sim  0.66$\tablenotemark{b}~ & bulge & Old merger (V) & 4 \\
MIPS8477  		& 1.84    & 19.44 &  1.55 & bulge & Advanced merger (IV) & 3 \\
MIPS8479  		& 1.30    & 20.51 &  3.25 & disk & Faint \& Compact & 5 \\
MIPS8493  		& 1.80    & 20.53 &  4.16 & disk & Triplet & 3 \\
MIPS8499  		& 0.60    & 17.27 &  5.92 & bulge & Advanced merger (IV) & 2 \\
MIPS8507  		& 0.76    & 17.98 &  4.40 & disk & Face-on spiral & \nodata \\
MIPS8521  		& 1.19    & 19.58 &  3.80 & disk & Pre-merger (III) & 3 \\
MIPS8532  		& 0.86    & 18.60 &  3.31 & disk & Face-on spiral & \nodata \\
MIPS8543  		& 0.65    & 18.09 &  4.05 & disk & Face-on spiral & \nodata \\
MIPS15690 		& 0.85    & 18.51 &  3.80 & disk & Edge-on spiral & \nodata \\
MIPS15755 		& 0.74    & 17.66 &  4.12 & disk & Face-on spiral & \nodata \\
MIPS15771 		& 2.20    & 21.07 &  1.76 & bulge & Pre-merger (III) & 4 \\
MIPS15776 		& 1.12    & 18.65 &  2.25 & bulge & Advanced merger (IV) & 2 \\
MIPS15840 		& 2.30    & 21.58 &  2.20 & disk & Pre-merger (III) & 4 \\
MIPS15871 		& \nodata & 20.54 & $\sim  0.64$\tablenotemark{b}~ & ambiguous & Old merger (V) & 5 \\
MIPS15880 		& 1.64    & 20.84 &  3.22 & disk & Pre-merger (III) & 3 \\
MIPS15897 		& 1.62    & 18.67 &  0.67 & bulge & Old merger (V) & 5 \\
MIPS15928 		& 1.50    & 19.59 &  2.51 & disk & Pre-merger (III) & 3 \\
MIPS15941 		& 1.23    & 20.11 &  1.85 & disk & Faint \& Compact & 4 \\
MIPS15949 		& 2.12    & 20.76 &  2.83 & disk & Advanced merger (IV) & 3 \\
MIPS15958 		& 1.97    & 21.14 &  0.81 & ambiguous & Old merger (V) & 4 \\
MIPS15967 		& 1.30    & 18.12 &  2.62 & bulge & Advanced merger (IV) & 4 \\
MIPS15977 		& 1.85    & 19.71 &  2.15 & bulge & Advanced merger (IV) & 3 \\
MIPS16037 		& 1.61    & 20.43 &  1.86 & bulge & Regular bulge & 6 \\
MIPS16047 		& 0.52    & 17.99 &  3.46 & disk & Edge-on spiral & \nodata \\
MIPS16059 		& 2.33    & 20.29 &  2.51 & disk & Pre-merger (III) & 1 \\
MIPS16080 		& 2.01    & 20.21 &  2.36 & disk & Advanced merger (IV) & 2 \\
MIPS16095 		& 1.81    & 20.01 &  1.75 & disk & Advanced merger (IV) & 3 \\
MIPS16099 		& 0.95    & 19.70 &  2.02 & bulge & Advanced merger (IV) & 3 \\
MIPS16113 		& 1.90    & 20.96 &  2.54 & disk & Pre-merger (III) & 3 \\
MIPS16118 		& 2.61    & 21.93 &  3.50 & disk & Pre-merger (III) & 4 \\
MIPS16122 		& 1.97    & 21.40 &  2.59 & disk & Advanced merger (IV) & 3 \\
MIPS16132 		& \nodata & 18.97 & $\sim  2.22$\tablenotemark{b}~ & disk & Pre-merger (III) & 4 \\
MIPS16135 		& 0.62    & 19.41 &  2.13 & disk & Edge-on spiral & \nodata \\
MIPS16144 		& 2.13    & 20.36 &  2.97 & disk & Advanced merger (IV) & 3 \\
MIPS16156 		& 0.72    & 18.16 &  3.86 & disk & Pre-merger (III) & 2 \\
MIPS16170 		& 0.32    & 17.72 &  2.35 & disk & Edge-on spiral & \nodata \\
MIPS16202 		& \nodata & 20.11 & $\sim  2.11$\tablenotemark{b}~ & disk & Advanced merger (IV) & 3 \\
MIPS16219 		& 2.72    & 19.68 &  1.17 & bulge & Old merger (V) & 5 \\
MIPS16249 		& 0.53    & 18.44 &  1.80 & disk & Pre-merger (III) & 2 \\
MIPS16267 		& 1.31    & 21.11 &  1.60 & ambiguous & Old merger (V) & 4 \\
MIPS22196 		& 0.80    & 18.73 &  0.96 & bulge & Advanced merger (IV) & 2 \\
MIPS22204 		& 1.97    & 20.22 &  1.05 & bulge & Pre-merger (III) & 2 \\
MIPS22248 		& 1.20    & 18.85 &  2.54 & disk & Faint \& Compact & 4 \\
MIPS22277 		& 1.77    & 19.68 &  2.02 & ambiguous & Pre-merger (III) & 4 \\
MIPS22303 		& 2.34    & 22.60 &  1.65 & disk & Pre-merger (III) & 3 \\
MIPS22307 		& 0.70    & 18.14 &  2.05 & bulge & Advanced merger (IV) & 2 \\
MIPS22352 		& 0.66    & 17.67 &  3.23 & bulge & Advanced merger (IV) & 3 \\
MIPS22371 		& 1.67    & 19.64 &  3.07 & disk & Pre-merger (III) & 3 \\
MIPS22379 		& 0.65    & 17.85 &  3.39 & bulge & Old merger (V) & 4 \\
MIPS22397 		& 1.83    & 19.76 &  2.78 & disk & Pre-merger (III) & 3 \\
MIPS22432 		& 1.59    & 19.80 &  4.18 & disk & Pre-merger (III) & 1 \\
MIPS22516 		& 1.35    & 19.67 &  1.20 & bulge & Advanced merger (IV) & 3 \\
MIPS22527 		& \nodata & 19.73 & $\sim  0.84$\tablenotemark{b}~ & bulge & Pre-merger (III) & 4 \\
MIPS22530 		& 1.95    & 20.70 &  3.54 & disk & Pre-merger (III) & 2 \\
MIPS22536 		& 1.59    & 19.04 &  2.73 & disk & Triplet & 2 \\
MIPS22549 		& 1.05    & 20.12 &  2.30 & disk & Faint \& Compact & 5 \\
MIPS22555 		& 1.88    & 20.82 &  1.99 & disk & Pre-merger (III) & 2 \\
MIPS22557 		& 0.79    & 18.03 &  3.17 & bulge & Advanced merger (IV) & 2 \\
MIPS22558 		& 3.20    & 21.61 &  0.60 & bulge & Advanced merger (IV) & 3 \\
MIPS22621 		& \nodata & 23.71 & $\sim  0.67$\tablenotemark{b}~ & ambiguous & Faint \& Compact & 6 \\
MIPS22635 		& 0.80    & 19.36 &  1.81 & disk & First contact (II) & 3 \\
MIPS22638 		& 0.98    & 19.21 &  4.26 & disk & Pre-merger (III) & 4 \\
MIPS22651 		& 1.73    & 19.33 &  2.50 & bulge & Pre-merger (III) & 3 \\
MIPS22661 		& 1.75    & 20.35 &  1.70 & bulge & Advanced merger (IV) & 4 \\
MIPS22690 		& 2.07    & 21.01 &  1.01 & disk & Pre-merger (III) & 4 \\
MIPS22699 		& 2.59    & 22.26 &  0.85 & ambiguous & Old merger (V) & 5 \\
MIPS22725 		& 1.40    & 19.37 &  1.57 & bulge & Old merger (V) & 5
\enddata
\tablenotetext{a}{MIPS42 is very close to a diffraction spike in our NICMOS image which causes all of our fits to fail.  Nevertheless, it is visually clear that it is a disk-dominated, \textquotedblleft faint \& compact\textquotedblright object.}
\tablenotetext{b}{Assuming $1" = 8$ kpc}
\tablecomments{The profile of some of our objects is listed as {\em ambiguous}.  These are objects whose profile is not well constrained either because they are very faint, barely resolved or possess a strong PSF component that hides the underlying galaxy. }
\end{deluxetable}

\end{document}